\documentclass[a4paper,12]{article}
\usepackage{amsmath, amsfonts, amssymb, amsthm, bm, graphics, bbm, color}
\usepackage{graphicx}
\usepackage{subfigure}
\usepackage{mathtools}
\usepackage[table]{xcolor}
\usepackage{url}
\usepackage{mathabx}
\usepackage{cancel}
\usepackage{multicol}
\usepackage{float}
\usepackage{caption}
\usepackage{mathalfa}
\usepackage{hyperref}
\usepackage{afterpage}
\DeclareMathAlphabet\mathbfcal{OMS}{cmsy}{b}{n}

\theoremstyle{definition}

\makeatletter
\newcommand\makebig[2]{%
  \@xp\newcommand\@xp*\csname#1\endcsname{\bBigg@{#2}}%
  \@xp\newcommand\@xp*\csname#1l\endcsname{\@xp\mathopen\csname#1\endcsname}%
  \@xp\newcommand\@xp*\csname#1r\endcsname{\@xp\mathclose\csname#1\endcsname}%
}
\makeatother

\makebig{biggg} {3.0}
\makebig{Biggg} {3.5}
\makebig{bigggg}{4.0}
\makebig{Bigggg}{4.5}
\makebig{biggggg}{5.0}
\makebig{Biggggg}{5.5}
\makebig{bigggggg}{6.0}
\makebig{Bigggggg}{6.5}

\newcommand{\dif}{\mathrm{d}}
\newcommand{\im}{\mathrm{i}}

\begin{document}
\title{Identification of Metallic Objects using Spectral MPT Signatures: Object Characterisation and Invariants}

\author{P.D. Ledger$^*$, B.A. Wilson$^\dagger$, A.A.S. Amad$^\dagger$ and W.R.B. Lionheart$^\ddagger$\\
$^*$School of Computing \& Mathematics, Keele University \\
$^\dagger$Zienkiewicz Centre for Computational Engineering, Swansea University \\
$^\ddagger$Department for Mathematics, The University of Manchester\\
Corresponding author: p.d.ledger@keele.ac.uk}

\maketitle

\section*{Abstract}
The early detection of terrorist threats, such as guns and knives, through improved metal detection, has the potential to reduce the number of attacks and improve public safety and security. To achieve this, there is considerable potential to use the fields applied and measured  by a metal detector to discriminate between different shapes and different metals since, hidden within the field perturbation, is object characterisation information. The magnetic polarizability tensor  (MPT) offers an economical characterisation of metallic objects that can be computed for different threat and non-threat objects and has an established theoretical background, which shows that the induced voltage is a function of the hidden object's MPT coefficients. In this paper, we describe the additional characterisation information that measurements of the induced voltage over a range of frequencies  offer compared to measurements at a single frequency. We call such object characterisations its MPT spectral signature. Then, we present a series of alternative rotational invariants for the purpose of classifying hidden objects using MPT spectral signatures. Finally, we include examples of computed MPT spectral signature characterisations of realistic threat and non-threat objects that can be used to train machine learning algorithms for classification purposes.

\noindent{\bf Keywords:} Finite element method; Magnetic polarizability tensor; Machine learning; Metal detection; Object classification; Reduced order model; Spectral; Validation.\\
\noindent{\bf MSC CLASSIFICATION:} 65N30; 65N21; 35R30; 35B30

%
%
%
%
%
%
%

\section{Introduction}\label{sect:intro}

The purpose of this paper is to provide a demonstration of  computing object characterisations for training machine learning algorithms, with the end goal being able to classify objects for metal detection. 
One key application is the  discrimination between threat and non-threat objects. With increased gun and knife crime in many countries (e.g. U.K. knife crime has seen a significant increase in the last 8-9 years~\cite{Commons}), the early detection of such weapons has the potential to reduce the number of related attacks and improve safety and security. The ability to discriminate between 
metallic clutter and threat objects is also important for improving the identification of hidden anti-personal mines and unexploded ordnance (UXO) in areas of former conflict in the developing world, offering improvements to the safety and security of local inhabitants and allowing the land to be returned to normal use.
As well as safety and security, the ability to detect and discriminate between different hidden metallic objects has important applications in archaeological searches and treasure hunts, non-destructive testing as well as detecting fake coins in vending machines and automatic checkouts.

The magnetic polarizability tensor (MPT)  has been shown to offer an economical characterisation of conducting permeable objects and explicit formulae for the computation of its $6$ independent complex coefficients, which are a function of  the exciting frequency, the object's size, its shape  as well as its conductivity and permeability, have been obtained ~\cite{Ammari2014,LedgerLionheart2015,LedgerLionheart2018,LedgerLionheart2019}. The behaviour of the MPT's coefficients as a function of frequency, known as its spectral signature, has been studied theoretically~\cite{LedgerLionheart2019} and an efficient method for the computation of the spectral signature using a reduced order model based on proper orthogonal decomposition (POD) has been proposed~\cite{ben2020}.  These computations employ the  established Open Source finite element (FE) package, \texttt{NGSolve}, and the recently derived alternative explicit expressions formulae for the MPT coefficients~\cite{LedgerLionheart2019}. The use of \texttt{NGSolve}~\cite{NGSolve,netgendet} ensures that the solutions to the underlying full order (eddy current type) transmission problems are accurately computed using high order $\bm{H}(\text{curl})$ conforming  (high order edge element) discretisations (see~\cite{ledgerzaglmayr2010,SchoberlZaglmayr2005,zaglmayrphd} and references therein) and the POD technique ensures their rapid computation over sweeps of frequency. Included in this approach~\cite{ben2020} are a-posteriori error bounds that can be computed at runtime with negligible additional cost, which
ensure reliability with the respect to the FE solutions and allow the MPT coefficients obtained with POD to be certified.

With the goal of identifying hidden metallic targets in mind, the MPT spectral signature has been previously used for simple library classification~\cite{Ammari2015,LedgerLionheartamad2019}, a $k$ nearest neighbours (KNN) classification algorithm~\cite{marsh2014b} and other machine learning approaches~\cite{WoutervanVerre2019}.  In addition, existing examples of practical MPT classification of objects  include in airport security screening~\cite{marsh2014,marsh2014b}, waste sorting~\cite{karimian2017} and anti-personal landmine detection~\cite{rehim2016}. 
In such situations, induced voltages are measured over a range of frequencies by a metal detector from which the MPT spectral signature of the hidden object is obtained and then a classifier applied~\cite{zhao2016,zhao2014,rehim2015}. With the exception of~\cite{Ammari2015,LedgerLionheartamad2019}, all previous studies used measured MPT spectral signature information to build the classifier.  As pointed out in~\cite{ben2020},  such libraries of measured MPT coefficients contain unavoidable errors if the object is placed in a non-uniform background field that varies significantly over the object as well as other errors and noise associated with capacitive coupling with other low-conducting objects or soil in the background. There will also be other generals noise (e.g. from amplifiers, parasitic voltages and filtering)~\cite{Makkonen2015}. This means the accuracy is about of measured MPT coefficients is about 1\% to 5\%~\cite{davidsoncoins, marsh2014b,Makkonen2015}, depending on the application. It is costly (time consuming) to produce a large library from measuring coefficients and using such libraries of measured MPT coefficients can limit the performance of machine learning classifiers (if they are used on a metal detector which has greater accuracy than that of MPT coefficients in the library). We will instead use the newly developed POD approach~\cite{ben2020} to build a library for object classification. This means that the MPT coefficients are obtained with higher accuracy than can currently be achieved from measurements (noise appropriate to the system can be added to the library during a classifier training if desired), the spectral signature is accurately computed for a large frequency range (up to the limit of the eddy current model) rather than obtained at a small number of discrete frequencies and, through scripting, it allows a much larger library of objects and variations of materials to be considered, which is all highly desirable for achieving greater fidelity and accuracy when training machine learning classifier

From previous studies of the simpler P\'oyla-Szeg\"o tensor characterisation of an object for a fixed conductivity contrast in electrical impedance tomography (EIT), it is known that shape and material contrast information cannot be separated~\cite{ammarikangbook} and, it is generally accepted, that using a MPT characterisation at a single frequency also only provides limited information.
By studying nanoparticles and their shape reconstruction from plasmonic resonances, Ammari {\it et al. }~\cite{ammari2019} show that such resonances are related to the Neumann-Poincar\'e operator, used in the computation of the P\'oyla-Szeg\"o tensor and generalised polarization (polarizability) tensors with 
contrasts $k(\omega)$ being a function of frequency $\omega$. In addition, for electro-sensing, where it has been postulated that electric fish characterise objects by P\'olya-Szeg\"o tensors~\cite{taufiq2016,taufiq2012}, spectral information, with frequency dependent contrasts $k(\omega)$, have been used for the successful classifications of objects~\cite{ammarishapeelectro}.  
Although there has been success in using the MPT spectral signature for object classification~\cite{Ammari2015,LedgerLionheartamad2019,marsh2014b,marsh2014,
karimian2017,rehim2016} and a theoretical study of the MPT spectral signature~\cite{LedgerLionheart2019} has been made, the additional information the spectral signature provides about the object's shape and its materials remains open. Furthermore, in such classifications, an MPT's eigenvalues are commonly used as the object's features, as they are invariant under object rotation, although there are other possibilities and the benefits of these will be explored in this work.


The main novelties of this work are, firstly, to establish that an MPT characterisation at a fixed (limiting) frequency only characterises an object upto an equivalent ellipsoid, which is not surprising given the known related result for the simpler P\'oyla-Szeg\"o tensor~\cite{ammarikangbook}. Secondly, that an MPT's spectral signature, provides a sequence of different equivalent ellipsoids at each frequency of excitation where, in general, using any single equivalent ellipsoid provides a different spectral signature to that of the MPT spectral signature of the original object. Thirdly, to list, and demonstrate, appropriate sets of tensor invariants, obtained from MPT coefficients, for the object features in object classification. Fourth, and finally, to provide a series of practically motivated MPT spectral signature characterisations computed for realistic threat and non-threat objects.

The paper is organised as follows: In Section~\ref{sect:MPTcharacter}, the MPT characterisation of a conducting permeable object is  briefly reviewed. This section also briefly reviews approaches for the efficient computation of their spectral signature and measurement of the tensor coefficients in metal detection. Then, in Section~\ref{sect:equivellip}, the relationship between an MPT characterisation of an object and an equivalent ellipsoid at a fixed frequency is discussed along with the added benefits of using an MPT spectral signature for object characterisation. Section~\ref{sect:classinv} considers alternative MPT spectral signature invariants that are invariant to object rotation for the purpose of training a machine learning algorithm for object classification. In Section~\ref{sect:trainset},  we present computational examples of MPT spectral signature characterisations of realistic exemplar threat and non-threat objects with the purpose being to build a training data set for object classification. The paper closes with some concluding remarks.


\section{Review of the MPT for object characterisation}\label{sect:MPTcharacter}

In this section we briefly recall the economical characterisation of a small conducting permeable isolated object $B_\alpha$ with conductivity $\sigma_*$ and permeability $\mu_*$ by an MPT if the background medium is non-conducting ($\sigma =0 \, \text{S/m}$) and has permeability of free space $\mu_0 = 4 \pi \times 10^{-7} \text{H/m}$. 
As mentioned above, this describes the situation of metal detection for applications including security screening e.g.,~\cite{marsh2014,marsh2014b}, waste sorting e.g.~\cite{karimian2017} and buried landmines and UXOs e.g.~\cite{rehim2016}  (where the soil's conductivity is much lower than that of object so that it can be neglected).
We use the description $B_\alpha = \alpha B + {\bm z}$, which means that the object is described by a size parameter $\alpha \ll 1$, a unit sized object $B$ containing the origin and a translation vector ${\bm z}$. Then, using the asymptotic formula obtained by Ammari, Chen, Chen, Garnier and Volkov~\cite{Ammari2014}, Ledger and Lionheart~ \cite{LedgerLionheart2015} have derived the simplified form 
\begin{align}
(\bm{H}_{\alpha}-\bm{H}_0)(\bm{x})_i=(\bm{D}_{\bm{x}}^2G(\bm{x},\bm{z}))_{ij}(\mathcal{M})_{jk}(\bm{H}_0(\bm{z}))_k+O(\alpha^4), \label{eqn:asymp}
\end{align}
for the magnetic field perturbation caused by the presence of the object, which holds as $\alpha\to 0$ and makes the MPT explicit.  In the above, ${\bm H}_\alpha$ is the magnetic interaction field, ${\bm H}_0$ is the background magnetic field (in absence of the object), $G(\bm{x} ,\bm{z}  ) := 1/ ({4\pi | \bm{x} -\bm{z} |})$ is the free space Laplace Green's function, $ (\bm{D}^2_x G)_{ij} = \partial_{x_j} \partial_{x_i} G$ denote the elements of the Hessian of $G$ and Einstein summation convention of the indices is implied in (\ref{eqn:asymp}). In addition, ${\mathcal M}=({\mathcal M})_{jk} {\bm e}_j \otimes {\bm e}_k$, where $\bm{e}_j$ denotes the $j$th orthonormal unit vector,  is the  complex symmetric rank 2 MPT, which describes the shape and material properties of the object $B_\alpha $ and is frequency dependent, but is independent of the object's position $\bm{z}$. 
%

We state below the explicit formulae for the computation of the coefficients $(\mathcal{M})_{ij}$ of ${\mathcal M}$ using the splitting, $(\mathcal{M})_{ij}:=(\mathcal{N}^0)_{ij}+(\mathcal{R})_{ij}+\im(  \mathcal{I})_{ij}$~\cite{LedgerLionheart2019} 
\begin{subequations}
\label{eqn:NRI}
\begin{align}
(\mathcal{N}^0[ \alpha B , \mu_r ] )_{ij}&:=\alpha^3\delta_{ij}\int_{B}(1-\mu_r^{-1})\dif \bm{\xi}+\frac{\alpha^3}{4}\int_{B\cup B^c}\tilde{\mu}_r^{-1}\nabla\times\tilde{\bm{\theta}}_i^{(0)}\cdot\nabla\times\tilde{\bm{\theta}}_j^{(0)}\ \dif \bm{\xi},\\
(\mathcal{R}[\alpha B, \omega,\sigma_*,\mu_r])_{ij}&:=-\frac{\alpha^3}{4}\int_{B\cup B^c}\tilde{\mu}_r^{-1}\nabla\times\bm{\theta}_j^{(1)}\cdot\nabla\times\overline{\bm{\theta}_i^{(1)}}\ \dif \bm{\xi},\\
(\mathcal{I}[\alpha B, \omega,\sigma_*,\mu_r])_{ij}&:=\frac{\alpha^3}{4}\int_B\nu\Big(\bm{\theta}_j^{(1)}+(\tilde{\bm{\theta}}_j^{(0)}+\bm{e}_j\times\bm{\xi})\Big)\cdot\Big(\overline{\bm{\theta}_i^{(1)}+(\tilde{\bm{\theta}}_i^{(0)}+\bm{e}_i\times\bm{\xi})}\Big)\ \dif \bm{\xi},
\end{align}
\end{subequations}
where $\im: = \sqrt{-1}$, $\omega$ is the angular frequency, 
$\nu:=\alpha^2\omega\mu_0\sigma_*$, $\delta_{ij}$ is the Kronecker delta,
${\bm \xi}:=(\xi_1, \xi_2,\xi_3) $ is measured from an origin in $B$, and, with the exception of the expression below (where it indicates the closure), the overbar denotes the complex conjugate throughout this work. Furthermore, the square brackets denote the tensors' dependence on size, geometry, materials and exciting frequency.
In addition, note that
\begin{align}
\tilde{\mu}_r ( \bm{\xi} ) := \left \{ \begin{array}{ll}  \mu_r :=\mu_*/\mu_0 & \bm{\xi} \in {B}\\
1 & \bm{\xi} \in B^c:= {\mathbb R}^3 \setminus \overline{B}  \end{array} \right . \nonumber.
\end{align}
 The computation of the tensor coefficients in (\ref{eqn:NRI}) rely on the real vector field solution ${\bm \theta}_i^{(0)}$, $i=1,2,3$, of the transmission problem~\cite{LedgerLionheart2019}
\begin{subequations}
\label{eqn:Theta0}
\begin{align}
\nabla\times\tilde{\mu}_r^{-1}\nabla\times\bm{\theta}_i^{(0)}&=\bm{0} &&\textrm{in }B\cup B^c,\\
\nabla\cdot\bm{\theta}_i^{(0)}&=0 &&\textrm{in }B\cup B^c,\\
[{\bm{n}}\times\bm{\theta}_i^{(0)}]_{\Gamma}&=\bm{0} &&\textrm{on }\Gamma,\\
[{\bm{n}}\times\tilde{\mu}_r^{-1}\nabla\times\bm{\theta}_i^{(0)}]_{\Gamma}&=\bm{0} &&\textrm{on }\Gamma,\\
\bm{\theta}_i^{(0)}-{\bm{e}}_i\times\bm{\xi}&=\bm{O}(|\bm{\xi}|^{-1}) &&\textrm{as }|\bm{\xi}|\rightarrow\infty,
\end{align}
\end{subequations}
where $\Gamma:=\partial B$ and the complex vector field solution ${\bm \theta}_i^{(1)}$, $i=1,2,3$, of the transmission problem
\begin{subequations}
\label{eqn:Theta1}
\begin{align}
\nabla\times {\mu}_r^{-1}\nabla\times\bm{\theta}_i^{(1)}-\im \nu  (\bm{\theta}_i^{(0)}+\bm{\theta}_i^{(1)})&=\bm{0}&&\textrm{in }B,\\
\nabla\times \nabla\times\bm{\theta}_i^{(1)} &=\bm{0}&&\textrm{in }B^c,\\
\nabla\cdot\bm{\theta}_i^{(1)}&=0&&\textrm{in }B^c,\\
[{\bm{n}}\times\bm{\theta}_i^{(1)}]_{\Gamma}&=\bm{0}&&\textrm{on }\Gamma,\\
[{\bm{n}}\times\tilde{\mu}_r^{-1}\nabla\times\bm{\theta}_i^{(1)}]_{\Gamma}&=\bm{0}&&\textrm{on }\Gamma,\\
\bm{\theta}_i^{(1)}&=\bm{O}(|\bm{\xi}|^{-1})&&\textrm{as }|\bm{\xi}|\rightarrow\infty.
\end{align}
\end{subequations}
Note also that we have chosen to introduce $\tilde{\bm{\theta}}_i^{(0)}\vcentcolon=\bm{\theta}_i^{(0)}-{\bm{e}}_i\times\bm{\xi}$, which can be shown to satisfy the same transmission problem as (\ref{eqn:Theta0}), except with a non-zero jump condition for $[{\bm{n}}\times\tilde{\mu}_r^{-1}\nabla\times\tilde{\bm{\theta}}_i^{(0)}]_{\Gamma}$ and the decay condition $\tilde{\bm{\theta}}_i^{(0)}(\bm{\xi})=\bm{O}(|\bm{\xi}|^{-1})$ as $|\bm{\xi}|\rightarrow\infty$. 

Furthermore, the rank 2 tensors $\mathcal{N}^0[ \alpha B , \mu_r ]$, $\mathcal{R}[\alpha B, \omega , \sigma_*, \mu_r ]$ and $\mathcal{I}[\alpha B, \omega, \sigma_*, \mu_r]$, which make up the splitting
$\mathcal{M}=\mathcal{N}^0+\mathcal{R} +\im  \mathcal{I}$, are real, symmetric and each have real eigenvalues~\cite{LedgerLionheart2019}. They are  related to the MPT by
\begin{align}
\text{Re}({\mathcal M}[\alpha B, \omega, \sigma_*, \mu_r]) =&\tilde{ \mathcal{R}}[\alpha B, \omega, \sigma_*, \mu_r]= \mathcal{N}^0[ \alpha B,\mu_r ] + \mathcal{R}[\alpha B, \omega, \sigma_*, \mu_r],  \nonumber\\
 \text{Im}({\mathcal M}[\alpha B, \omega, \sigma_*, \mu_r]) =&  \mathcal{I}[\alpha B, \omega, \sigma_*, \mu_r].
\nonumber
\end{align}
 We call the MPT's coefficients as a function of $\omega$ its spectral signature.
Note that the above formulation for the MPT and associated transmission problems is for the case of a single homogenous object $B$, the extension to multiple inhomogeneous objects can be found in~\cite{LedgerLionheartamad2019,LedgerLionheart2019}.

\subsection{Efficient computation of the MPT spectral signature} \label{sect:rommethod}
An efficient procedure for computing the MPT spectral signature of a conducting permeable object has been proposed in~\cite{ben2020} and will be used throughout the work to generate the numerical results. In this approach, discrete approximations to $\bm{\theta}_i^{(1)}(\omega_n)$, $i=1,2,3$ are computed accurately  at a small number of logarithmically spaced frequencies $\omega_n$, $n =1,\ldots,N$ using the high order finite element solver \href{https://ngsolve.org}{\texttt{NGSolve}} and a ${\bm H} (\text{curl})$ conforming discretisation~\cite{SchoberlZaglmayr2005, NGSolve, zaglmayrphd, netgendet} on unstructured tetrahedral grids. These solutions are called the representative full order model solution snapshots. A proper orthogonal decomposition approach using projection (PODP)~\cite{hesthaven2016} is then applied to predict the discrete approximations to $\bm{\theta}_i^{(1)}(\omega)$ at other frequencies $\omega=\omega_m$, $m=1, \ldots,M$ with $M \gg N$ and then resulting MPT coefficients $({\mathcal M}^{PODP}[\alpha B , \omega, \sigma_*, \mu_r ])_{ij}$, which make up the MPT's spectral signature,  follow by simple post-processing.  The a-posteriori error estimates  
\begin{subequations}
\begin{align}
| ({\mathcal R}^{PODP}[\alpha B , \omega, \sigma_*, \mu_r  ] )_{ij} - ({\mathcal R} [\alpha B , \omega, \sigma_*, \mu_r  ] )_{ij} | \le &  (\Delta [\omega] )_{ij},\\
| ({\mathcal I}^{PODP}[\alpha B , \omega, \sigma_*, \mu_r  ] )_{ij} - ({\mathcal I} [\alpha B , \omega , \sigma_*, \mu_r ] )_{ij} | \le & (\Delta [\omega] )_{ij},
\end{align}
\end{subequations}
which bound the accuracy of the real and imaginary parts of the PODP MPT predication with respect to the full order solution, have been derived and can be  computed at low-computational cost during the online stage of the reduced order model for each  $\omega=\omega_m$, $m=1, \ldots,M$. They allow the reduced order predictions to be certified  without having compute additional full order model solutions and, if desired, can be used to adaptively choose additional frequencies for the representative full order model solution snapshots so as to improve the accuracy of the PODP prediction of the MPT. Note that in subsequent results we focus on PODP solutions and drop the PODP superscript unless confusion may arise. We refer to~\cite{ben2020} for details of the approach.

 \subsubsection{Limiting frequency of the MPT spectral signature}\label{sect:irregtetexp}
 
 The eddy current model, on which the MPT description is based, is a low frequency approximation to the full Maxwell system and is commonly accepted to be valid if the quasi-static assumption applies (dimension $ D \approx \alpha$ of the object $ B_\alpha $ is small compared to the wavelength) and the conductivities are high ($\sigma_{max}  \gg \omega \epsilon_{max}$, where $\epsilon_{max}$ denotes the object's maximum permittivity, here assumed to be $\epsilon_{max}=\epsilon_0 \approx 8.854 \times 10^{-12} \text{F/m}$). However, the topology of the object has important role to play in determining the limiting frequency at which the approximation remains valid.  Schmidt, Sterz and Hiptmair~\cite{schmidteddycurrent} have obtained the following estimates
 \begin{subequations} \label{eqn:eddylimit}
 \begin{align}
 C_1 \epsilon_{max} \mu_{max} \omega^2 D^2 & \ll 1, \\
 C_2 \epsilon_{max}  \omega \sigma_{min}^{-1} & \ll 1,
 \end{align}
\end{subequations}
that are required to hold to ensure the validity of the eddy current model. In the above, $C_1=C_1(B)$ and $C_2=C_2(B)$ are constants that depend on the object's topology.
  In particular, an object with a long thin extension or with a small gap  (eg hoarse-shoe shaped conductor) lead to capacitive coupling and have large $C_1(B)$ and $C_2(B)$ limiting the frequency $\omega_{limit}$ at which the MPT spectral signature remains valid compared to  using the quasi-static and high conductivity conditions alone.
Schmidt {\it et al.} have describe a numerical procedure that allows the constants to be estimated numerically for different objects by solving a low-dimensional eigenvalue problem.  Once the constants have been found, the limiting the frequency $\omega_{limit}$ can be estimated from (\ref{eqn:eddylimit}). We apply this procedure to the numerical examples in this work.

\subsubsection{Illustrative example of the MPT spectral signature for an irregular tetrahedron} \label{sect:irregulartet}

To illustrate the method proposed in~\cite{ben2020} for computing the MPT spectral signature, as well the procedure for obtaining $\omega_{limit}$ described in Section~\ref{sect:irregtetexp}, we consider the case where $B$ is the irregular tetrahedron with vertices
\begin{equation}
{\bm v}_1 = \begin{pmatrix} 
0 \\
0 \\
0 
\end{pmatrix}, \qquad 
{\bm v}_2= \begin{pmatrix} 
7 \\
0 \\
0 
\end{pmatrix}, \qquad  
{\bm v}_3 =  \begin{pmatrix} 
5.5 \\
4.6 \\
0.0 
\end{pmatrix}, \qquad 
{\bm v}_4=  \begin{pmatrix} 
3.3 \\
2.0 \\
5.0 
\end{pmatrix},
\label{eqn:tetvertices}
\end{equation}
size $\alpha=0.01 \text{ m}$ and material properties $\mu_r= 2$ and $\sigma_* = 5.96\times 10^{6} \text{ S/m}$.  This object has been chosen as it does not have rotational or reflectional symmetries and so the associated MPT has $6$ independent complex coefficients at each frequency.

\begin{figure}
\begin{center}
$\begin{array}{cc}
\includegraphics[width=0.5\textwidth]{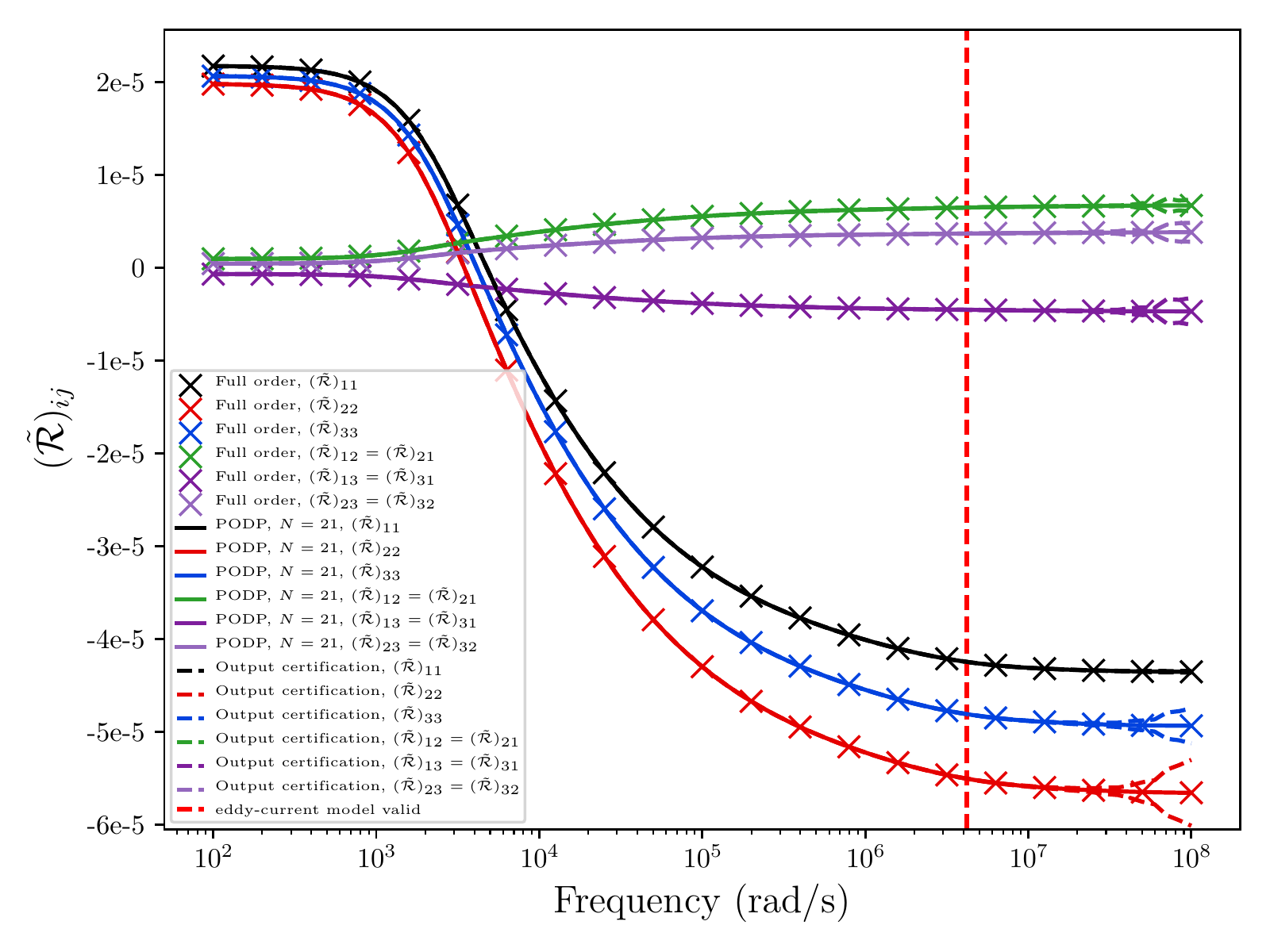} &
\includegraphics[width=0.5\textwidth]{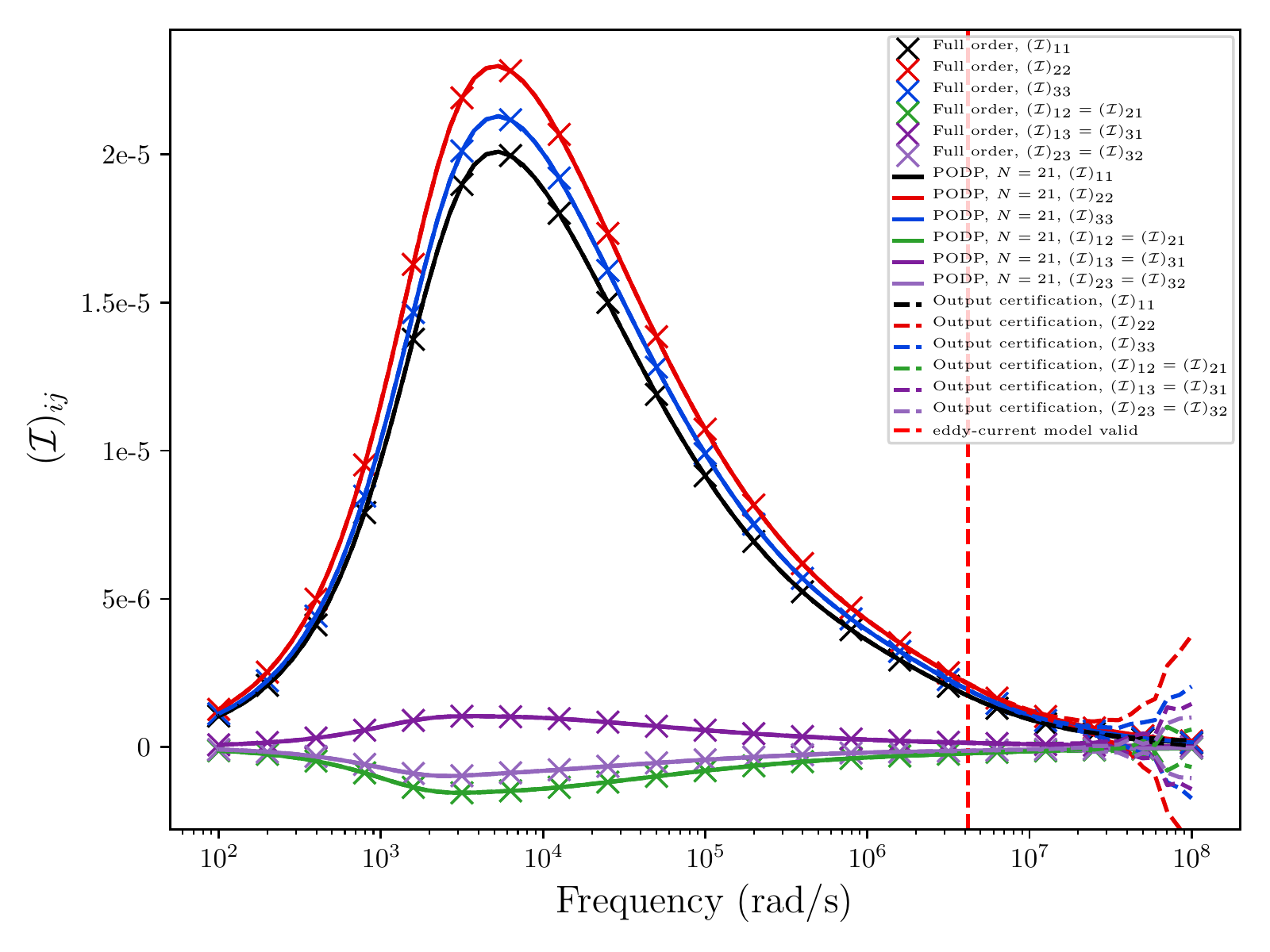} \\
(\tilde{\mathcal R}[\alpha B, \omega, \sigma_*, \mu_r ])_{ij} & ({\mathcal I}[\alpha B , \omega, \sigma_*, \mu_r])_{ij}
\end{array}$
\end{center}
\caption{Irregular tetrahedron $B$ with vertices as in (\ref{eqn:tetvertices}), $\alpha=0.01 \text{ m}$ $\mu_r= 2$ and $\sigma_* = 5.96\times 10^{6} \text{ S/m}$: PODP prediction of the spectral signature for $(\tilde{\mathcal R}[\alpha B, \omega, \sigma_*, \mu_r ])_{ij} $ and $({\mathcal I}[\alpha B , \omega, \sigma_*, \mu_r])_{ij}$ including the certification of the output $(\cdot) \pm (\Delta[\omega])_{ij}$ and the limiting frequency $\omega_{limit}$. } \label{fig:irregtetpodsweep}
\end{figure}

To describe the object and surrounding truncated unbounded region (set as the $[-100,100]^3$ box), a mesh of $21\, 427$ unstructured tetrahedra was generated and elements of order $p=3$ were used so that the representative full order model solution  snapshots (corresponding to discrete approximations of ${\bm \theta}_i^{(1)}(\omega_n)$, $i=1,2,3$), obtained at $N=21$ logarithmically spaced frequencies, $\omega_n$, were found to be converged. The PODP approach was then applied with a tolerance of $TOL=10^{-4}$ to obtain the approximations to the solutions ${\bm \theta}_i^{(1)}(\omega)$, $i=1,2,3$, at other frequencies and, hence, obtain the spectral signature shown in Figure~\ref{fig:irregtetpodsweep}. In this plot, the crosses are used to indicate the frequencies $\omega_n$ used to compute the representative full order model solution snapshots, the dashed lines show the certification of the output $(\cdot ) \pm (\Delta[\omega])_{ij}$  and vanish to the predicted MPT coefficients, at all but the highest frequencies, indicating the high accuracy of the PODP solution with respect to the full order solution. Note that acceptable PODP solutions can also be obtained with $N=13$ representative full order model solutions, but $N=21$ was chosen to ensure  $(\Delta[\omega])_{ij}$ is small.
The plot also includes a vertical dashed line, which indicates the limiting frequency $\omega_{limit}$. For $\omega > \omega_{limit}$ we expect the MPT to no longer provide a valid characterisation of the object due to the eddy current model being invalid. 
This figure illustrates the typical characteristic behaviour of the MPT spectral signature of an object with homogeneous material parameters in which  the diagonal coefficients of $\tilde{\mathcal{R}} [\alpha B, \omega,\sigma_*,\mu_r ]$ and $ \mathcal{I} [\alpha B, \omega,\sigma_*,\mu_r]$ dominate over their off-diagonal coefficients. The coefficients of    $\tilde{\mathcal{R}} [\alpha B, \omega,\sigma_*,\mu_r ]$ illustrate the typical sigmoid behaviour with $\log \omega$ and the coefficients of  $ \mathcal{I} [\alpha B, \omega,\sigma_*,\mu_r]$  have a single local maximum with $\log \omega$, which is also well understood theoretically~\cite{LedgerLionheart2019} (See also Section~\ref{sect:spectral}). Note that behaviour of these coefficients is different for objects with inhomogeneous materials~\cite{LedgerLionheartamad2019}.

\subsection{Measurement of MPT coefficients}


In the electrical engineering community, the measurement of MPT coefficients as a function of frequency is sometimes called MPT spectroscopy and procedures  have been developed for
anti-personal landmine detection~\cite{norton2001, ambrus, dekdouk,abdel} as well as for MPT measurement and real time classification  for security screening~\cite{marsh2013,marsh2014,marsh2014b,marsh2015,zhao2016} and other applications.
The basic idea is that background fields ${\bm H}_0^{(m)}$, $m=1,\ldots,M_e$, from $M_e$ different exciters are generated, in turn, and measurements of the resulting magnetic field perturbation $({\bm H}_\alpha - {\bm H}_0^{(m)})({\bm x})$ caused by the presence of a hidden conducting permeable  object are made at  sufficiently many positions ${\bm x}$ around the object for a range of exciting frequency. Considering a single frequency excitation, the perturbed magnetic field is usually measured as an induced voltage in the form
\begin{equation}
{\rm V}_{nm}^{\text{ind}}  = \int_{C^{(n)}} {\bm n} \cdot ({\bm H}_\alpha - {\bm H}_0^{(m)} )({\bm x})  \dif {\bm x} \label{eqn:vnm},
\end{equation}
where $n=1,\ldots,M_r$ correspond to the different receiver coils and $C^{(n)}$ is an appropriate surface related to coil $n$~\cite{LedgerLionheart2018}. In light of (\ref{eqn:asymp}), and the MPT object characterisation,  we see that
the leading order term gives an approximation to ${\rm V}_{nm}^{\text{ind}} $ and the accuracy of the approximation will depend on the uniformity of ${\bm H}_0({\bm z}) $ as well as the object size. As explained in Section~\ref{sect:intro}, the measurements ${\rm V}_{nm}^{\text{ind, measured}} $ will also have unavoidable errors and noise from a range of sources.
Accepting these errors and noise, the task of determining the MPT of a hidden object separates in to first determining the position of the object ${\bm z}$, which can be done for example using a MUSIC approach~\cite{Ammari2014,Ammari2015} or some other localisation based approach, and secondly determining the MPT's coefficients from an over determined linear system of the form
\begin{equation}
{\mathbf A} {\mathbf u} ={\mathbf b} , \label{eqn:overdetsys}
\end{equation}
using least squares where bold Roman fonts are used for matrices and vectors in linear systems and, for  $i = (n -1)M_e +m$, $m=1,\ldots,M_e$, $n=1,\ldots,M_r$ $j = 3(q-1)+r$, $q,r=1,2, 3$, they have the entries
\begin{align}
({\mathbf u})_j =&  {\rm u}_j    =  ({\mathcal M})_{qr} \nonumber , \\ 
({\mathbf A})_{ij} = & {\rm A}_{ij} =  ( {\bm H}_0^{(m)} ( {\bm z} ))_r  \int_{C^{(n)}}   ({\bm D}_{\bm x}^2  G({\bm x},{\bm z} ) {\bm n} )_{q} \dif {\bm x}   \nonumber , \\
({\mathbf b})_i = &  {\rm b}_i =  \int_{C^{(n)}}  {\bm n} \cdot  ({\bm H}_\alpha - {\bm H}_0^{(m)} )({\bm x}) \dif {\bm x} \approx V_{nm}^{\text{ind, measured}}  \nonumber .
\end{align}
In addition, the known symmetry of ${\mathcal M}$ can be used to reduce the number of complex unknowns in (\ref{eqn:overdetsys}) from $9$ to $6$. Clearly, we need the product $M_eM_r  > 6$ in order for the system to be over determined, however, it is not only important that we have sufficiently many measurements, but also that the location of emitting and receiving coils are correctly chosen so that all of the coefficients of ${\mathcal M}$ can be determined. For further details, and an algorithm of how this can be automated, see~\cite{LedgerLionheart2018}. The discrete, measured spectral signature of ${\mathcal M}$ follows by repeating the above process using
${\rm V}_{mn}^{\text{ind, measured}}$  at different excitation frequencies $\omega_m$, $m=1,\ldots,M$ and depending on the system and application errors of around 1-5\% can be expected with current systems~\cite{davidsoncoins, marsh2014b,Makkonen2015}. Note that using a higher order expansion of  (\ref{eqn:asymp}), derived in~\cite{LedgerLionheart2018g}, will lead to an improved approximation of ${ \rm V}_{nm}^{\text{ind}} $ in (\ref{eqn:vnm}), particularly if ${\bm H}_0({\bm z})$ is non-uniform, and allow the coefficients of generalised MPTs to be found.

\section{MPTs and equivalent ellipsoids} \label{sect:equivellip}

In this section, we examine the extent to which a MPT uniquely characterises an object. First, we consider eigenvalue decompositions of the real and imaginary parts of the MPT for a fixed frequency and, secondly, for spectral data. Thirdly, we prove how equivalent ellipsoids can be constructed at limiting frequencies and discuss the construction of equivalent ellipsoids at other frequencies.

\subsection{Eigenvalue decomposition at fixed frequency}\label{sect:fixedfreq}
Consider the characterisation of an object $B_\alpha=\alpha B$ by an MPT (recall the characterisation is independent of ${\bm z}$),  which can be expressed by the splitting
\begin{align}
\mathcal{M}[\alpha B, \omega,\sigma_*,\mu_r ] = & \mathcal{N}^0[\alpha B,\mu_r] +\mathcal{R} [\alpha B, \omega,\sigma_*,\mu_r ] +\im  \mathcal{I} [\alpha B, \omega,\sigma_*,\mu_r ], \nonumber\\
= & \tilde{\mathcal{R}} [\alpha B, \omega,\sigma_*,\mu_r ] +\im  \mathcal{I} [\alpha B, \omega,\sigma_*,\mu_r ],  \label{eqn:freqdepsplit}
\end{align}
where our interest in this section is for a fixed frequency $ 0 \le  \omega=\Omega < \infty $.
For simplicity in this section, we assume that the parameter dependent coefficients of the rank 2 tensors are arranged as $3 \times 3$ matrices and use the same notation for both. 
Thus, the associated matrices 
$\tilde{\mathcal{R}} [\alpha B, \Omega,\sigma_*,\mu_r ] $ and ${\mathcal I} [\alpha B,\Omega,\sigma_*,\mu_r]$ are symmetric, ${\mathcal N}^0 [\alpha B,\mu_r]$ is positive definite if $\mu_r>1$,  
${\mathcal R}  [\alpha B,\Omega,\sigma_*,\mu_r]$ is negative definite and ${\mathcal I} [\alpha B,\Omega,\sigma_*,\mu_r]$ is
positive definite and they can be diagonalised as follows
\begin{subequations} \label{eqn:eigdecomp}
\begin{align}
 \tilde {\mathcal R}  [\alpha B,\Omega, \sigma_*,\mu_r  ] =& {\mathbf Q}^{\tilde{\mathcal R} [\alpha B,\Omega, \sigma_*,\mu_r  ]} {\mathbf \Lambda}^{\tilde{\mathcal R} [\alpha B,\Omega, \sigma_*,\mu_r   ]} ({\mathbf Q}^{\tilde{\mathcal R} [\alpha B,\Omega, \sigma_*,\mu_r    ] })^T , 
\\
 {\mathcal I}  [\alpha B,\Omega, \sigma_*,\mu_r   ] = & {\mathbf Q}^{ {\mathcal I} [\alpha B, \Omega, \sigma_*,\mu_r   ]} {\mathbf \Lambda}^{{\mathcal I} [\alpha B,\Omega, \sigma_*,\mu_r  ]} ({\mathbf Q}^{{\mathcal I} [\alpha B,\Omega, \sigma_*,\mu_r  ]} )^T ,
 \end{align}
 \end{subequations}
 where ${\mathbf Q}^{\tilde{\mathcal R} [\alpha B,\Omega, \sigma_*,\mu_r  ]} $ is an orthogonal matrix whose columns are the eigenvectors of $ \tilde {\mathcal R}  [\alpha B,\Omega, \sigma_*,\mu_r  ] $ and ${\mathbf \Lambda}^{\tilde{\mathcal R} [\alpha B,\Omega, \sigma_*,\mu_r   ]}$ is a diagonal matrix whose diagonal entries are the eigenvalues of $ \tilde {\mathcal R}  [\alpha B,\Omega, \sigma_*,\mu_r  ] $ and $T$ denotes the transpose.  The matrices ${\mathbf Q}^{ {\mathcal I} [\alpha B,\Omega, \sigma_*,\mu_r  ]} $ and ${\mathbf \Lambda}^{{\mathcal I} [\alpha B, \Omega, \sigma_*,\mu_r]} $ contain the eigenvectors and eigenvalues of  ${\mathcal I}  [\alpha B,\Omega,  \sigma_*, \mu_r ]$, respectively.
 Furthermore, if the object has reflectional or rotational symmetries, the number of independent coefficients in $ \tilde{\mathcal R} [\alpha B,\Omega, \sigma_*,\mu_r]$ and ${\mathcal I} [\alpha B,\Omega, \sigma_*,\mu_r]$ are reduced. In the case that  $ \tilde{\mathcal R} [\alpha B,\Omega, \sigma_*,\mu_r]$ and ${\mathcal I} [\alpha B,\Omega, \sigma_*,\mu_r]$ have at most $3$ independent coefficients, this reduction means that
  ${\mathbf Q }^{\tilde{\mathcal R} [\alpha B,\Omega, \sigma_*,\mu_r] }  ={\mathbf Q }^{{\mathcal I} [\alpha B,\Omega, \sigma_*,\mu_r]} ={\mathbf Q}(B)$  where we emphasise that ${\mathbf Q}$ only depends  on $B$. Moreover, when $ \tilde {\mathcal R}  [\alpha B,\Omega, \sigma_*,\mu_r] $ and ${\mathcal I}  [\alpha B,\Omega, \sigma_*,\mu_r]$ are diagonal, due to canonical choice of $B$ and the object's reflectional and rotational symmetries~\cite{LedgerLionheart2015}, then ${\mathbf Q}={\bm {\mathbb I}}$ is the identity matrix.

\subsection{Eigenvalue decomposition in the spectral case} \label{sect:spectralobject}

We now consider the case of the characterisation of an object by an MPT with varying frequency  $\omega$  so that the real and imaginary parts of an MPT expressed by (\ref{eqn:freqdepsplit}) are available at discrete frequencies $0\le \omega_m< \infty$, $m=1\ldots,M$. A similar decomposition to (\ref{eqn:eigdecomp}) again applies, except ${\mathbf Q}^{\tilde{\mathcal R}}$, ${\mathbf \Lambda}^{\tilde{\mathcal R}}$, ${\mathbf Q}^{{\mathcal I}}$ and ${\mathbf \Lambda}^{{\mathcal I}}$ are functions of $\omega_m$. If $B$ has reflectional or rotational symmetries, ${\mathbf Q}^{\tilde{\mathcal R} [\alpha B,\omega_m, \sigma_*,\mu_r] }  ={\mathbf Q}^{{\mathcal I} [\alpha B,\omega_m, \sigma_*,\mu_r]} ={\mathbf Q}(B)$  and, in  the limiting case where $ \tilde {\mathcal R}  [\alpha B,\omega_m, \sigma_*,\mu_r] $ and ${\mathcal I}  [\alpha B,\omega_m, \sigma_*,\mu_r]$ are diagonal, ${\mathbf Q}={\mathbb I}$ is the identity matrix. If ${\mathbf Q}^{\tilde{\mathcal R} [\alpha B,\omega_m, \sigma_*,\mu_r] }  ={\mathbf Q}^{{\mathcal I} [\alpha B,\omega_m, \sigma_*,\mu_r]} ={\mathbf Q}(B)$ , the only dependence of the MPT's coefficients on $\alpha$, $\omega_m$, $\mu_r$ and $\sigma_*$ is through
 ${\mathbf \Lambda}^{\tilde{\mathcal R} [\alpha B,\omega_m, \sigma_*,\mu_r] } $ and ${\mathbf \Lambda}^{{\mathcal I} [\alpha B,\omega_m, \sigma_*,\mu_r]} $.
 

\subsection{Equivalent Ellipsoids}

For an ellipsoidal object $E_\alpha = \alpha E$ of size $\alpha$  with material parameters $\mu_r$, $\sigma_*$ and aligned with coordinate axes such that $E$ is defined by
\begin{align}
\left ( \frac{\xi_1}{a} \right )^2 + \left ( \frac{\xi_2}{b} \right )^2 + \left ( \frac{\xi_3}{c} \right )^2 =1, \label{eqn:ellipsoid}
\end{align}
with $a\ge b\ge b \ge c$ then, for a fixed frequency $\omega=\Omega$, its MPT $\mathcal{M}[\alpha E, \Omega,\sigma_*,\mu_r ]$, as well as its real and imaginary parts, are diagonal. Similarly, in the spectral case, the MPTs $\mathcal{M}[\alpha E, \omega_m,\sigma_*,\mu_r ]$, $m=1,\ldots,M$ are diagonal. 

We now show that for the cases of $\omega=0$ or $\omega\to\infty$ that equivalent ellipsoids $E(0)$ and $E(\infty)$ can be found, which have the same MPT as $\mathcal{M}[\alpha B, 0,\sigma_*,\mu_r ]$ and $\mathcal{M}[\alpha B, \infty,\sigma_*,\mu_r ]$, respectively, and comment on the construction of ellipsoids for other fixed frequencies and the  spectral case. 

\subsubsection{Equivalent ellipsoid $E(0)$} \label{sect:equivellip0}
For the limiting case of $\omega=0$ it is known that~\cite{LedgerLionheart2019}
\begin{equation}
{\mathcal M} [\alpha B, \omega=0,\sigma_*, \mu_r] = {\mathcal N}^0 [ \alpha B,  \mu_r ] = {\mathcal T} [\alpha B, \mu_r ], \label{eqn:lowfreq}
\end{equation}
where ${\mathcal T}[\alpha B,  k ] =( {\mathcal T}[\alpha B, k ])_{ij} {\bm e}_i \otimes {\bm e}_j$
is the P\'olya-Szeg\"o tensor~\cite{ammarikangbook}. 
This tensor is simpler than the MPT and characterises small homogeneous conducting objects with shape $B$ in electrical impedance tomography ($k=\sigma_*/\sigma_0$) and small permeable homogeneous objects with shape $B$ in magnetostatics ($k=\mu_r=\mu_*/ \mu_0)$, it is  symmetric and is positive (negative) definite provided that the contrast $k>1$ ($0\le k <1$)~\cite{ammarikangbook}.  
Given $\alpha$ and $\mu_r$, we wish to show there is an equivalent ellipsoid $E(0)$ such that
\begin{align}
{\mathbf \Lambda}^{\tilde{\mathcal R} [\alpha B,0, \sigma_*,\mu_r ]} = & \tilde{\mathcal{R}} [\alpha E(0), 0 ,\sigma_*,\mu_r ]= {\mathcal T} [\alpha E(0),  \mu_r ],  \nonumber \\
{\mathbf \Lambda}^{{\mathcal I} [\alpha B,0,  \sigma_*,\mu_r ]} = & {\mathcal{I}} [\alpha E(0), 0 ,\sigma_*,\mu_r ] = 0, \nonumber
\end{align}
holds and to determine its radii $a$, $b$ and $c$. Note that the second expression does not provide any additional information and is  automatically satisfied. 

We recall that for the ellipsoid defined by (\ref{eqn:ellipsoid}) an analytical expression is available for ${\mathcal T}[\alpha E,k]$ in the form
\begin{equation}
{\mathcal T}[\alpha E, k] = \alpha^3 (k-1) |E| \left (
\begin{array}{ccc}
\frac{1}{1-A_1 +kA_1} & 0 & 0\\
0 & \frac{1}{1-A_2 +kA_2}  & 0\\
0 & 0 & \frac{1}{1-A_3 +kA_3} \end{array}
 \right ),
\end{equation}
where $|E|:= \frac{4}{3}\pi a b c$ and
$A_1$, $A_2$, $A_3 $ are the elliptical integrals 

\begin{align}
A_1:=&  \frac{bc}{a^2} \displaystyle \int_1^\infty \frac{1}{t^2 \sqrt{t^2 -1 + \left ( \frac{b}{a} \right )^2 } \sqrt{t^2 - 1 + \left ( \frac{c}{a} \right )^2 }} \dif t \nonumber , \\
A_2:=&  \frac{bc}{a^2} \displaystyle \int_1^\infty \frac{1}{\left ( t^2 -1 + \left ( \frac{b}{a} \right )^2 \right )^{3/2} \sqrt{t^2 - 1 + \left ( \frac{c}{a} \right )^2 }} \dif t \nonumber,  \\
A_3:=&  \frac{bc}{a^2} \displaystyle \int_1^\infty \frac{1}{\sqrt{ t^2 -1 + \left ( \frac{b}{a} \right )^2  }\left (  t^2 - 1 + \left ( \frac{c}{a} \right )^2 \right ) ^{3/2} } \dif t \nonumber ,
\end{align}
that are a function of $a,b,c$~\cite{ammarikangbook}. These integrals can also be shown to be equivalent to the alternative expressions in terms of incomplete elliptic integrals given by Osborn~\cite{osborn}, which can be computed using standard libraries.

\begin{figure}
\begin{center}
\includegraphics[width=0.5\textwidth]{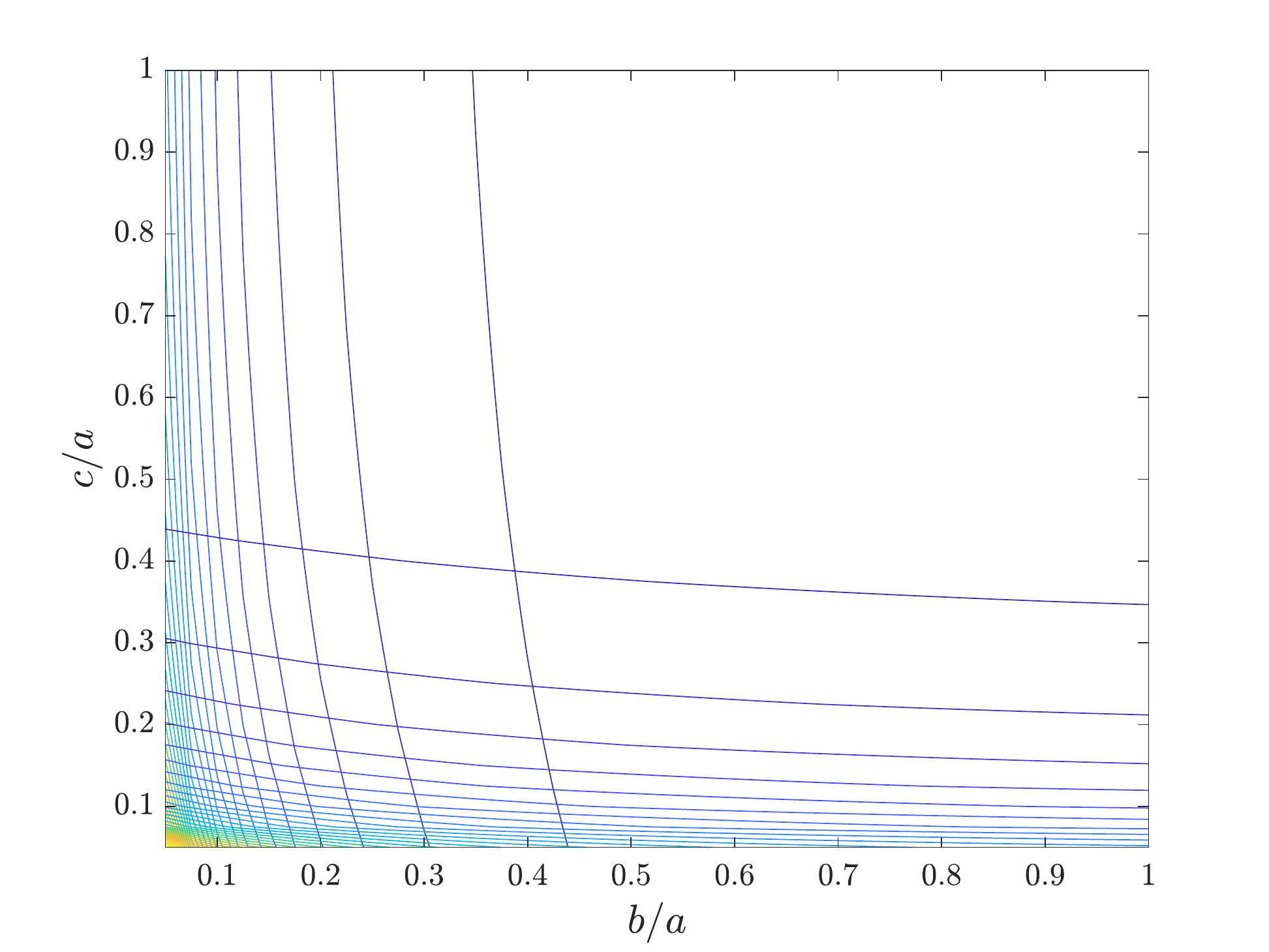}
\end{center}
\caption{Contours of $A_2/A_1$ and $A_3/A_1$ as function of $b/a$ and $c/a$.}\label{fig:ellipticintinvert}
\end{figure}

 Given that we know  $A_1+A_2+A_3=1$ from~\cite{osborn}, then, if $\alpha$ and $\mu_r$ are known (or at least if $\frac{1}{\alpha^3} \left ( 1+ \frac{3}{\mu_r-1} \right )$ is known), we can determine $|E(0)|$ from
\begin{align}
|E(0)| = \frac{1}{\alpha^3 L } \left ( 1 + \frac{3}{\mu_r-1} \right ) , \qquad L =\sum_{i=1}^3 \frac{1}{({\mathbf \Lambda}^{{\mathcal N}^0 [\alpha B,\mu_r]})_{ii}},
\end{align}
using the eigenvalues of ${\mathcal N}^0 [\alpha B,\mu_r]$. As  $|E(0)|$ represents the volume of the unit sized equivalent ellipsoid, we expect $|E(0)|= O( 1)$  as $\alpha \to 0$. Once $|E(0)|$ is known, then we can determine
\begin{align}
A_i= \frac{\alpha^3|E(0)|}{({\mathbf \Lambda}^{{\mathcal N}^0 [\alpha B,\mu_r]})_{ii}} -\frac{1}{\mu_r-1} , \qquad i=1,2,3,
\end{align}
and compute  $A_2/A_1$ and $A_3/A_1$.
Figure~\ref{fig:ellipticintinvert}, which was obtained by numerical evaluation of the elliptic integrals in MATLAB, shows that $(A_2/A_1, A_3/A_1) \to (b/a,c/a)$ is injective and shows the range of the map (numerical evaluation of Osborn's alternative expressions in terms of incomplete elliptic integrals gives the same result). Thus, given $(A_2/A_1, A_3/A_1)$, we know the solution $(b/a,c/a)$ exists and is unique. Hence, we can determine $(b/a,c/a)$ and  find a unique equivalent ellipsoid defined by $a$, $b$ and $c$ using  $|E(0)|= \frac{4}{3}\pi a b c$.
In addition, by ordering $({\mathbf \Lambda}^{{\mathcal N}^0 [\alpha B,\mu_r]})_{11}\ge( {\mathbf \Lambda}^{{\mathcal N}^0 [\alpha B,\mu_r]})_{22}\ge ({\mathbf \Lambda}^{{\mathcal N}^0 [\alpha B,\mu_r]})_{33}$, we will obtain $A_3 \ge A_2\ge A_1$. Since $0 < A_i < 1$ then,  $A_3/A_1 \ge A_2/A_1$ and, hence from Figure~\ref{fig:ellipticintinvert}, we find $b/a \ge c/a$. Thus, $a\ge b\ge c$ defines the equivalent ellipsoid. This means that ${\mathcal M} [\alpha B, \omega=0,\sigma_*, \mu_r]$ does not provide a unique object characterisation as there is an equivalent ellipsoid that has the same MPT.



Note that a numerical approach to finding the equivalent ellipsoid has been previously proposed by~\cite{taufiq} and involves solving the  minimisation problem 
\begin{equation}
\min_{\mathbf u}\left  | \sum_{i=1}^3 ({\mathbf \Lambda}^{{\mathcal N}^0 [\alpha B,\mu_r]})_{ii} -( {\mathcal T} [ \alpha E(0,{\mathbf{u} }),\mu_r   ] )_{ii} \right |^2, \label{eqn:minproblem}
\end{equation}
for ${\mathbf u} :=(a,b,c)$ 
so as to find the equivalent ellipsoid $E(0)$. Since $a>0$, $b>0$ and $c>0$ it can be shown that 
the associated non-linear system is continuous with respect to the unknown variables, hence the approximate solution of this non-linear system is well-posed~\cite{taufiq}, but it was unclear if the solution $a$, $b$ and $c$ exists and is unique. Our proof and alternative approach above overcomes this difficulty.  In practice, applying (\ref{eqn:minproblem}) gives the same result as our procedure for the examples presented in Section~\ref{equiv:ellips}.

\subsubsection{Equivalent ellipsoid $E(\infty)$}
For the case of $\sigma_* \to \infty$ it is known that~\cite{LedgerLionheart2019}
\begin{equation}
\lim_{\sigma_* \to \infty} {\mathcal M} [\alpha B, \omega ,\sigma_*, \mu_r ] = {\mathcal M}^\infty [ \alpha B] \equiv \lim_{\omega \to \infty} {\mathcal M} [\alpha B, \omega ,\sigma_*, \mu_r ] ,\label{eqn:pecmpt}
\end{equation}
where, importantly, the latter equivalence must be viewed with care as $\omega \to \infty$ would violate the eddy current assumption and, instead, this limit should be viewed as the limiting frequency for which the eddy current model is valid. If the topology of $B$ is such that its Betti number $\beta_1(B)=0$ then~\cite{LedgerLionheart2019}
\begin{equation}
\lim_{\sigma_* \to \infty} {\mathcal M} [\alpha B , \omega ,\sigma_*, \mu_r ] = {\mathcal M}^\infty [ \alpha B] = {\mathcal T}[\alpha B,0 ]  \equiv \lim_{\omega \to \infty} {\mathcal M} [\alpha B, \omega ,\sigma_*, \mu_r ] ,  \label{eqn:highfreq}
\end{equation}
and the unique equivalent ellipsoid $E(\infty)$ can also be found analogously to the approach in Section~\ref{sect:equivellip0}.
Similarly, we find that  $\lim_{\omega \to \infty} {\mathcal M} [\alpha B, \omega,\sigma_*, \mu_r]$ does not provide a unique object characterisation. 
 For the definition of Betti numbers and their implications for MPTs see~\cite{LedgerLionheart2016} and references therein.

\subsubsection{Equivalent ellipsoids $E(\Omega)$ and $E(\omega_m)$}
For the non-limiting fixed frequency case $\omega =\Omega$,  $\mathcal{I} [\alpha B, \omega,\sigma_*,\mu_r ]$ is non-vanishing and we can no longer express ${\mathbf \Lambda}^{\tilde{\mathcal R} [\alpha B,0, \sigma_*,\mu_r]}$ in terms ${\mathcal T}[\alpha E, k]$. However, a semi-analytical solution is available for the MPT of an ellipsoid~\cite{barrowes2008, barrowes2004, aospheroid}, which would allow an analogous numerical procedure to (\ref{eqn:minproblem}) to be applied to find an equivalent ellipsoid. But, given the non-explicit nature of this solution, it is not possible to show existence or uniqueness in this case. Nonetheless,  we conjecture that for $\beta_1(B)=0$ such an equivalent ellipsoid exists so that the MPT characterisation at a fixed frequency does not uniquely  characterise the object's shape and materials. To be able do this alot more data would be required.

If this is extended to the spectral case then, as results for $E(0)$ and $E(\infty)$ illustrate, there is no longer a single equivalent ellipsoid, but instead a sequence of equivalent ellipsoids $E(\omega_m)$, $m=1,\ldots,M$, which have different dimensions and, consequently, the MPT spectral signature constructed from an equivalent ellipsoid  at one frequency does not match the object's MPT eigenvalue spectral signature. To illustrate this, we consider the practical application to the earlier irregular tetrahedron described in Section~\ref{sect:irregulartet}.

\subsubsection{Equivalent ellipsoid for an irregular tetrahedron}\label{equiv:ellips}
\begin{figure}[!h]
\begin{center}
$\begin{array}{cc}
\includegraphics[width=0.5\textwidth, keepaspectratio]{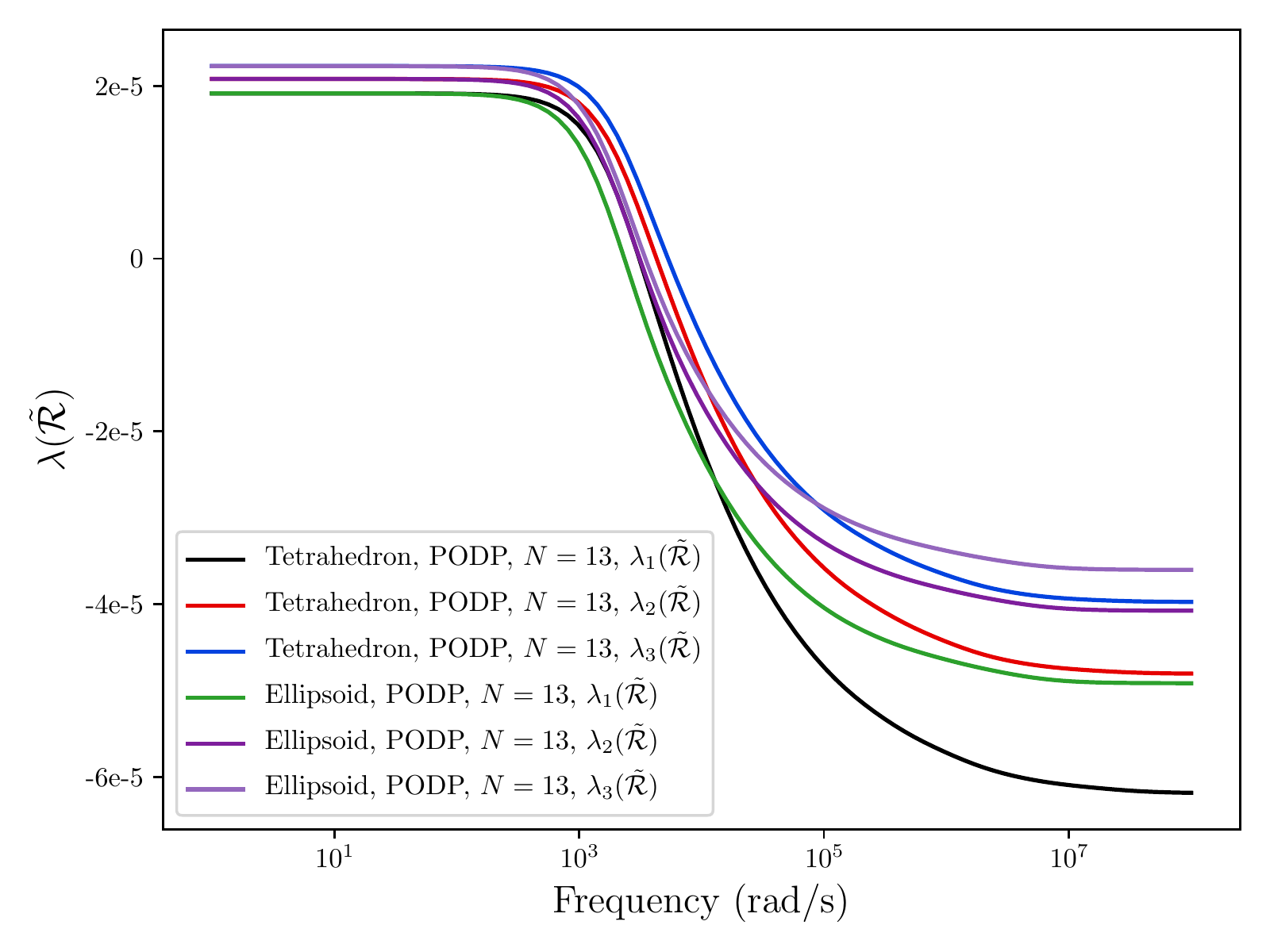} &
\includegraphics[width=0.5\textwidth, keepaspectratio]{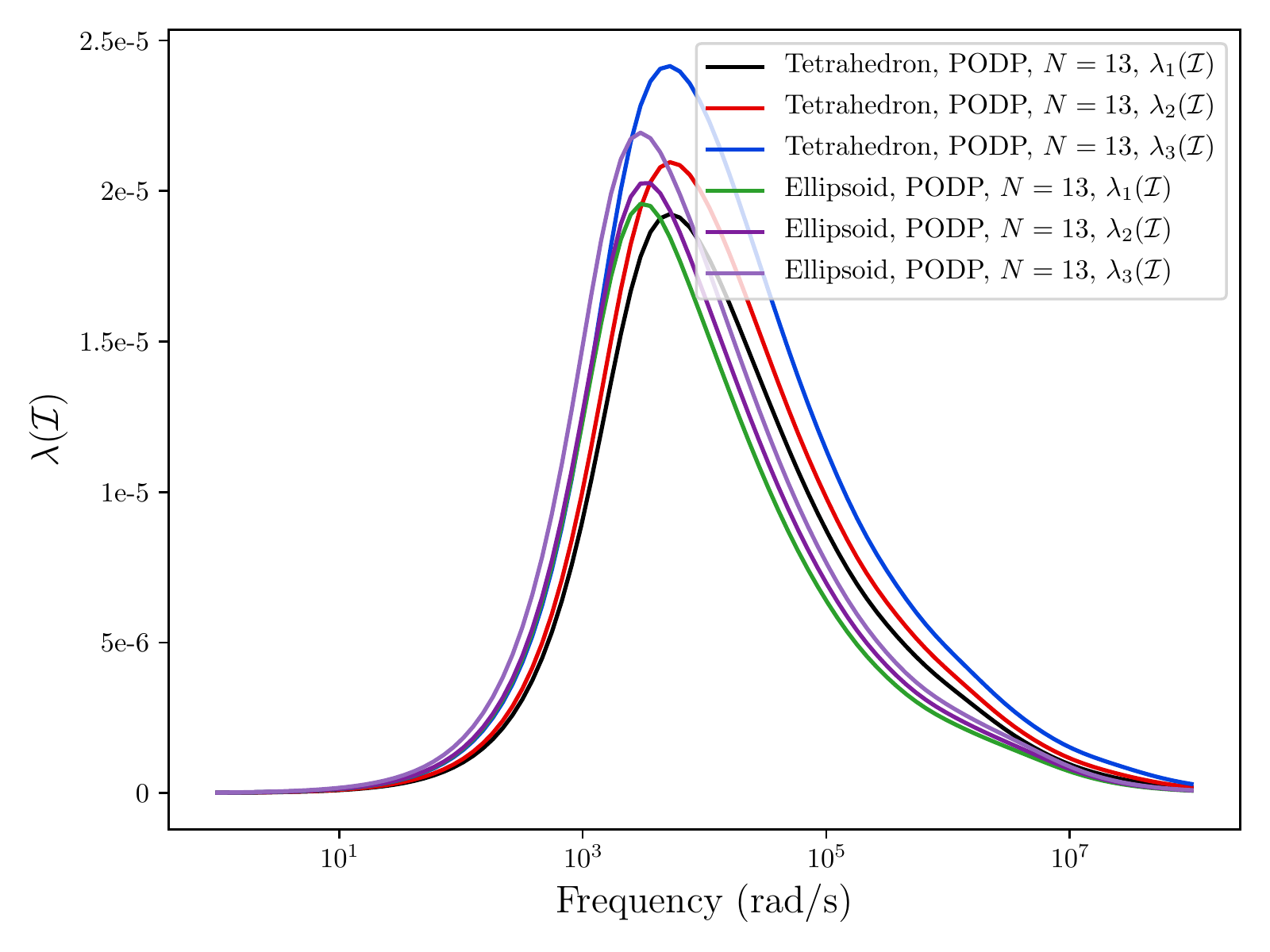}\\
\lambda_i ( \tilde{\mathcal R}) & \lambda_i({\mathcal I})
\end{array}$
\caption{Irregular tetrahedron $B$ with vertices as stated in (\ref{eqn:tetvertices}),  $\alpha=0.01 \text{ m}$, $\mu_r= 2$ and $\sigma_* = 5.96\times 10^{6} \text{ S/m}$. Comparison of $\lambda_i(\tilde {\mathcal R}[\alpha B, \omega, \sigma_*, \mu_r ] )$ and
$\lambda_i(\tilde {\mathcal R}[\alpha E(0), \omega, \sigma_*, \mu_r ] )$
as well as
 $\lambda_i ( {\mathcal I}[\alpha B, \omega, \sigma_*, \mu_r] )$ and
$ \lambda_i(  {\mathcal I}[\alpha E(0), \omega, \sigma_*, \mu_r ] )$
  using an equivalent ellipsoid $E(0)$.}
\label{fig:tetramurnot1}
\end{center}
\end{figure}

To illustrate that the spectral signature of the MPT for an object contains richer information than the spectral signature of an equivalent ellipsoid obtained at a fixed frequency, we compare ${\mathcal M}[\alpha B, \omega, \sigma_*, \mu_r ] $ and ${\mathcal M}[\alpha E(0) , \omega, \sigma_*, \mu_r ] $ in Figure~\ref{fig:tetramurnot1} using method summarised in Section~\ref{sect:rommethod} for the case where $B$ is the irregular tetrahedron
as described in Section~\ref{sect:irregtetexp}. 
The resulting equivalent ellipsoid $E(0)$ has $a= 1.4426$, $b= 1.8797$ and $c= 2.4243$ (to 4dp).  We observe that the eigenvalues of the real and imaginary parts of  ${\mathcal M}[\alpha E(0), \omega, \sigma_*, \mu_r ] $ agrees well with those of ${\mathcal M}[
\alpha B, \omega, \sigma_*, \mu_r] $ for small $\omega$, but the spectral signature differs for large $\omega$.

\begin{figure}[!h]
\begin{center}
$\begin{array}{cc}
\includegraphics[width=0.5\textwidth, keepaspectratio]{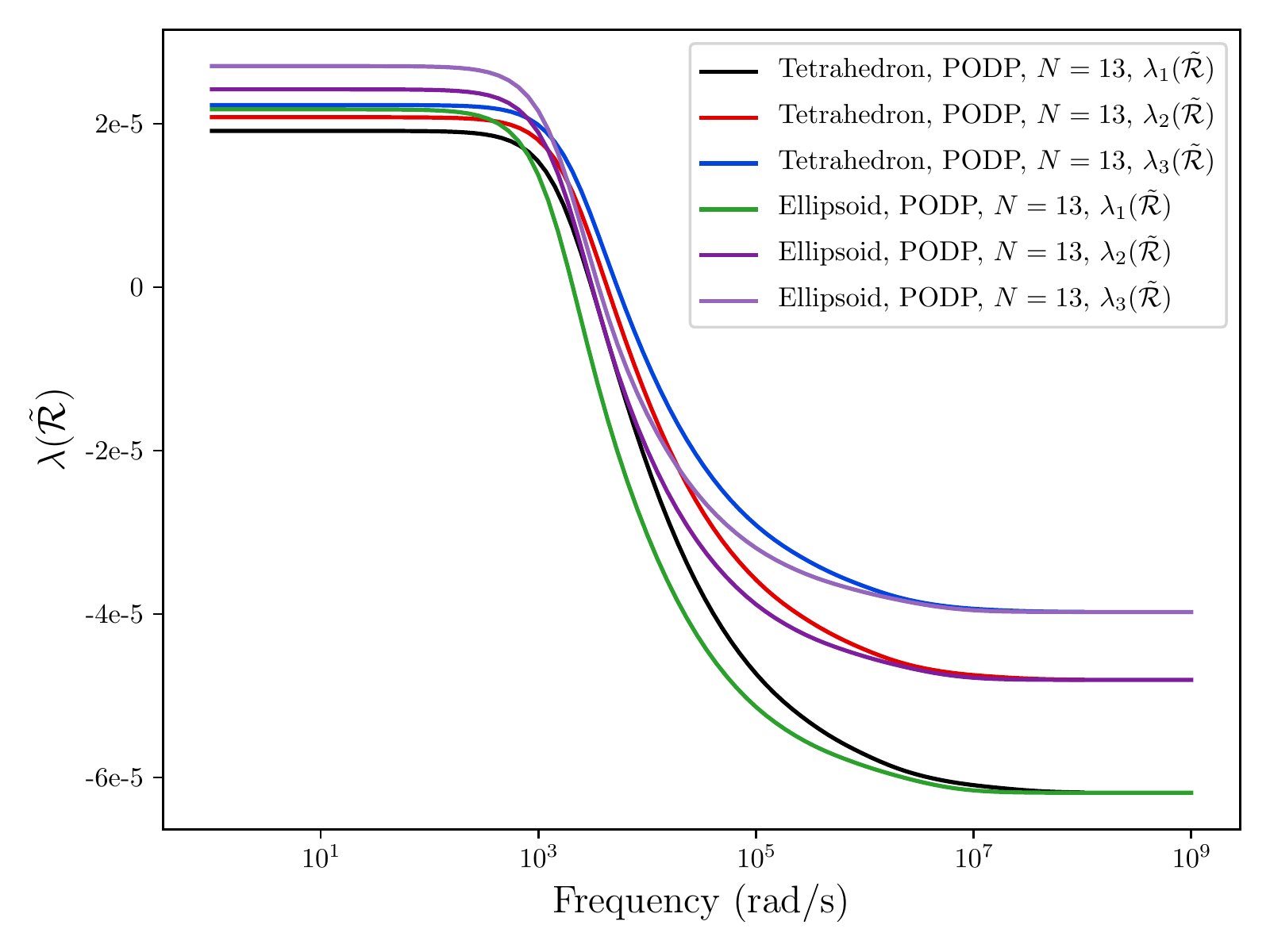} &
\includegraphics[width=0.5\textwidth, keepaspectratio]{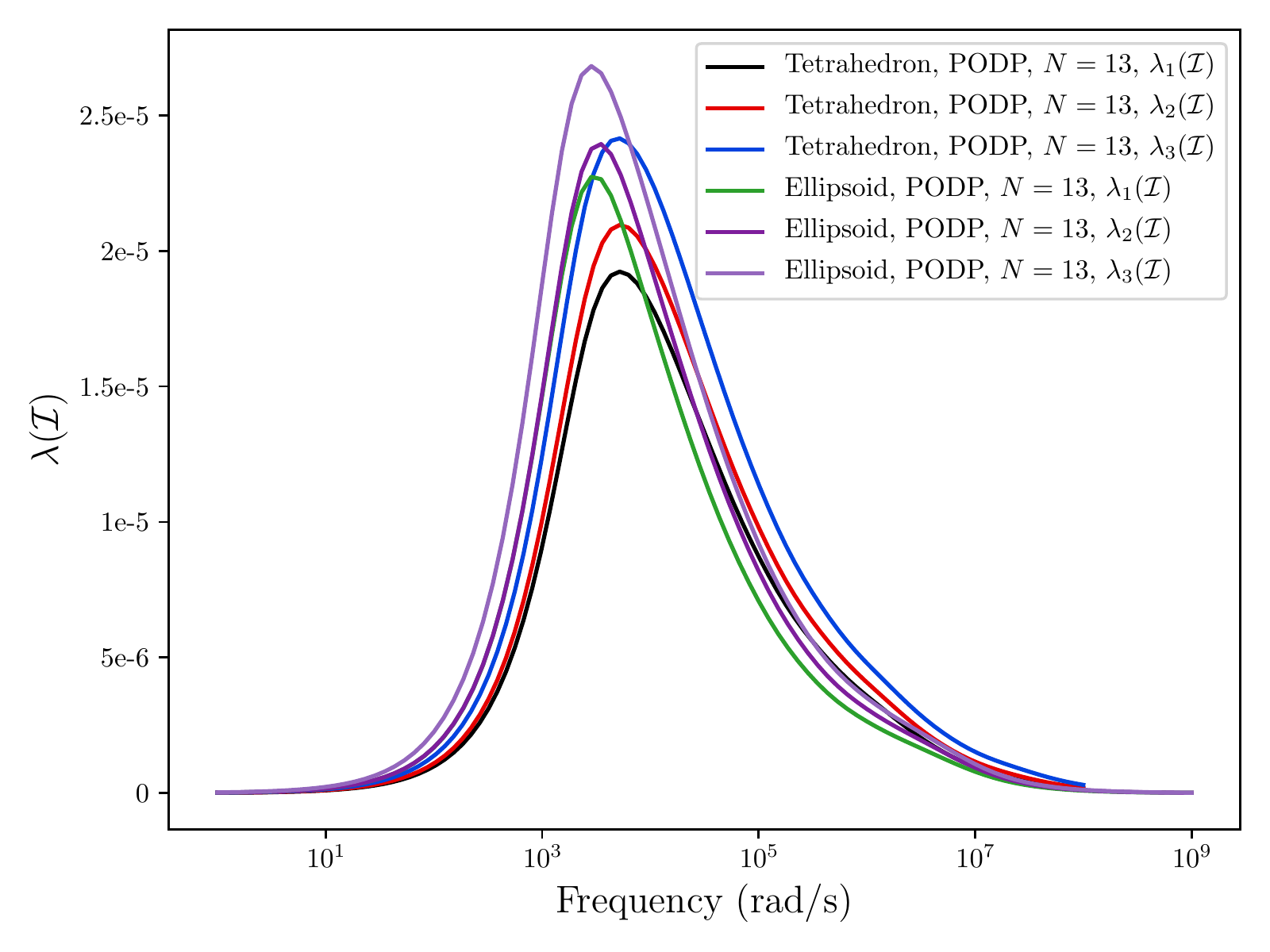} \\
\lambda_i ( \tilde {\mathcal R}) & \lambda_i({\mathcal I})
\end{array}$
\caption{Irregular tetrahedron $B$ with vertices as stated in  (\ref{eqn:tetvertices}), $\alpha=0.01 \text{ m}$ $\mu_r= 2$ and $\sigma_* = 5.96\times 10^{6} \text{ S/m}$: Comparison of $\lambda_i(\tilde{\mathcal R}[\alpha B, \omega, \sigma_*, \mu_r ] )$ and
$\lambda_i(\tilde{\mathcal R}[\alpha E(\infty), \omega, \sigma_*, \mu_r ] )$
as well as
 $\lambda_i ( {\mathcal I}[\alpha B, \omega, \sigma_*, \mu_r] )$ and
$ \lambda_i(  {\mathcal I}[\alpha E(\infty), \omega, \sigma_*, \mu_r ] )$
  using an equivalent ellipsoid $E(\infty)$.}
\label{fig:case2:tetramurnot1}
\end{center}
\end{figure}

The corresponding results comparing ${\mathcal M}[\alpha B , \omega, \sigma_*, \mu_r] $ and ${\mathcal M}[\alpha E(\infty) , \omega, \sigma_*, \mu_r] $ are shown in Figure~\ref{fig:case2:tetramurnot1}, where the equivalent ellipsoid $E(\infty)$ is defined by  $a= 1.3693$, $b= 1.9090$ and $c= 2.9404$ (to 4dp).  We observe that the eigenvalues of the real and imaginary parts of ${\mathcal M}[\alpha E(\infty) , \omega, \sigma_*, \mu_r ] $ agree well with those of ${\mathcal M}[\alpha B, \omega, \sigma_*, \mu_r ] $ for small and large $\omega$, but the spectral signature differs considerably for other $\omega$.

These results motivate the advantages of using spectral MPT signature over using MPT information from a single frequency, since, rather than a single equivalent ellipsoid, there are multiple equivalent ellipsoids corresponding to different $\omega$ providing greater information for the classification.

\subsection{Understanding the spectral signature of MPTs} \label{sect:spectral}

Building on the above, we now provide a theoretical justification as to why an object's MPT spectral signature offers an improved object characterisation over an MPT obtained at a single frequency. Recalling that $\nu: = \alpha^2 \omega \mu_0 \sigma_* $, 
then for an object $B$,  the following spectral representations have been derived~\cite{LedgerLionheart2019} for the tensor coefficients
\begin{subequations} \label{eqn:spectralrandi}
\begin{align}
({\mathcal R}[\alpha B , \omega, \sigma_*, \mu_r])_{ij} & = \frac{\alpha^3}{4} \sum_{n=1}^\infty \text{Re}( \beta_n^B(\nu))\lambda_n^B  \left < {\bm \phi}_n^B , {\bm \theta}_i^{(0)}  \right >_{L^2(B)} \left < {\bm \phi}_n^B , {\bm \theta}_j ^{(0)}   \right >_{L^2(B)} , \\
({\mathcal I}[\alpha B, \omega, \sigma_*, \mu_r ])_{ij} & = \frac{\alpha^3}{4} \sum_{n=1}^\infty \text{Im}( \beta_n^B(\nu) )\lambda_n^B  \left < {\bm \phi}_n^B , {\bm \theta}_i ^{(0)}  \right >_{L^2(B)} \left < {\bm \phi}_n^B , {\bm \theta}_j^{(0)}  \right >_{L^2(B)} ,
\end{align}
\end{subequations}
where the notation $\left < {\bm u} , {\bm v} \right >_{L^2(B)} : = \int_B {\bm u} \cdot  \overline{\bm v} \dif {\bm \xi}$ denotes the $L^2$ inner product over $B$ and
\begin{equation}
\beta_n^B(\nu) = \text{Re} ( \beta_n^B(\nu) + \im \text{Im}(\beta_n^B(\nu))  = - \frac{\nu^2}{\nu^2+(\lambda_n^B)^2} + \im \frac{\nu \lambda_n^B}{\nu^2 + (\lambda_n^B)^2} ,
\end{equation}
are functions  whose real part is sigmoid with $\log \nu$ (or $\nu$) and whose imaginary part has a single local maximum with $\log \nu$ (or $\nu$). In the above,
 $(\lambda_n^B, {\bm \phi}_n^B)$ are real eigen--solution pairs for the problem
\begin{subequations}
\begin{align}
\nabla \times \mu_r^{-1} \nabla \times {\bm \phi}_n^B & = \lambda_n {\bm \phi}_n^B && \text{in $B$} \\
\nabla \times \nabla \times {\bm \phi}_n^B & = {\bm 0} && \text{in $B^c$} \\
[ {\bm n} \times {\bm \phi}_n ^B] & = {\bm 0} && \text{on $\partial B$} \\
[ {\bm n} \times \mu_r \nabla \times {\bm \phi}_n ^B ]  & = {\bm 0} && \text{on $\partial B$} \\
{\bm \phi}_n^B & = O( | {\bm \xi } |^{-1} | ) && \text{as $| {\bm \xi} | \to \infty$}.
\end{align}
\end{subequations}
and the real weights $\lambda_n^B \left < {\bm \phi}_n^B , {\bm \theta}_i^{(0)}  \right >_{L^2(B)} \left < {\bm \phi}_n^B , {\bm \theta}_j ^{(0)}   \right >_{L^2(B)} $ together with the complex functions
$\beta_n^B(\nu)$
ultimately determine the shape of an MPT's spectral signature. For small $\omega$, the shape of the MPT spectral signature is well captured by a single dominant mode in the summations (\ref{eqn:spectralrandi}) so that, if the dominant mode is the first mode,
\begin{subequations} 
\begin{align}
({\mathcal R}[\alpha B , \omega, \sigma_*, \mu_r])_{ij} & \approx \frac{\alpha^3}{4}  \text{Re}( \beta_1^B(\nu))\lambda_1^B  \left < {\bm \phi}_1^B , {\bm \theta}_i^{(0)}  \right >_{L^2(B)} \left < {\bm \phi}_1^B , {\bm \theta}_j ^{(0)}   \right >_{L^2(B)} , \\
({\mathcal I}[\alpha B, \omega, \sigma_*, \mu_r ])_{ij} & \approx \frac{\alpha^3}{4}  \text{Im}( \beta_1^B(\nu) )\lambda_1^B  \left < {\bm \phi}_1^B , {\bm \theta}_i ^{(0)}  \right >_{L^2(B)} \left < {\bm \phi}_1^B , {\bm \theta}_j^{(0)}  \right >_{L^2(B)} ,
\end{align}
\end{subequations}
which makes the role of  $\text{Re} ( \beta_1^B(\nu))$  $\text{Im}(\beta_1^B(\nu))$ explicit in determining the shape of the MPT frequency spectral signature. For further details see~\cite{LedgerLionheart2019}.


Similar representations to (\ref{eqn:spectralrandi}) can be obtained for $ ({\mathcal R}[\alpha E, \omega, \sigma_*, \mu_r ])_{ij}$ and  $ ({\mathcal I}[\alpha E, \omega, \sigma_*, \mu_r, ])_{ij}$ for an ellipsoidal object $E_\alpha=\alpha E$, which is aligned with the coordinate axes such that the associated tensor is diagonal.  The non-zero coefficients for $i=j$ are expressed
 in terms of real eigen--solutions $(\lambda_n^E, {\bm \phi}_n^E)$ where
\begin{subequations}
\begin{align}
\nabla \times \mu_r^{-1} \nabla \times {\bm \phi}_n^E & = \lambda_n {\bm \phi}_n^E && \text{in $E$}, \\
\nabla \times \nabla \times {\bm \phi}_n^E & = {\bm 0} && \text{in $E^c$}, \\
[ {\bm n} \times {\bm \phi}_n ^E] & = {\bm 0} && \text{on $\partial E$}, \\
[ {\bm n} \times \mu_r \nabla \times {\bm \phi}_n ^E ]  & = {\bm 0} && \text{on $\partial E$}, \\
{\bm \phi}_n^E & = O( | {\bm \xi } |^{-1} | ) && \text{as $| {\bm \xi} | \to \infty$}.
\end{align}
\end{subequations}

For simplicity, we consider objects $B$ where the MPT is diagonal such that the eigenvalues of ${\mathcal N^0}, {\mathcal R}$ and ${\mathcal I}$ coincides with the corresponding non-zero tensor coefficients in the following two subsections.

\subsubsection{Equivalent ellipsoids at $\omega=0$}

Comparing ${\mathcal M}[\alpha B, \omega,\sigma_*, \mu_r ]$ and ${\mathcal M}[\alpha E(0), \omega,\sigma_*, \mu_r ]$ we have, for an equivalent ellipsoid constructed at $\omega =0$, that
\begin{equation}
{\mathcal M}[\alpha B, \omega=0 ,\sigma_*, \mu_r ] = {\mathcal N}^ 0 [\alpha B, \mu_r] ={\mathcal N}^ 0 [\alpha E(0), \mu_r]  = {\mathcal M}[\alpha E(0),\omega= 0,\sigma_*, \mu_r ],
\end{equation}
but, there is no equivalence between the eigensolutions  $(\lambda_n^B, {\bm \phi}_n^B)$ and $(\lambda_n^E, {\bm \phi}_n^E)$ and, hence, we expect $\tilde{\mathcal R}[\alpha B , \omega, \sigma_*, \mu_r ]$, ${\mathcal I}[  \alpha B , \omega, \sigma_*, \mu_r ]$ to be different from $\tilde{\mathcal R}[ \alpha E(0) , \omega, \sigma_*, \mu_r ]$, ${\mathcal I}[ \alpha E(0), \omega, \sigma_*, \mu_r ]$ away from $\omega=0$.

\subsubsection{Equivalent ellipsoids as $\omega\to \infty$}
Comparing ${\mathcal M}[\alpha B, \omega,\sigma_*, \mu_r ]$ and ${\mathcal M}[\alpha E (\infty) , \omega,\sigma_*, \mu_r ]$  we have, for an equivalent ellipsoid constructed for $\omega \to \infty$, that
\begin{align}
{\mathcal M}^\infty  [\alpha B] = &{\mathcal N}^0[\alpha B, \mu_r ] + \lim_{\omega \to \infty} ( {\mathcal R} [ \alpha B , \omega, \sigma_*, \mu_r ]  + \im{\mathcal I} [ \alpha B, \omega, \sigma_*, \mu_r ] ) 
\nonumber \\
=& {\mathcal N}^0[\alpha B, \mu_r] + \lim_{\omega \to \infty}  {\mathcal R} [ \alpha B, \omega, \sigma_*, \mu_r]
\nonumber \\
= &  {\mathcal N}^0[\alpha E (\infty) , \mu_r] + \lim_{\omega \to \infty}  {\mathcal R} [ \alpha E (\infty) , \omega, \sigma_*, \mu_r ] =  {\mathcal M}^ \infty [\alpha E(\infty) ]  .
\end{align}
In addition,
\begin{align} 
\lim_{\omega \to \infty}  ({\mathcal R} [\alpha B, \omega, \sigma_*, \mu_r ])_{ij}  = - \frac{\alpha^3}{4} \sum_{n=1}^\infty  \lambda_n^B  \left < {\bm \phi}_n^B , {\bm \theta}^{(0)}_i \right >_{L^2(B)} \left < {\bm \phi}_n^B , {\bm \theta}^{(0)} _j \right >_{L^2(B)} , \nonumber \\
\lim_{\omega \to \infty}  ({\mathcal R} [\alpha E(\infty),  \omega, \sigma_*, \mu_r ])_{ij}  = - \frac{\alpha^3}{4} \sum_{n=1}^\infty  \lambda_n^E  \left < {\bm \phi}_n^E , {\bm \theta}^{(0)}_i \right >_{L^2( E(\infty) )} \left < {\bm \phi}_n^E , {\bm \theta}^{(0)}_j \right >_{L^2( E(\infty))} \nonumber .
\end{align}

\paragraph{Case of $\mu_r=1$}
If $\mu_r=1$ then $({\mathcal N}^0[\alpha B, \mu_r ] )_{ij}= ({\mathcal N}^0[\alpha E ( \infty) , \mu_r ] )_{ij} =0$ and, thus,  for $i=j$ we  have
\begin{equation}
 \sum_{n=1}^\infty  \lambda_n^B  \left < {\bm \phi}_n^B , {\bm \theta}^{(0)}_i \right >_{L^2(B)} ^2
 = 
\sum_{n=1}^\infty  \lambda_n^E  \left < {\bm \phi}_n^E , {\bm \theta}^{(0)}_i \right >_{L^2(E(\infty))}^2 \label{eqn:equivspecmu1oinf}.
\end{equation}
However, only in the case where  $B= E(\infty)$ can we expect the solutions ${\bm \theta}^{(0)}_i $ to the transmission problems
(\ref{eqn:Theta0}) for the objects $B$ and $E(\infty)$ to be same. Otherwise, we expect the solutions ${\bm \theta}^{(0)}_i $  to be different and 
$(\lambda_n^B, {\bm \phi}_n^B) \ne (\lambda_n^{E}, {\bm \phi}_n^{E })$. Expression
  (\ref{eqn:equivspecmu1oinf}) only guarantees  that the sum is the same and, consequently,  ${\mathcal M}[\alpha B, \omega,\sigma_*, \mu_r ]$ will be different to ${\mathcal M}[\alpha E(\infty) , \omega,\sigma_*, \mu_r ]$ away from the limiting cases of $\omega=0$ and $\omega \to \infty$ for $\mu_r=1$.

\paragraph{Case of $\mu_r\ne1$}
If $\mu_r\ne 1$ then $({\mathcal N}^0[\alpha B, \mu_r ] )_{ij} \ne ({\mathcal N}^0[\alpha E(\infty) , \mu_r] )_{ij} $ and, therefore,  for $i=j$ we have
\begin{equation}
 \sum_{n=1}^\infty  \lambda_n^B  \left < {\bm \phi}_n^B , {\bm \theta}^{(0)}_i \right >_{L^2(B)} ^2
 \ne 
\sum_{n=1}^\infty  \lambda_n^E  \left < {\bm \phi}_n^E , {\bm \theta}^{(0)}_i \right >_{L^2(E)}^2.
 \end{equation}
Thus, $\lambda_n^E \ne  \lambda_n^B$ and  $\left < {\bm \phi}_n^B , {\bm \theta}^{(0)}_i \right >_{L^2(B)} ^2
\ne  \left < {\bm \phi}_n^E , {\bm \theta}^{(0)} _i \right >_{L^2(E(\infty) )}^2$ for all $n$. Consequently,  ${\mathcal M}[\alpha B, \omega,\sigma_*, \mu_r ]$ will be different to ${\mathcal M}[\alpha E(\infty) , \omega,\sigma_*, \mu_r ]$ away from the limiting cases of $\omega \to \infty$ for $\mu_r\ne 1$.

\section{MPT Spectral signature invariants for object classification} \label{sect:classinv}
Bishop~\cite{bishopbook} describes the process of classification  as taking an input vector ${\mathbf x}$ and assigning it to one of $K$ discrete classes $C_k$, $k=1,\ldots,K$. For example, in security screening, the simplest form of classification with $K=2$ involves only the classes {\em threat} ($C_1$) and {\em non-threat} ($C_2$), and one with a higher level of fidelity might include the classes  of metallic objects such as key ($C_1$), coin ($C_2$), gun ($C_3$), knife ($C_4$) ... where the class numbers are assigned as desired. He recommends that it is convenient (in probabilistic methods of classification) to use a 1-of-$K$ coding system in which the entries in a  vector ${\mathbf t}\in {\mathbb R}^K$ take the form
\begin{equation}
{\rm t}_i := \left \{ \begin{array}{ll} 1 & \text{if $i=k$} \\
0 & \text{otherwise} \end{array} \right . ,
\nonumber 
\end{equation}
 if the correct class is $C_k$.
 Requiring that we always have $\sum_{k=1}^K {\rm t}_k=1$, then this approach has the advantage that ${\rm t}_k$ can be interpreted as the probability that the correct class is $C_k$.  
In this section, we focus on alternative choices of the $F$  features in the input vector ${\mathbf x}\in{\mathbb R}^F$ for the classifier. In future work we will compare the performance of different classifiers based for these alternatives.
We focus on suitable features that are invariant to rotation of the object. Note that the rank 2 MPT, and hence the invariants considered below, are invariant to the position of the object.

%

\subsection{Tensor eigenvalues}
Recall that the diagonal matrices 
${\mathbf \Lambda}^{\tilde{\mathcal R} [\alpha B,\omega_m,  \sigma_*, \mu_r  ]}$ and
${\mathbf \Lambda}^{{\mathcal I} [\alpha B,\omega_m , \sigma_*, \mu_r ]}$ contain the eigenvalues of \\ $\tilde{\mathcal R} [\alpha B,\omega_m,  \sigma_*, \mu_r  ]$ and ${\mathcal I} [\alpha B,\omega_m , \sigma_*, \mu_r ]$, respectively, and satisfy the object rotation invariant property
\begin{align}
\lambda_i ( \tilde{\mathcal R} [\alpha B,\omega_m,  \sigma_*,\mu_r  ]) = ({\mathbf \Lambda}^{\tilde{\mathcal R} [\alpha B,\omega_m, \sigma_*,\mu_r  ]} )_{ii}= &  
({\mathbf \Lambda}^{\tilde{\mathcal R} [\alpha R(B),\omega_m, \sigma_*, \mu_r  ]})_{ii} 
=\lambda_i ( \tilde{\mathcal R} [\alpha R(B),\omega_m, \sigma_*, \mu_r  ])
\nonumber, \\
\lambda_i ( {\mathcal I} [\alpha B,\omega_m,  \sigma_*, \mu_r   ]) =
({\mathbf \Lambda}^{{\mathcal I} [\alpha B, \omega_m, \sigma_*,\mu_r  ]})_{ii} = &  
({\mathbf \Lambda}^{{\mathcal I} [\alpha R(B), \omega_m, \sigma_*, \mu_r   ]})_{ii} =\lambda_i ( {\mathcal I} [\alpha R(B),\omega_m,  \sigma_*, \mu_r  ])  \nonumber  ,
\end{align}
at each discrete frequencies $\omega_m$, $m=1,\ldots,M$ in the MPT spectral signature.
Thus, one option is to select the features for the classifier as
\begin{equation}
({\mathbf x})_i = {\rm x}_i = \left \{ \begin{array}{ll} \lambda_j ( \tilde{\mathcal R} [\alpha B,\omega_m,  \sigma_*,\mu_r  ]), & i =  j+ (m-1)M  \\
\lambda_j ( {\mathcal I} [\alpha B,\omega_m,  \sigma_*, \mu_r   ]), & i = j+ (m+2) M
\nonumber  \end{array} \right . , \label{eqn:feateig}
\end{equation}
where $j=1,2,3$  and  $m=1,\ldots,M$ so that $F=6M$.
This is particularly attractive, since any hidden object is likely to be in some unknown rotated configuration compared to canonical choice of the corresponding object in the library and, as the eigenvalues are invariant under object rotation, we do not need knowledge of the orthogonal rotation matrix ${\mathbf R}$ to perform the classification. Furthermore, 
in practice, {measurements lead to} noisy tensor coefficients in the form ${\mathcal M} + {\mathcal E}_r+\im {\mathcal E}_\im$ where ${\mathcal E}_r+\im {\mathcal E}_\im$ is a complex symmetric rank 2 tensor and represents the noise. 
To understand the effects of noise, consider for simplicity a symmetric real matrix ${\mathbf A}
$ corrupted by a real symmetric  ${\mathbf E}$, applying results on eigenvalue perturbations~\cite{golub} we find that
\begin{equation}
\sum_{i=1}^3 (\lambda_i( {\mathbf A} +{\mathbf E}) - \lambda_i({\mathbf A} ))^2 \le \| {\mathbf E} \|_F^2 \nonumber ,
\end{equation}
so that the  eigenvalues $\lambda_i$  of ${\mathbf A}$ are similar to those of ${\mathbf A}+{\mathbf E}$ provided ${\mathbf E}$ represents the low-moderate noise. However, for an eigenvalue-eigenvector pair $\lambda_1,\mathbf{q}_1$~\cite{golub}
\begin{equation}
\text{dist}(\mathbf{q}_1( {\mathbf A} ) , \mathbf{q}_1( {\mathbf A}+{\mathbf E})) \le \frac{4}{d} C({\mathbf E}) \nonumber ,
\end{equation}
where  $d=\min_{\mu \in\lambda_i({\mathbf A})} | \lambda_i -\mu | >0$ and $C({\mathbf E})$ is a constant depending on ${\mathbf E}$.
 In other words, if the eigenvalues are close (so $d$ is small), the eigenvectors will be badly effected by the noise. The same applies to the real and imaginary parts of ${\mathcal M} + {\mathcal E}_r+\im {\mathcal E}_\im$  when the coefficients are arranged as matrices.
\subsection{Tensor invariants}
While $\lambda_i ( \tilde{\mathcal R} [\alpha B,\omega_m,\mu_r , \sigma_* ])$, $\lambda_i ( {\mathcal I} [\alpha B,\omega_m, \mu_r , \sigma_* ]) $, $m=1,\ldots,M$,  are invariant under object rotation, their behaviour is well understood and they behave well for noisy measurements, classifying objects on the basis of these may still cause practical issues. Firstly, care is needed  with the ordering of the eigenvalues since choosing  a simple rule such as  $\lambda_1 \ge \lambda_2 \ge \lambda_3 $ may lead to confusing results. For example, if the object has rotational and/or reflectional symmetries, we might find there are only $2$ independent eigenvalues at each frequency in the real and imaginary parts of the MPT, then, applying  the aforementioned rule   independently to
$\tilde{\mathcal R} [\alpha B, \omega_m , \sigma_*, \mu_r  ]$ and ${\mathcal I} [\alpha B,\omega_m,  \sigma_*, \mu_r  ]$
could lead to $\lambda_2( \tilde{\mathcal R} [\alpha B, \omega_m,  \sigma_* , \mu_r  ]) = \lambda_3 ( \tilde{\mathcal R} [\alpha B, \omega_m , \sigma_*, \mu_r   ])$ and 
$\lambda_1( {\mathcal I} [\alpha B,\omega_m,  \sigma_*, \mu_r  ]) = \lambda_2 ( {\mathcal I} [\alpha B,\omega_m, \sigma_*, \mu_r  ])$. Secondly, there is a danger  that different ordering rules are applied in the creation of the training library for the classifier compared to that used for testing some new candidate object. To overcome this,  tensor invariants can be used, which  are independent of how $\lambda_1$, $\lambda_2$ and $\lambda_3$ are assigned.
One possibility are the principal tensor invariants, which, for a rank 2 tensor ${\mathcal A}$, are (e.g.~\cite{bonet})
\begin{subequations}\label{eqn:princpinv}
\begin{align}
I_1 ({\mathcal A}) := & \hbox{tr} \, ({\mathcal A}) = \lambda_1 ({\mathcal A}) +  \lambda_2 ({\mathcal A}) +  \lambda_3 ({\mathcal A}) , \\
I_2 ({\mathcal A}) := &\frac{1}{2} \left (  \hbox{tr}\,  ({\mathcal A})^2 - \hbox{tr}\,   ( {\mathcal A}^2 ) \right ) = \lambda_1 ({\mathcal A})  \lambda_2 ({\mathcal A}) + \lambda_1 ({\mathcal A})  \lambda_3 ({\mathcal A}) + \lambda_2 ({\mathcal A})  \lambda_3 ({\mathcal A}) , \\
I_3 ({\mathcal A}) := &\hbox{det}\, ({\mathcal A}) = \lambda_1 ({\mathcal A})  \lambda_2 ({\mathcal A})  \lambda_3 ({\mathcal A}) ,
\end{align} 
\end{subequations}
which contain the same information as the tensor's eigenvalues $\lambda_i({\mathcal A})$ and can also be computed from (\ref{eqn:princpinv}). 
They satisfy\begin{align}
\lambda^3 - I_1  ({\mathcal A}) \lambda^2 + I_2 ({\mathcal A}) \lambda - I_3 ({\mathcal A}) = 0,
\end{align}
are rotationally invariant and, like the eigenvalues, are less-susceptible to noise than the tensor's eigenvectors. 

{Borrowing notation from continuum mechanics (e.g.~\cite{bonet}), $I_1 ({\mathcal A})$ is related to the {\em hydrostatic part} of ${\mathcal A}$ given by ${\mathcal H}= \frac{1}{3} \hbox{tr} ({\mathcal A} ) {\mathbb I}$ and is associated with the extent to which the operation ${\mathcal H} {\bm v}$ {\em stretches} or {\em shrinks} the magnitude of ${\bm v}$. The invariant  $I_2 ( {\mathcal A}  )$ is often, but not exclusively, related to the {\em deviatoric part} of $ {\mathcal A}  $ given by ${\mathcal S} = {\mathcal A} - {\mathcal H}$
describing the  extent to which 
${\mathcal S} {\bm v}$ distorts the components of ${\bm v}$. The invariant $I_3({\mathcal A})$ describes the extent of coupling of the two aforementioned cases and whether or not the tensor ${\mathcal A}$, when arranged as a $3 \times 3 $ matrix, is singular or not.  In addition,  when applied to (limiting cases) of $\tilde{\mathcal R} [\alpha B,\omega_m,  \sigma_*,\mu_r  ] $ and ${\mathcal I} [\alpha B,\omega_m,  \sigma_*,\mu_r  ] $, it has a further physical interpretation: Recall that the product $\alpha B$ implies that there are an infinite number of ways to choose $\alpha \ll 1 $ and $B$, which still result in the same $\alpha B$. For example, if $|B|$ is chosen such that $\hbox{det}\, ({\mathcal N}^0 [\alpha B,  \mu_r] ) = I_3 ({\mathcal N}^0 [\alpha B,  \mu_r] )= \alpha^9 $ then this invariant provides object size information, while, in general $I_3 ({\mathcal N}^0 [\alpha B,  \mu_r] )$, will be a function of $|B|$, $\alpha$ and $\mu_r$. Similarly,  $I_3 ( \tilde{\mathcal R} [\alpha B,\omega_m, \sigma_*, \mu_r  ])$ and $I_3 ( {\mathcal I} [\alpha B,\omega_m, \sigma_* , \mu_r  ])$ will be functions of $|B|$, $\alpha$, $\omega_m$, $\sigma_*$ and $\mu_r$.  Thus,  $I_3 ( \tilde{\mathcal R} [\alpha B,\omega_m, \sigma_*, \mu_r  ])$ and $I_3 ( {\mathcal I} [\alpha B,\omega_m, \sigma_* , \mu_r  ])$, for fixed $\alpha$, $\omega_m$, $\sigma_*$ and $\mu_r$, will scale like $|B|$.}
Using principal invariants, we could then select the features as
\begin{equation}
{\rm x}_i = \left \{ \begin{array}{ll} I_j ( \tilde{\mathcal R} [\alpha B,\omega_m,  \sigma_*,\mu_r  ]), & i= j+ (m-1)M  \\
I_j ( {\mathcal I} [\alpha B,\omega_m,  \sigma_*, \mu_r   ]), & i=  j+ (m+2)M
 \end{array} \right . , \label{eqn:featinv}
\end{equation}
where $j=1,2,3$  and  $m=1,\ldots , M$ so that $F=6 M$.


As an alternative to the principal invariants stated in (\ref{eqn:princpinv}), the alternative set of invariants (e.g.~\cite{bonet})
\begin{subequations} \label{eqn:altinv}
\begin{align}
I_1 ({\mathcal A}) := & \hbox{tr} \, ({\mathcal A}) = \lambda_1 ({\mathcal A}) +  \lambda_2 ({\mathcal A}) +  \lambda_3 ({\mathcal A}) , \\
J_2 ({\mathcal A}) := &\frac{1}{2}\hbox{tr} \, ({\mathcal S}^2) = \frac{1}{3}  I_1 ({\mathcal A}) ^2 - I_2 ({\mathcal A})  = \frac{1}{2} ( s_1({\mathcal S}) ^2 +s_2({\mathcal S})^2 +s_3({\mathcal S})^2) \nonumber \\
= & \frac{1}{2} ( (\lambda_1({\mathcal A}) - I_1({\mathcal A})/3)^2 +(\lambda_2({\mathcal A}) - I_1({\mathcal A})/3)^2 +(\lambda_3( {\mathcal A}) - I_1({\mathcal A})/3)^2),\\
J_3 ({\mathcal A}) := &\hbox{det} \, ({\mathcal S}) =  \frac{2}{27} I_1 ({\mathcal A}) ^3  - \frac{1}{3} I_1 ({\mathcal A}) I_2 ({\mathcal A})  + I_3 ({\mathcal A})   =\frac{1}{3} s_1({\mathcal S})  s_2({\mathcal S})  s_3({\mathcal S})  \nonumber  \\
= &(\lambda_1({\mathcal A}) - I_1({\mathcal A})/3) (\lambda_2({\mathcal A}) - I_1({\mathcal A})/3)(\lambda_3({\mathcal A}) - I_1({\mathcal A})/3),
\end{align} 
\end{subequations}
where  $\lambda_i ({\mathcal A})=s_i({\mathcal A}) + I_1 ({\mathcal A})/3$ can be used. These invariants  satisfy 
\begin{equation}
s^3 - J_2 ({\mathcal A})s  - J_3 ({\mathcal A})  =0,
\nonumber 
\end{equation}
and the roots of this equation are the eigenvalues $s_i$ of ${\mathcal S}$. {The invariants  $J_2 ( {\mathcal A}  )$ and   $J_3 ( {\mathcal A}  )$ are both related to the extent to which ${\mathcal S} {\bm v}$ distorts the components of ${\bm v}$.}
In this case, we can select the features as 
\begin{equation}
{\rm x}_i = \left \{ \begin{array}{ll} I_1 ( \tilde{\mathcal R} [\alpha B,\omega_m,  \sigma_*,\mu_r  ]), & i= 1+ (m-1)M \\
J_j ( \tilde{\mathcal R} [\alpha B,\omega_m,  \sigma_*,\mu_r  ]), & i= j+ (m-1)M \\
I_1 ( {\mathcal I} [\alpha B,\omega_m,  \sigma_*, \mu_r   ]), & i=  1+ (m+2)M  \\
J_j ( {\mathcal I} [\alpha B,\omega_m,  \sigma_*, \mu_r   ]), & i=   j+ (m+2)M
 \end{array} \right . , \label{eqn:featinv2}
\end{equation}
where $j=2,3$  and  $m=1,\ldots,M$ and $F=6 M$.
 One potential advantage of using (\ref{eqn:altinv}) as a set of features is that, for the case where 
$ \tilde{\mathcal R} [\alpha B,\omega_m,  \sigma_*,\mu_r  ]$ and ${\mathcal I} [\alpha B,\omega_m,  \sigma_*, \mu_r   ]$
  are both multiples of identify (such as for the MPT characterisation of a cube or sphere), $J_2$ and $J_3$ vanish.  

\subsection{Eigenvalues of the commutator of $\tilde{\mathcal R} [\alpha B,\omega,  \sigma_*, \mu_r ]$ and ${\mathcal I} [\alpha B,\omega,  \sigma_*, \mu_r ]$}
\begin{figure}
\begin{center}
$\begin{array}{cc}
\includegraphics[width=3in]{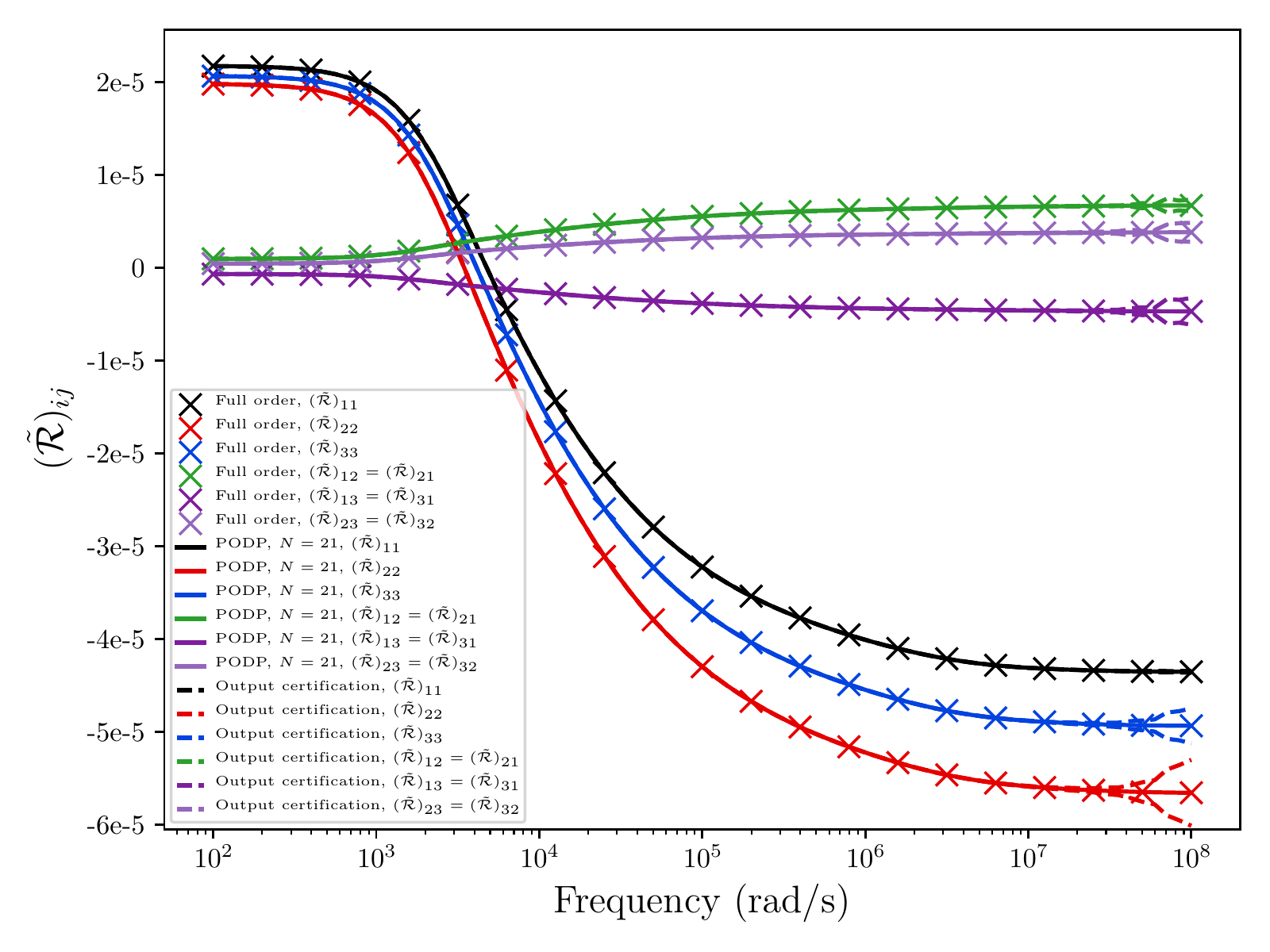} &
\includegraphics[width=3in]{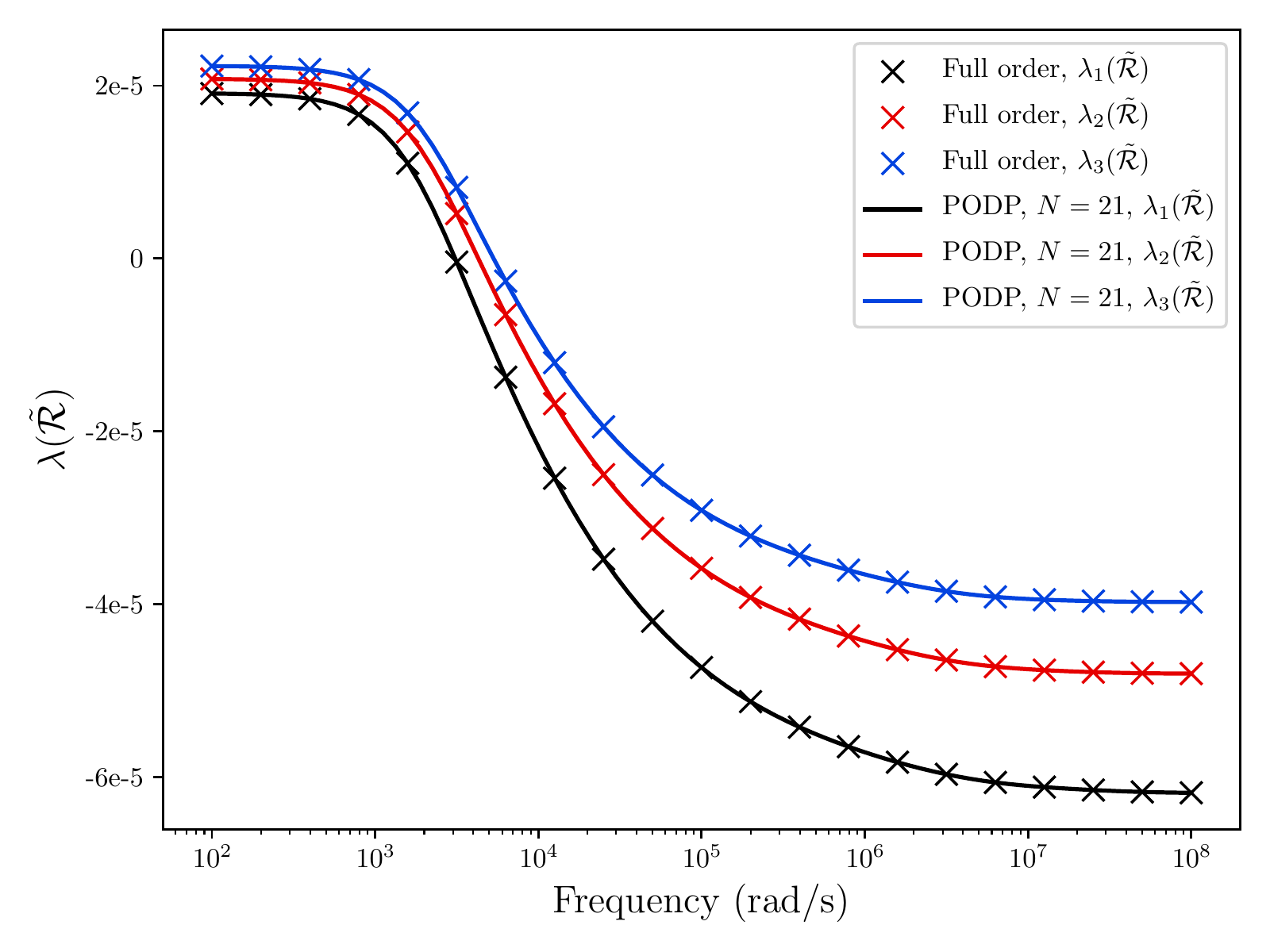}\\
(\tilde{\mathcal R}[\alpha B, \omega, \sigma_*, \mu_r ])_{ij} & \lambda_i(\tilde{\mathcal R}[\alpha B , \omega, \sigma_*, \mu_r])
\end{array}$
\end{center}
\caption{Irregular tetrahedron $B$ with vertices  (\ref{eqn:tetvertices}), $\alpha=0.01 \text{ m}$ $\mu_r= 2$ and $\sigma_* = 5.96\times 10^{6} \text{ S/m}$: Comparison of $(\tilde{\mathcal R}[\alpha B, \omega, \sigma_*, \mu_r ])_{ij} $ and $\lambda_i(\tilde{\mathcal R}[\alpha B , \omega, \sigma_*, \mu_r]) $.} \label{fig:tetrealcoefeig}
\end{figure}

The off-diagonal enteries $( \tilde{\mathcal R} [\alpha B,\omega,\sigma_*, \mu_r ])_{ij}$ and $( {\mathcal I} [\alpha B,\omega , \sigma_*,\mu_r ])_{ij}$ with $i\ne j$ 
are much smaller than those on the diagonal with $i=j$ as the results
\begin{subequations}
\begin{align}
| (\tilde{\mathcal R} [\alpha B, \omega, \sigma_*,\mu_r ])_{ij}| \le & | \text{tr} ( \tilde{\mathcal R} [\alpha B,\omega, \sigma_*,\mu_r ]) | = \left | \sum_{k=1}
 ^3 \lambda_k ( \tilde{\mathcal R} [\alpha B,\omega , \sigma_*,\mu_r ]) \right | , \\
 ( {\mathcal I} [\alpha B,\omega , \sigma_*,\mu_r ])_{ij}  \le &  \text{tr} ( {\mathcal I} [\alpha B,\omega , \sigma_*, \mu_r ])= \sum_{k=1}
 ^3 \lambda_k ( {\mathcal I} [\alpha B,\omega , \sigma_*, \mu_r ]) ,
 \end{align}
 \end{subequations}
  obtained  in Lemma 6.1 of~\cite{LedgerLionheart2019} show. This implies that  the tensor's eigenvalues, and hence its principal invariants (\ref{eqn:princpinv}) (as well as the alternative invariants (\ref{eqn:altinv})), are dominated by the behaviour of its diagonal coefficients. 
    To illustrate this, we show, in Figure~\ref{fig:tetrealcoefeig}, the comparison between $\lambda_i( \tilde{\mathcal R} [\alpha B,\omega, \sigma_*,\mu_r ])$ and $ ( \tilde{\mathcal R} [\alpha B,\omega, \sigma_*, \mu_r ])_{ij} $ for the irregular tetrahedron discussed in
 Section~\ref{sect:spectralobject}. We observe that the behaviour of the eigenvalues is dominated by the diagonal coefficients of ${\tilde{\mathcal R}} [\alpha B,\omega , \sigma_*, \mu_r ]$, similar arguments also apply to $\lambda_i( {{\mathcal I} [\alpha B,\omega, \sigma_*,\mu_r ]})$ and $({\mathcal I} [\alpha B,\omega, \sigma_*, \mu_r ])_{ij}$.

To improve the discrimination between objects whose tensors 
$ \tilde{\mathcal R} [\alpha B, \omega, \sigma_* , \mu_r]$ and ${\mathcal I} [\alpha B, \omega,  \sigma_*, \mu_r ] $  have different eigenvectors,
 we consider their commutator, which has coefficients
\begin{align}
({\mathcal Z}[ \alpha B,\omega,  \sigma_*, \mu_r])_{ij} : = (\tilde{\mathcal R} [\alpha B,\omega,  \sigma_*, \mu_r ])_{ik}
({\mathcal I} [\alpha B,\omega,  \sigma_*, \mu_r ])_{kj} -
( {\mathcal I} [\alpha B,\omega, \sigma_*, \mu_r ])_{ik}
(\tilde{\mathcal R} [\alpha B,\omega,  \sigma_* , \mu_r])_{kj}
\end{align}
where Einstein summation convention of the indices is implied.
The commutator measures how different the eigenspaces of $\tilde{\mathcal R} [\alpha B,\omega, \sigma_*, \mu_r ]$ and ${\mathcal I} [\alpha B,\omega, \sigma_*, \mu_r ]$ are. It  vanishes when the 
$\tilde{\mathcal R} [\alpha B,\omega, \sigma_*, \mu_r ]$ and ${\mathcal I} [\alpha B,\omega, \sigma_*, \mu_r ]$ are simultaneously diagonalisable (i.e. the
eigenvectors of $ {\mathbf Q}^{\tilde{\mathcal R} [\alpha B,\omega, \sigma_*, \mu_r ]}= {\mathbf Q}^{{\mathcal I} [\alpha B,\omega, \sigma_*, \mu_r ]}= {\mathbf Q}(B)$  are the same). In Lemma 8.11 of ~\cite{LedgerLionheart2019}, $|({\mathcal Z} [ \alpha B,\omega,  \sigma_*, \mu_r] )_{ij}|$  has been shown to grow at most linearly with $\omega$. 
In addition, the coefficients of ${\mathcal Z}$ transform as a rank 2 tensor and so the eigenvalues of ${\mathcal Z}[\alpha B]$ and ${\mathcal Z}[\alpha {\mathbf R} (B)]$ are the same.

It is easy to show that, since $\tilde{\mathcal R} [\alpha B,\omega, \sigma_*, \mu_r ]$ and ${\mathcal I} [\alpha B,\omega,  \sigma_*, \mu_r ]$
are symmetric,  ${\mathcal Z}[\alpha B,\omega, \sigma_*, \mu_r]$
has vanishing diagonal coefficients and is skew symmetric. Then, by arranging the coefficients of ${\mathcal Z}[\alpha B,\omega, \sigma_*, \mu_r]$ as a $ 3 \times 3$ matrix, 
we find that its eigenvalues are zero or purely imaginary
\begin{align}
\lambda_i ({\mathcal Z}) \in \left \{ 0, \pm \im \sqrt{({\mathcal Z})_{12}^2 + ({\mathcal Z})_{13}^2 + ({\mathcal Z})_{23}^2 }  \right \} , \nonumber
\end{align}
and, thus, $\sqrt{({\mathcal Z})_{12}^2 + ({\mathcal Z})_{13}^2 + ({\mathcal Z})_{23}} = \sqrt{I_2 ( {\mathcal Z}[ \alpha B,\omega,  \sigma_*, \mu_r]  )}$ is useful as an additional classifier 
for situations where the off-diagonal coefficients of the tensors are amongst its independent coefficients~\footnote{ Note that $I_1(   {\mathcal Z}[ \alpha B,\omega,  \sigma_*, \mu_r])= I_3(  {\mathcal Z}[ \alpha B,\omega_m,  \sigma_*, \mu_r])=0 $}.  For an object where the only independent coefficients
$( \tilde{\mathcal R} [\alpha B,\omega, \sigma_*, \mu_r ])_{ij}$ and $( {\mathcal I} [\alpha B,\omega, \sigma_*, \mu_r ])_{ij}$ are associated with  $i=j$  then $ \sqrt{({\mathcal Z})_{12}^2 + ({\mathcal Z})_{13}^2 + ({\mathcal Z})_{23}} $ vanishes. This invariant can easily be added to the list of features in (\ref{eqn:feateig}), (\ref{eqn:featinv}) or (\ref{eqn:featinv2}) as
\begin{equation}
{\rm x}_i = \sqrt{I_2 ( {\mathcal Z}[ \alpha B,\omega_m,  \sigma_*, \mu_r])}, \qquad i= 6M + m, \label{eqn:featz}
\end{equation}
for $m=1,\ldots, M$ bringing the total number of features to $F=7M$.

\section{Building a training data set for classification} \label{sect:trainset}

In this section, we provide a series of illustrative examples to demonstrate how the PODP approach~\cite{ben2020} to compute the MPT spectral signatures described in Section~\ref{sect:rommethod} can be combined with an appropriate choice of eigenvalues or tensor invariants in Section~\ref{sect:classinv} and sampling at $M$ frequencies to form a training dataset for object classification. Forming this dictionary involves considering different physical objects $B_\alpha^{(p)}$, $p=1,\ldots,P$, and  provides the pairs (${\mathbf x}_p  \in {\mathbb R}^{F}$, ${\mathbf t}_p \in {\mathbb R}^K $) for each object where the entries ${\mathbf x}_p$ are of the form of   (\ref{eqn:feateig}) (or (\ref{eqn:featinv}) or (\ref{eqn:featinv2})) and (\ref{eqn:featz}) and the entries of ${\mathbf t}_p$ are all $0$ except for one entry with value $1$, corresponding to the class of the object $B_\alpha^{(p)}$.
For the numerical examples considered, we use the author's \href{https://github.com/BAWilson94/MPT-Calculator}{\texttt{MPT-Calculator}}  program (commit number b861dfb) to generate the MPT spectral signatures, which, in turn, uses the \href{https://ngsolve.org/}{\texttt{NGSolve}} package (version 6.2.2004) developed by the group led by Sch\"oberl for the underlying finite element computations~\cite{SchoberlZaglmayr2005,NGSolve,zaglmayrphd,netgendet}. We note that this data set can be easily enriched by using the scaling results obtained in Lemma 5.1 and Lemma 5.2 of~\cite{ben2020}, which enable the MPT spectral signatures of objects with the same underlying geometry, but with different sizes and different conductivities to be obtained at no cost.

We separate the results into four subsections the first two relating to non-threat items (house keys and British coins) and the second two to threat items (TT-33 semi automatic pistol and knives).

\subsection{Non-threat items:  Keys for pin-tumbler locks}

\begin{figure}[htbp]
\begin{center}
$\begin{array}{cc}
    \includegraphics[width=0.4\textwidth]{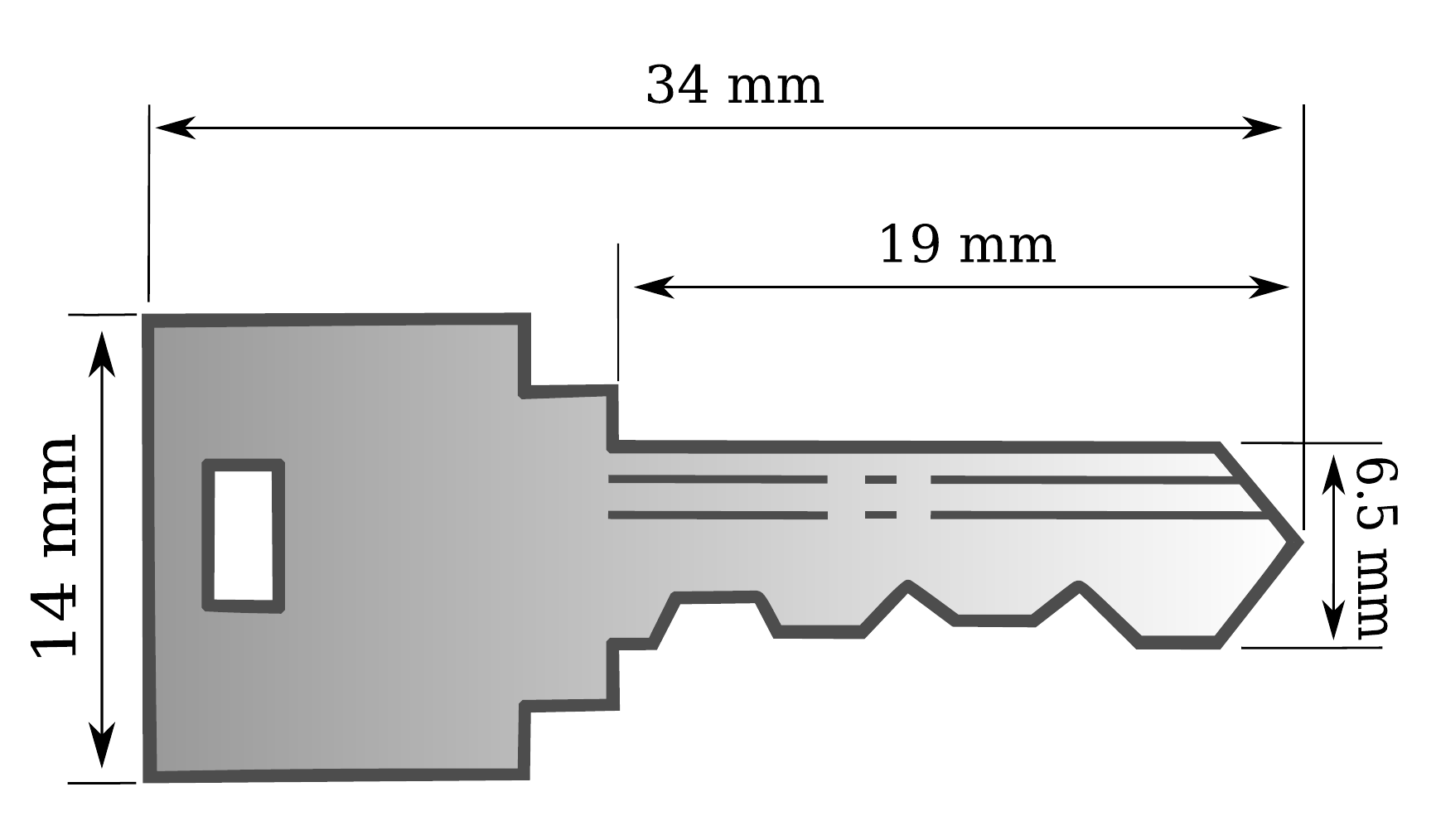}  &
    \includegraphics[width=0.4\textwidth]{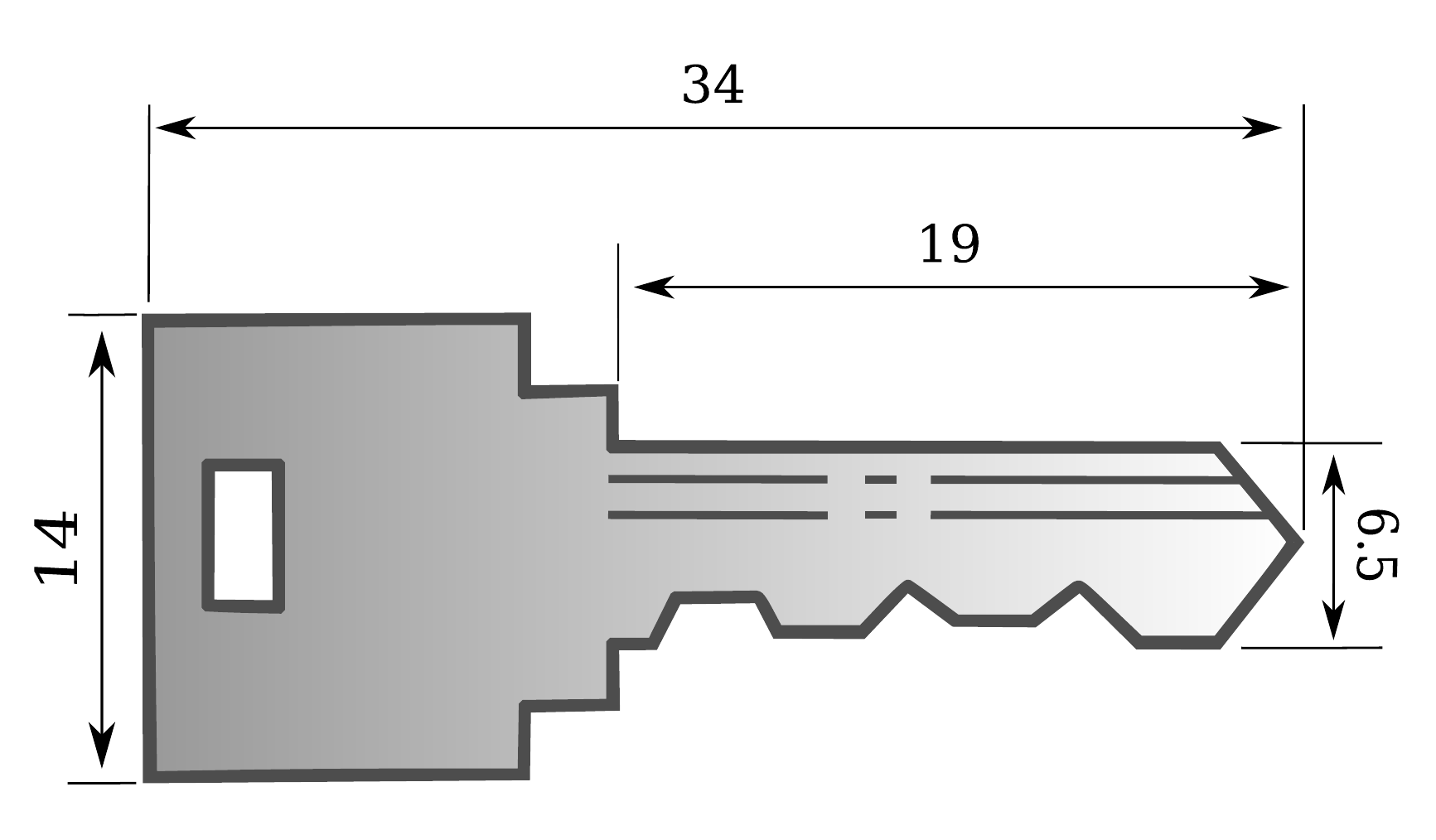}  \\
 \text{(a)} \    B_\alpha & \text{(b)} \  B
\end{array}$
\end{center}
  \caption{Typical 
  key for a pin-tumbler lock: (a) Physical object $B_\alpha$ and  (b) non-dimensional object $B$.}
        \label{fig:diagramkey}
\end{figure}

Common materials for keys for pin-tumbler locks include brass, plated brass, nickel silver, and steel. Amongst these, brass is often chosen due to its low cost, ease of cutting and its self lubricating characteristics, which avoids the key getting stuck in a lock. Therefore, for this study, we restrict ourselves to keys made of brass and we select the material parameters to be $\mu_r=\mu_*/\mu_0 = 1$ and $\sigma_* = 1.5\times 10^{7} \text{ S/m}$ that correspond to its conductivity value being at $26 \%$ of the value for copper in the International Annealed Copper Standard (IACS)~\cite{conductivity}.
An illustration of the cross-section of a key for a pin-tumbler lock is included in Figure~\ref{fig:diagramkey} where the dimensions are similar to a house key and we indicate the physical key $B_\alpha$ as well as the non-dimensional object $B$ used in the computations.

\begin{figure}[!h]
\centering
    \subfigure[Key 1]{\includegraphics[scale=0.12]{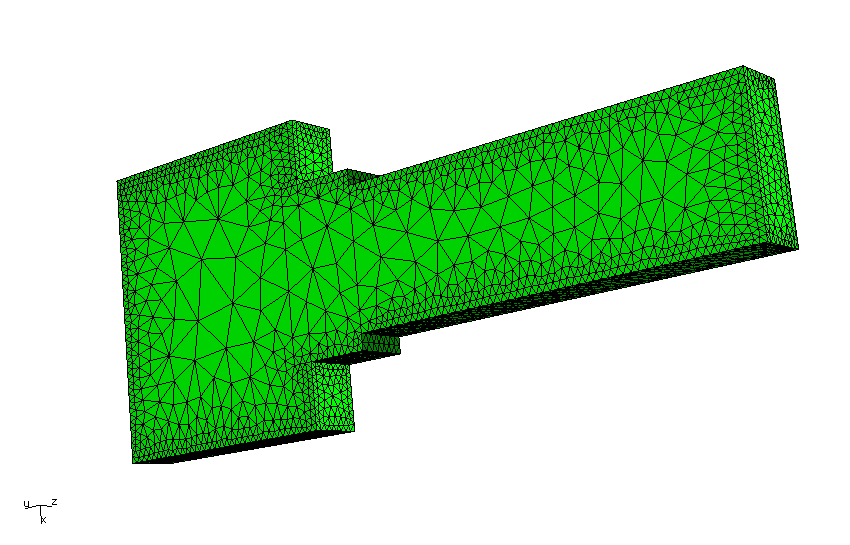}
    \label{fig:set1:fineMesh:case1}
    }
    \subfigure[Key 2]{\includegraphics[scale=0.12]{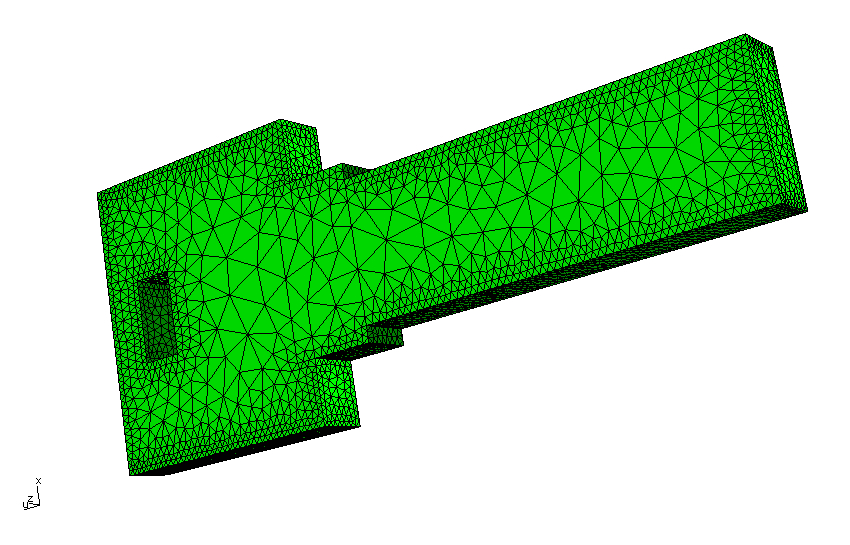}
    \label{fig:set1:fineMesh:case2}
    }
    \subfigure[Key 3]{\includegraphics[scale=0.12]{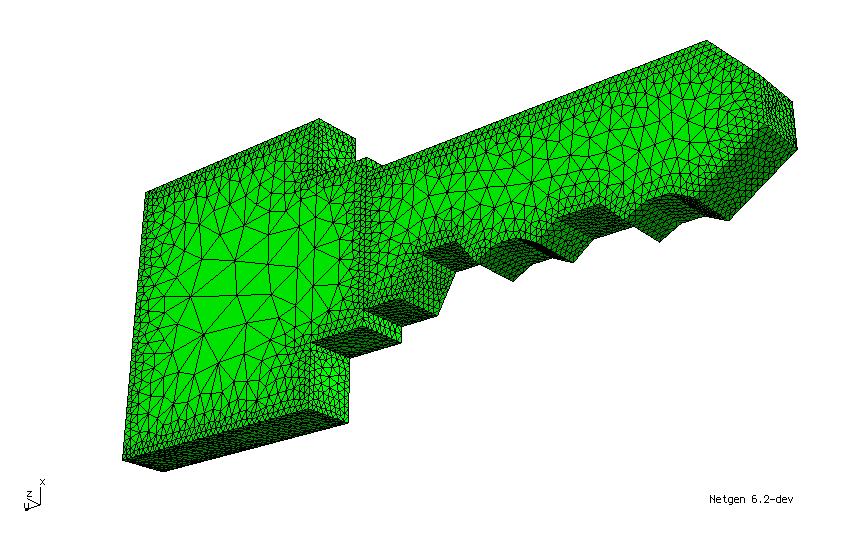}
    \label{fig:set1:fineMesh:case3}
    }
    \subfigure[Key 4]{\includegraphics[scale=0.12]{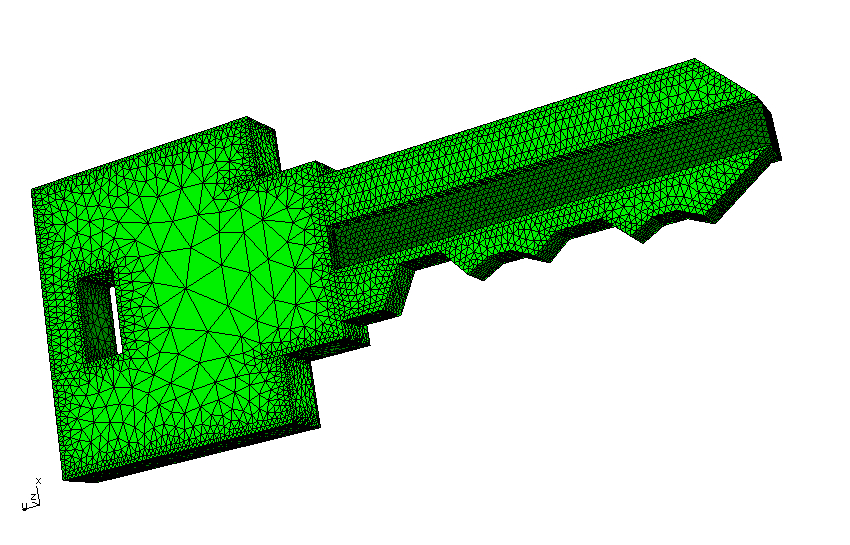}
    \label{fig:set1:fineMesh:case4}
    }
    \subfigure[Key 9]{\includegraphics[scale=0.12]{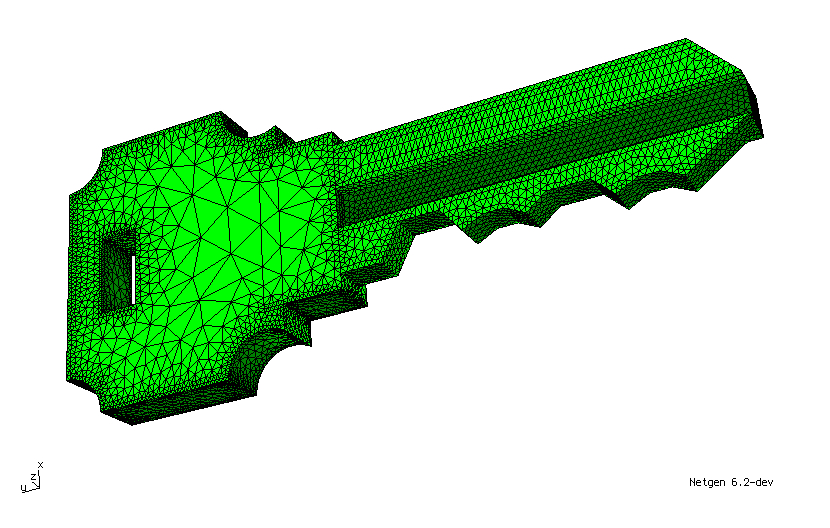}
    \label{fig:set1:fineMesh:case9}
    }
  \caption{Set 1 of brass house keys : Surface distribution of elements  of the keys in the meshes cases 1-4 and 9.}
        \label{fig:set1:fineMesh:objects}
\end{figure}

\begin{figure}[!h]
\centering
    \subfigure[Key 5]{\includegraphics[scale=0.12]{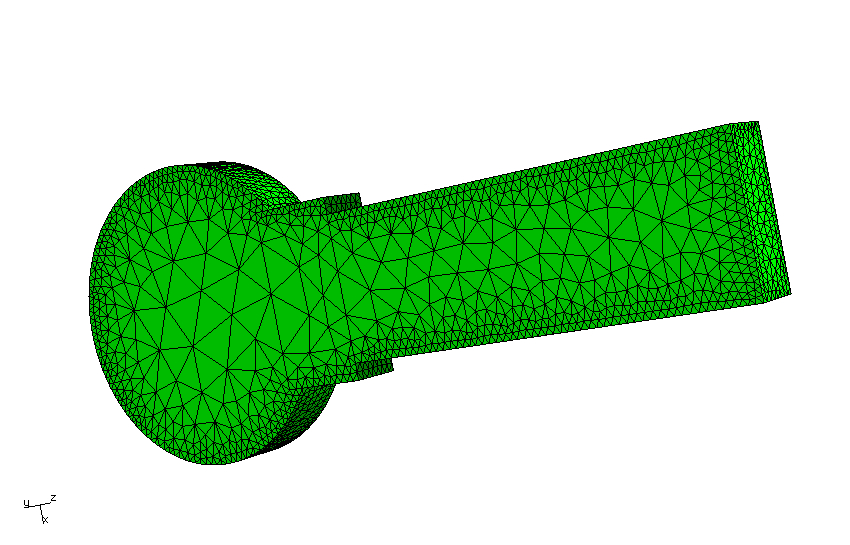}
    \label{fig:set2:fineMesh:case5}
    }
    \subfigure[Key 6]{\includegraphics[scale=0.12]{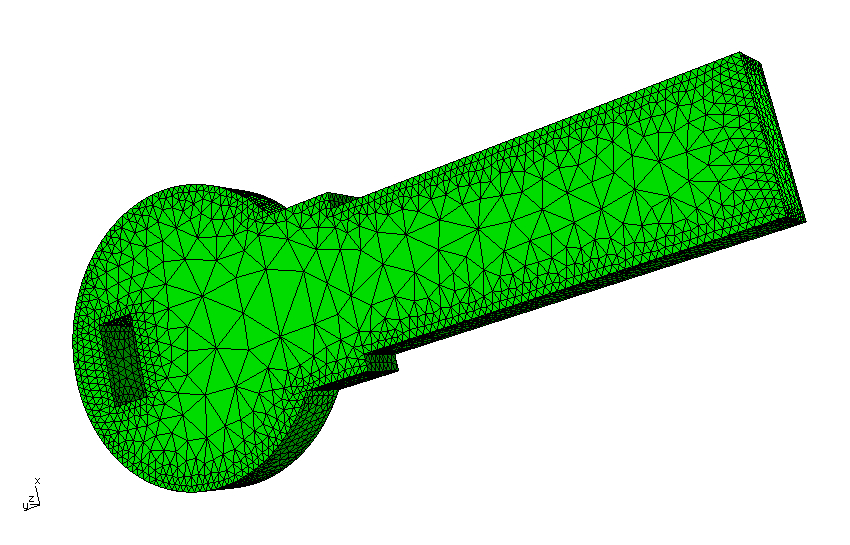}
    \label{fig:set2:fineMesh:case6}
    }
    \subfigure[Key 7]{\includegraphics[scale=0.12]{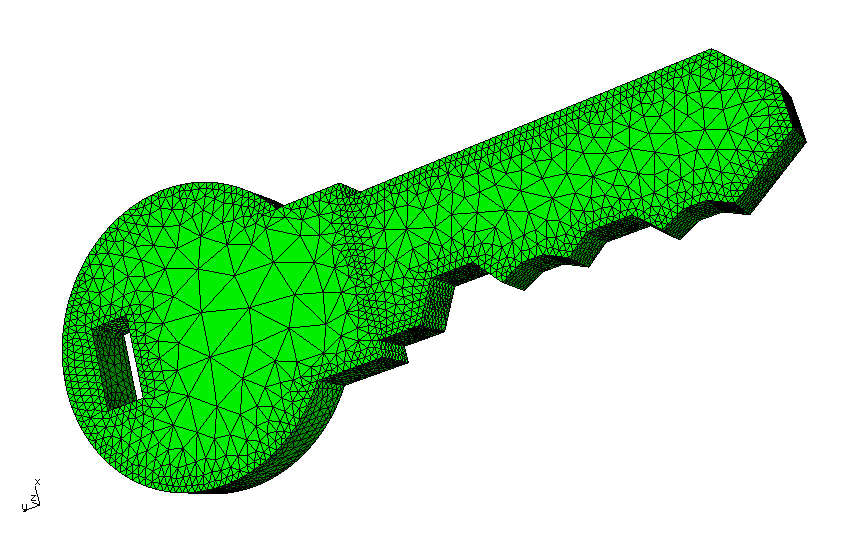}
    \label{fig:set2:fineMesh:case7}
    }
    \subfigure[Key 8]{\includegraphics[scale=0.12]{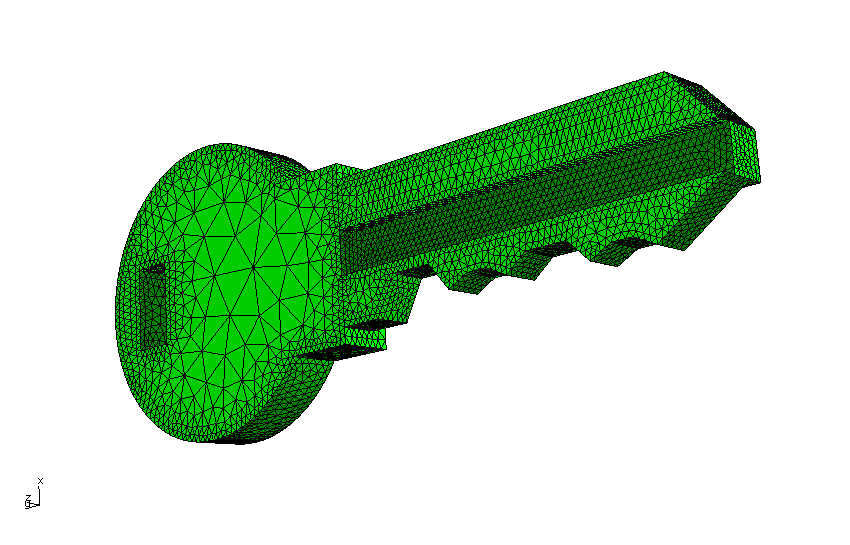}
    \label{fig:set2:fineMesh:case8}
    }
  \caption{Set 2 of brass house keys : Surface distribution of elements  of the keys in the meshes cases 5-8 (see Figure~\ref{fig:set1:fineMesh:objects} for key 9 also in this set).}
        \label{fig:set2:fineMesh:objects}
\end{figure}

To understand how the effects of small changes in the key geometry affect the MPT frequency spectral signature for frequencies in the range $10^2 \le \omega \le 10^8 \text{rad/s}$,  a sequence of $9$ different key geometries were produced. In each case, we set $\alpha =0.001$m and specify the dimensions of the different cases for $B$ to be non-dimensional. For example, for key 1, $B$ is of length  $34$, has a  width of 6.5 (min)-14 (max)  and a thickness of $2.5$ whereas $B$ for key 4 has a maximum thickness of 2.5 and a deep blade cut of 0.75 and notches of maximum size 1.75. The meshes of the two sets of keys are shown in Figures~\ref{fig:set1:fineMesh:objects} and~\ref{fig:set2:fineMesh:objects}. These meshes have local refinement towards the edges of the keys and each case the mesh extends out to a truncation boundary in the form of the $[-1000,1000]^3$  rectangular box and comprise of between $51\, 726$ and $108\, 523$ unstructured tetrahedra.  Importantly, note that the connectedness of the different key types. Of the different keys, keys 2, 4,  6, 7, 8 and 9 are multiply connected and have Betti numbers $\beta_0(B)=\beta_1(B)=1$ and $\beta_2(B)$ the remaining keys are simply connected with $\beta_0(B)=1$ and $\beta_1(B)=\beta_2(B)$.

\subsubsection{Set 1 of brass house keys}
\begin{figure}[!h]
\centering
\hspace{-1.cm}
    \subfigure[ $\lambda_{1} ( \tilde{\mathcal{R}} ) =\lambda_{2} ( \tilde{\mathcal{R}} )  $ ]{\includegraphics[scale=0.5]{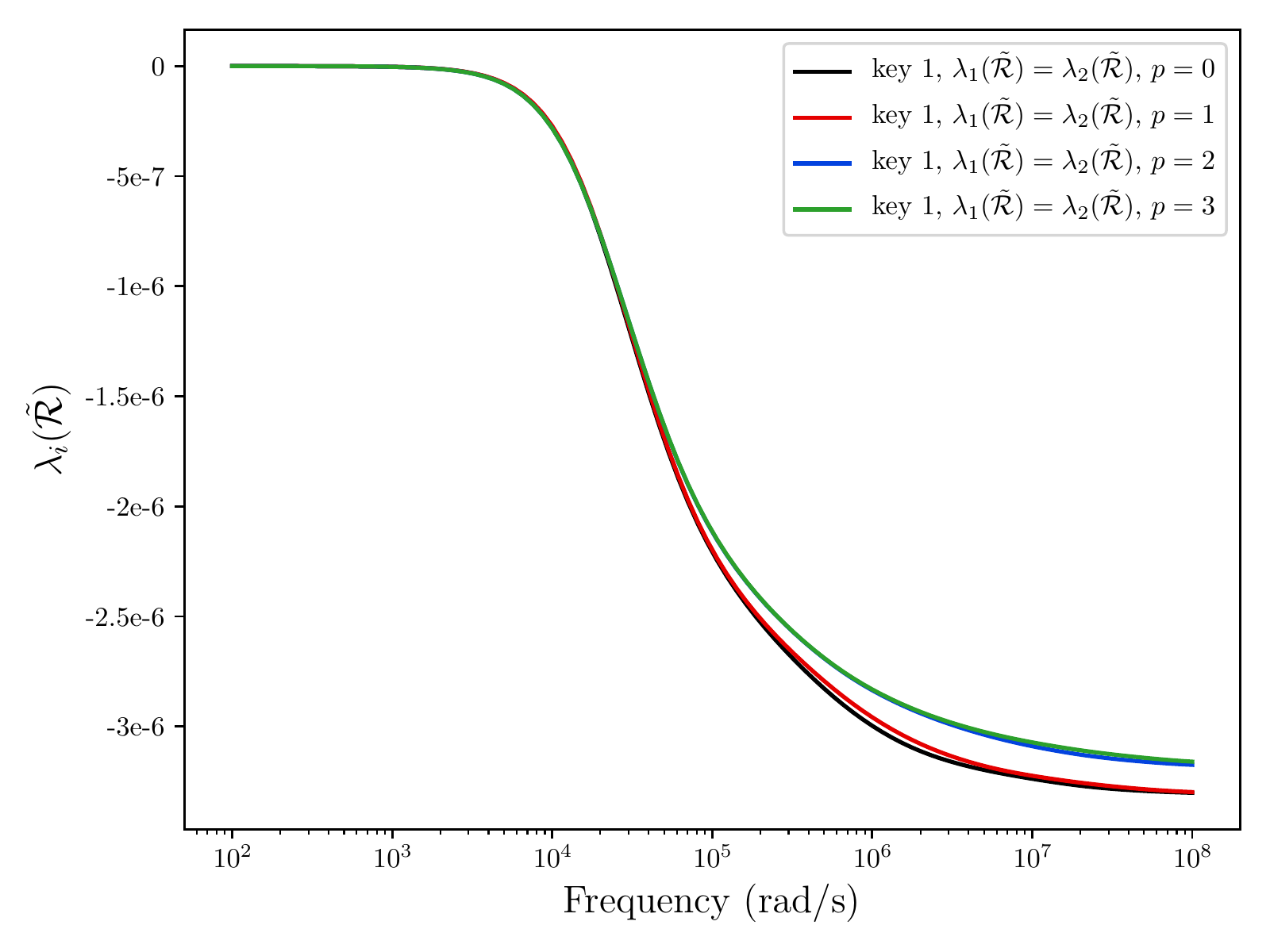} 
    }
    \subfigure[ $ \lambda_{3} ( \tilde{\mathcal{R}} ) $ ]{\includegraphics[scale=0.5]{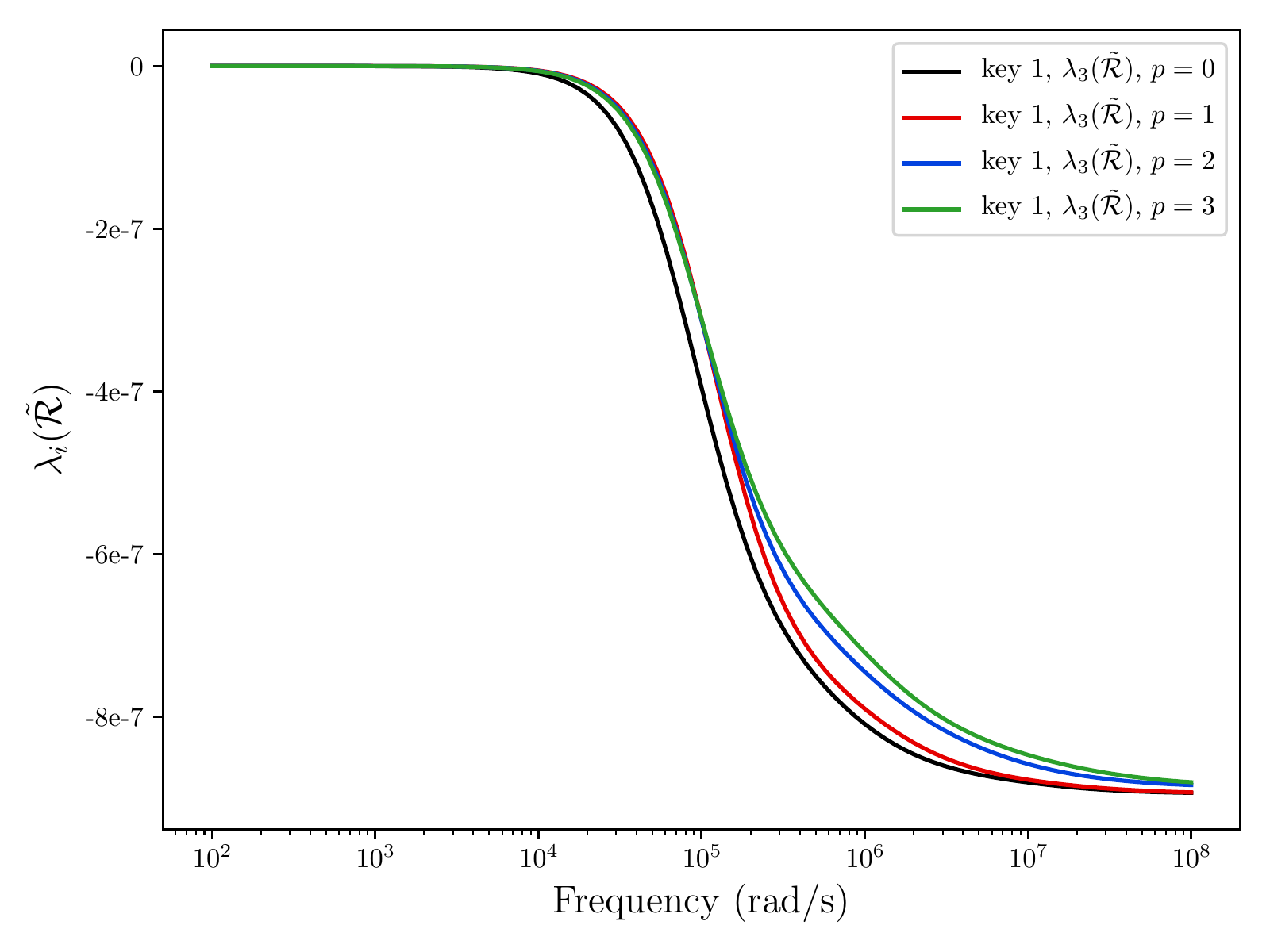} 
    }

  \caption{Key 1 from set 1 of brass house keys: $p$-refinement study using $p=0,1,2,3$ order elements for (a) $\lambda_{1} ( \tilde{\mathcal{R}} ) = \lambda_{2} ( \tilde{\mathcal{R}} )  $, (b) $ \lambda_{3} ( \tilde{\mathcal{R}} ) $.}
        \label{fig:set1:key1:fineMesh:pConvergence:Re}
\end{figure}

\begin{figure}[!h]
\centering
\hspace{-1.cm}
     \subfigure[$\lambda_{1} ( \mathcal{I} ) =\lambda_{2} ( \mathcal{I} ) $]{\includegraphics[scale=0.5]{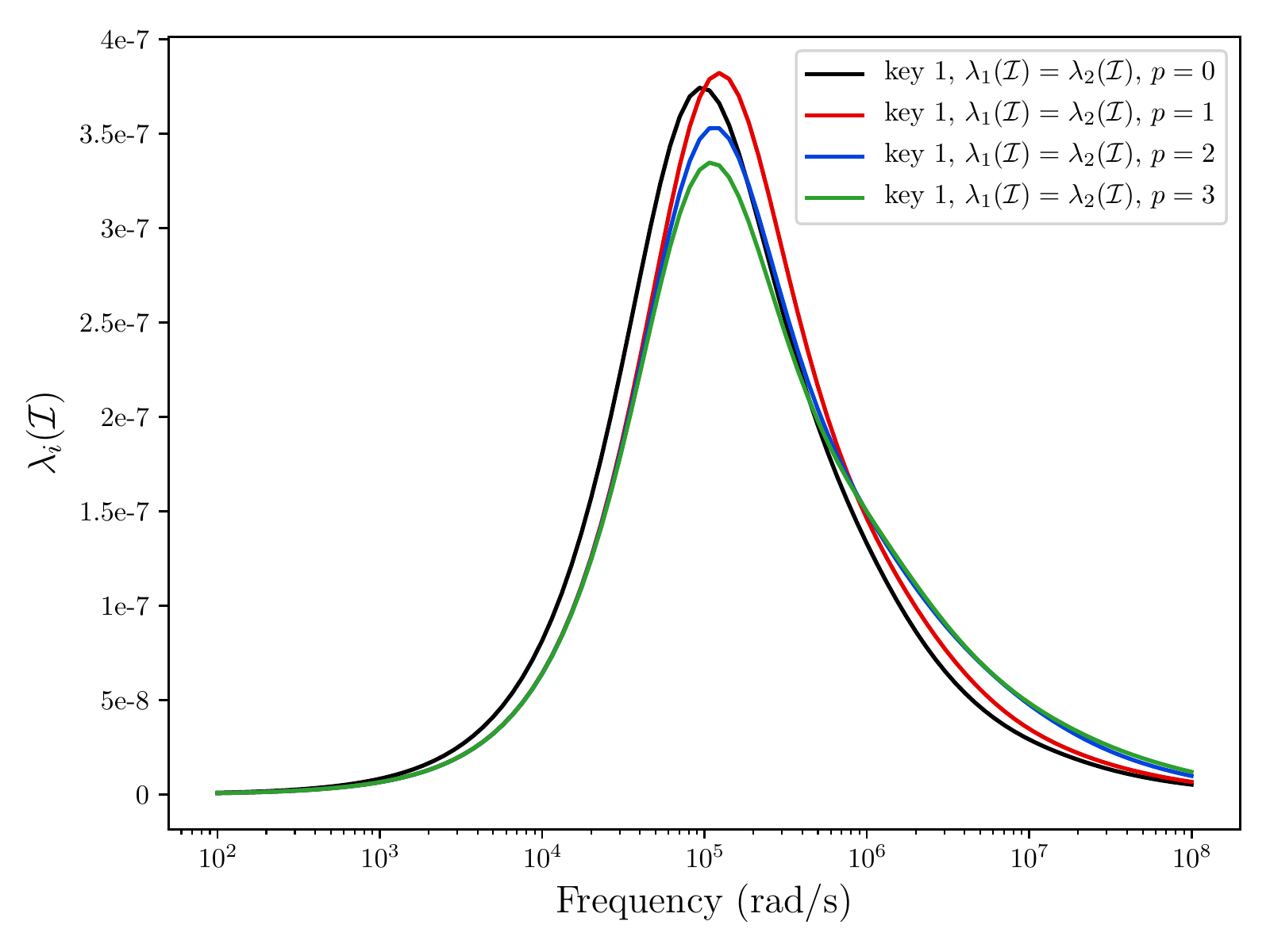} 
         }
    \subfigure[$\lambda_{3} ( \mathcal{I} ) $]{\includegraphics[scale=0.5]{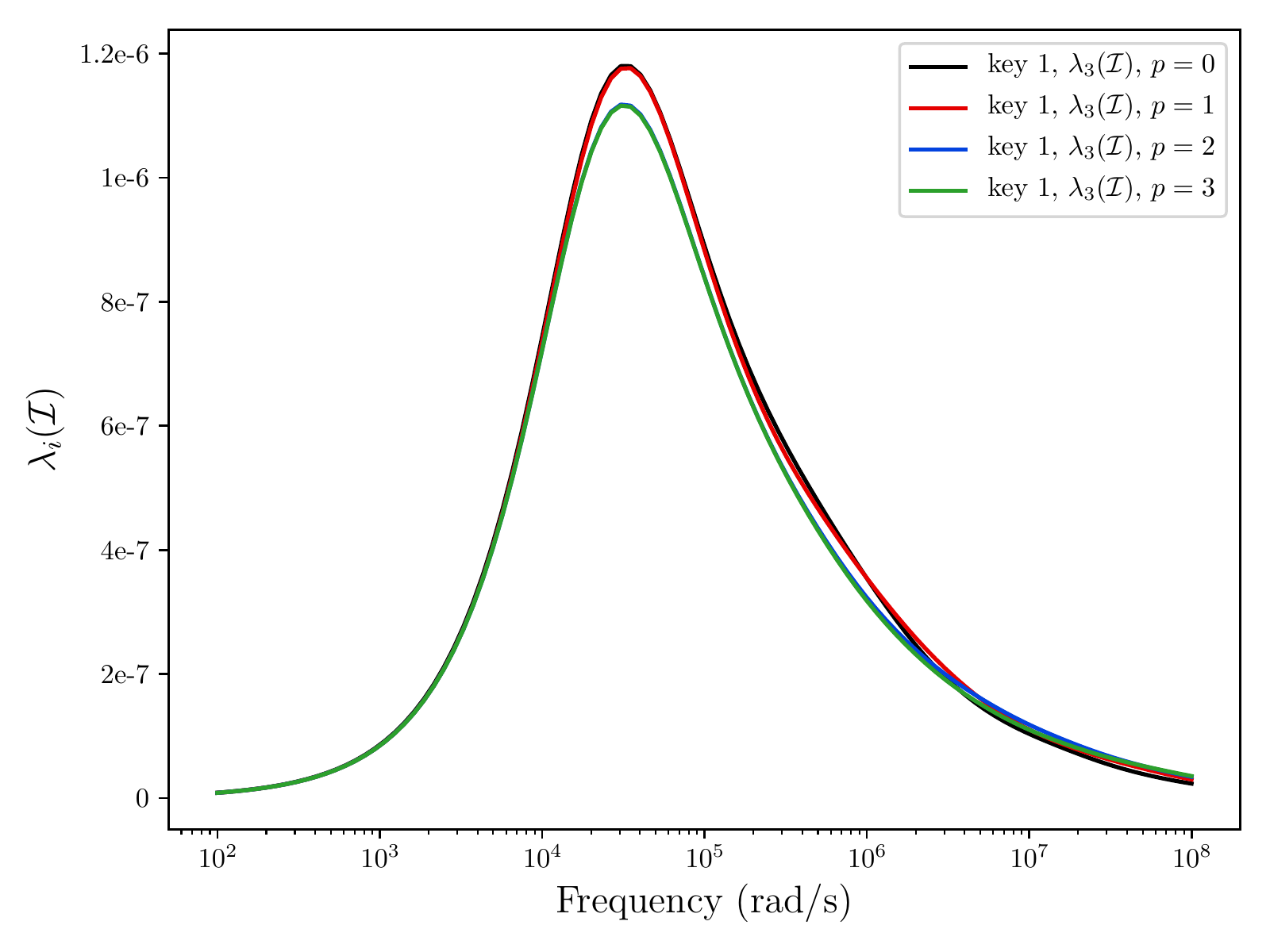} 
    }
  \caption{Key 1 from set 1 of brass house keys: $p$-refinement study using $p=0,1,2,3$  order elements for (a) $\lambda_{1} ( \mathcal{I} ) = \lambda_{2} ( \mathcal{I}  ) $ and (b) $\lambda_{3} ( \mathcal{I}  ) $.}
        \label{fig:set1:key1:fineMesh:pConvergence:Im}
\end{figure}

Restricting consideration to the set 1 of house keys, we begin by illustrating that $p$-refinement of the mesh of $56\, 241$ unstructured tetrahedra for key 1 using $p=0,1,2,3$ order elements leads to a rapid convergence of the MPT spectral signature presented in the form of the eigenvalues of $\tilde{\mathcal R}[\alpha B, \omega,\sigma_*,\mu_r]$~\footnote{Note that the coefficients of ${\mathcal N}^0$ vanish in this case as $\mu_r=1$, but we keep to the notation of  $\tilde{\mathcal R}={\mathcal N}^0+ {\mathcal R}$ for ease of comparison with later results} and ${\mathcal I}[\alpha B, \omega,\sigma_*,\mu_r]$, namely $\lambda_{i} ( \tilde{\mathcal{R}} ) $ and   $\lambda_{i}\left( \mathcal{I} \right) $, $i=1,2,3$, as illustrated in Figures~\ref{fig:set1:key1:fineMesh:pConvergence:Re} and \ref{fig:set1:key1:fineMesh:pConvergence:Im}. Note that, due to the reflectional symmetries for key 1, there are only two independent coefficients each in $\tilde{\mathcal{R}} $ and ${\mathcal I}$ and hence only two distinct eigenvalues in each case.  Also, we have chosen to order the eigenvalues so that those corresponding to $i=1$ and $i=2$ are the same in these plots.

\begin{figure}
\begin{center}
$\begin{array}{cc}
\includegraphics[scale=0.5]{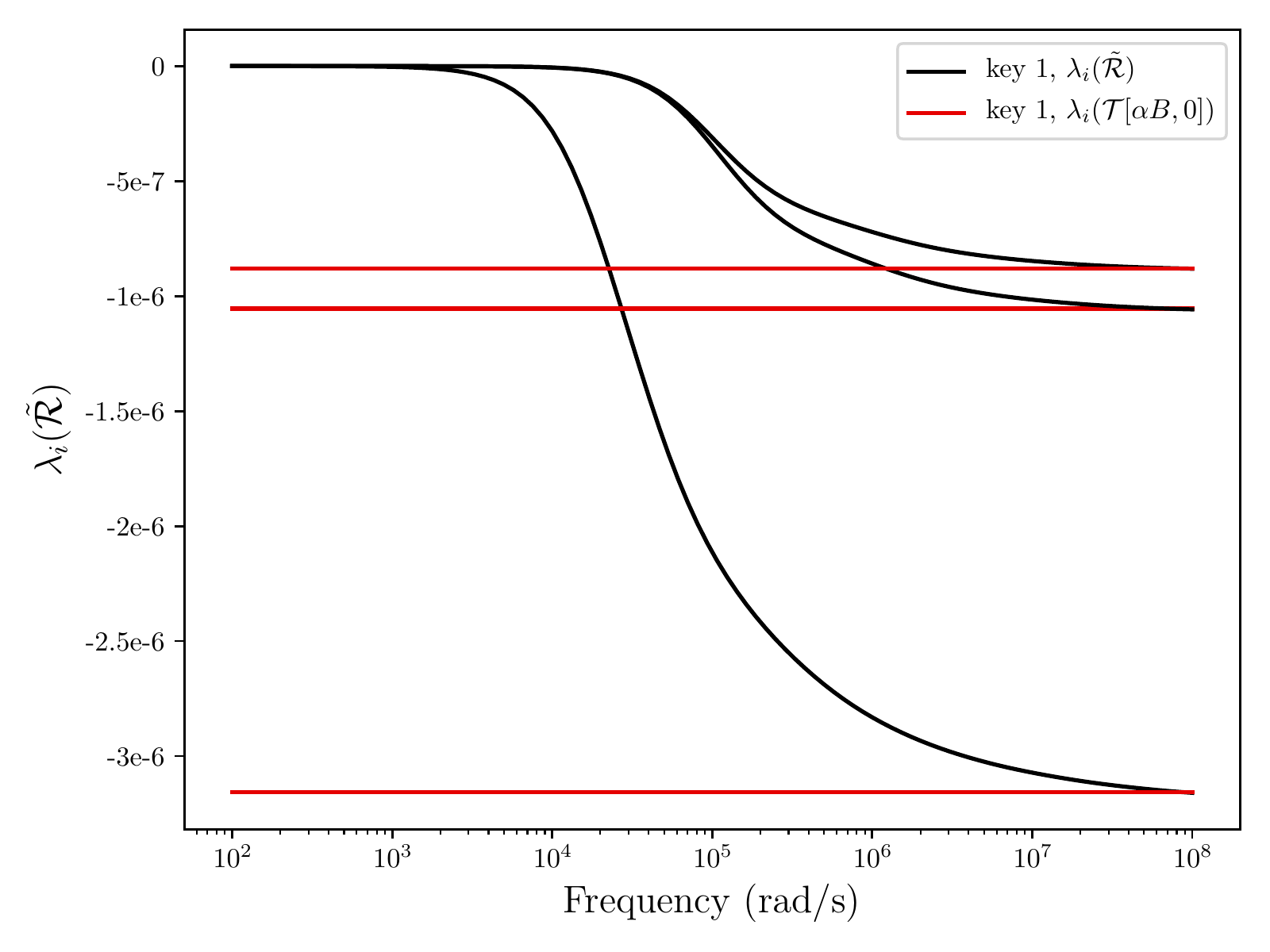} &
\includegraphics[scale=0.5]{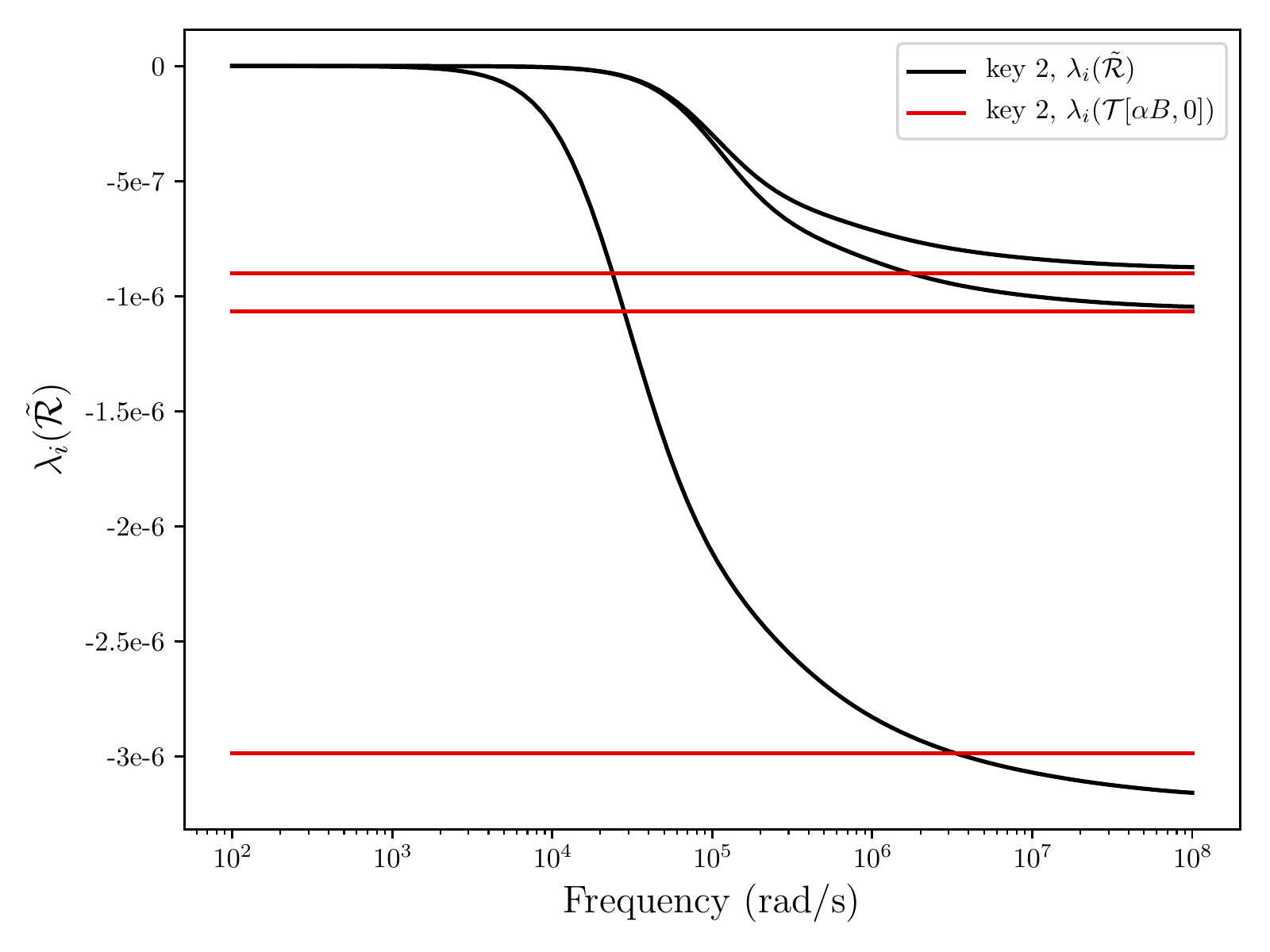}  \\
\text{(a) Key 1} & 
\text{(a) Key 2} 
\end{array}$
\end{center}
\caption{Set 1 of brass house keys: Comparison of $\lambda_i ( \tilde{\mathcal R})$ and $\lambda_i({\mathcal T}[\alpha B,0])$ for (a) Key 1 and (b) Key 2} \label{fig:bettinumber}
\end{figure}

The role played by a key's topology and its equivalent ellipsoid at a fixed frequency is now considered.  Previously, in (\ref{eqn:highfreq}), the equivalence between ${\mathcal M}^\infty [ \alpha B]$ and ${\mathcal T}[ \alpha B, 0]$ for the situation where $\beta_1(B)=0$ was highlighted. In Figure~\ref{fig:bettinumber}, we compare $\lambda_i ( \tilde{\mathcal R})$ and $\lambda_i ({\mathcal T}[\alpha B,0])$ for key 1 and key 2, the former having $\beta_1(B)=0$ and the latter having $\beta_1(B)=1$. As expected, since $\lim_{\omega \to \infty} ({\mathcal I})_{ij} =0$, we see good agreement in the limiting case as $\omega \to \infty$ (up to the limit of the eddy current model) between $\lambda_i (\tilde{\mathcal R})$ and $\lambda_i({\mathcal T}[\alpha B,0])$ 
for key 1, but not for key 2. Thus, highlighting the important role that an object's topology plays.

\begin{figure}[!h]
\centering
\hspace{-1.cm}
\subfigure[$( \tilde{\mathcal{R}} )_{ij} $]{\includegraphics[scale=0.5]{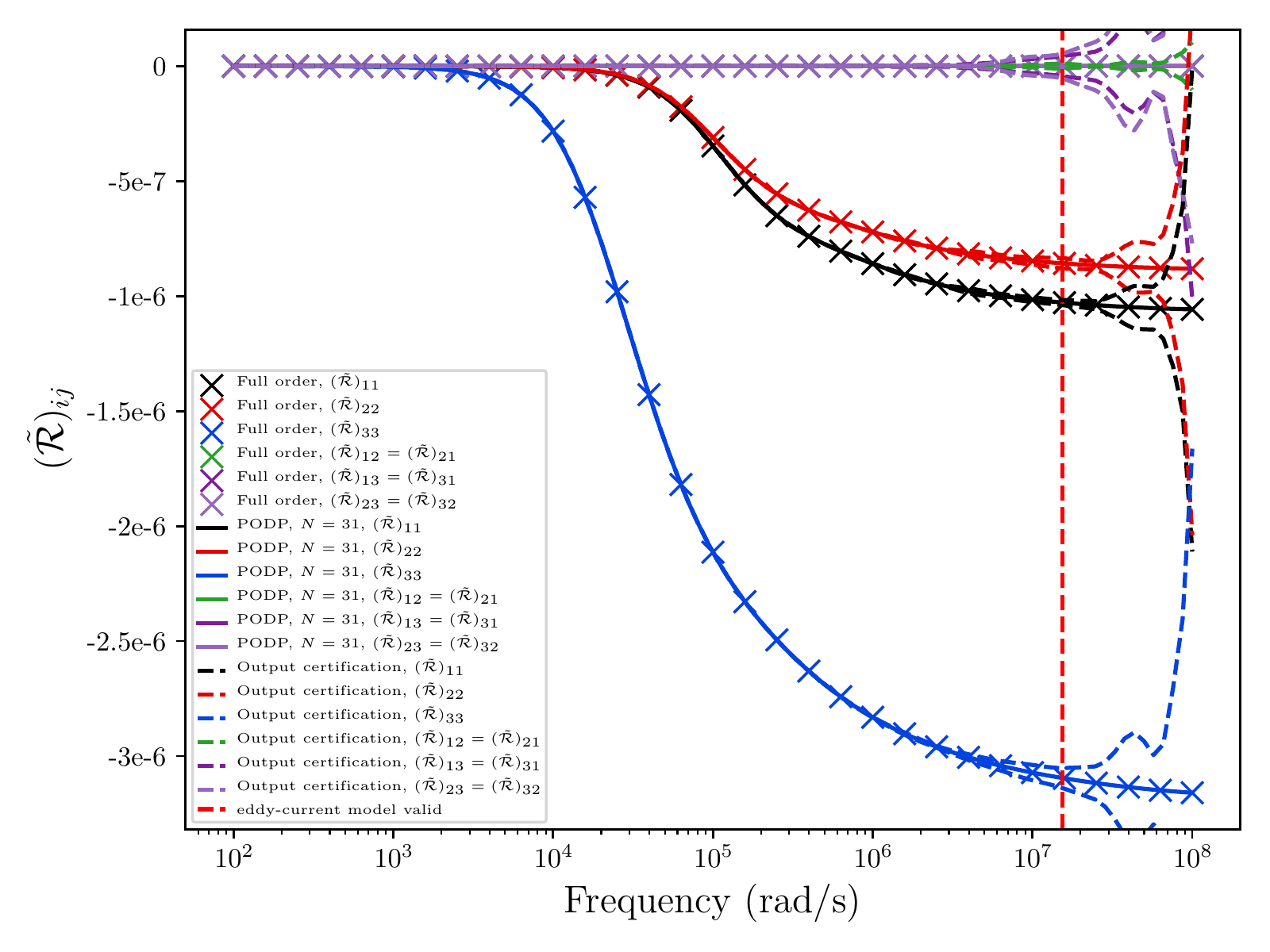}
    }
    \subfigure[$( \mathcal{I} )_{ij} $]{\includegraphics[scale=0.5]{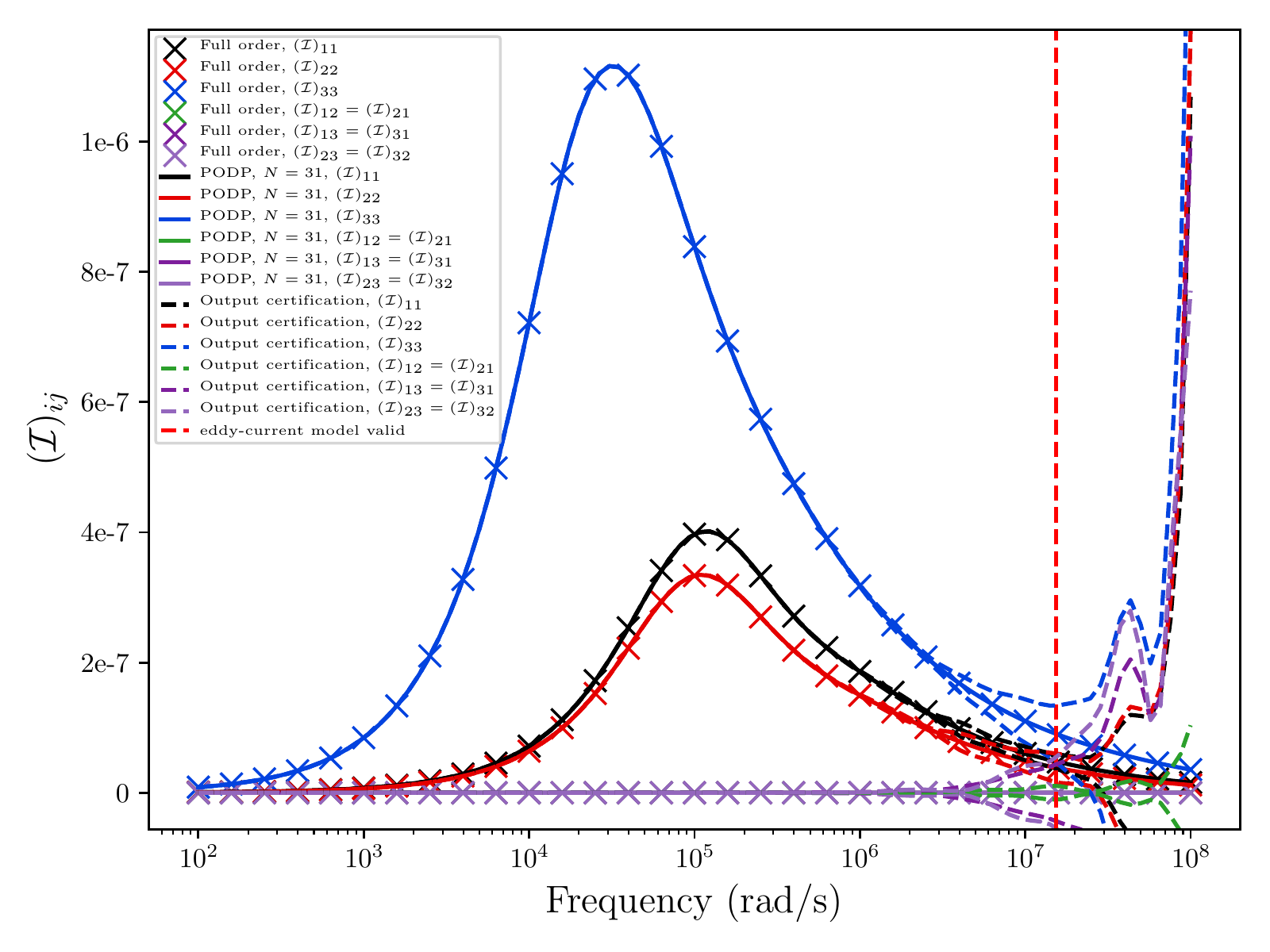} 
    }
  \caption{Key 1 from set 1 of brass house keys: PODP prediction of the spectral signature  showing also the frequencies used for the representative full order solution snapshots and limiting frequency for  (a) $ ( \tilde{\mathcal{R}} )_{ij} $ and (b) $ ( \mathcal{I} )_{ij} $.  }
        \label{fig:set1:key1:certificates}
\end{figure}

In a similar manner to the results shown in Figures~\ref{fig:set1:key1:fineMesh:pConvergence:Re} and \ref{fig:set1:key1:fineMesh:pConvergence:Im}, by performing $p$-refinement on the meshes for the other keys, and considering snapshot frequencies, the MPT coefficients were found converge using $p=3$ elements. However, to accelerate the computation of the MPT spectral signature for the keys, the PODP approach described in~\cite{ben2020}, and outlined in Section~\ref{sect:rommethod}, was followed. This involves computing solutions at  $N$ representative full order model solutions at logarithmically spaced frequencies and  then extracting a basis using a tolerance of $TOL=10^{-8}$ and solving reduced sized problems to approximate ${\bm \theta}_i^{(1)}(\omega)$ at other frequencies and, henceforth, predict the MPT coefficients at other frequencies. We illustrate the process in Figure~\ref{fig:set1:key1:certificates} for key 1 using $N=31$. The a-posteriori error estimates $ (\Delta [\omega] )_{ij}$ that are obtained at low-computational cost at run-time during the online stage of  the reduced order model,  are used to certify the reduced order model solutions that have been obtained,  are also shown in this figure. These illustrate that, in this case, the reduced order model predictions are reliable with respect to the full order model prediction of the MPT. Note that the PODP solutions are also very acceptable using $N=13$ representative full order model solution snapshots,
 but we have used $N=31$ to ensure small $ (\Delta [\omega] )_{ij}$  is small at all but the highest frequencies.
  Still further, the frequency $\omega_{limit}$, obtained using the method described in Section~\ref{sect:irregtetexp},  at which the eddy current approximation for this object is predicted to break down is shown. Tighter certificates bounds could be obtained by increasing $N$, however, this was not deemed to be necessary as the bounds, which provide confidence that the PODP predictions are accurate, are already tight for $\omega \le \omega_{limit}$.

\begin{figure}[!h]
\centering
\hspace{-1.cm}
$\begin{array}{cc}
\includegraphics[scale=0.5]{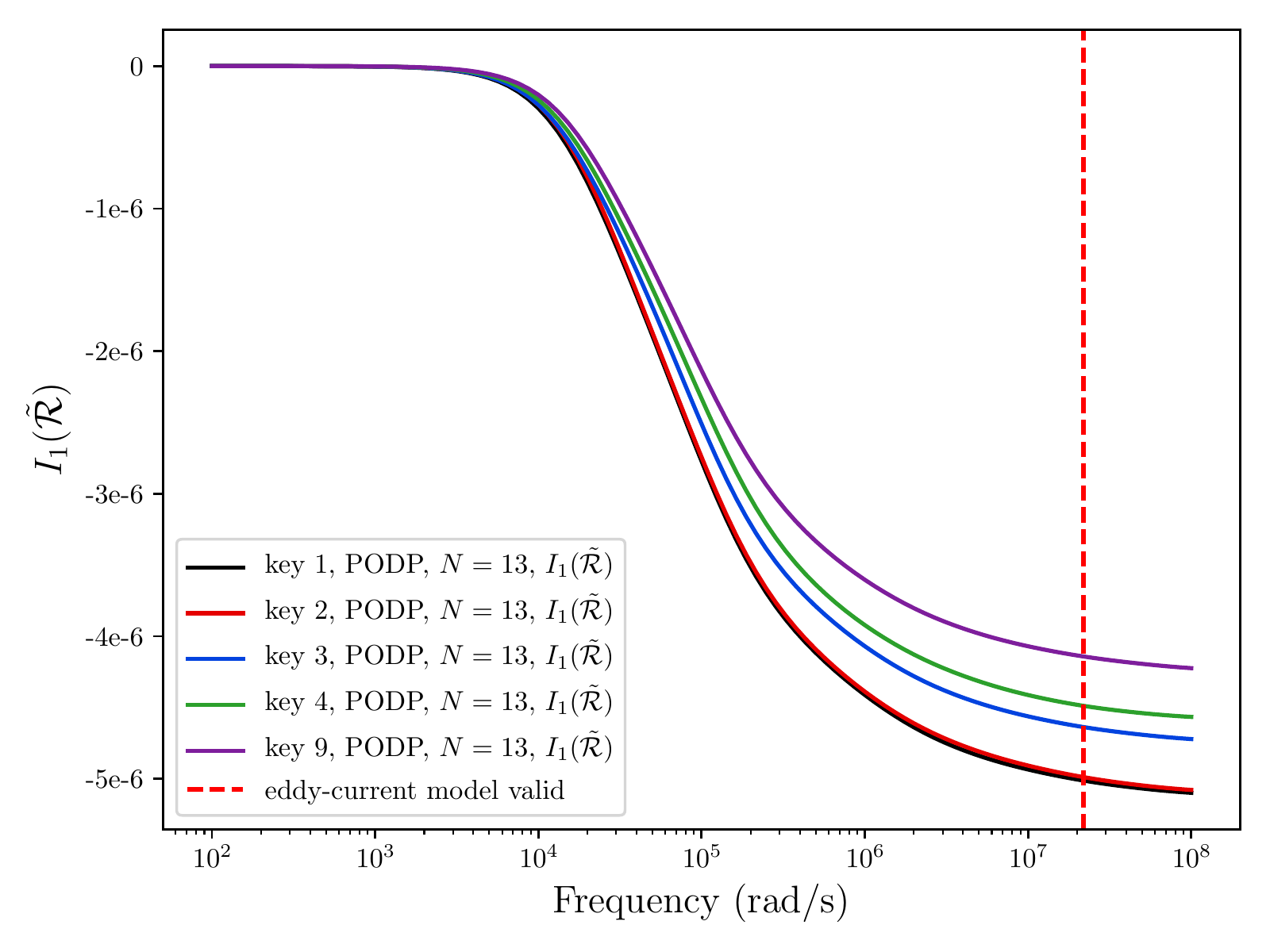}  &
 \includegraphics[scale=0.5]{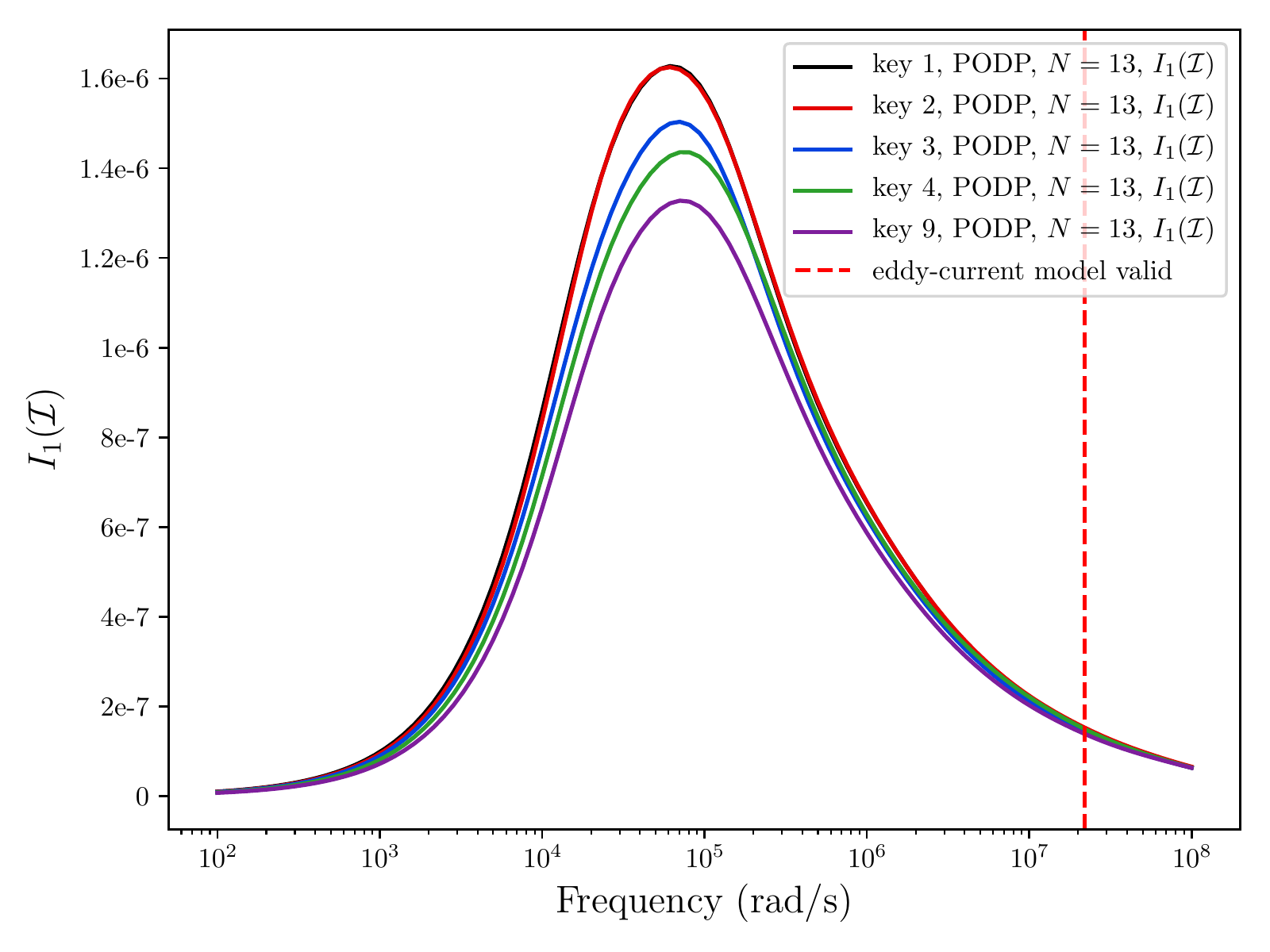}  \\
\text{(a) } I_{1} ( \tilde{\mathcal{R}} ) & 
\text{(b) } I_{1} ( \mathcal{I} )  \\
\includegraphics[scale=0.5]{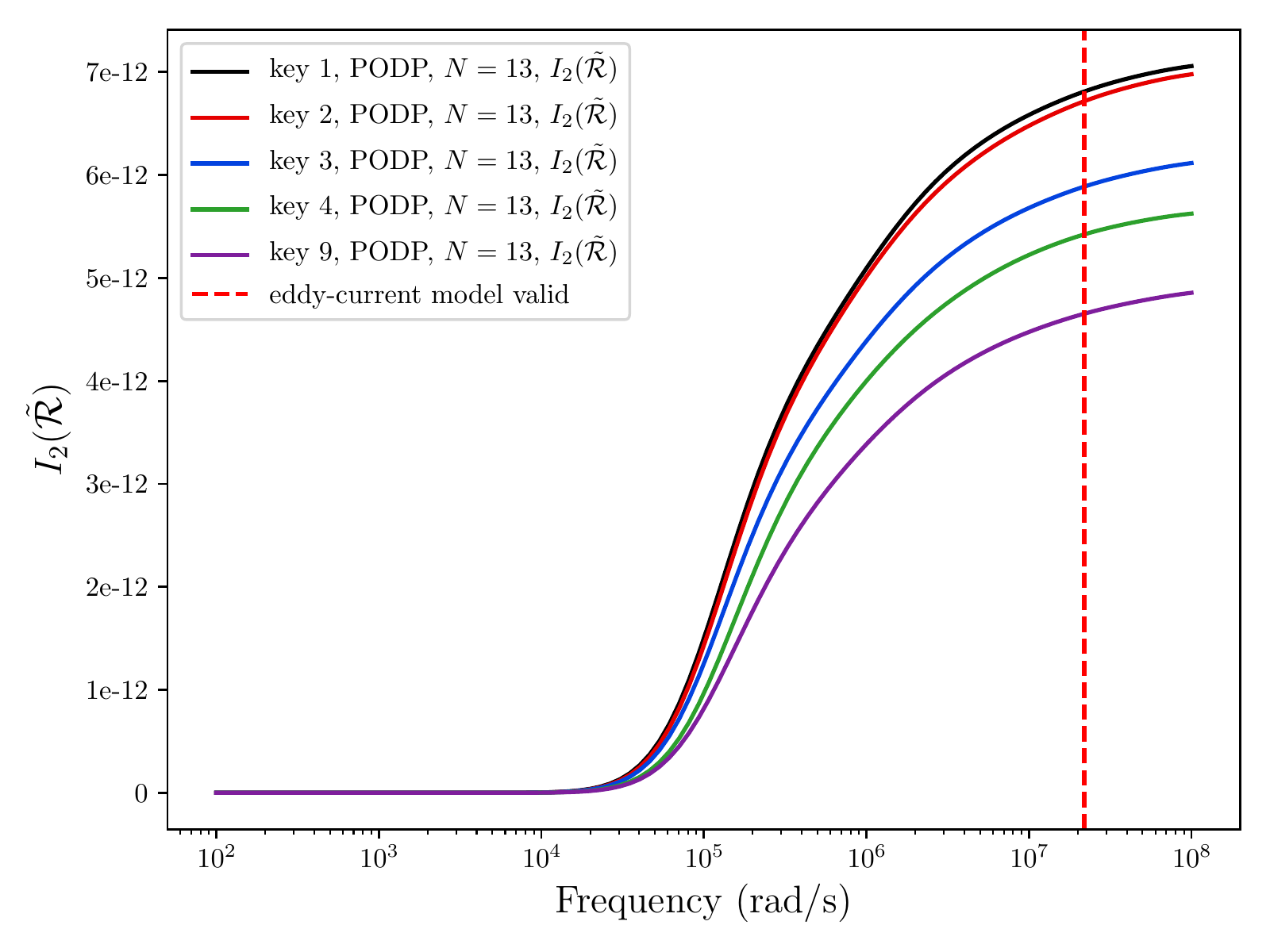}  &
 \includegraphics[scale=0.5]{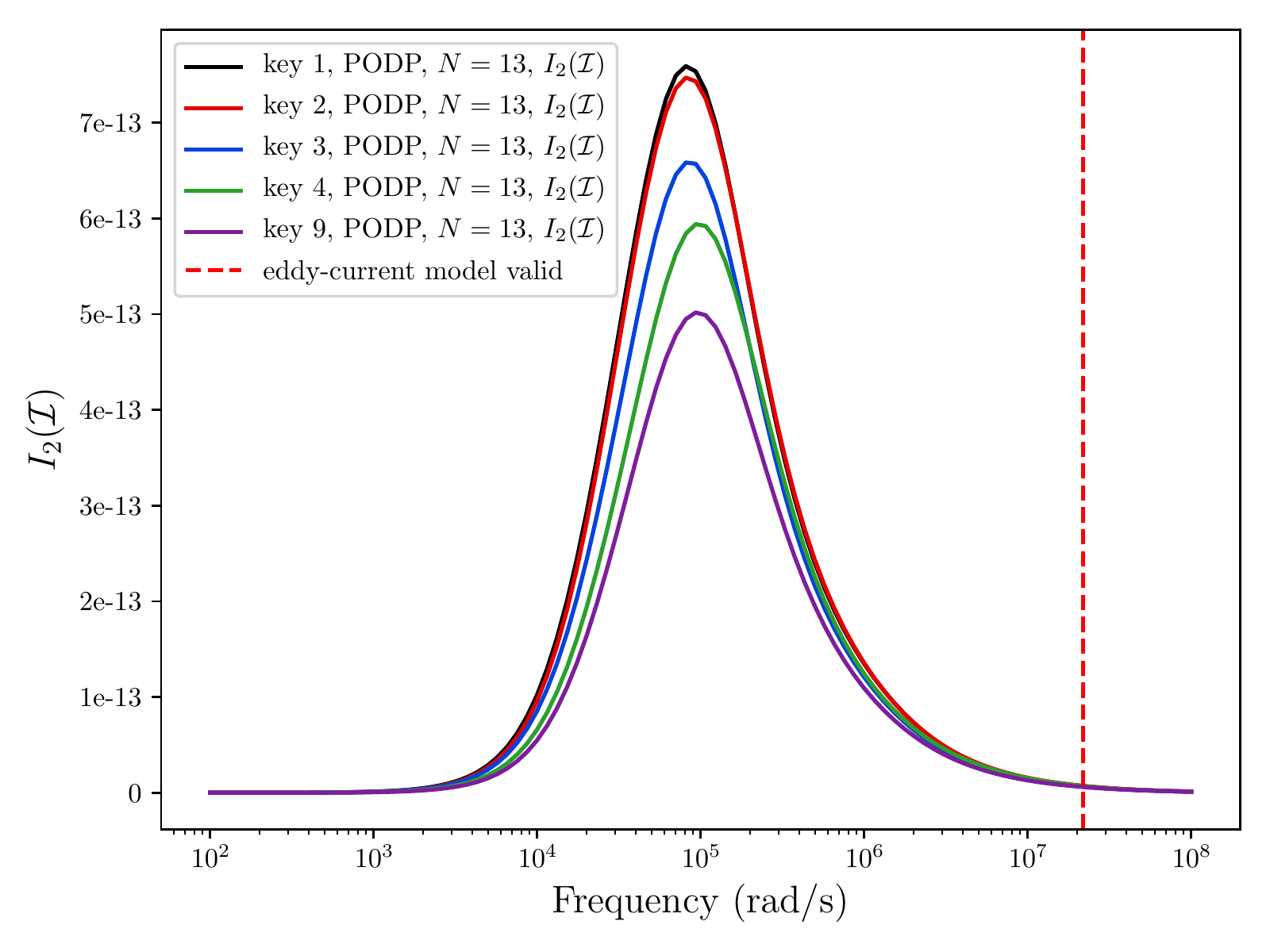}  \\
\text{(c) } I_{2} ( \tilde{\mathcal{R}} ) & 
\text{(d) } I_{2 } ( \mathcal{I} )  \\
\includegraphics[scale=0.5]{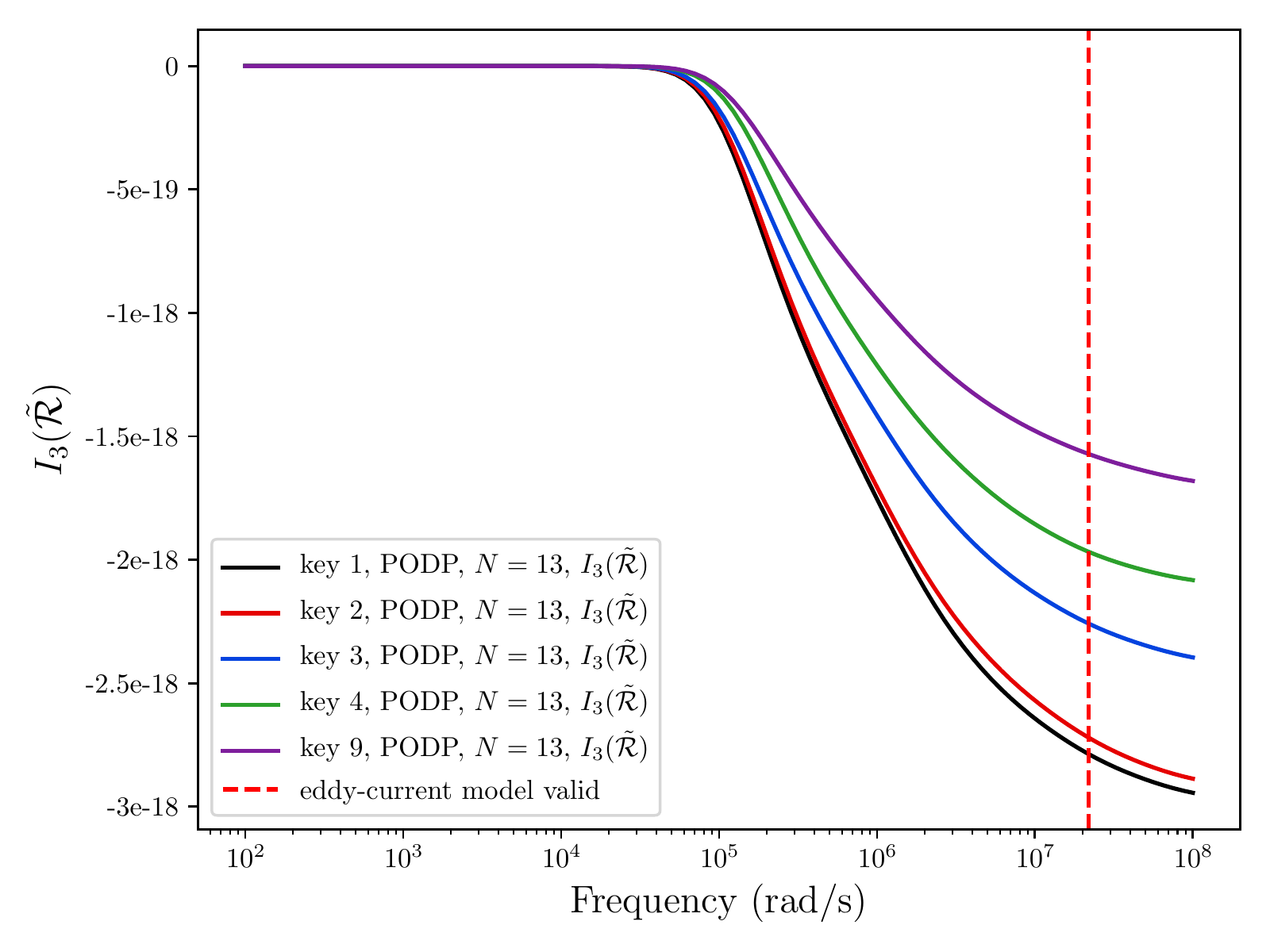}  &
 \includegraphics[scale=0.5]{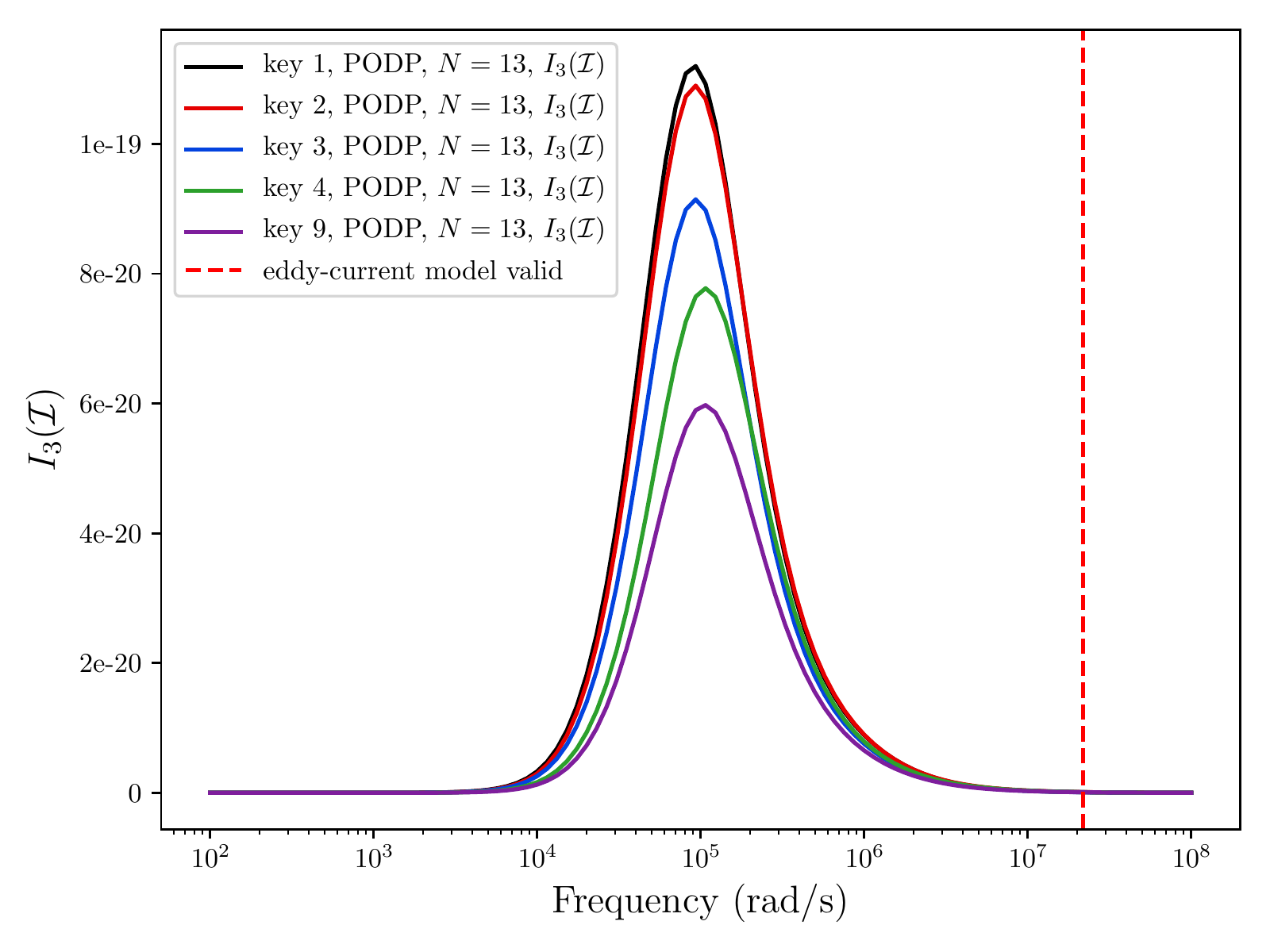}  \\
\text{(e) } I_{3} ( \tilde{\mathcal{R}} ) &
\text{(f) }  I_{3} ( \mathcal{I} )  
\end{array}$
  \caption{Set 1 of brass house keys: Comparison of tensor invariants. (a) $I_{1} ( \tilde{\mathcal{R}} ) $, (b) $I_{1} ( \mathcal{I} ) $
  (c) $I_{2} ( \tilde{\mathcal{R}} ) $, (d) $I_{2} ( \mathcal{I} ) $,
  (e) $I_{3} ( \tilde{\mathcal{R}} ) $ and (f)  $I_{3} ( \mathcal{I} ) $.}
        \label{fig:set1:fineMesh:I1_I3}
\end{figure}

To compare the results for different keys in set 1, we present the MPT spectral signature using the principal invariants $I_i$, $i=1,2,3$, for $\tilde{\mathcal R} [\alpha B,\omega,\sigma_*, \mu_r  ]$ and ${\mathcal I} [\alpha B,\omega, \sigma_*, \mu_r   ]$ that have been obtained using the PODP approach in Figure~\ref{fig:set1:fineMesh:I1_I3}.
In this figure, we observe a family of curves that each show a similar behaviour for all the keys in the set. 

The invariant $I_1 (\tilde{\mathcal R})$ is monotonically decreasing with $\log \omega$, implying the hydrostatic part of $\tilde{\mathcal R}$ is associated with a maximum response at high frequencies, while
the invariant $I_2(  \tilde{\mathcal R})$, which  is monotonically increasing with $\log \omega$, implies the deviatoric part of $\tilde{\mathcal R}$ begin to plays a significant role for $\omega > 10^5 \text{rad/s}$.
 The invariant $I_3(\tilde{\mathcal R})$, which is monotonically decreasing with $\log \omega$,  implies the interaction between the hydrostatic and deviatoric part of $\tilde{\mathcal R}$ begin to plays a significant role for $\omega > 10^5 \text{rad/s}$. 
 The invariants $I_i({\mathcal I})$, $i=1,2,3$, each have a single local maximum and are greater or equal to $0$ for all $\omega$. The invariant $I_1({\mathcal I})$ implies that hydrostatic part of ${\mathcal I}$ is associated with a maximum response at $\omega\approx 10^5 \text{rad/s}$   and has a broad response over the frequency range $10^2 \le \omega \le 10^8 \text{rad/s}$ while the invariant $I_2 ({\mathcal I})$ implies the deviatoric part of  ${\mathcal I}$ is associated with a maximum response at $\omega\approx 10^5 \text{rad/s}$ , but its effects are more limited to the narrower frequency band $10^4 \le \omega \le 10^7 \text{rad/s}$. Finally, the invariant $I_3( {\mathcal  I})$ has a maximum at $\omega\approx 10^5 \text{rad/s}$, although interaction between hydrostatic and deviatoric parts are more limited to the $10^4 \le \omega \le 10^6 \text{rad/s}$.
Comparing the keys, the effects are diminished from keys 1,2,3,4 and 9, in turn and, for example, the peak value of $I_3({\mathcal I})$ reduces in sequence of the volumes of the keys, as expected. Furthermore, the results for $I_i$, $i=2,3$ applied to $\tilde{\mathcal R}$ and ${\mathcal I}$ are similar when comparing the keys 1 and 2.
%
%
%


\begin{figure}[!h]
\centering
\hspace{-1.cm}
$\begin{array}{cc}
\includegraphics[scale=0.5]{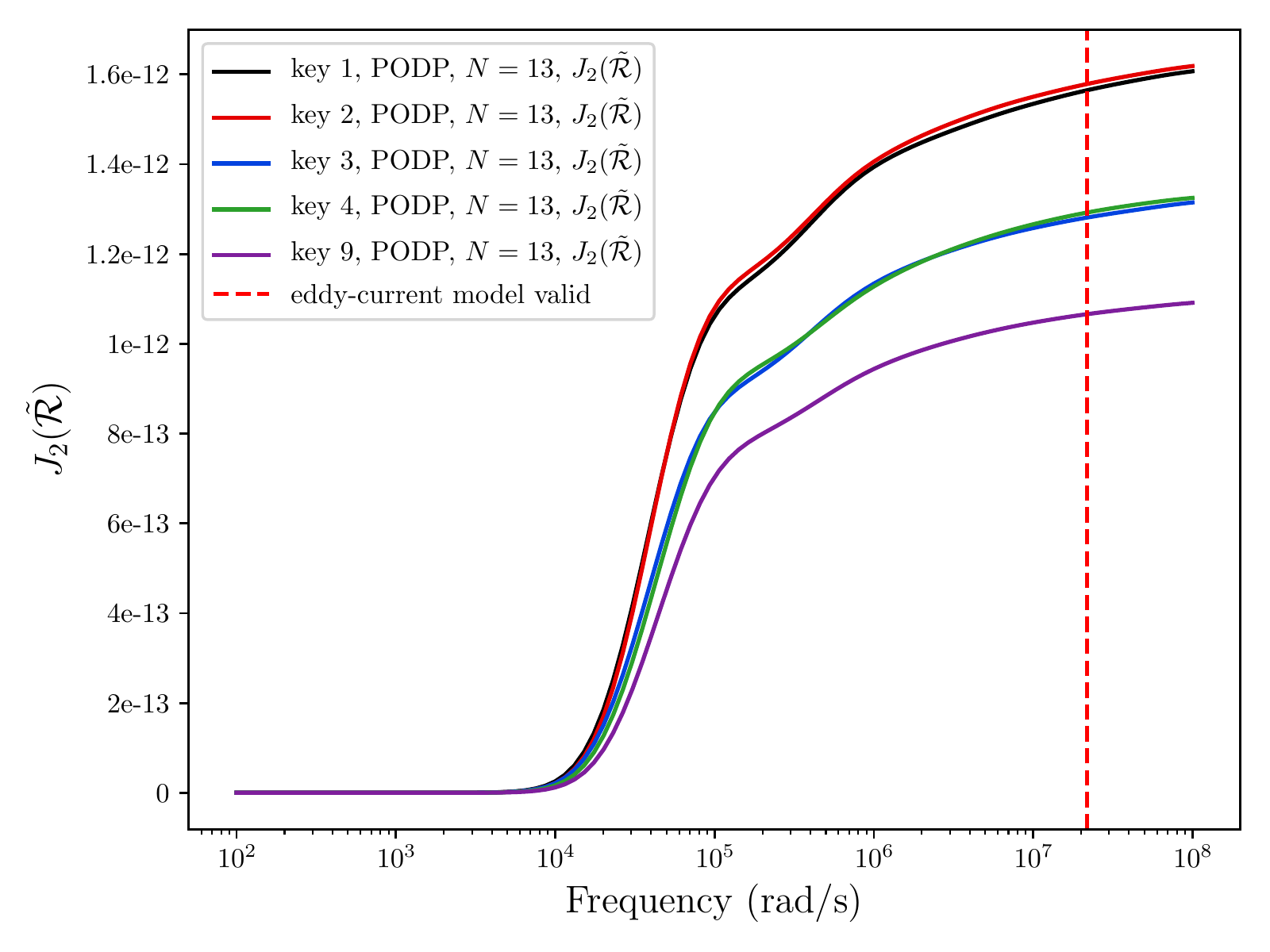}  & \includegraphics[scale=0.5]{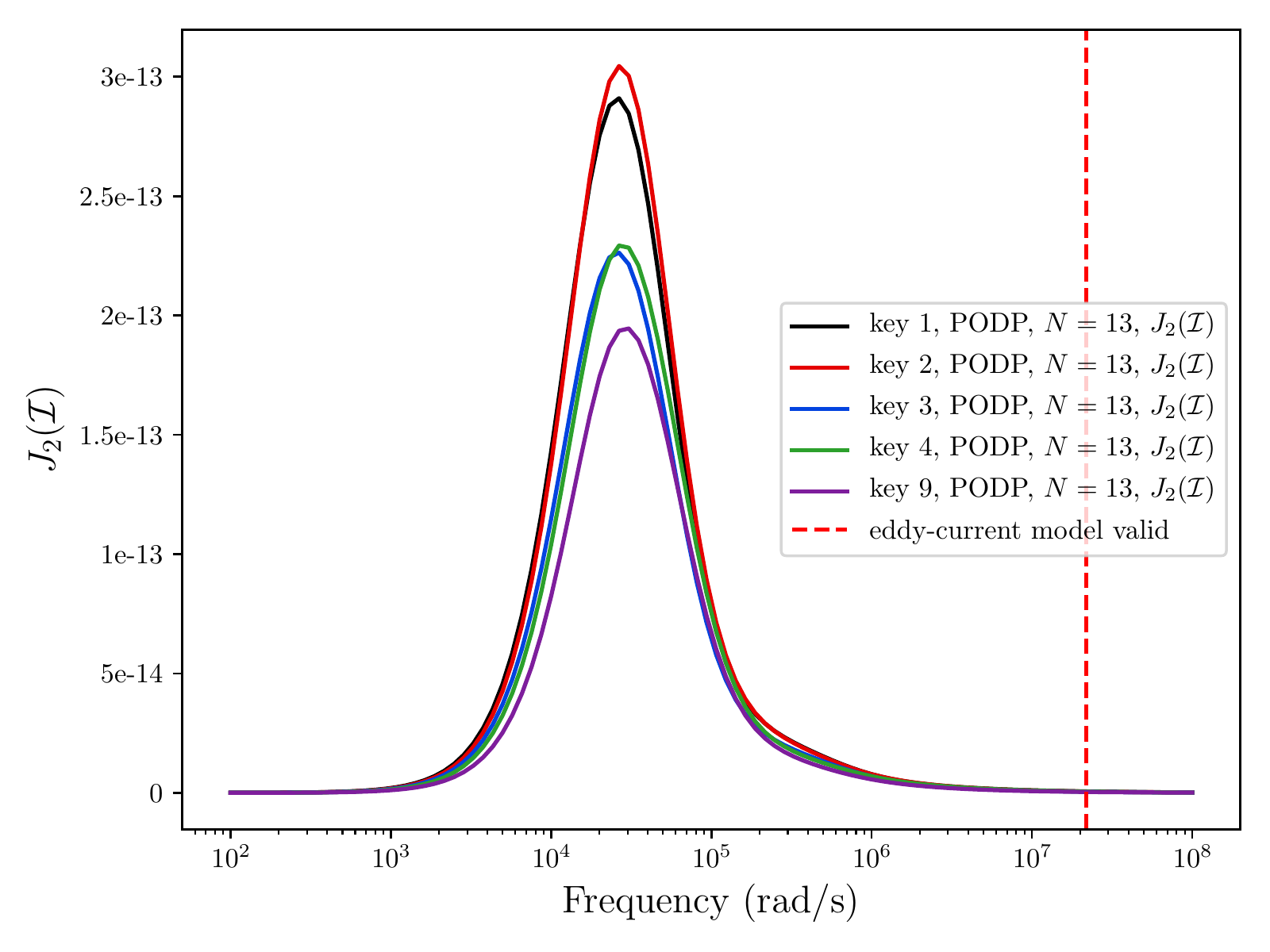}  \\
\text{(a) } J_{2} ( \tilde{\mathcal{R}} ) & 
\text{(b) } J_{2 } ( \mathcal{I}  )  \\
\includegraphics[scale=0.5]{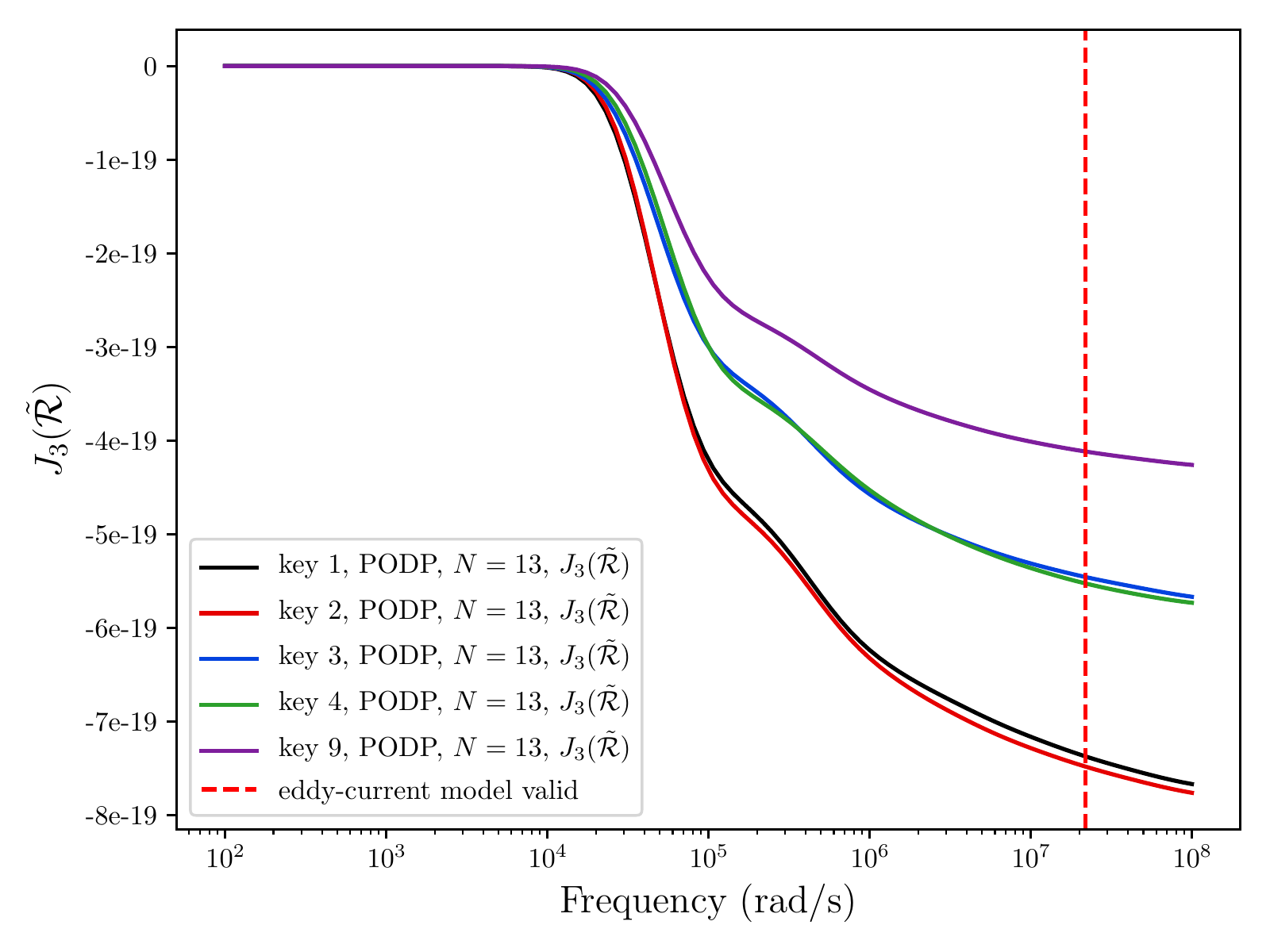}  & \includegraphics[scale=0.5]{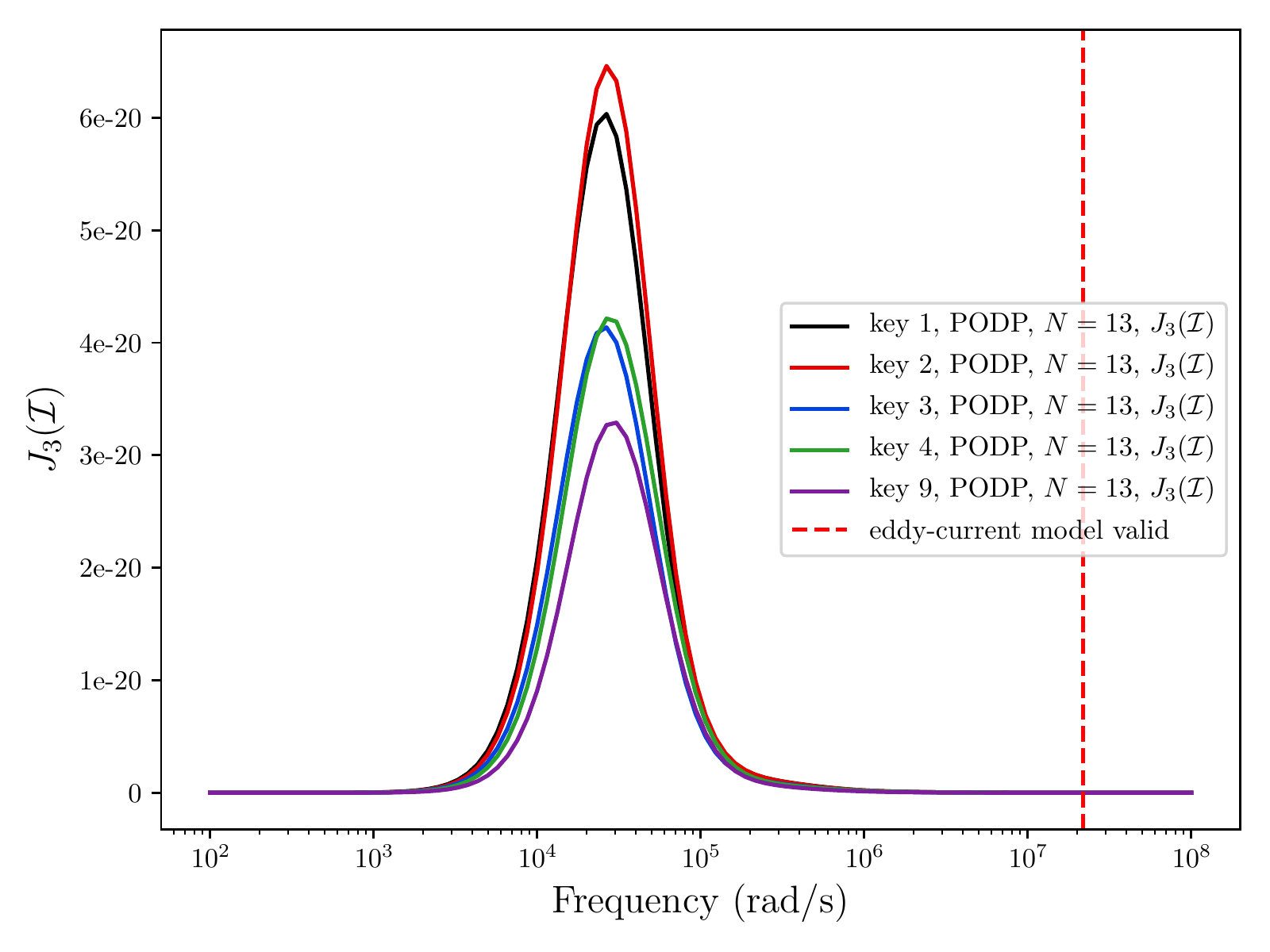} 
\\
\text{(c) } J_{3}  ( \tilde{\mathcal{R}} ) &
\text{(d) }  J_{3} ( \mathcal{I} )  
\end{array}$
  \caption{Set 1 of brass house keys: Comparison of tensor invariants.   (a) $J_{2} ( \tilde{\mathcal{R}} ) $, (b) $J_{2} ( \mathcal{I} ) $,
  (c) $J_{3}  ( \tilde{\mathcal{R}} ) $ and (d)  $J_{3} ( \mathcal{I} ) $.}
        \label{fig:set1:fineMesh:J2_J3}
\end{figure}

Next,  in Figure~\ref{fig:set1:fineMesh:J2_J3},  we present the MPT spectral signature using the alternative invariants $J_i$, $i=2,3$, for $\tilde{\mathcal R} [\alpha B,\omega, \sigma_*, \mu_r   ]$ and ${\mathcal I} [\alpha B,\omega, \sigma_*, \mu_r  ]$ that have been obtained using the PODP approach. Note that we do not reproduce the invariant $I_1$, which also forms part of this set, as it has already been shown in Figure~\ref{fig:set1:fineMesh:I1_I3}. We observe that, for the different keys making up set 1, a family of similar curves is obtained and there are similarities to the behaviour of the invariants $I_i$, $i=2,3$, with frequency for these tensors.
 {However, the  following differences are noteworthy, firstly,  the monotonic increase and decrease of $J_2( \tilde{\mathcal R})$ and $J_3( \tilde{\mathcal R})$ with $\log \omega$, respectively, is initially much more rapid than that of $I_2( \tilde{\mathcal R})$ and $I_3( \tilde{\mathcal R})$. Secondly, the curves for  $J_i (\tilde{\mathcal R})$ and $J_i( {\mathcal I} )$, $i=2,3$, are very similar when comparing the keys 3 and 4, whereas the corresponding curves for $I_i (\tilde{\mathcal R})$ and $I_i( {\mathcal I} )$, $i=2,3$, for these keys are different. On the other hand, the similarities previously  observed between the invariants $I_i$, $i=2,3$, for $\tilde{\mathcal R}$ and ${\mathcal I}$ keys 1 and 2 are also reflected by the invariants $J_i$, $i=2,3$, for these tensors and keys.}


\begin{figure}[!h]
\centering
\hspace{-1.cm}
  \subfigure[$\text{Key 1}$]{ \includegraphics[scale=0.5]{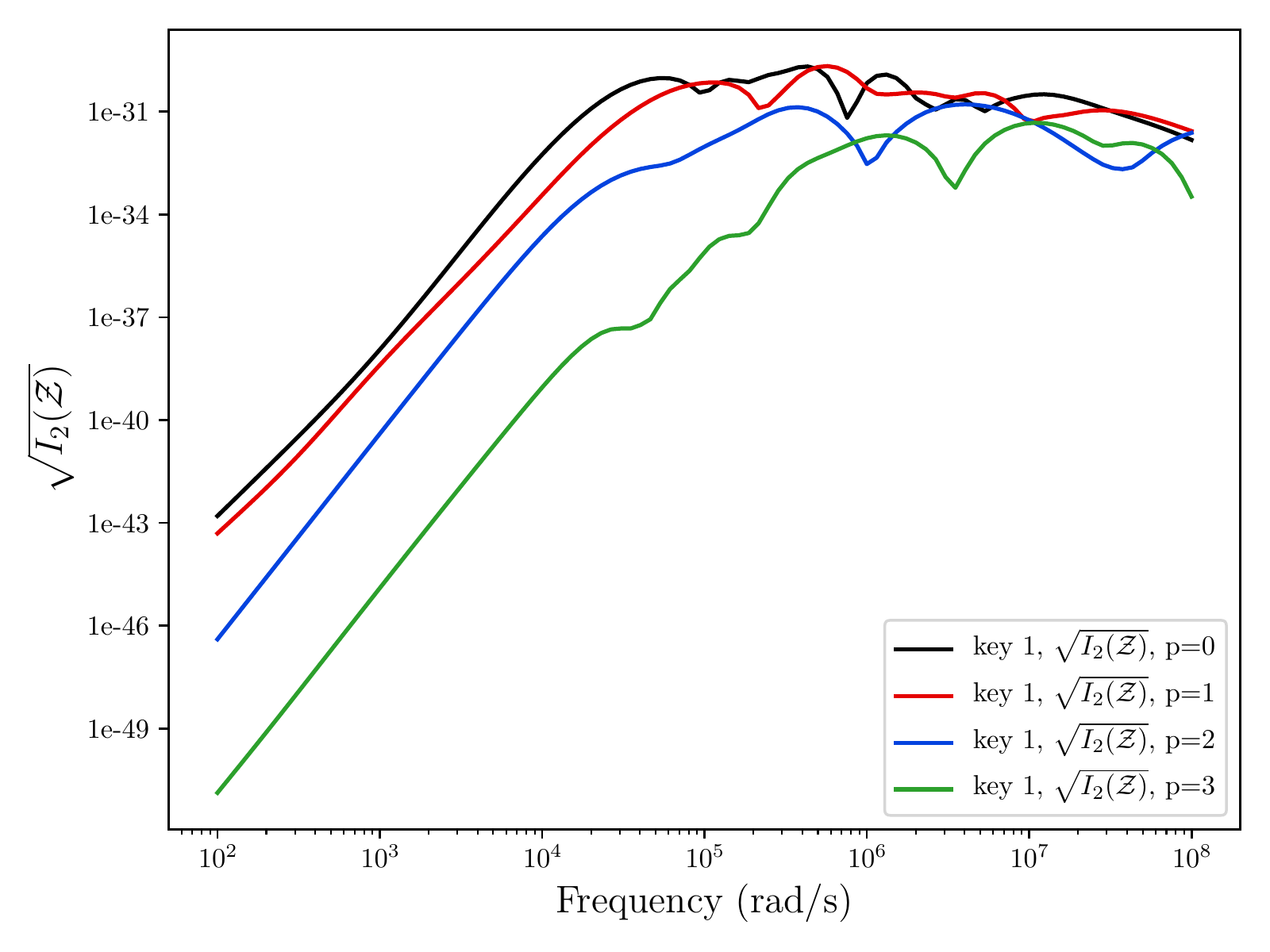} 
    }
    \subfigure[$\text{Key 4}$]{ \includegraphics[scale=0.5]{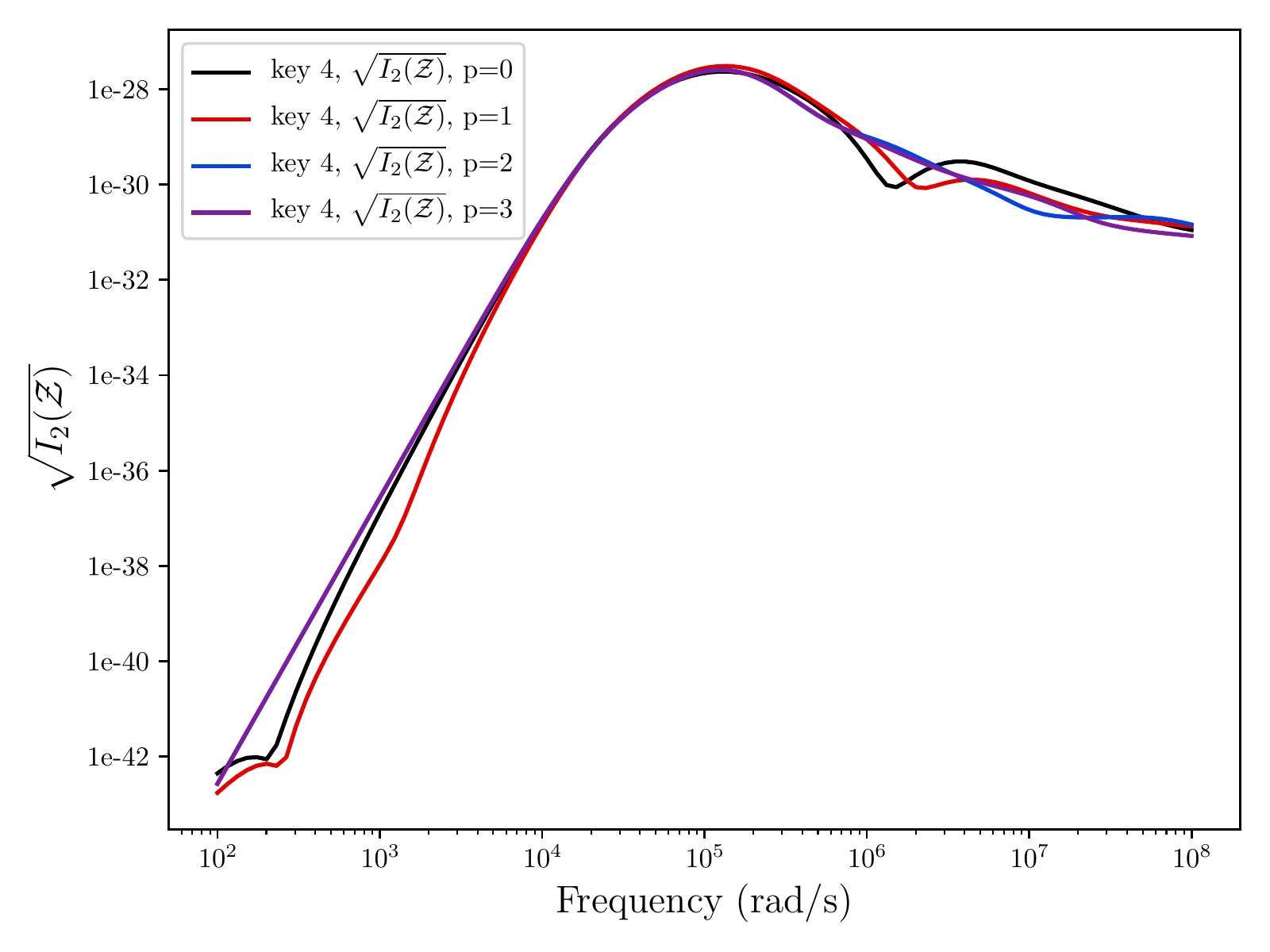} 
    }
  \caption{Set 1 of brass keys: $p$-refinement study for $\sqrt{I_2( {\mathcal Z})}$ for (a) Key 1 and (b) Key 4.}
        \label{fig:set1:key1:zCurve}
\end{figure}

\begin{figure}[!h]
\centering
    \includegraphics[scale=0.5]{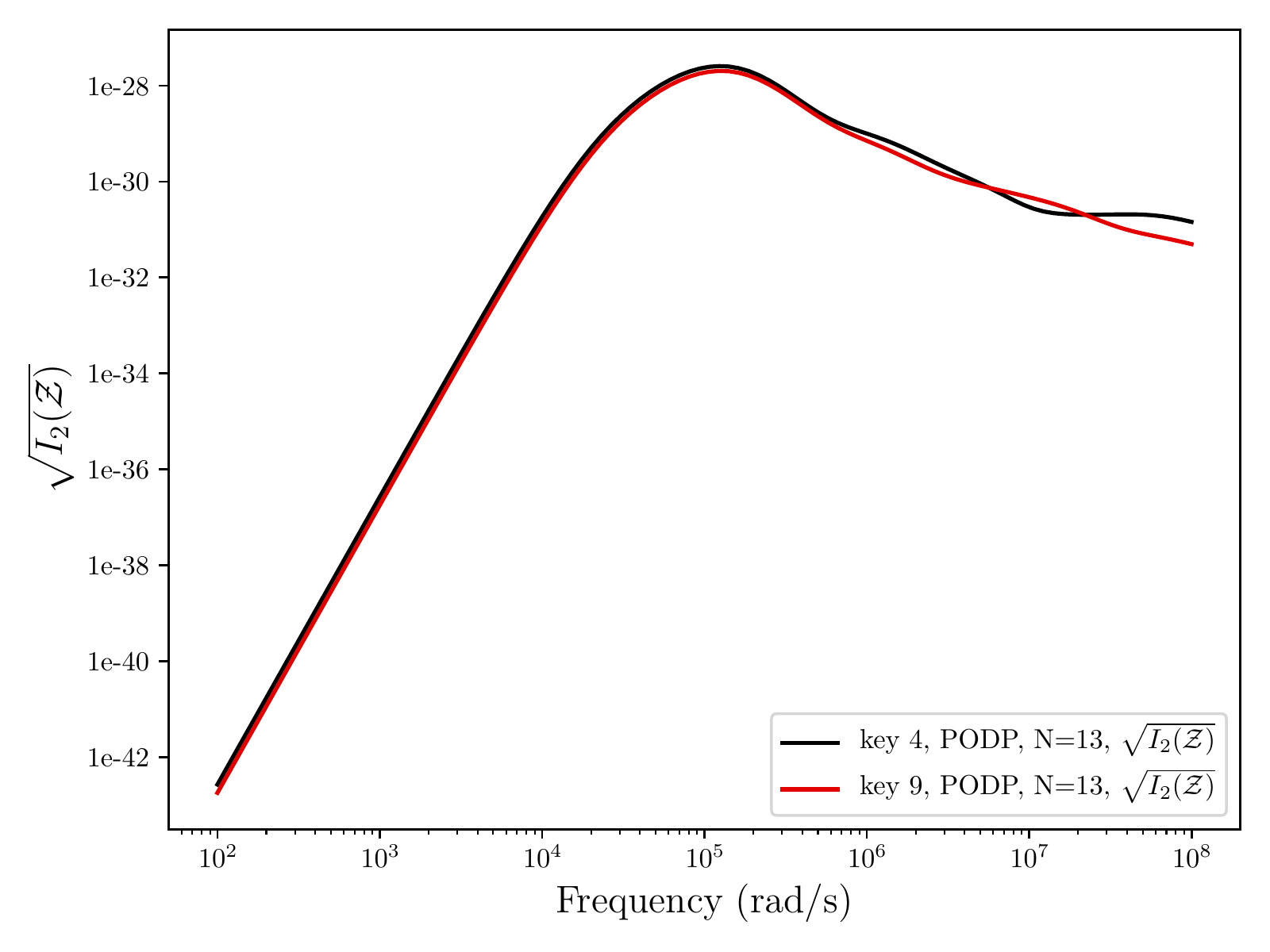} 
  \caption{Set 1 of brass keys: Comparison of $\sqrt{I_2( {\mathcal Z})}$ for keys  4 and 9.}
        \label{fig:set1:fineMesh:zCurve}
\end{figure}

Finally, we consider comparisons of the  invariant $ \sqrt{I_2 ( {\mathcal Z}[ \alpha B,\omega,  \sigma_*, \mu_r])} $ as a function of $\omega$ for keys in set 1, which  can provide additional information about the object's characterisation if the independent coefficients of the MPT are not only associated with its diagonal coefficients. We begin by 
showing the convergence of $ \sqrt{I_2 ( {\mathcal Z} )} $ to $0$ under $p$-refinement for key 1 in Figure~\ref{fig:set1:key1:zCurve}. The mirror symmetries for this object imply that
 $\tilde{\mathcal R} [\alpha B,\omega, \sigma_*, \mu_r  ]$ and ${\mathcal I} [\alpha B,\omega, \sigma_*, \mu_r  ]$
  each have  only $3$ independent coefficients  (at each frequency)
  that lie on the diagonal of the tensors and so we expect $ \sqrt{I_2 ( {\mathcal Z} )} $ to vanish for exact computations. Alongside this, in the same figure, we show the $p$-convergence of key 4, which has $6$ independent coefficients each in  $\tilde{\mathcal R} $ and ${\mathcal I} $ (at each frequency) and exhibits rapid convergence of the invariant $ \sqrt{I_2 ( {\mathcal Z})} $ to the shown curve as a function of $\omega$. The behaviour of $ \sqrt{I_2 ( {\mathcal Z})}$ for keys 4 and 9 is shown in Figure~\ref{fig:set1:fineMesh:zCurve}. For keys $1-3$ we henceforth set $\sqrt{I_2 ( {\mathcal Z} )} =0$.

\subsubsection{Set 2 of brass house keys}

\begin{figure}[!h]
\centering
\hspace{-1.cm}
$\begin{array}{cc}
\includegraphics[scale=0.5]{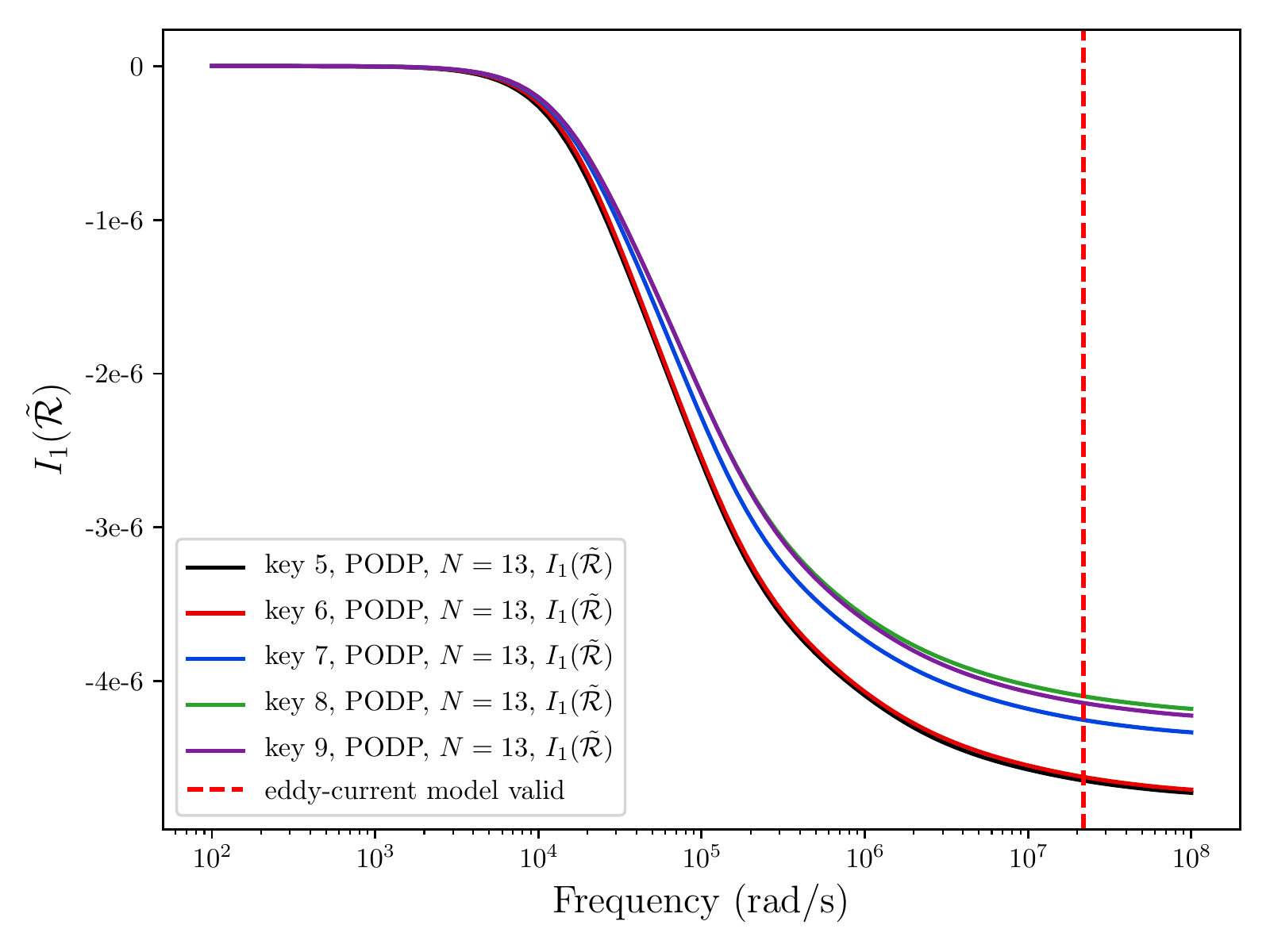}  &
 \includegraphics[scale=0.5]{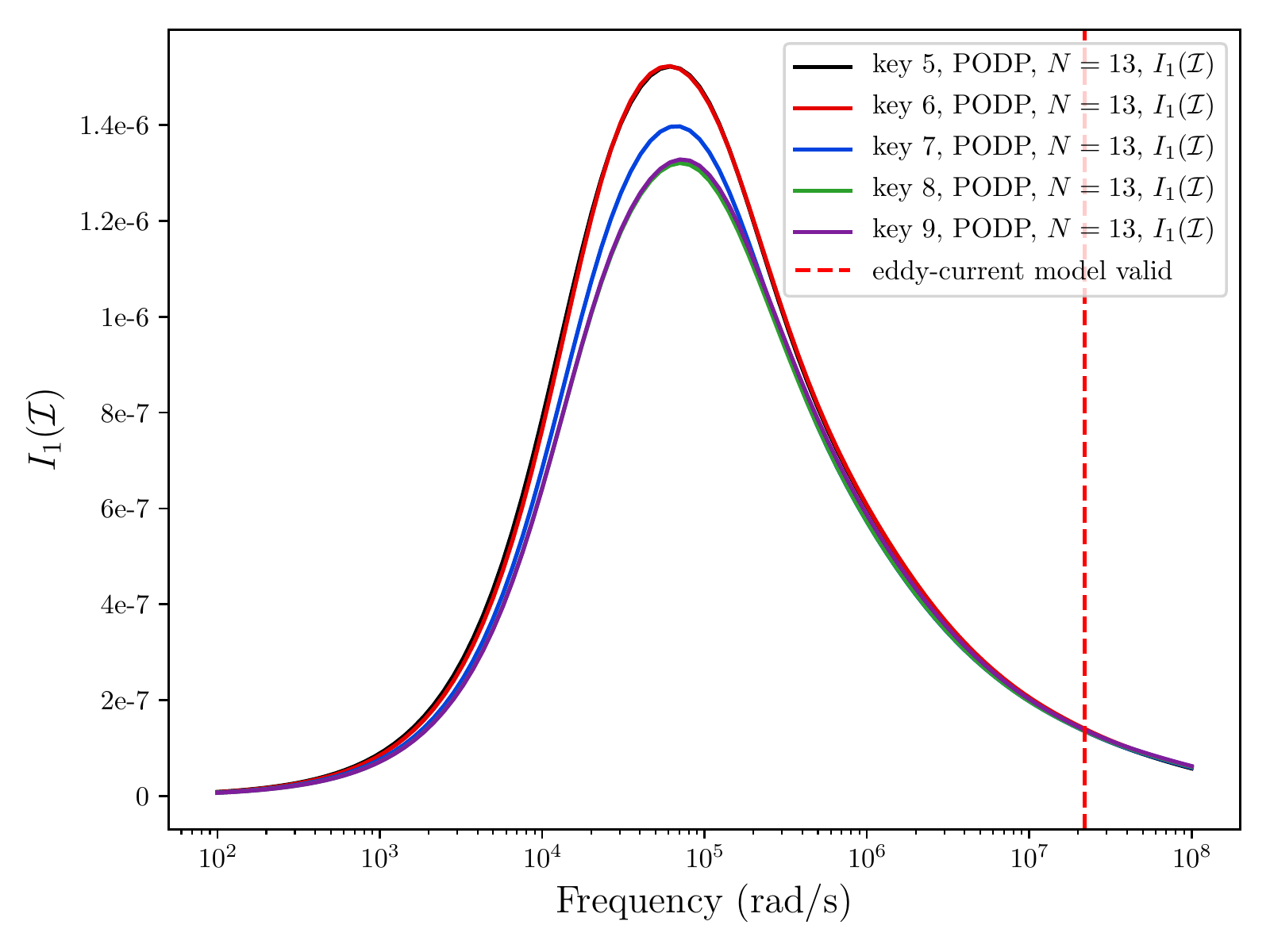}  \\
\text{(a) } I_{1} ( \tilde{\mathcal{R}} ) & 
\text{(b) } I_{1} ( \mathcal{I} )  \\
\includegraphics[scale=0.5]{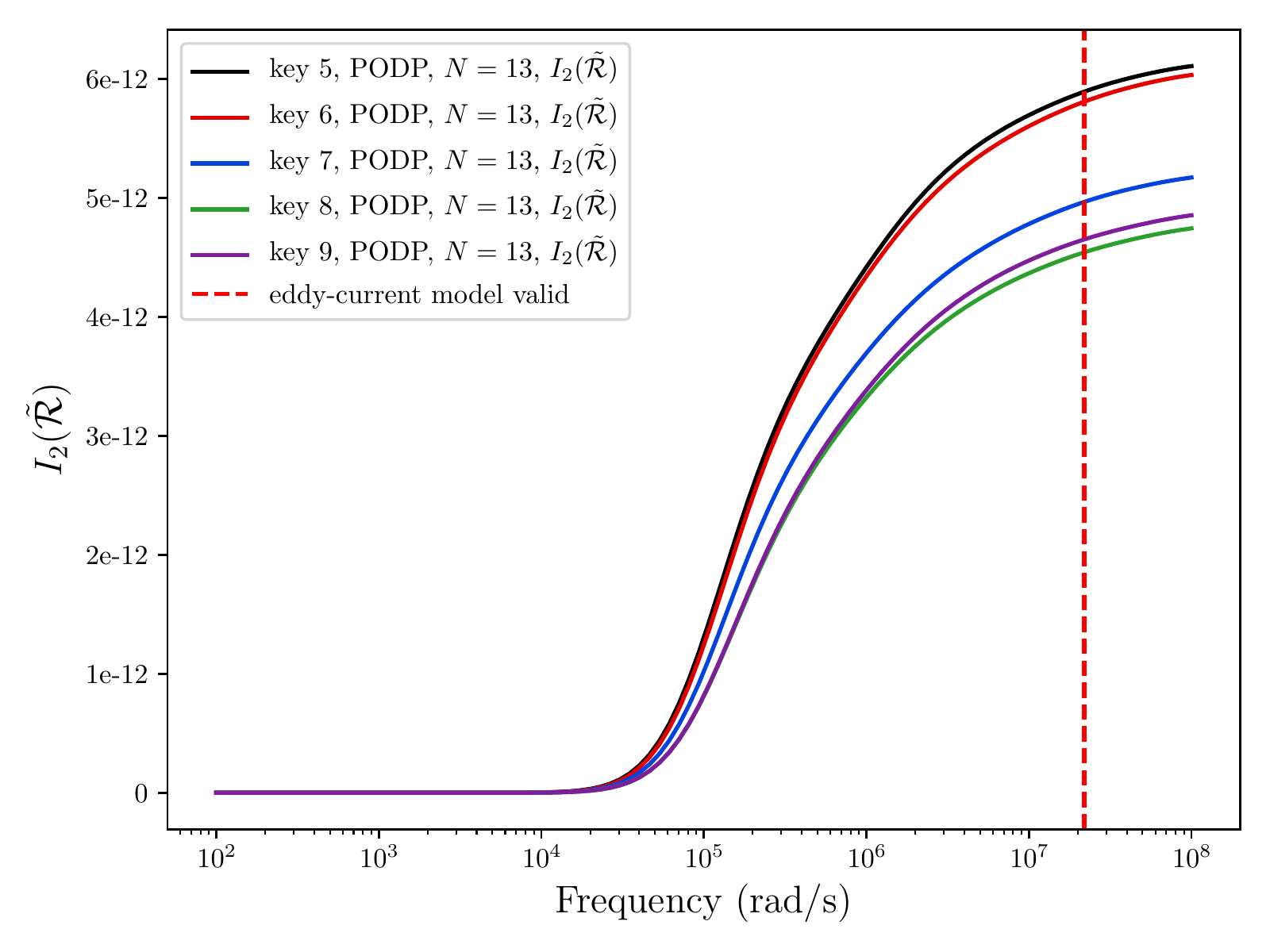}  &
 \includegraphics[scale=0.5]{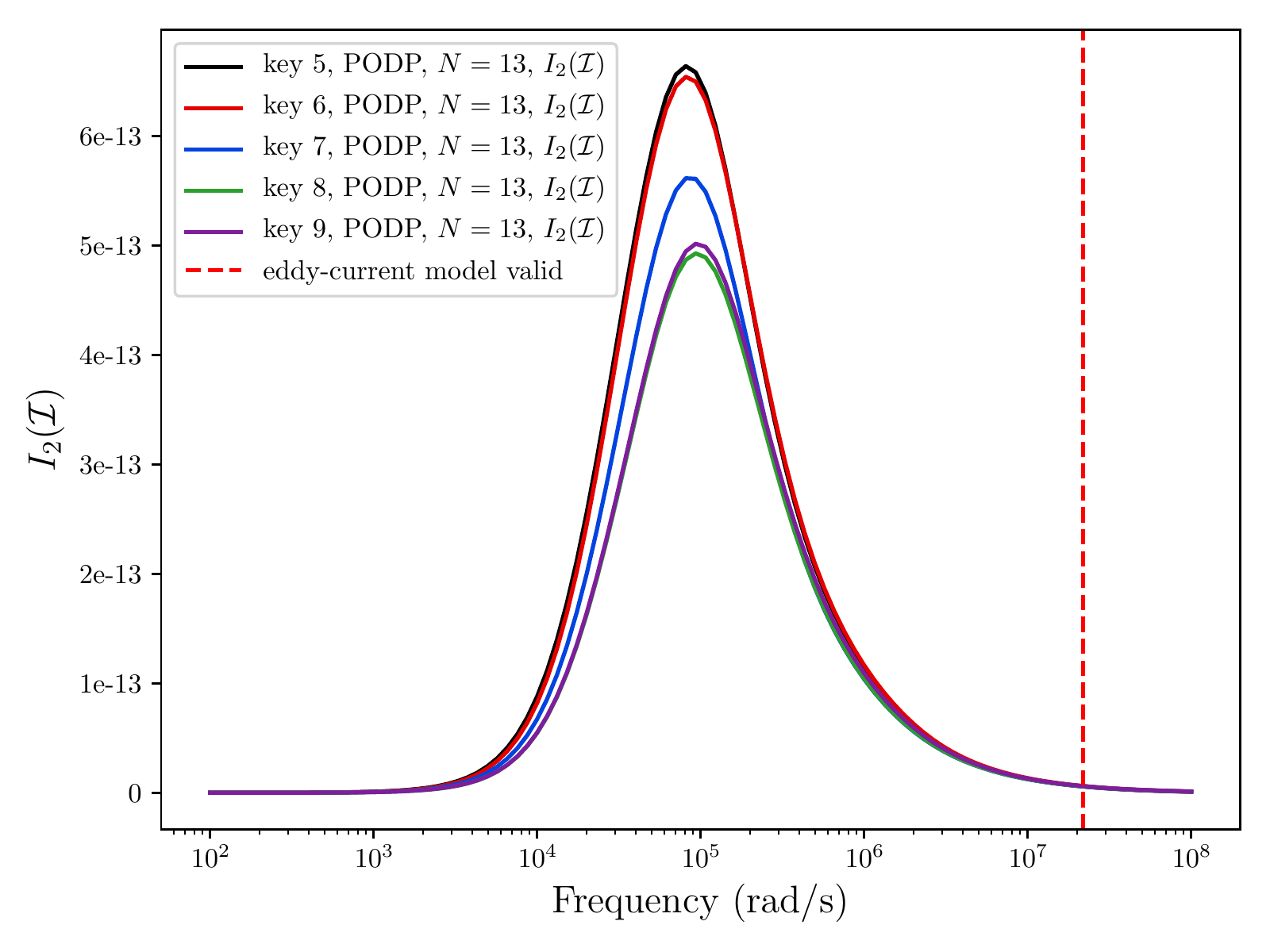}  \\
\text{(c) } I_{2} ( \tilde{\mathcal{R}} ) & 
\text{(d) } I_{2 } ( \mathcal{I} )  \\
\includegraphics[scale=0.5]{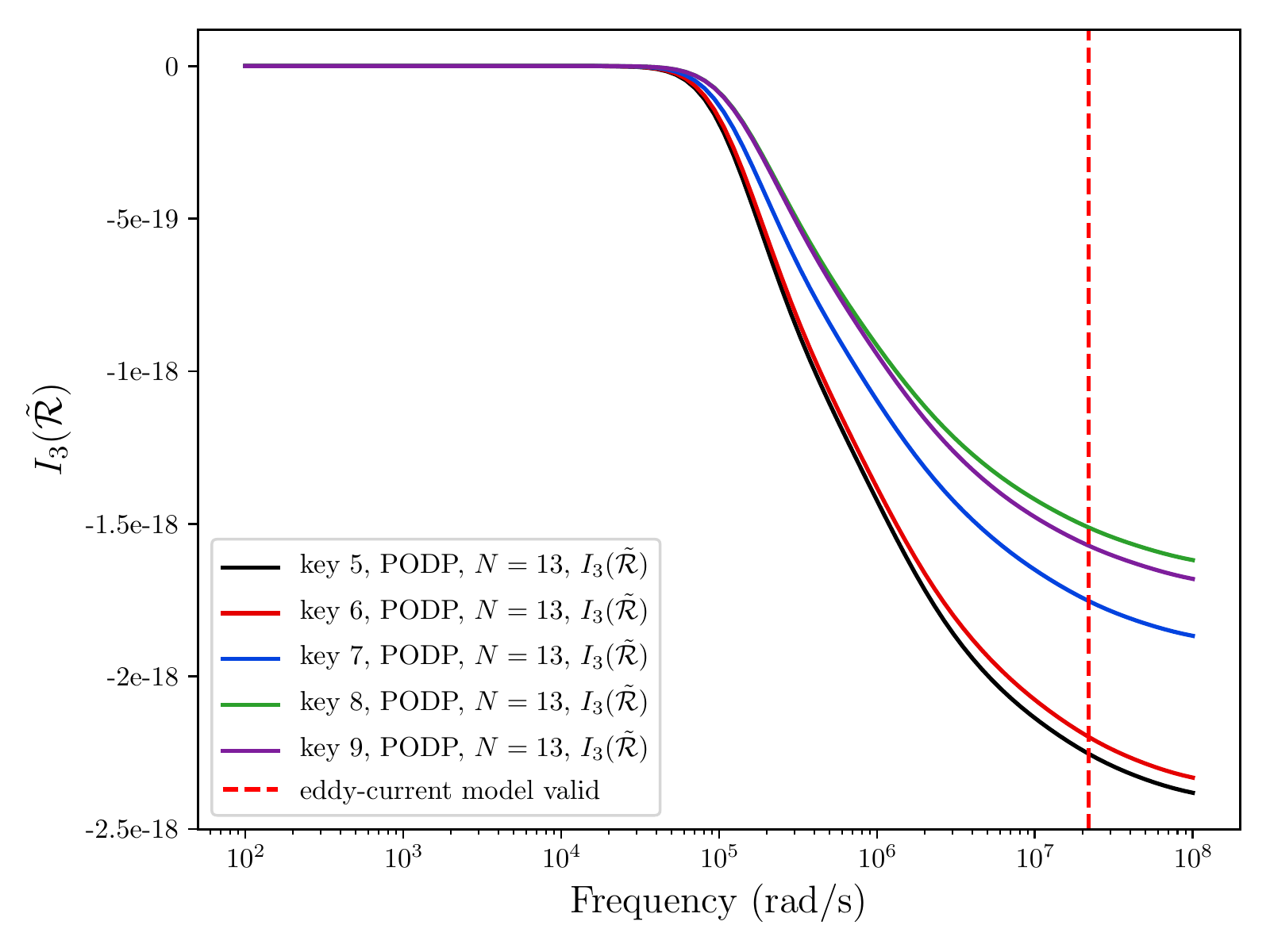}  &
 \includegraphics[scale=0.5]{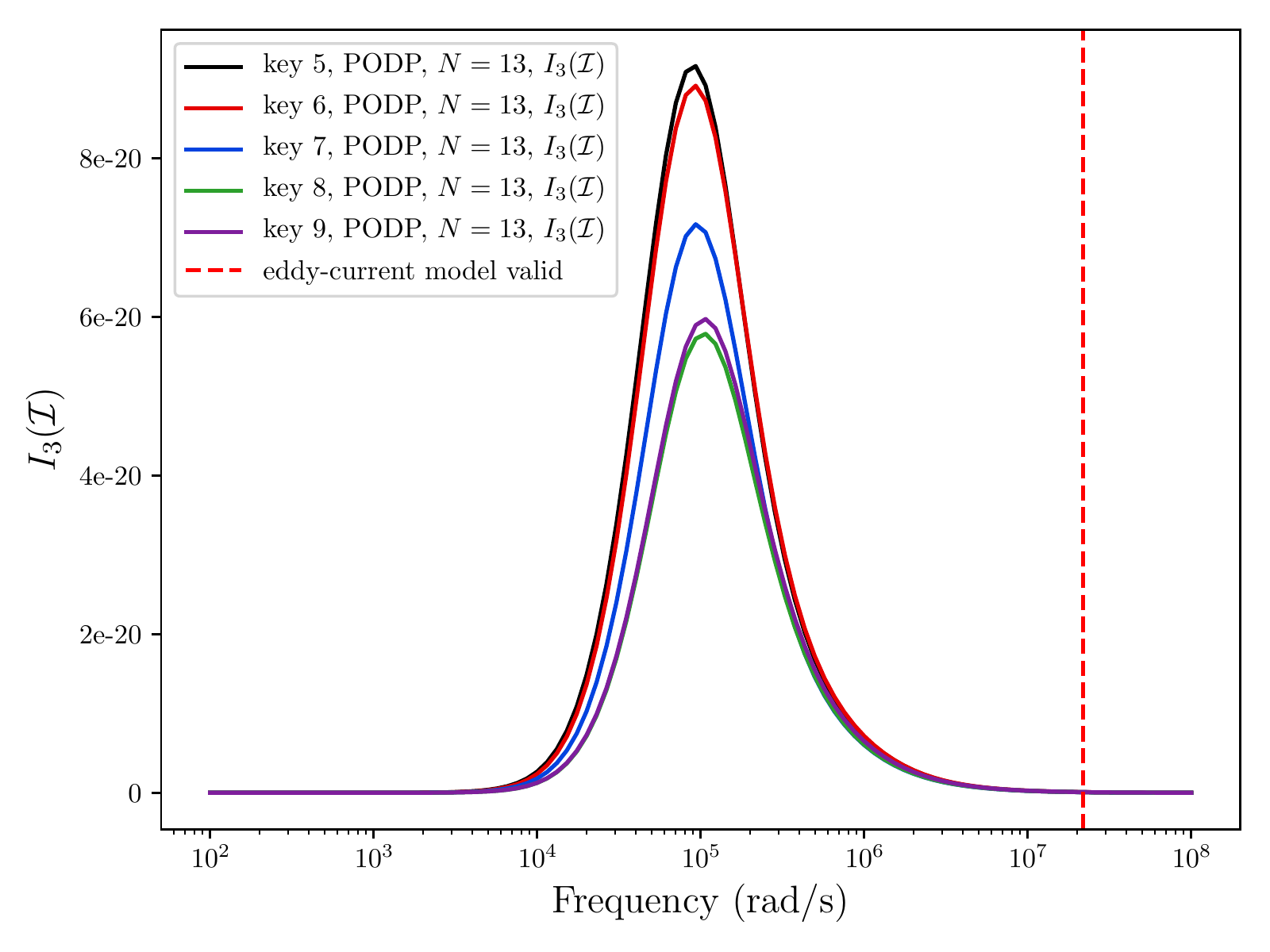}  \\
\text{(e) } I_{3}  ( \tilde{\mathcal{R}} ) &
\text{(f) }  I_{3} ( \mathcal{I} )  
\end{array}$
  \caption{Set 2 of brass house keys: Comparison of tensor invariants. (a) $I_{1} ( \tilde{\mathcal{R}} ) $, (b) $I_{1} ( \mathcal{I} ) $
  (c) $I_{2} ( \tilde{\mathcal{R}} ) $, (d) $I_{2} ( \mathcal{I} ) $,
  (e) $I_{3} ( \tilde{\mathcal{R}} ) $ and (f)  $I_{3} ( \mathcal{I} ) $}
        \label{fig:set2:fineMesh:I1_I3}
\end{figure}

Turning our attention to set 2 of the brass house keys, we reproduce the results previously shown in Figure~\ref{fig:set1:fineMesh:I1_I3}  for the invariants $I_i$, $i=1,2,3$, for $\tilde{\mathcal R} [\alpha B,\omega,  \sigma_*, \mu_r  ]$~\footnote{Note that the coefficients of ${\mathcal N}^0$ vanish as $\mu_r=1$, but we keep to the notation of  $\tilde{R}={\mathcal N}^0+ {\mathcal R}$ for ease of comparison with later results} and ${\mathcal I} [\alpha B,\omega, \sigma_*, \mu_r  ]$  for the second set of keys and show these in~Figure~\ref{fig:set2:fineMesh:I1_I3}.
In a similar manner to Figure~\ref{fig:set1:fineMesh:I1_I3}, we see that the results included in~Figure~\ref{fig:set2:fineMesh:I1_I3}  form  a family of similar curves {and that their behaviour follows a similar pattern to that previous described for the keys in set 1. The results for $I_i$, $i=1,2,3$ for the tensor characterisations of keys 5 and 6 are similar, which is not surprising given the similarities in these geometries. In addition, there are only small differences in $I_i$, $i=1,2,3$ for the tensor characterisations of keys 8 and 9. Note that key 8 has a circular head and key 9 a polygonal head, but the volume of material is similar and the symmetries of the objects and the number of independent coefficients in $\tilde{\mathcal R} $ and ${\mathcal I} $  (for each frequency) are otherwise the same for these two keys. Keys $5-8$ are associated with a gradual reduction in the volume of the material for the key and we can see that the magnitude of the associated $I_3(  {\mathcal I} )$ and $I_3(\tilde{\mathcal R})$, curves for these cases reduces as expected. }

\begin{figure}[!h]
\centering
\hspace{-1.cm}
$\begin{array}{cc}
\includegraphics[scale=0.5]{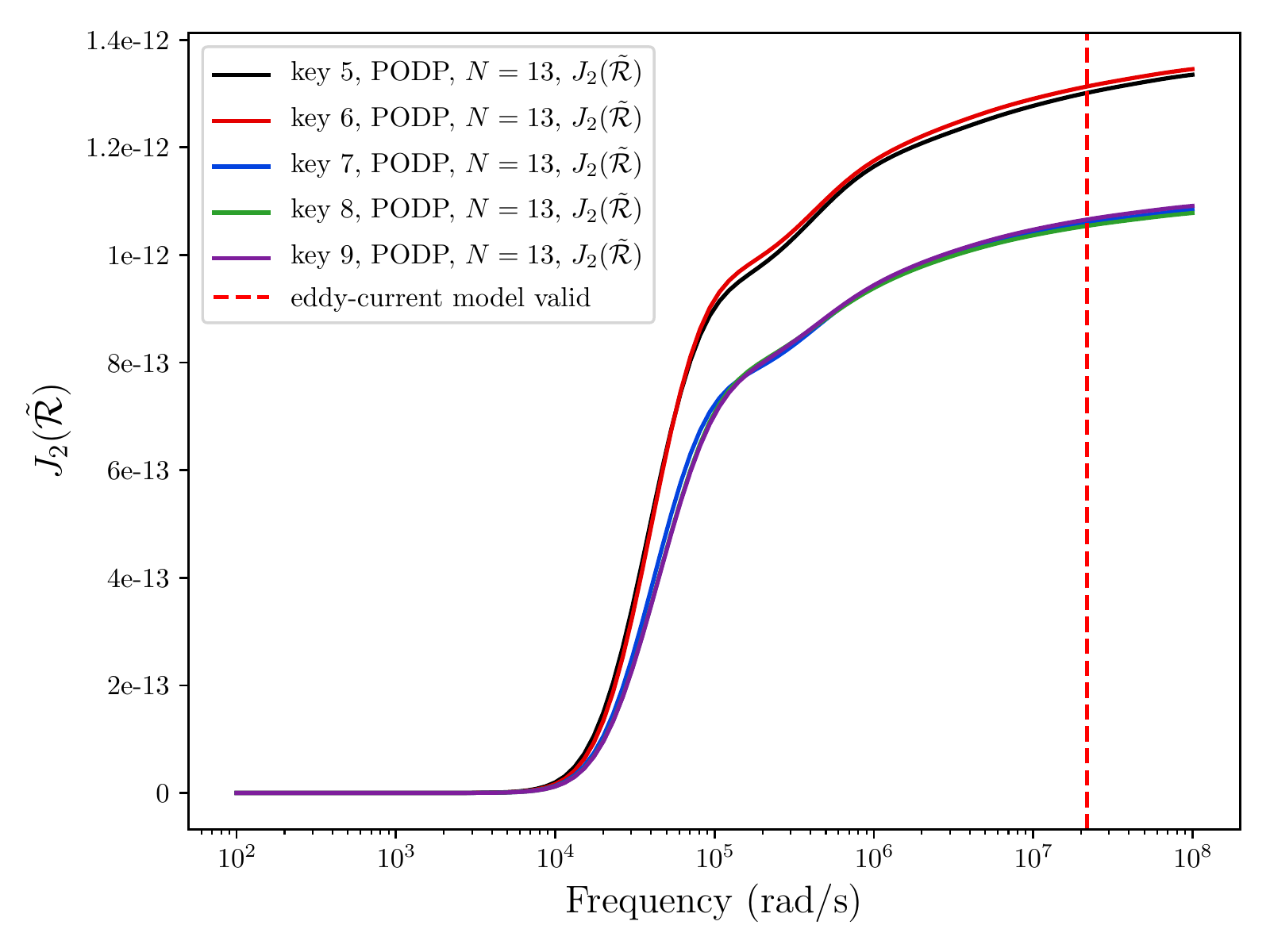}  & \includegraphics[scale=0.5]{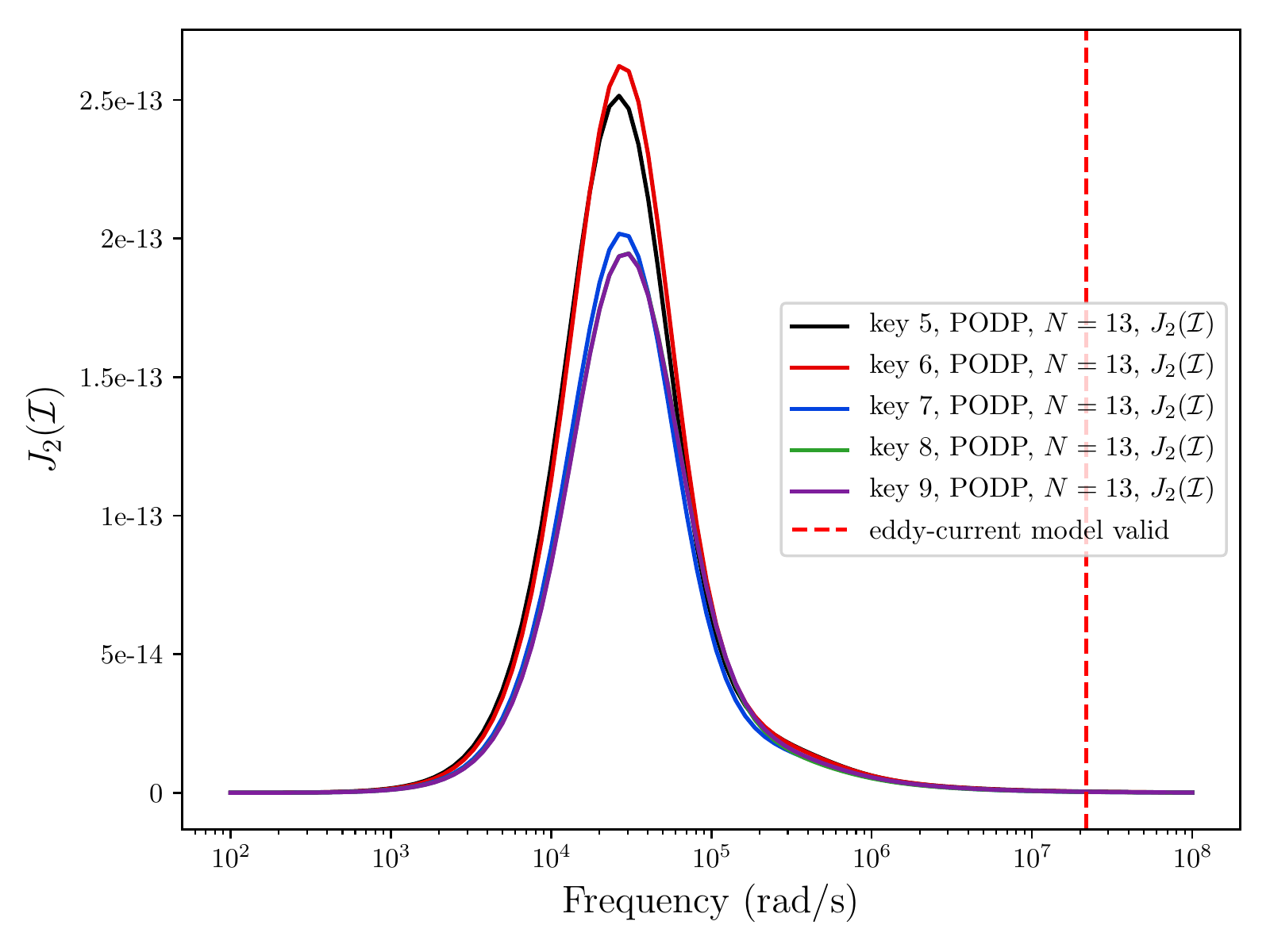} \\
\text{(a) } J_{2} ( \tilde{\mathcal{R}} ) & 
\text{(b) } J_{ 3}( \mathcal{I} )  \\
\includegraphics[scale=0.5]{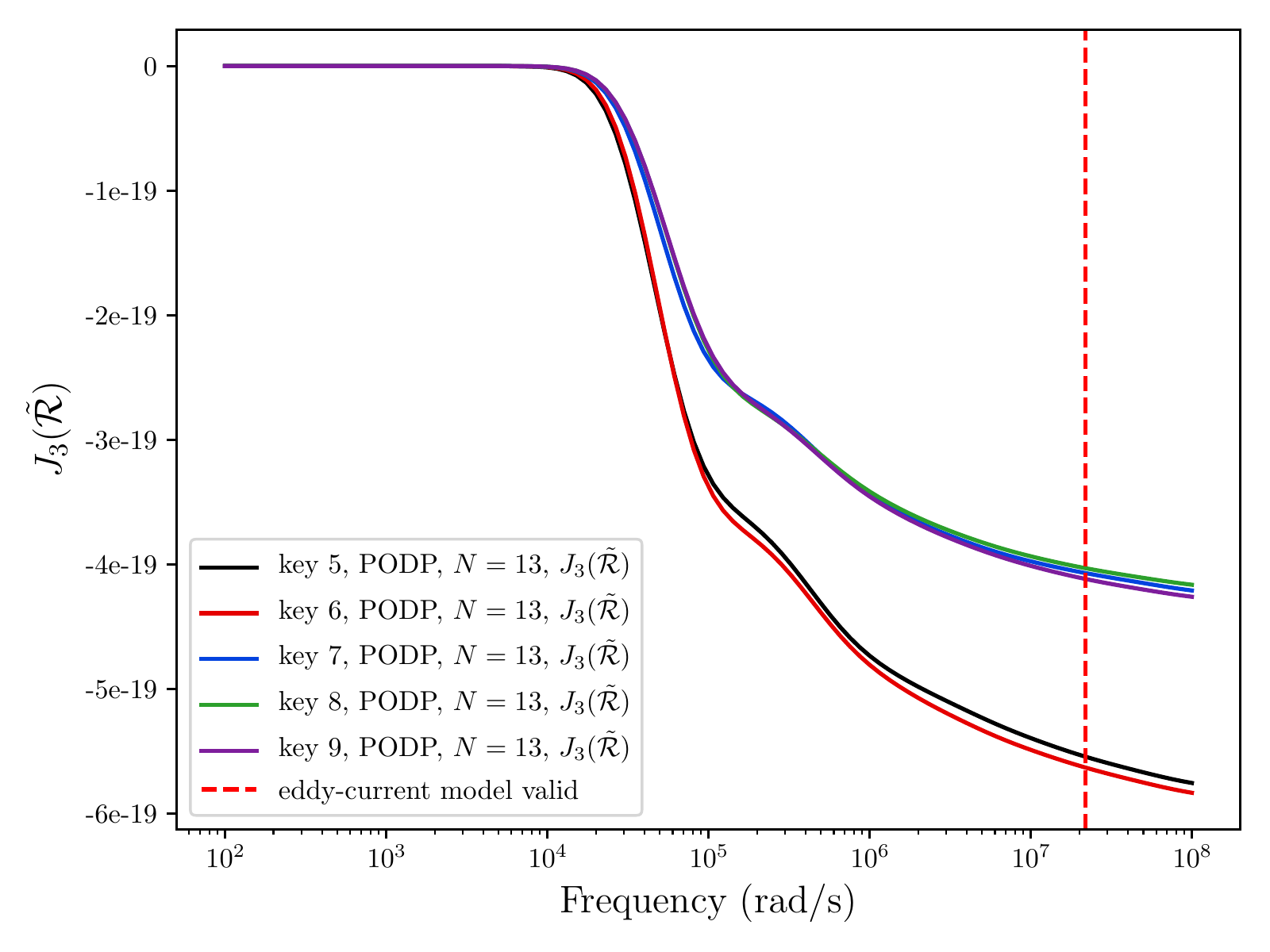}  & \includegraphics[scale=0.5]{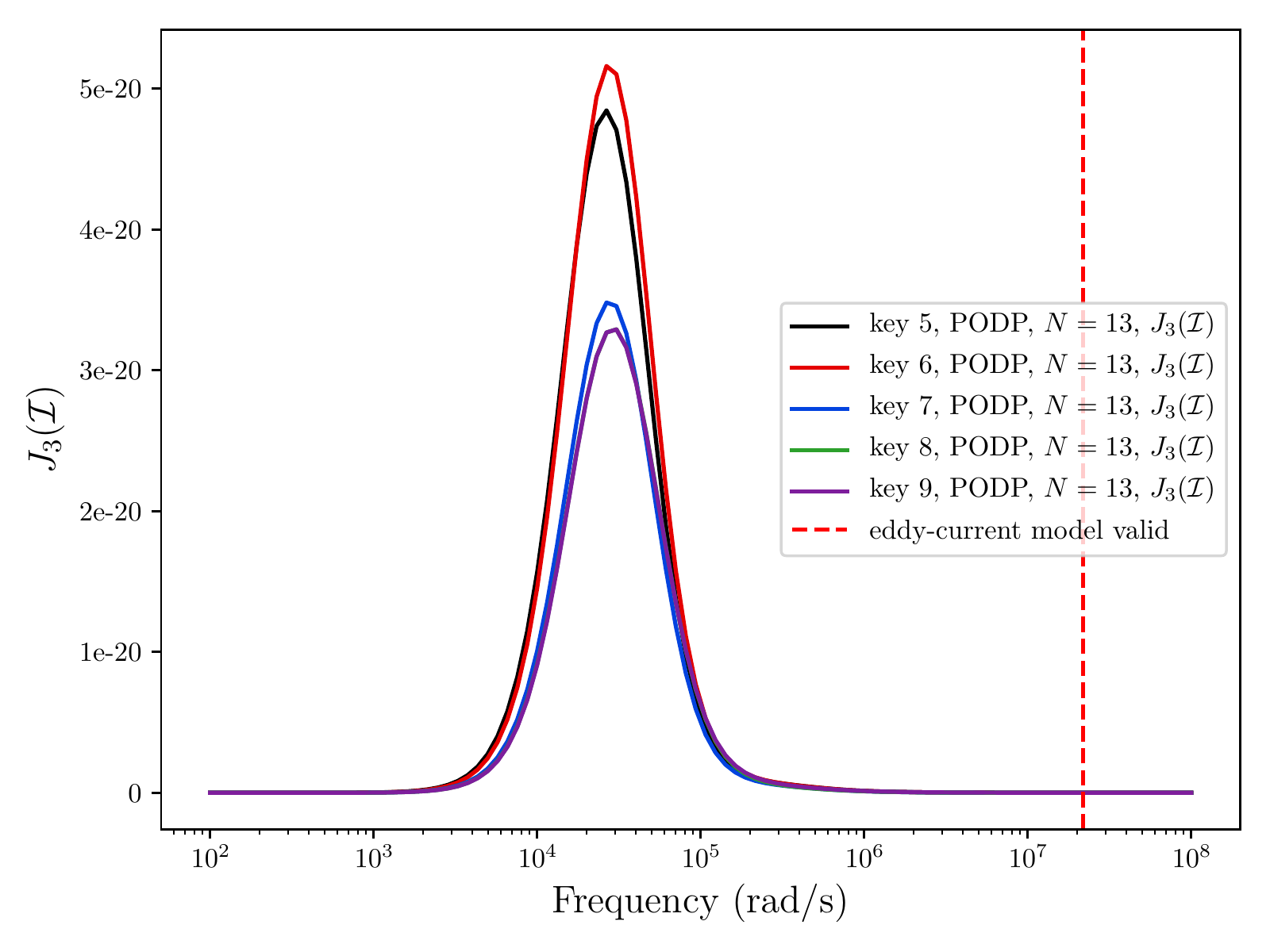} \\
\text{(c) } J_{3} ( \tilde{\mathcal{R}}) &
\text{(d) }  J_{3} ( \mathcal{I} )  
\end{array}$
  \caption{Set 2 of brass house keys: Comparison of tensor invariants. (a) $J_{2} ( \tilde{\mathcal{R}} ) $, (b) $J_{2} ( \mathcal{I} ) $,
  (c) $J_{3}  ( \tilde{\mathcal{R}} ) $ and (d)  $J_{3}( \mathcal{I} ) $}
        \label{fig:set2:fineMesh:J2_J3}
\end{figure}

Similarly, in~Figure~\ref{fig:set2:fineMesh:J2_J3}, we see the results for the invariants $J_i$, $i=2,3$ form a family of curves {with the behaviour of the invariants similar to that described for the keys in set 1. Note that the results for the keys in set 2 for $J_i$, $i=2,3$ can be grouped into keys 5, 6 and keys 7, 8 and 9 where the results for keys 8 and 9 for $J_2({\mathcal I})$ (and $J_3({\mathcal I})$) are indistinguishable on this scale. The former group does not contain the notches or the blade cut while the latter set all have the same notches, keys 8 and 9 have the deep blade cut and key 9 differs from the others by having a polygonal head rather than a circular head (although has a similar volume to keys 7 and 8).}

\begin{figure}[!h]
\centering
    \includegraphics[scale=0.5]{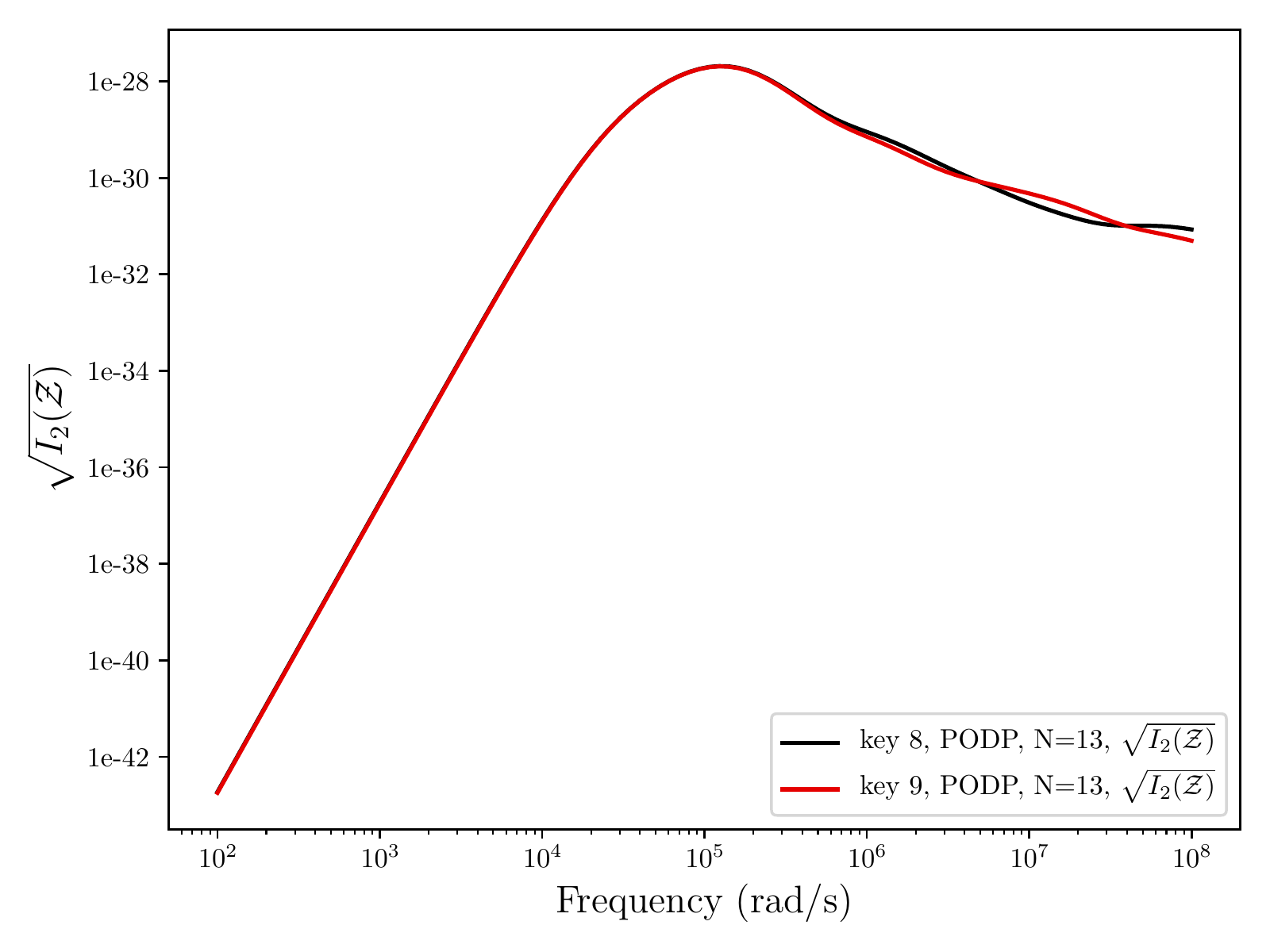} 
      \caption{Set 2 of brass house keys: Comparison of the tensor invariant    $\sqrt{I_2 ( {\mathcal Z}  )} $.}
        \label{fig:set2:fineMesh:zCurve}
\end{figure}

Of the keys in set 2, only keys 8 and 9 have independent coefficients in ${\mathcal R} [\alpha B,\omega,  \sigma_*, \mu_r  ]$ and ${\mathcal I} [\alpha B,\omega,  \sigma_*, \mu_r  ]$  
 that are not only associated with the diagonal entries of the tensor. The behaviour of $\sqrt{I_2 ( {\mathcal Z}[ \alpha B,\omega,  \sigma_*, \mu_r])} $ for these keys is shown in Figure~\ref{fig:set2:fineMesh:zCurve}. For the other keys in set 2 we set $\sqrt{I_2 ( {\mathcal Z} )}=0$.

\subsection{Non-threat Items: British coins}
In this section, inspired by the previous article on MPT characterisations of US coins~\cite{davidsoncoins},
we present MPT spectral signature characterisations for British coins in the denominations 1 pence (p), 2p, 5p, 10p, 20p, 50p, \pounds 1, \pounds 2. We use the 1982 (20p), 1992 (1p, 2p), 1997 (50p), 1998 (\pounds 2), 2012 (5p,10p), 2017 (\pounds 1)  issues of these denominations as listed in Table~\ref{tab:Coins}, which also summarises the shape, diameter, thickness, composition based on the information available from the Royal Mint~\cite{royalmint}.  The table also sets out the electrical properties, where the conductivity values for the different material compositions have been obtained from~\cite{coinconductivity} at room temperature. For the quoted compositions, we have assumed that $\mu_r=1$, however, in practice, some Copper-Nickel mixtures with a high iron content can have a $\mu_r$ slightly above $1$ (e.g~\cite{coinpermeability}). The later issues of the 1p, 2p, 5p and 10p coins have a significantly different composition to that presented in Table~\ref{tab:Coins} and, instead of a high copper content, they are instead copper plated steel. Note that each of the coins considered are simply connected.

With the exception of the \pounds 1 and \pounds 2 denominations, the coins are modelled as homogeneous conductors while the former are each modelled as an annulus with two different materials. The majority of the coins have a circular face and only the 20p and 50p differ, being Reuleaux heptagonal discs. The coins with a circular face are modelled so that their circular region lies in the plane spanned by ${\bm e}_1$ and ${\bm e}_2$  and, hence, they have rotational symmetry about the ${\bm e}_3$ axis (for any angle).  Consequently, the 
 independent coefficients
 of $\tilde{\mathcal R} [\alpha B,\omega,\mu_r , \sigma_* ]$~\footnote{Note that the coefficients of ${\mathcal N}^0$ vanish as $\mu_r=1$, but we keep to the notation of  $\tilde{R}={\mathcal N}^0+ {\mathcal R}$ for ease of comparison with later results} and ${\mathcal I} [\alpha B,\omega,\mu_r , \sigma_* ]$  
  for such coins are $(\tilde{\mathcal R}  )_{11}=(\tilde{\mathcal R}  )_{22} $,
 $(\tilde{\mathcal R})_{33} $
  and $({\mathcal I}  )_{11} = ({\mathcal I}  )_{22} $ and
 $({\mathcal I} )_{33} $ 
   (for each frequency). The Reuleaux heptagonal discs are modelled in a similar way, with a 51.428 (4dp) degree rotational symmetry about the ${\bm e}_3$ and, consequently, it also follows that their independent coefficients of the MPT for such coins are associated with the same entries.

\begin{table}[!h]
\begin{center}
\begin{tabular}{!\vrule p{2.5cm}!\vrule p{1.7cm}!\vrule p{1.7cm}!\vrule p{1.7cm}!\vrule p{2.3cm}!\vrule p{2cm}!\vrule p{2cm}!\vrule}
\hline
Coin & Shape & Diameter in mm & Thickness in mm & Composition & Relative Permeability ($\mu_r$) & Conductivity ($\sigma_*$) in S/m \\\hline
 \raisebox{0.0\totalheight}{ \includegraphics[width=0.11\textwidth, keepaspectratio, angle=-90]{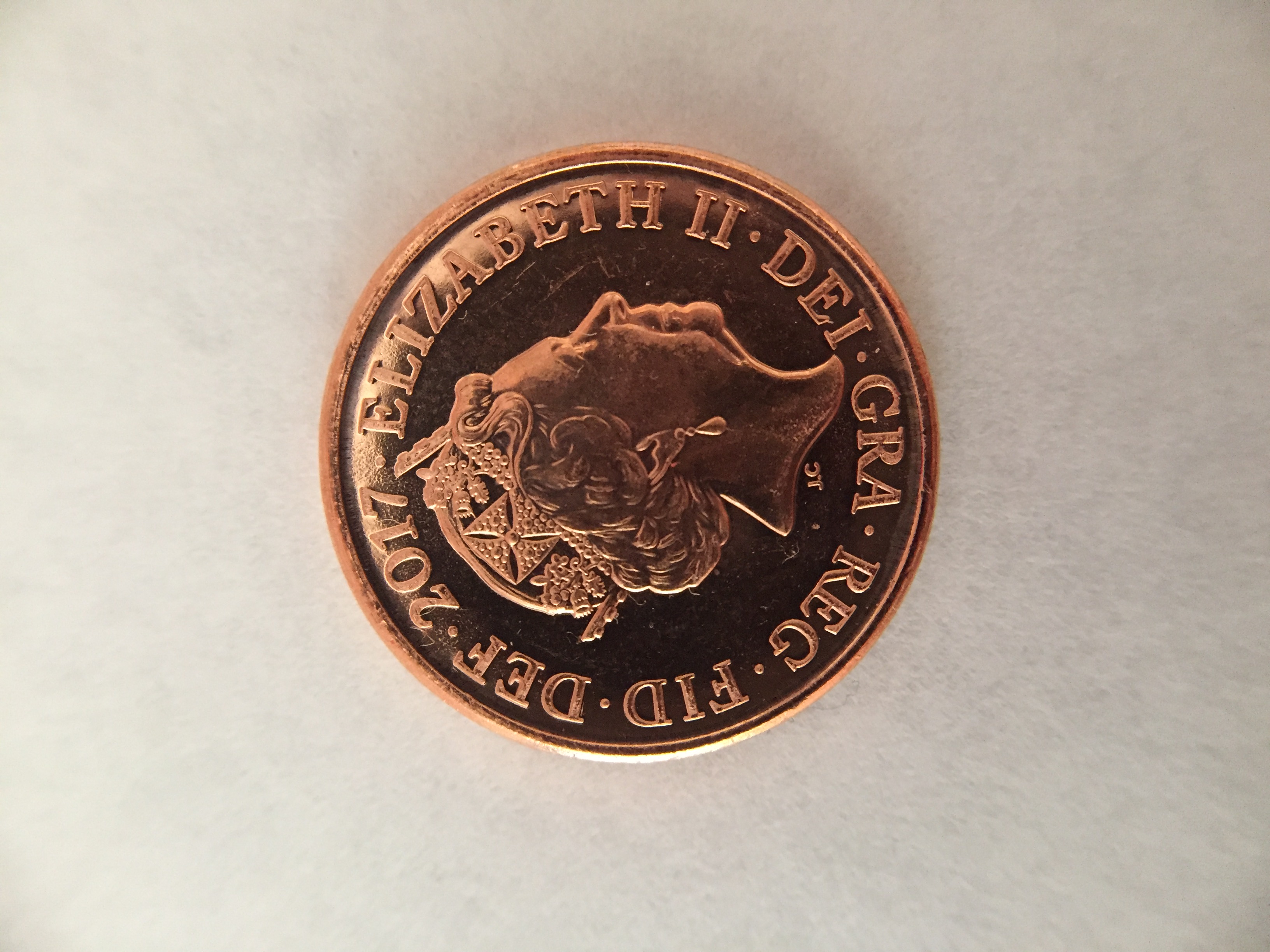}}
1p  (1971-Date) &
Circular Disc & 20.3 & 1.52 & 97\% Copper, 2.5\% Zinc and 0.5\% Tin & 1 & 4.03$\times 10^7$\\\hline
\raisebox{0.0\totalheight}{\includegraphics[width=0.11\textwidth, keepaspectratio, angle=-90]{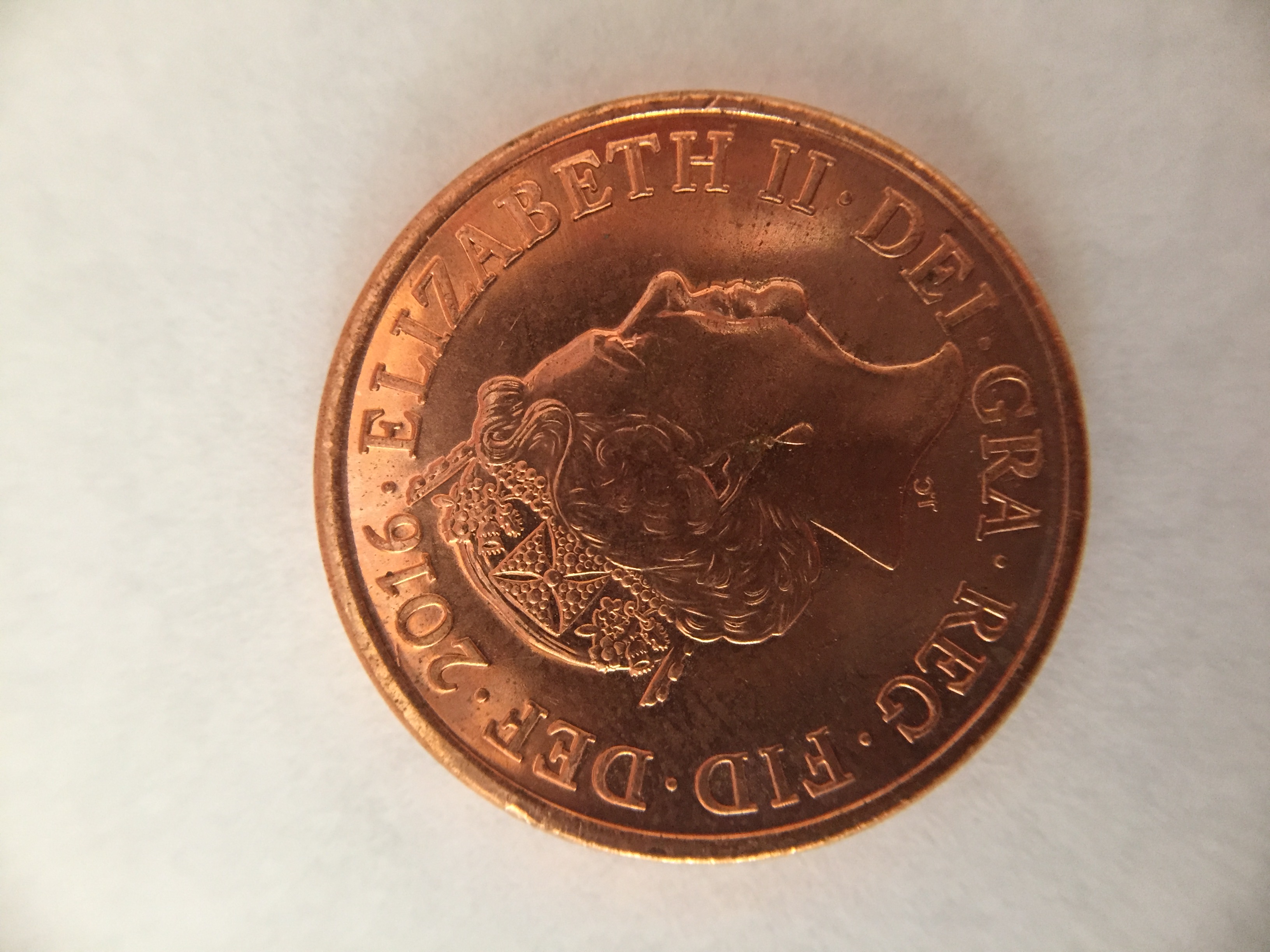}}  
2p (1971-Date) & Circular Disc& 25.9 & 2.03 & 97\% Copper, 2.5\% Zinc and 0.5\% Tin & 1 & 4.03$\times 10^7$\\\hline
 \raisebox{0.0\totalheight}{\includegraphics[width=0.11\textwidth, keepaspectratio,angle=-90]{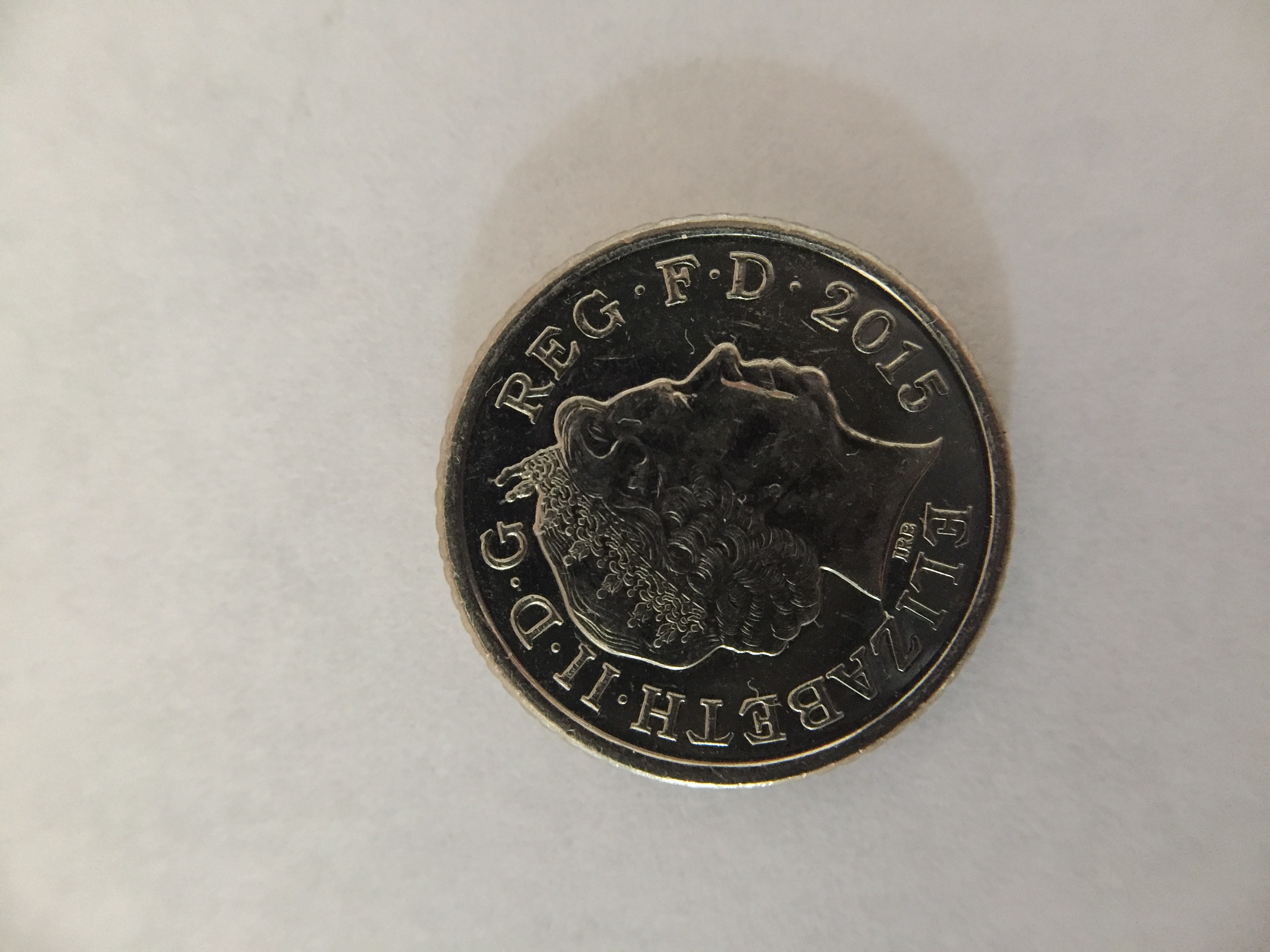}} 
5p (1990-Date) & Circular Disc &18 & 1.7 & 75\% Copper and 25\% Nickel & 1 & 2.91$\times 10^6$\\\hline
\raisebox{0.0\totalheight}{\includegraphics[width=0.11\textwidth, keepaspectratio, angle=-90]{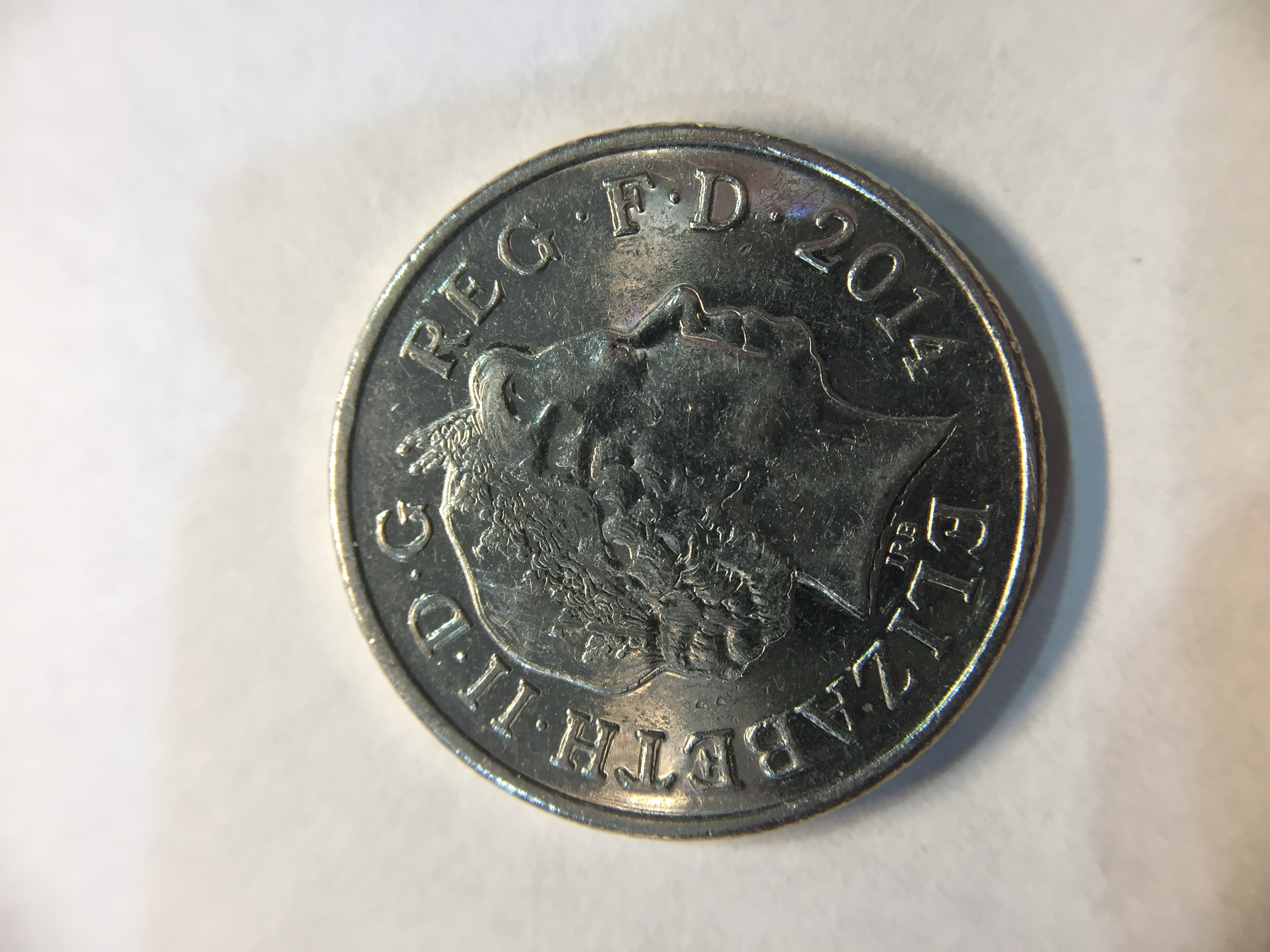}} \ \ \ \
10p (1990-Date) & Circular Disc &  24.5 & 1.85 & 75\% Copper and 25\% Nickel & 1 & 2.91$\times 10^6$\\\hline
\raisebox{0.0\totalheight}{\includegraphics[width=0.11\textwidth, keepaspectratio, angle=-90]{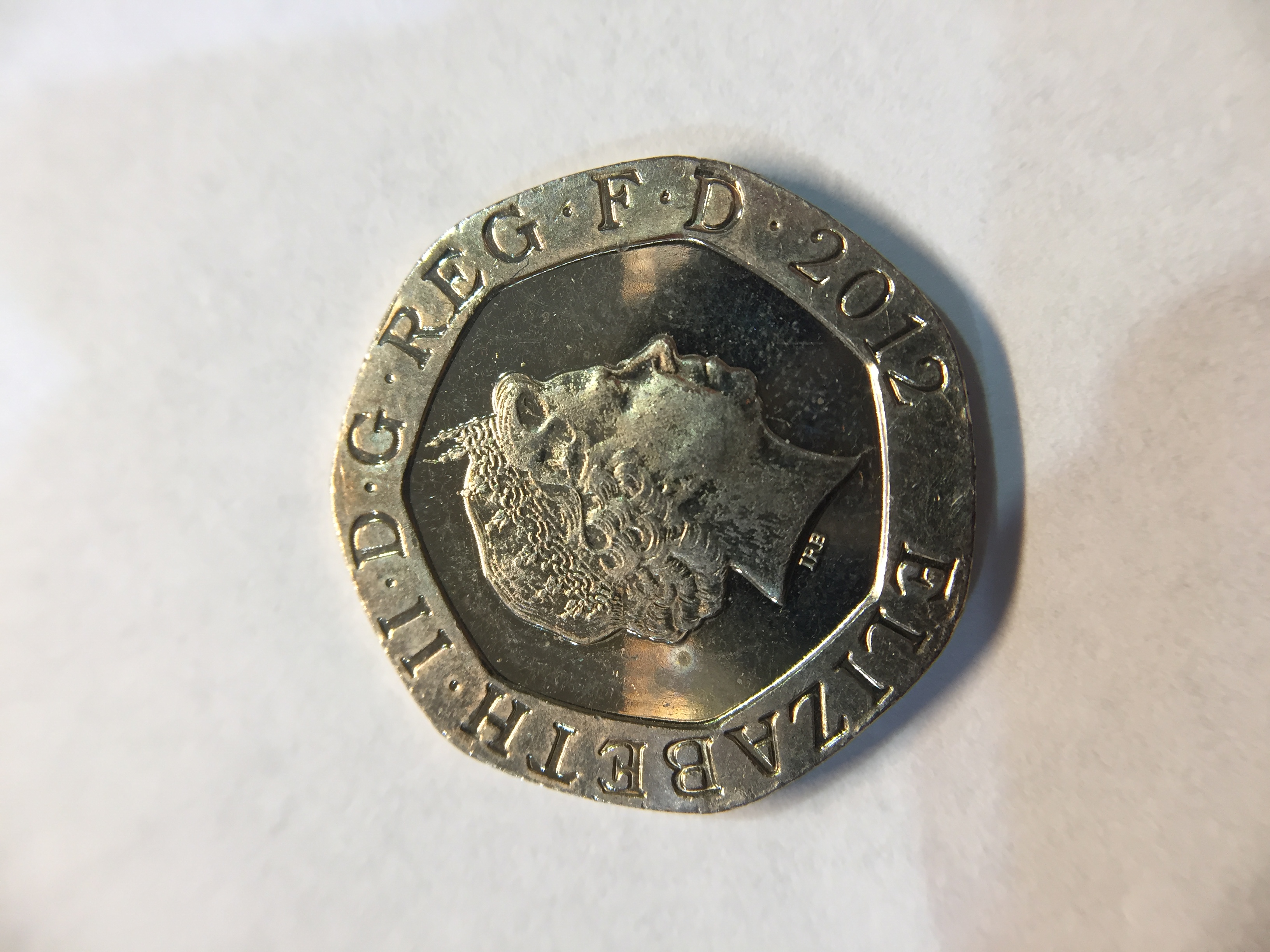}} \ \ \ \
20p (1982-Date) & Reuleaux Heptagonal Disc & 21.4 & 1.7 & 84\% Copper and 16\% Nickel & 1 & 5.26$\times 10^6$\\\hline
\raisebox{0.0\totalheight}{\includegraphics[width=0.11\textwidth, keepaspectratio, angle=-90]{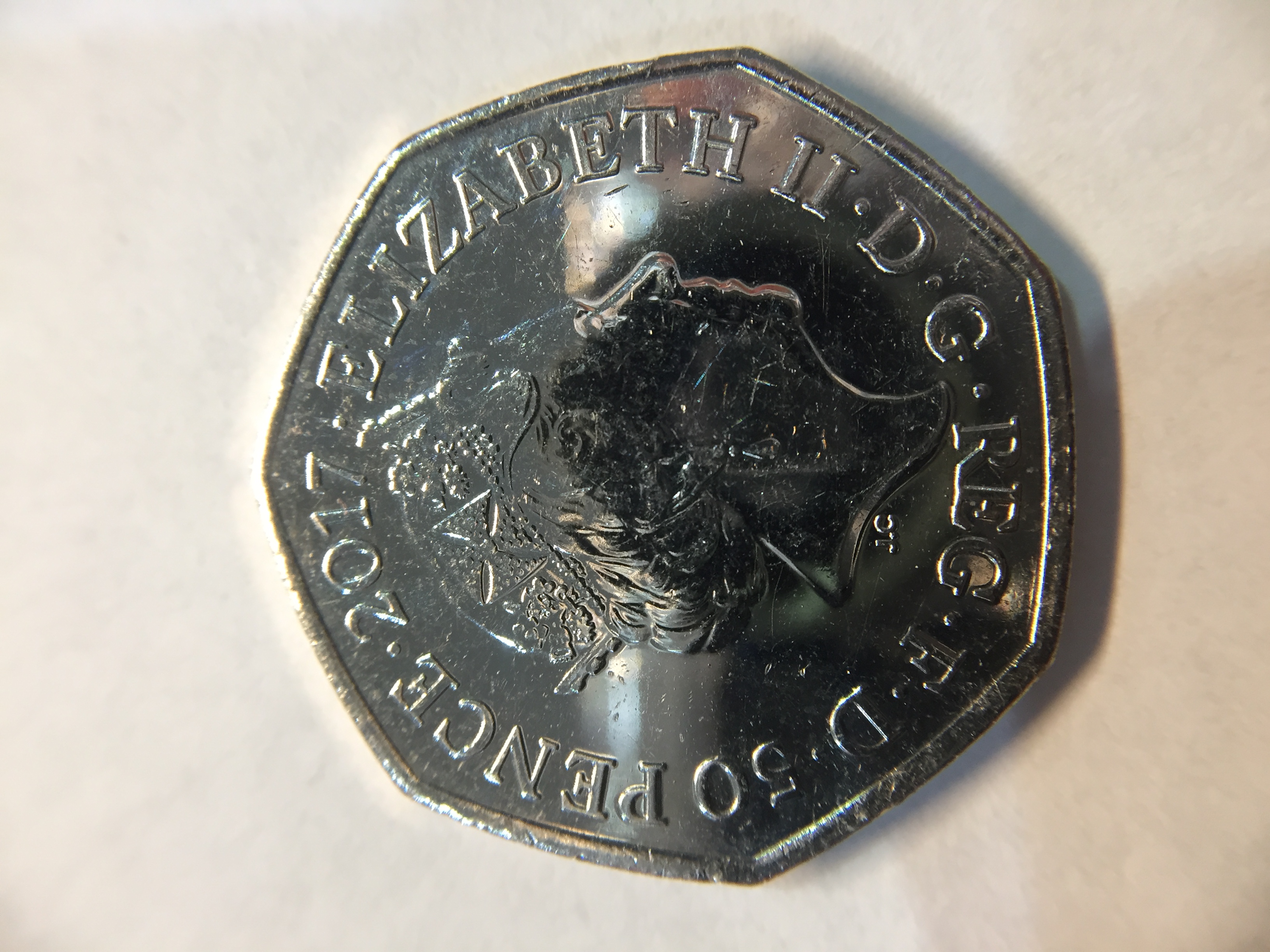}}  \ \ \ 
50p (1997-Date)
 & Reuleaux Heptagonal Disc & 27.3 & 1.78 & 75\% Copper and 25\% Nickel & 1 & 2.91$\times 10^6$\\\hline
\raisebox{0.0\totalheight}{ \includegraphics[width=0.11\textwidth, keepaspectratio, angle=-90]{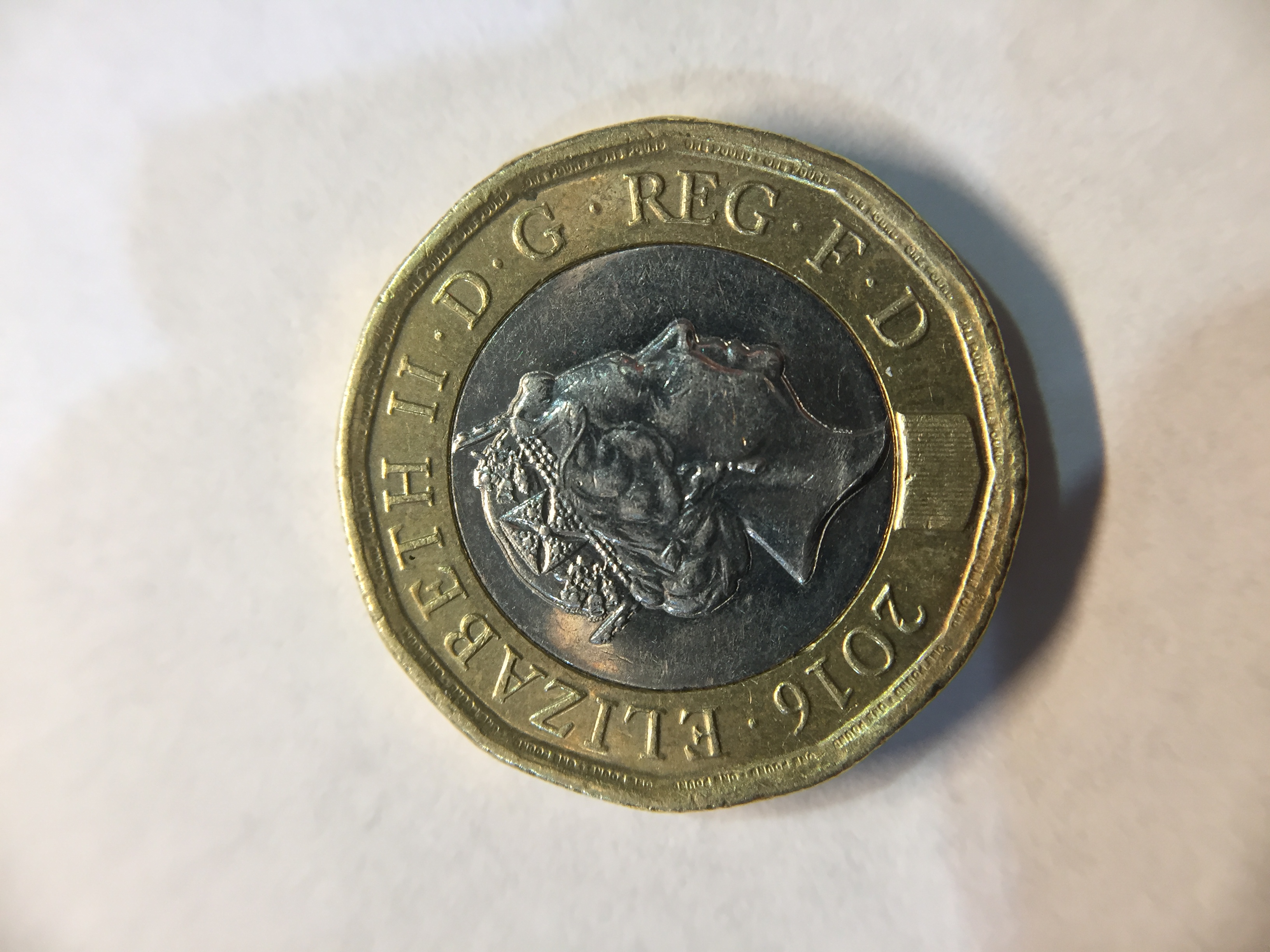} } \ \ \ \ 
\pounds 1 (2017-Date) & Annulus & 15.2 /23.45 (in/out) & 2.8 / 2.8 (in/out) & Nickel Plated Brass / 70\% Copper, 24.5\% Zinc and 5.5\% Nickel & 1 /1 (in/out)& 1.63$\times 10^7$ / 5.26$\times 10^6$ (in/out) \\\hline
\raisebox{0.0\totalheight}{ \includegraphics[width=0.11\textwidth, keepaspectratio, angle=-90]{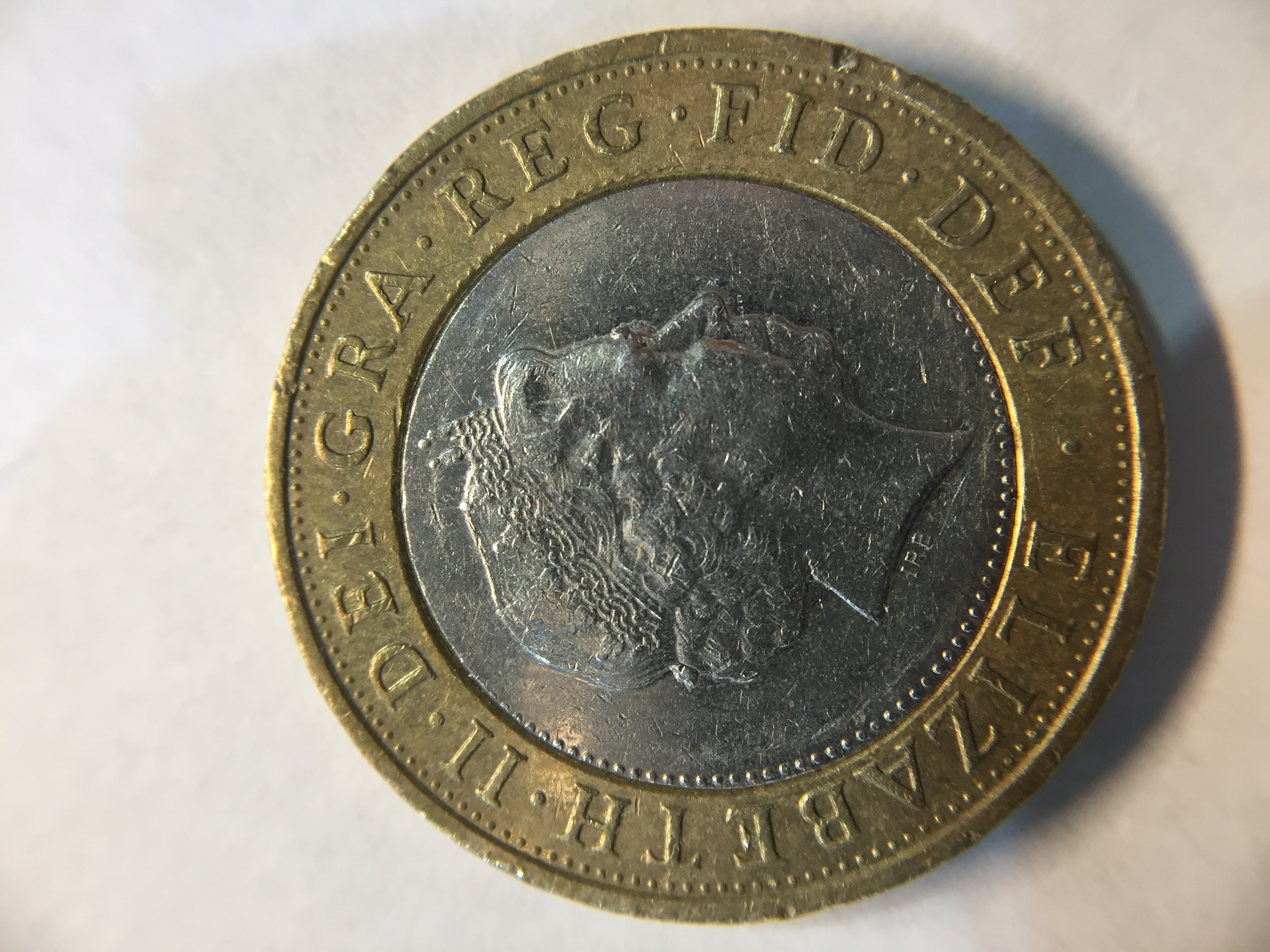}}   \ \ \ \ 
\pounds 2 (1998-Date) & Annulus &  21 /28.4  (in/out) & 2.5 /2.5 (in/out) & 75\% Copper and 25\% Nickel/ 97\% Copper, 2.5\% Zinc and 0.5\% Tin & 1 /1 (in/out) & 2.91$\times 10^6$ / 1.93$\times 10^7$ (in/out)\\\hline
\end{tabular}\\\bigskip
\caption{Set of British Coins 1p, 2p, 5,10p, \pounds 1 and \pounds 2 : Coin shape, dimensions and electrical properties.} \label{tab:Coins}
\end{center}
\end{table}

To model the 1p coin, we considered $B$ to be a circular disc of diameter 20.3 and thickness 1.52 and set $\alpha=0.001$ mm.  An unstructured mesh of $33\, 351$ unstructured tetrahedra was generated to model the object and the region surrounding it out to a truncation boundary in the form of the rectangular box $[-1000,1000]^3$. In a similar way, unstructured meshes of between $24\, 963$ and $36\, 957$ tetrahedra were generated to model the other coins. On these meshes, $p=4$ elements were found to be satisfactory for accurately computing the representative full order model solution snapshots. In order to produce the MPT spectral signature for the coins, $N=13$ representative full order solution snapshots were obtained at logarithmically spaced frequencies over the range $10^1 \le \omega \le 10^{10}  \text{rad/s}$ were used in combination with the PODP approach and a tolerance of $TOL =10^{-6}$. 

\begin{figure}[!h]
\centering
\hspace{-1.cm}
     \subfigure[$( \tilde{\mathcal{R}} )_{ii} $]{\includegraphics[scale=0.5]{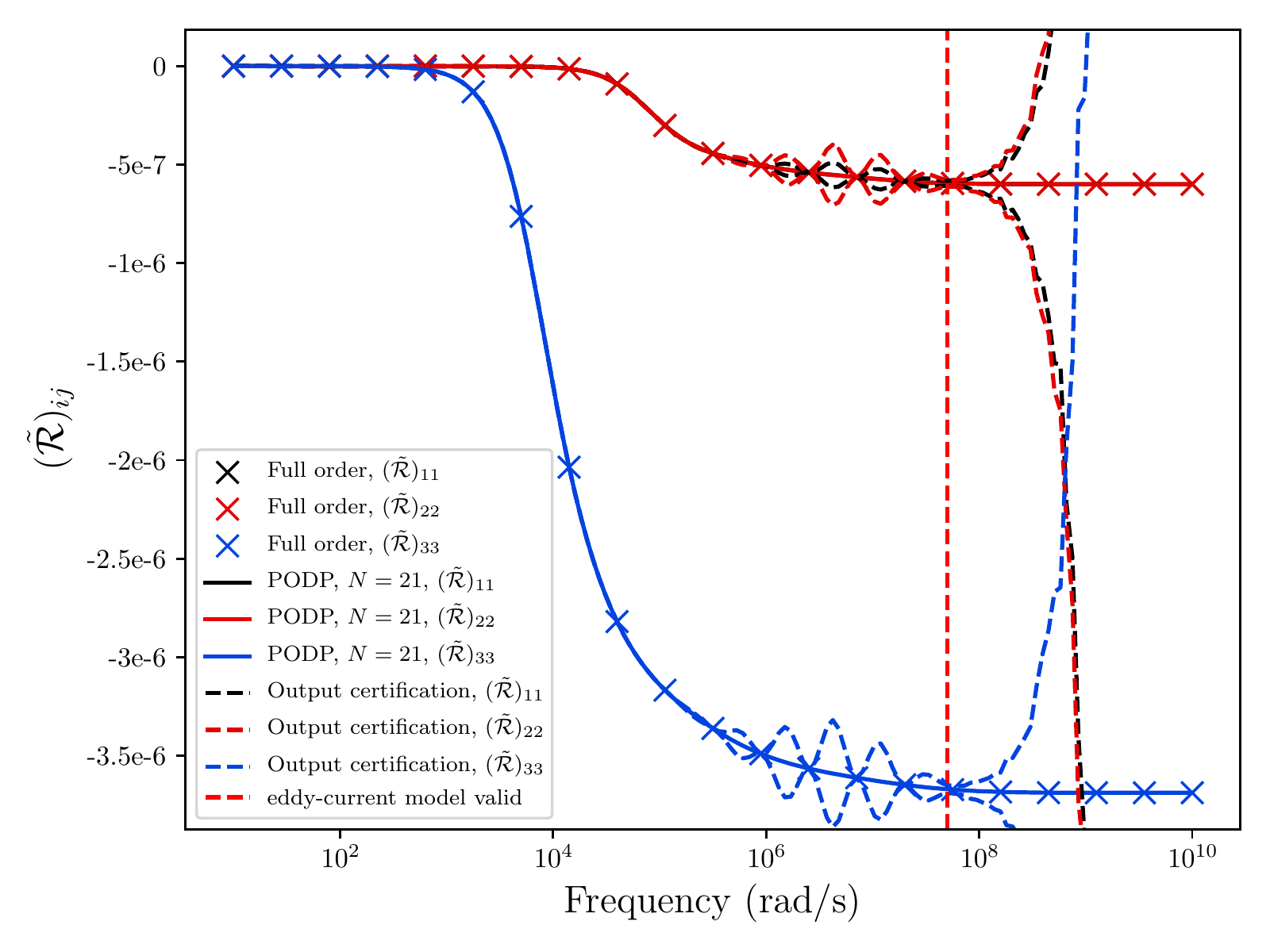}
    }
    \subfigure[$( \mathcal{I} )_{ii} $]{\includegraphics[scale=0.5]{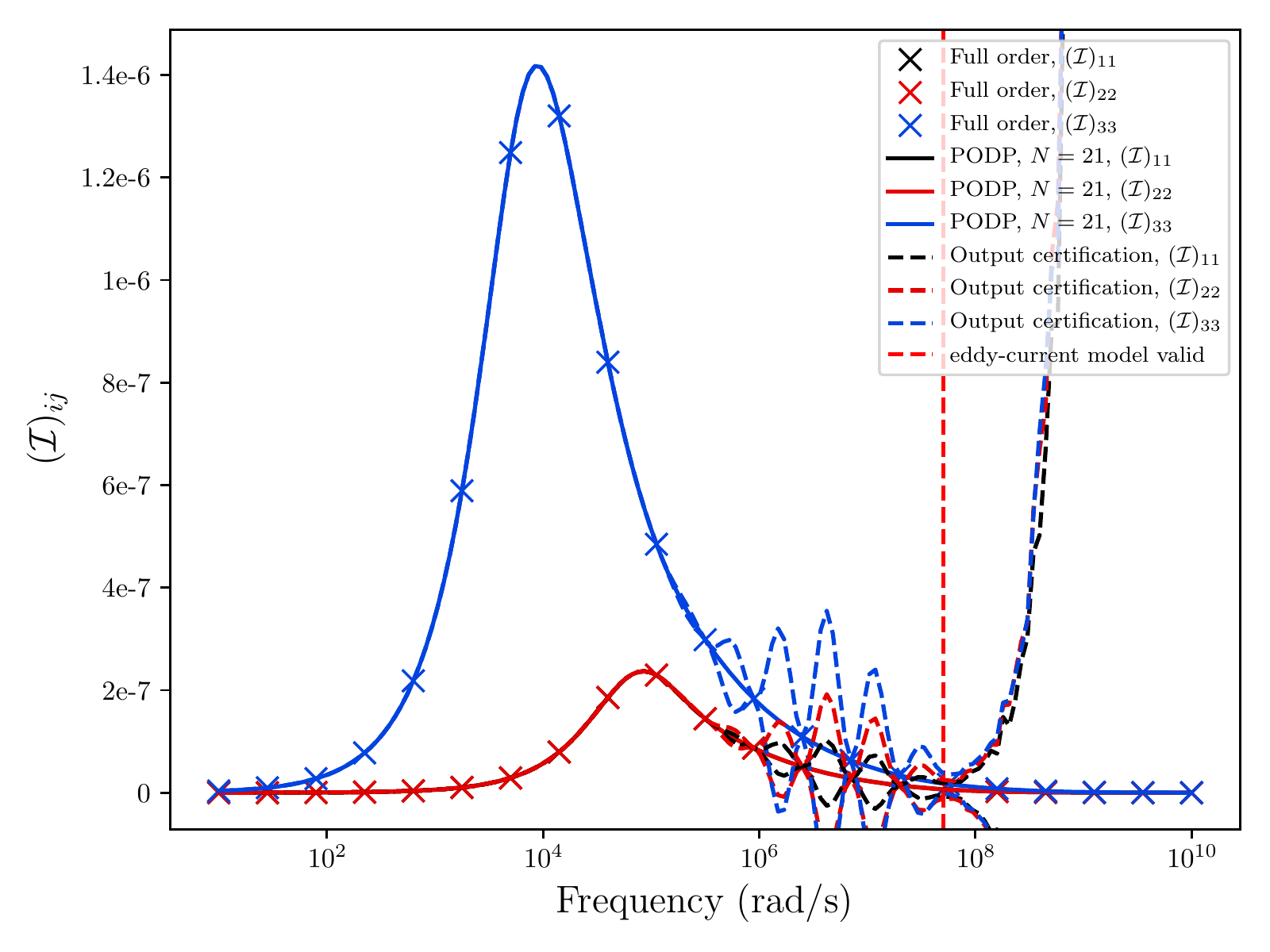}
    }

     \caption{1p Coin from set of British coins:  PODP prediction of the spectral signature showing also the frequencies used for the representative full order  solution snapshots and limiting frequency for  (a) $ ( \tilde{\mathcal{R}}  )_{ii} $ and (b) $ ( \mathcal{I}  )_{ii} $.}
\label{fig:ErrorBars}
\end{figure}

Although the PODP solutions are very acceptable using $N=13$ representative full order model solution snapshots, in order to achieve smaller a-posteriori error estimates,  results obtained with $N=21$  and $TOL=10^{-8}$ are considered and shown in Figure~\ref{fig:ErrorBars}. Also included in this figure is the limiting frequency $\omega_{limit}$ predicted by following the approach in Section~\ref{sect:irregtetexp}. 
The rotational symmetry of the object implies that the object has just two independent coefficients each in $  \tilde{\mathcal R}[\alpha B,\omega,  \sigma_*, \mu_r   ]$ and  $ {\mathcal I}  [\alpha B,\omega, \sigma_*, \mu_r  ]$, which lie on the diagonal of the tensors. Of these $( \tilde{\mathcal R}    )_{33} $ and  $ ( \mathcal{I}   )_{33}$ have the largest magnitude in a direction that is perpendicular to the plane of the disc, which is as expected for a non-magnetic disc~\cite{davidsoncoins}. Note that the 1p coin issued after 1992, which has a high $\mu_r$ value, would have dominant components $ ( \tilde{\mathcal  R}    )_{11} =  ( \tilde{\mathcal R}   )_{22} $ and  $ ( \mathcal{I}   )_{11}= ( \mathcal{I} )_{22}$ in the plane of the disc, as expected for a magnetic disc~\cite{davidsoncoins}.
A similar study was performed for each of the coins listed in Table~\ref{tab:Coins} in~\cite{thesisben} in order to ensure the results were accurate.

\begin{figure}[!h]
\centering
\hspace{-1.cm}
$\begin{array}{cc}
\includegraphics[scale=0.5]{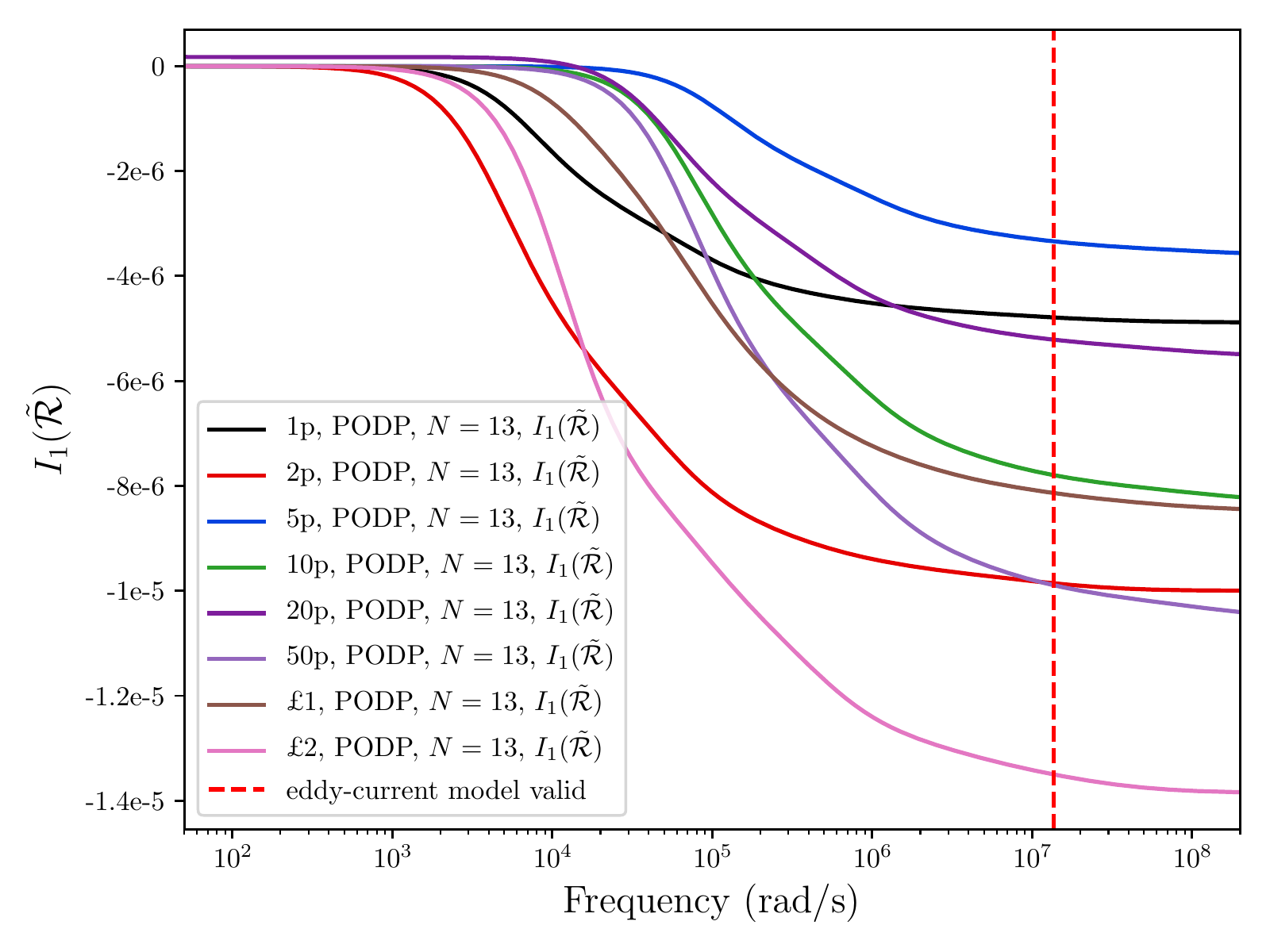} &
 \includegraphics[scale=0.5]{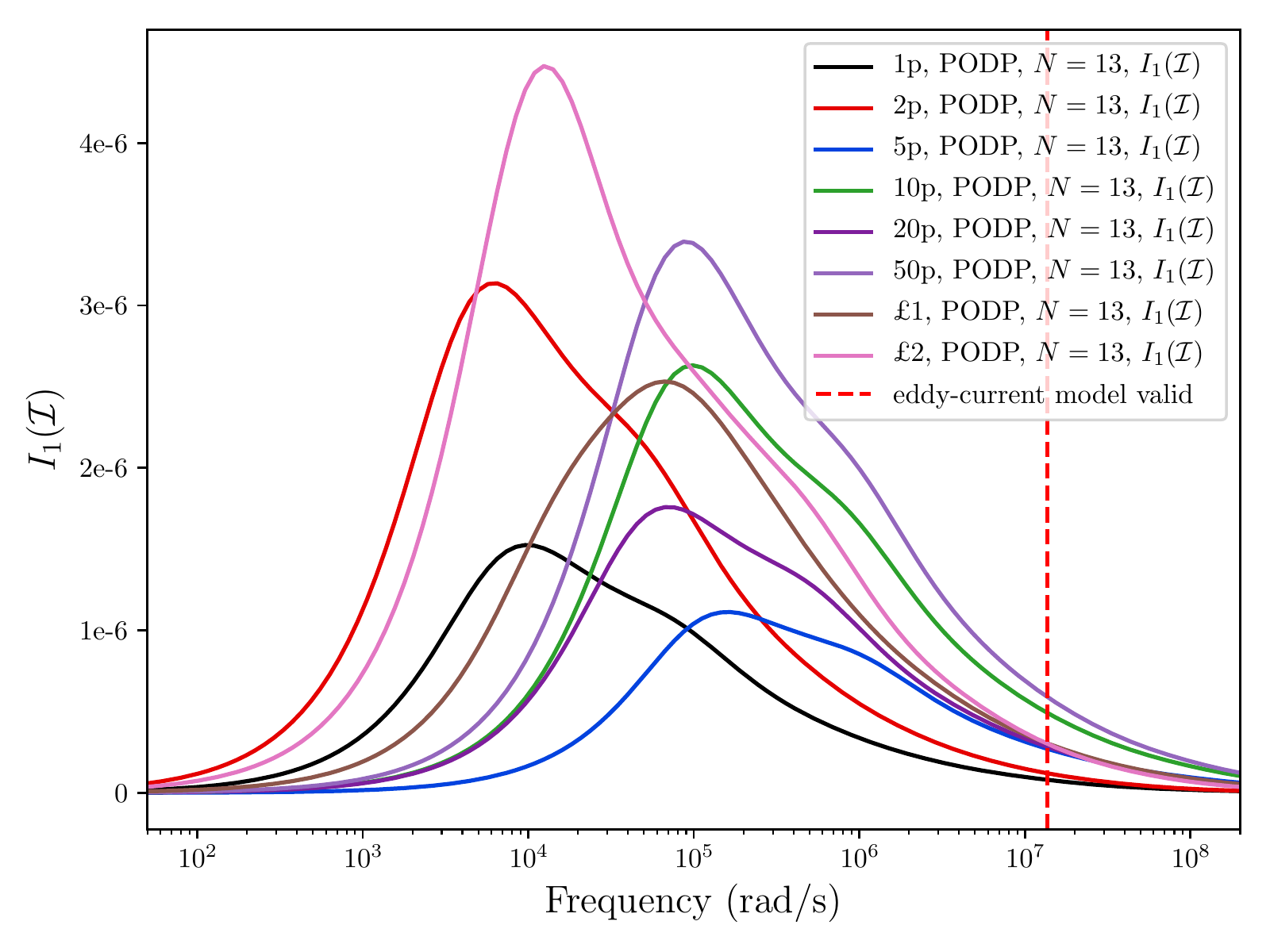}  \\
\text{(a) } I_{1} ( \tilde{\mathcal{R}} ) & 
\text{(b) } I_{1}( \mathcal{I} )  \\
\includegraphics[scale=0.5]{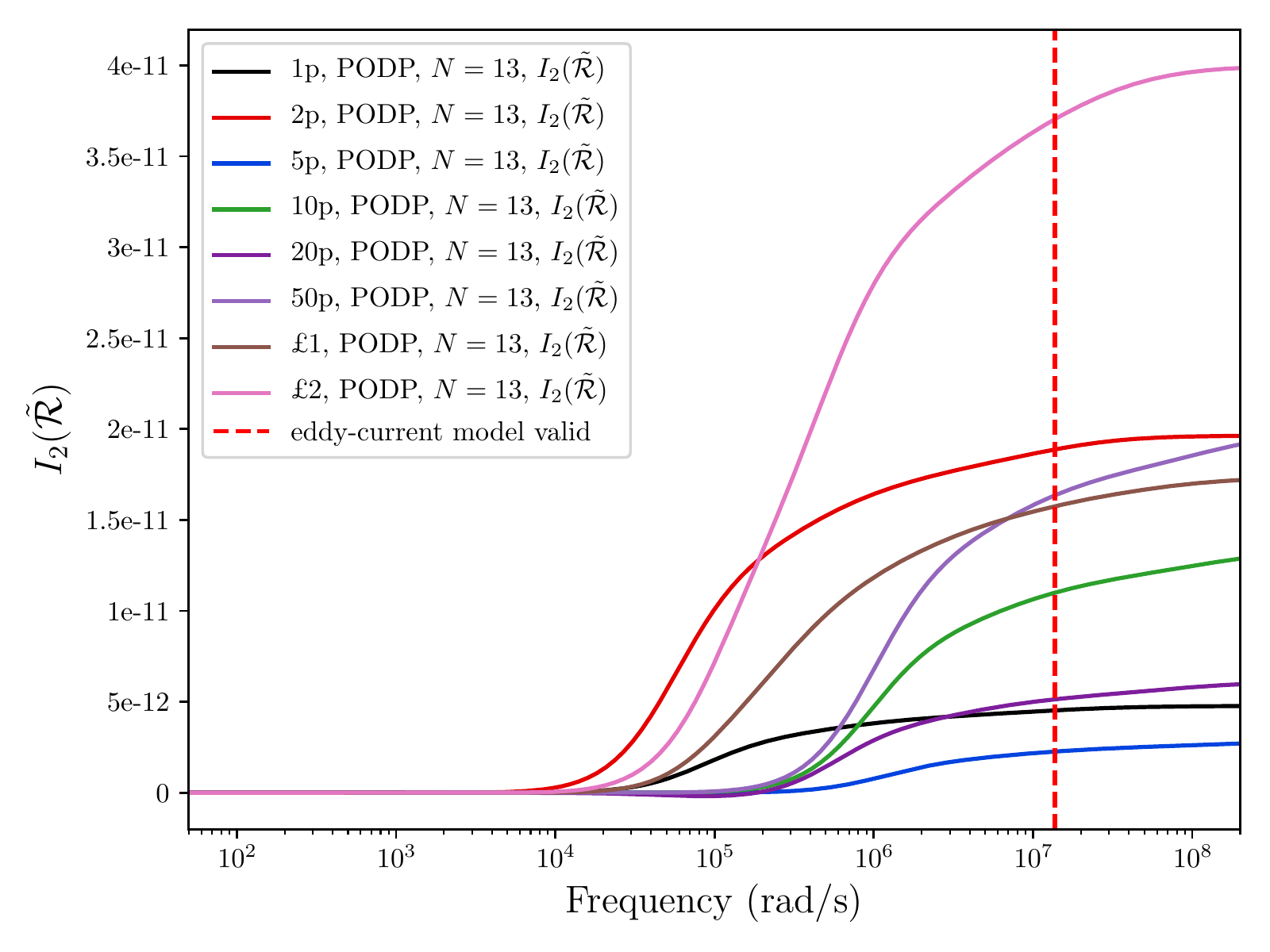} &
 \includegraphics[scale=0.5]{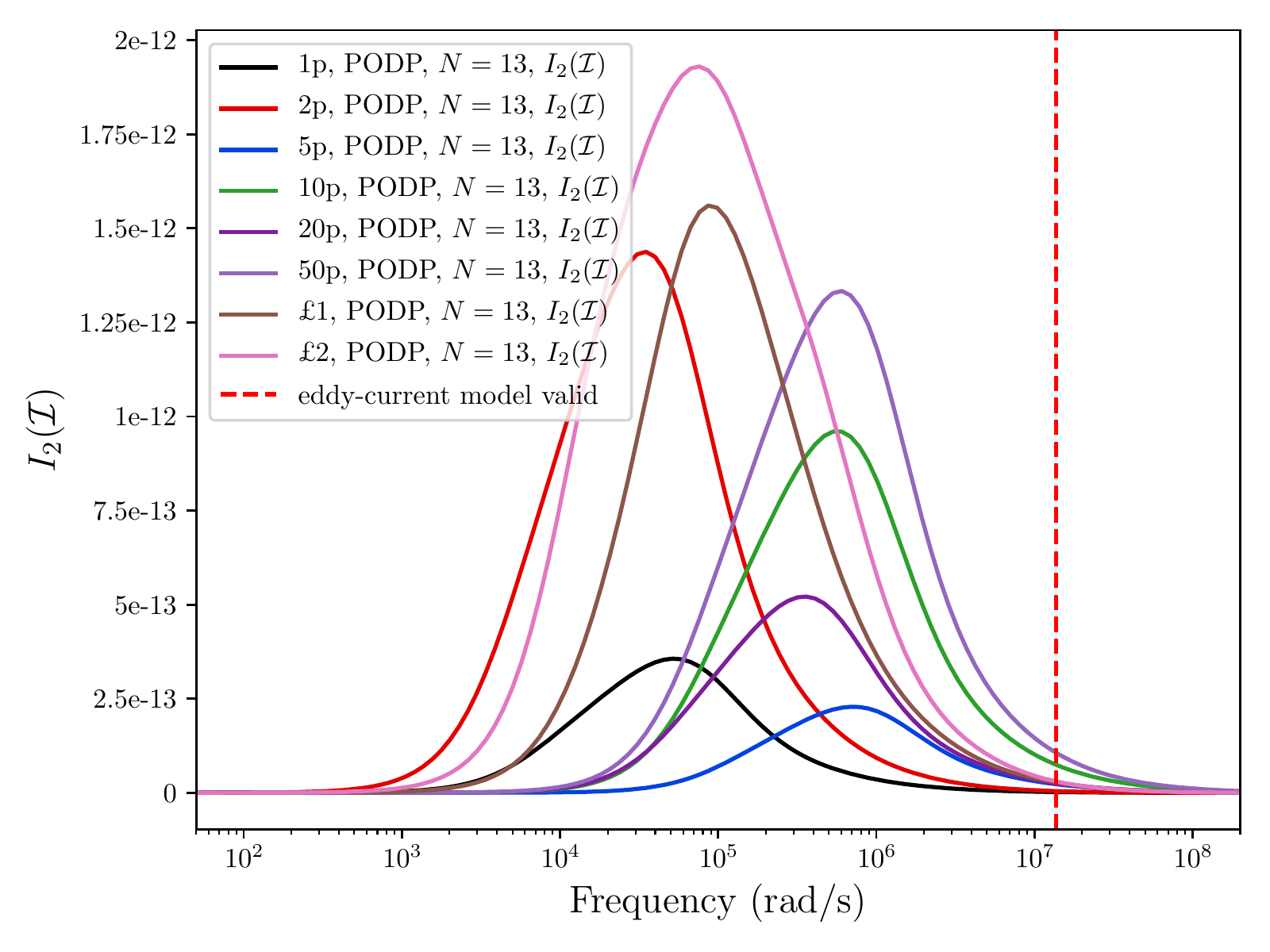}  \\
\text{(c) } I_{2} ( \tilde{\mathcal{R}} ) & 
\text{(d) } I_{2 }( \mathcal{I} )  \\
\includegraphics[scale=0.5]{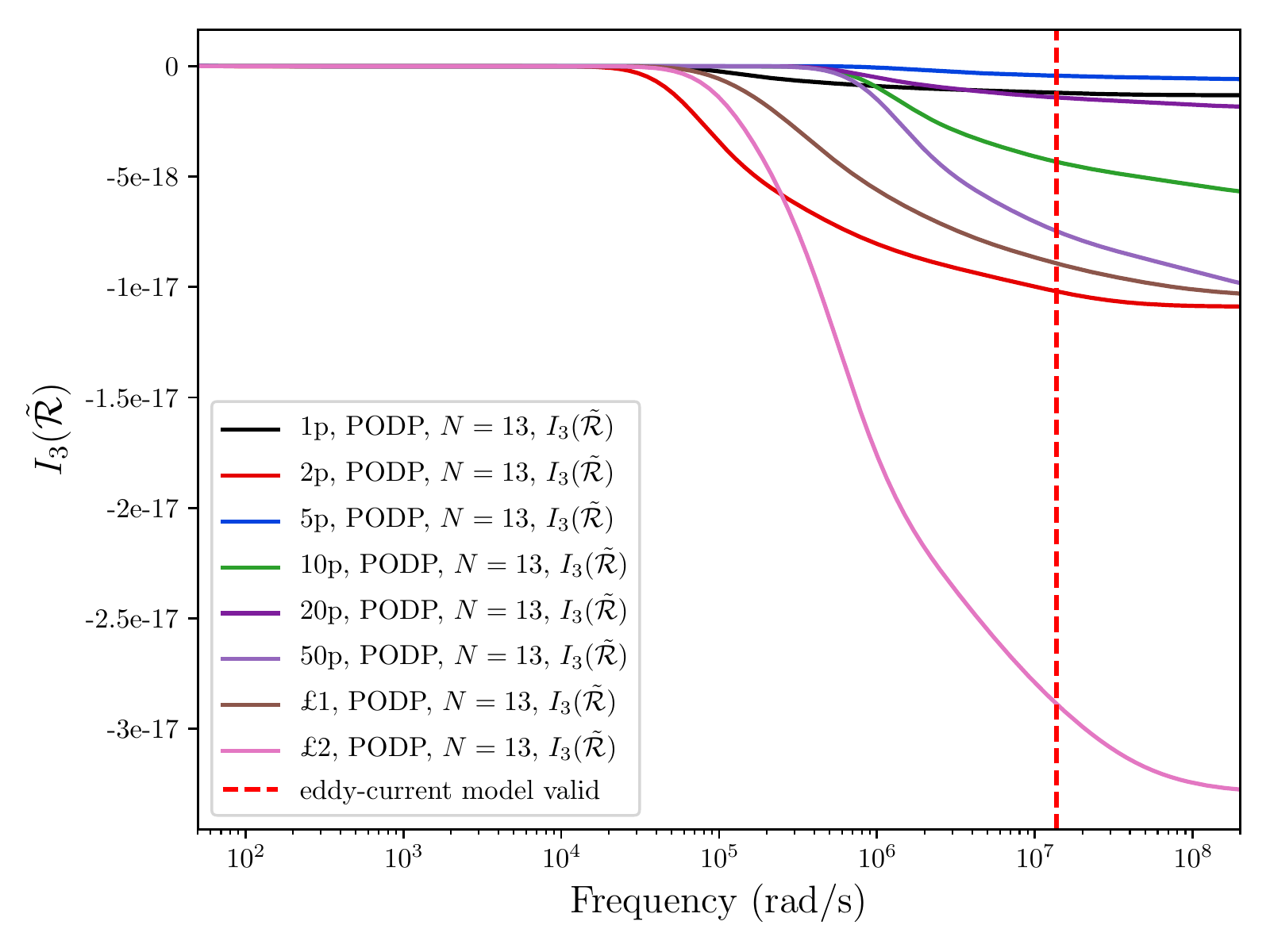} &
 \includegraphics[scale=0.5]{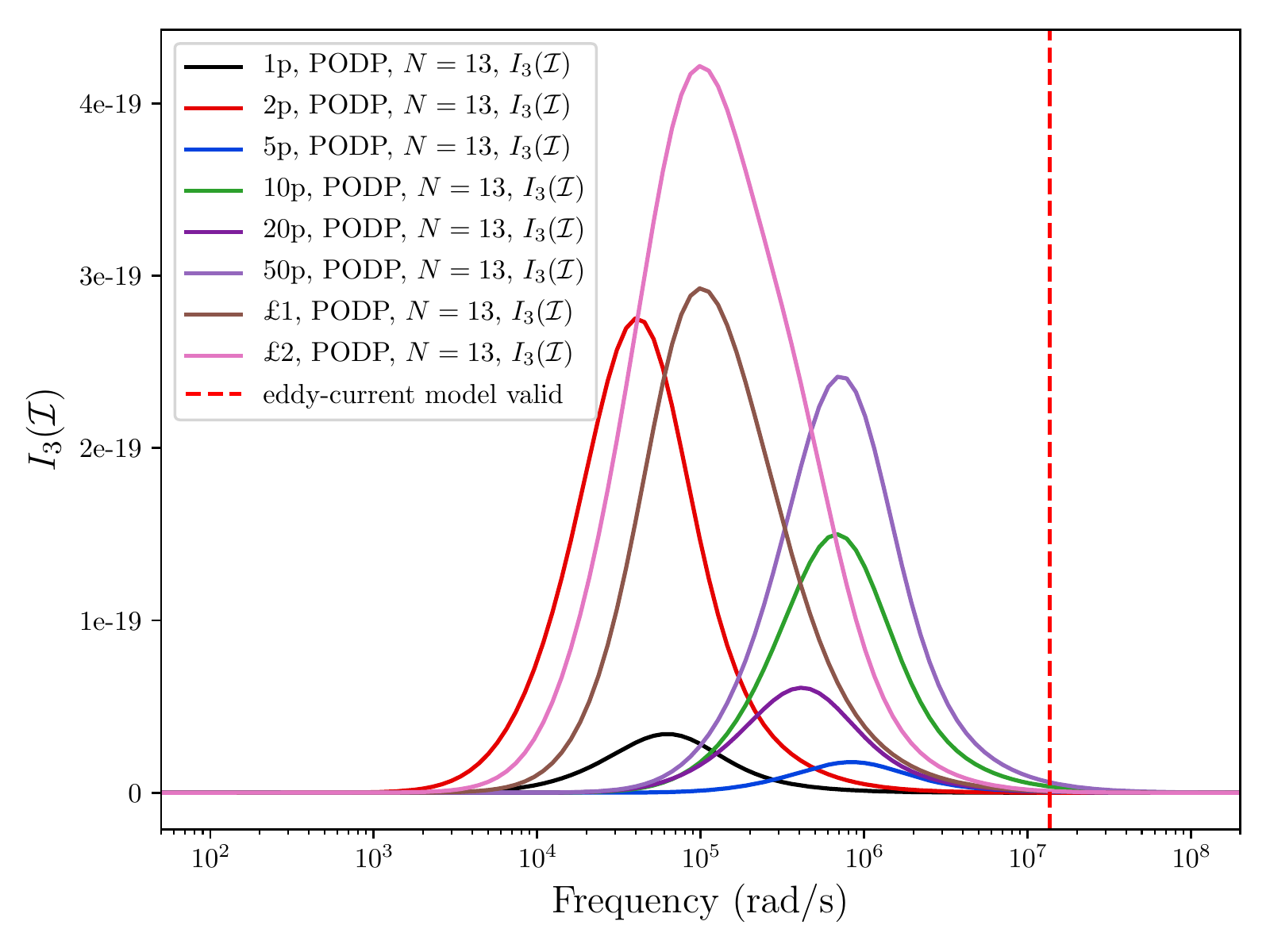}  \\
\text{(e) } I_{3} ( \tilde{\mathcal{R}} ) &
\text{(f) }  I_{3} ( \mathcal{I} )  
\end{array}$
  \caption{British coins: Comparison of tensor invariants. (a) $I_{1} ( \tilde{\mathcal{R}} ) $, (b) $I_{1}( \mathcal{I} ) $
  (c) $I_{2} ( \tilde{\mathcal{R}} ) $, (d) $I_{2}( \mathcal{I} ) $,
  (e) $I_{3} ( \tilde{\mathcal{R}} ) $ and (f)  $I_{3}( \mathcal{I} ) $.}
        \label{fig:CompInv}
\end{figure}

To compare the results for the different coins, we present the MPT spectral signature using the principal invariants $I_i$, $i=1,2,3$, for $\tilde{\mathcal R} [\alpha B,\omega,  \sigma_*, \mu_r   ]$ and ${\mathcal I} [\alpha B,\omega, \sigma_*, \mu_r  ]$ that have been obtained using the PODP approach in Figure~\ref{fig:CompInv}. In this figure, we have restricted consideration to frequencies such that $10^2 \le \omega \le 10^8 \text{rad/s}$ in order to allow comparisons with the earlier key results. In practice, the eddy current model brakes down at a frequency of $\omega_{limit} < 10^8 \text{rad/s}$  (or greater) for all the coins considered and so higher frequencies are physically invalid in any case.
{Unlike the corresponding results for the house keys shown in Figures~\ref{fig:set1:fineMesh:I1_I3} and~\ref{fig:set2:fineMesh:I1_I3}, the results obtained for the coins shown in Figure~\ref{fig:CompInv} do not form a family of similar curves since both the volumes and materials of the coins vary significantly motivating the ability to discriminate between different coins, however, some of the trends previously observed cary over to this case also. The curves for $I_1( \tilde{\mathcal R})$ and $I_3( \tilde{\mathcal R})$ are monotonically decreasing with $\log \omega$, while 
$I_2(\tilde {\mathcal R})$  is monotonically increasing with $\log \omega$.  The curves for $I_i({\mathcal I} ) $, $i=1,2,3$, each have a single local maximum, although the peaks appear at different frequencies for different coins and the different invariants, however, the width of the frequency band reduces for all cases, when considering $I_2({\mathcal I} ) $ and $I_3({\mathcal I} ) $ compared to $I_1({\mathcal I} ) $.
On considering the different coins, we see similarities between the MPT spectral signatures of the 1p, 2p coins, the 5p, 10p coins and the 20p and 50p coins.  This can be explained as follows: the composition of the coins in these groups is the same and their dimensions can be approximately obtained by a simple scaling, hence, the scaling results in Lemma 5.2 of~\cite{ben2020} predict that the tensor coefficients of the larger sized coin can be obtained from the smaller object by a translation and scaling, which is also observed in the invariants. The 50p coin has the largest volume and also the highest peak value in $I_3({\mathcal I})$, the magnitude of the peaks reduce in sequence of the volumes of the coins, as expected. The multiple local maxima in the coefficients of ${\mathcal I}$  and the multiple points of inflection in the coefficients of  $\tilde{\mathcal R} $, which are known to be associated with objects with inhomogeneous conductivity~\cite{LedgerLionheartamad2019}, are not easily distinguished on the invariants for the \pounds 1 and \pounds 2 coins, 
probably due to the difference in conductivities being approximately 1 order of magnitude or less.}

\begin{figure}[!h]
\centering
\hspace{-1.cm}
$\begin{array}{cc}
\includegraphics[scale=0.5]{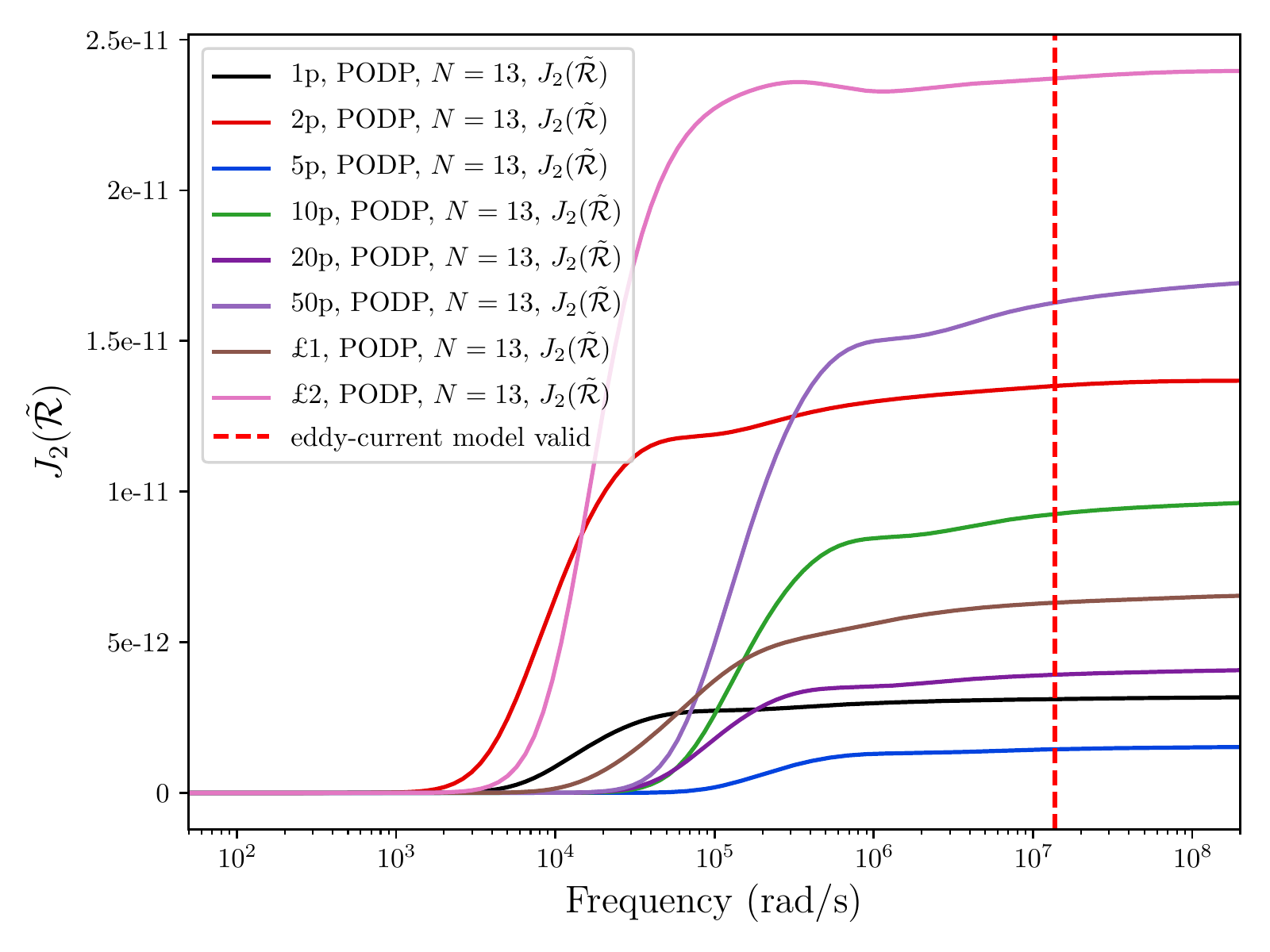} &
 \includegraphics[scale=0.5]{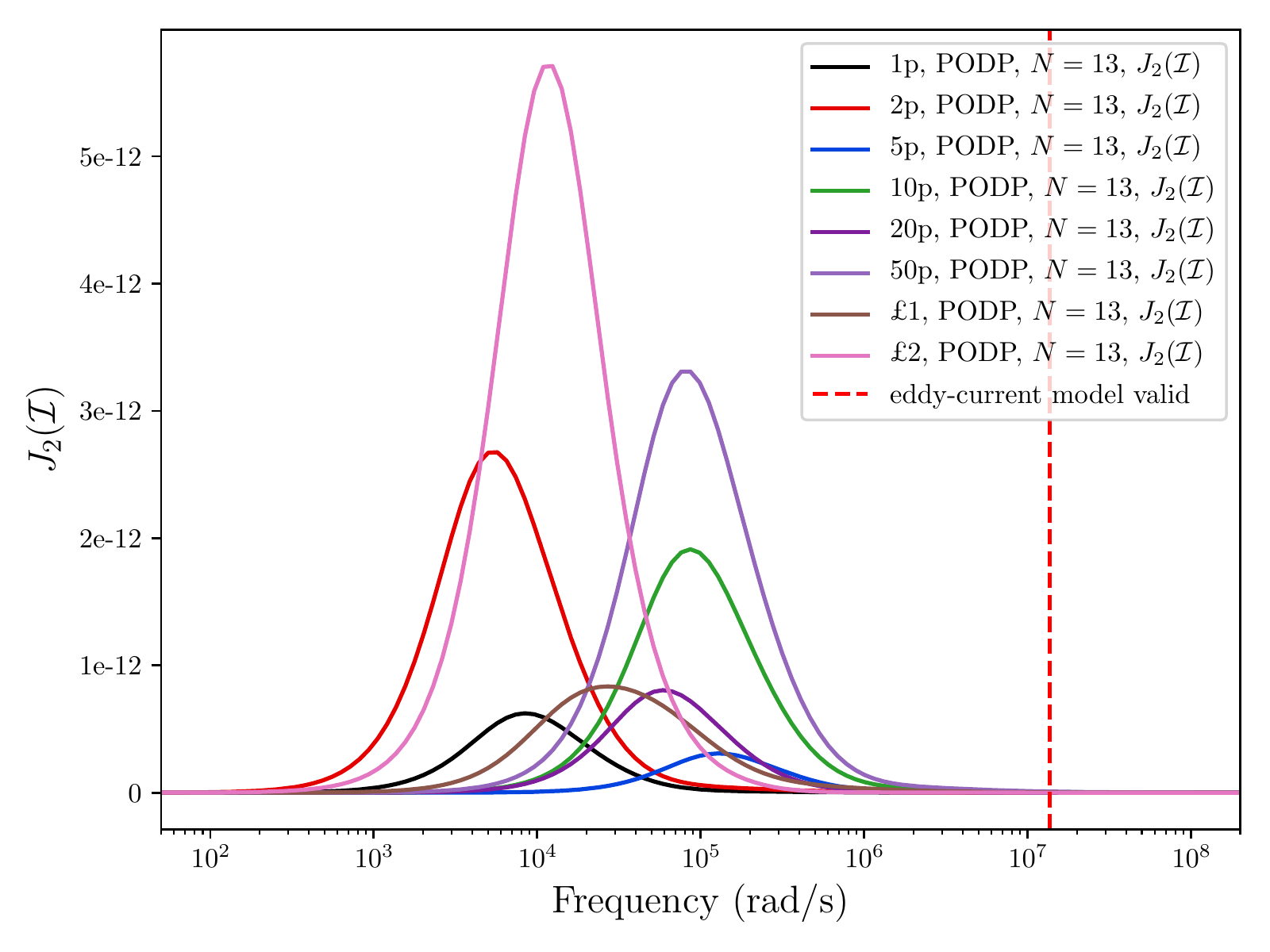}  \\
\text{(a) } J_{2}  ( \tilde{\mathcal{R}}  ) & 
\text{(b) } J_{2 } ( \mathcal{I} )  \\
\includegraphics[scale=0.5]{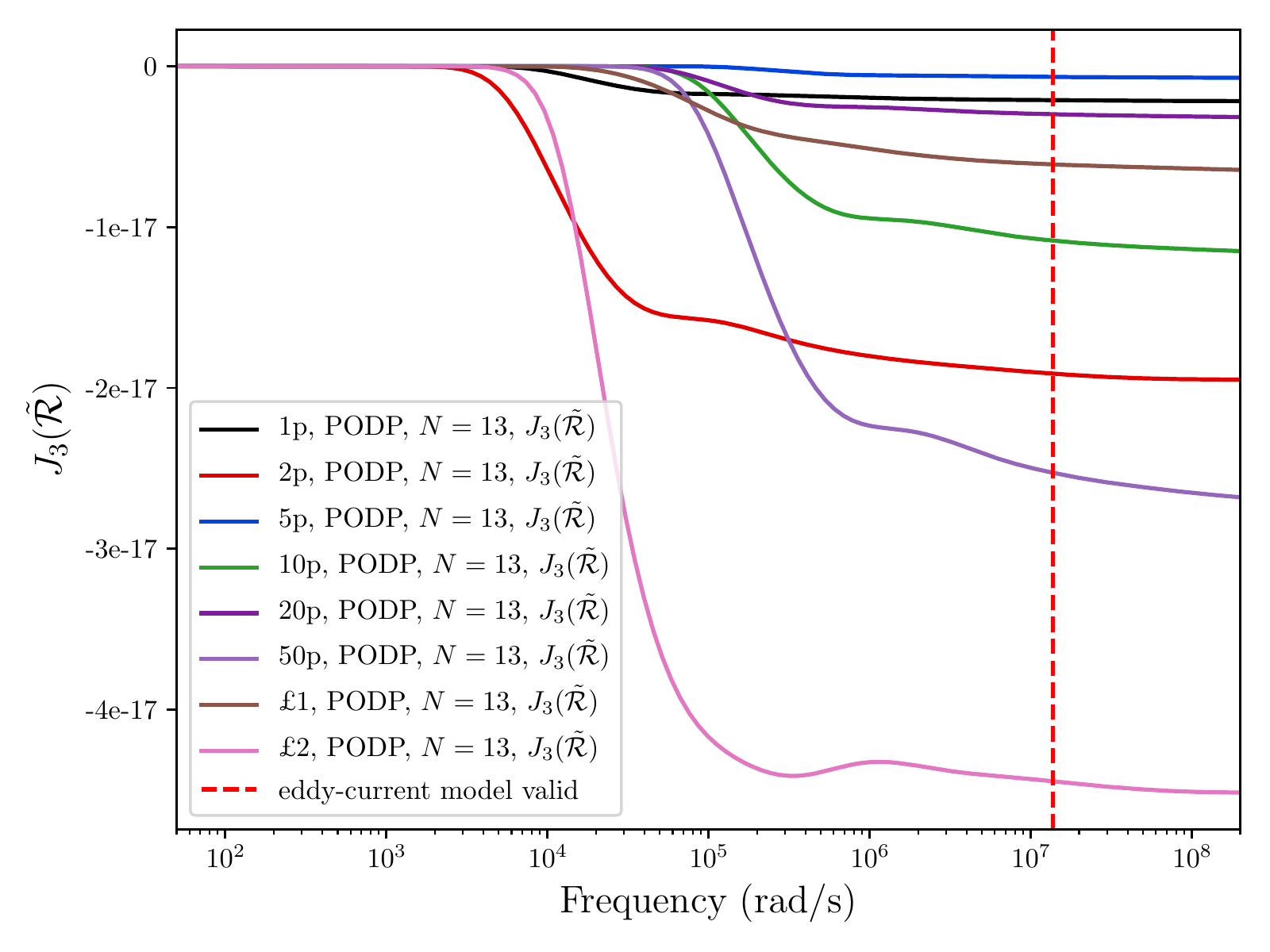} &
 \includegraphics[scale=0.5]{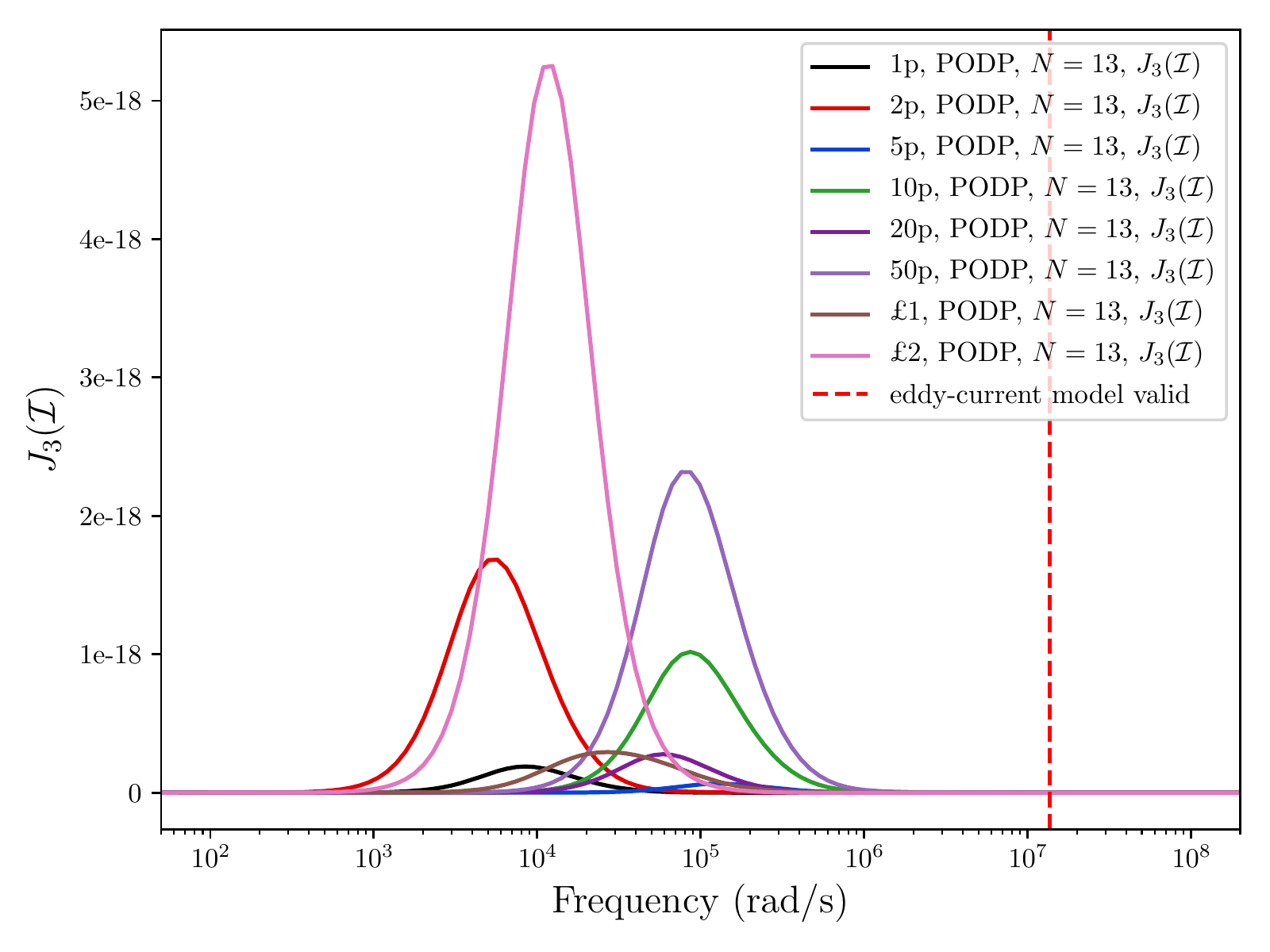}  \\
\text{(c) } J_{3}  ( \tilde{\mathcal{R}} ) &
\text{(d) }  J_{3} ( \mathcal{I} )  
\end{array}$
  \caption{British coins: Comparison of tensor invariants.  (a) $J_{2} ( \tilde{\mathcal{R}} ) $, (b) $J_{2} ( \mathcal{I} ) $,
  (c) $J_{3}  ( \tilde{\mathcal{R}} ) $ and (d)  $J_{3} ( \mathcal{I} ) $.}
        \label{fig:CompDevInv}
\end{figure}

The corresponding results obtained for the alternative invariants  $J_i$, $i=2,3$, for $ \tilde{\mathcal R} [\alpha B,\omega, \sigma_*, \mu_r   ]$ and ${\mathcal I} [\alpha B,\omega, \sigma_*, \mu_r   ]$ are presented in Figure~\ref{fig:CompDevInv}. 
Again, unlike the keys, we see that the plots of these invariants do not form a family of similar curves as both the volumes and conductivities of the different coins are different. 
{While  for most coins $J_2(\tilde{\mathcal R})$ is monotonically increasing with $\log \omega$ and $J_3(\tilde{\mathcal R})$ is monotonically decreasing with $\log \omega$, we see that there are exceptions, most notably with the \pounds 2 coin, which can be explained by its inhomogeneous materials. One might expect a similar behaviour with the inhomogeneous \pounds 1 coin, but it is difficult to observe on this scale. The behaviour of $J_2({\mathcal I})$ and  $J_3({\mathcal I})$ with $\log \omega$ shows a single local maximum for each coin where the presence of multiple local maxima for the \pounds 1 and \pounds 2 can't be observed on this scale. }
Of the coins considered, the curves associated with the 2p, 50p and \pounds 2 cases have the largest magnitude, indicating that they have the largest deviatoric component, which is also expected, given the geometries of these objects.


\begin{figure}[!h]
\begin{center}
$\begin{array}{ccc}
\includegraphics[width=0.5\textwidth, keepaspectratio]{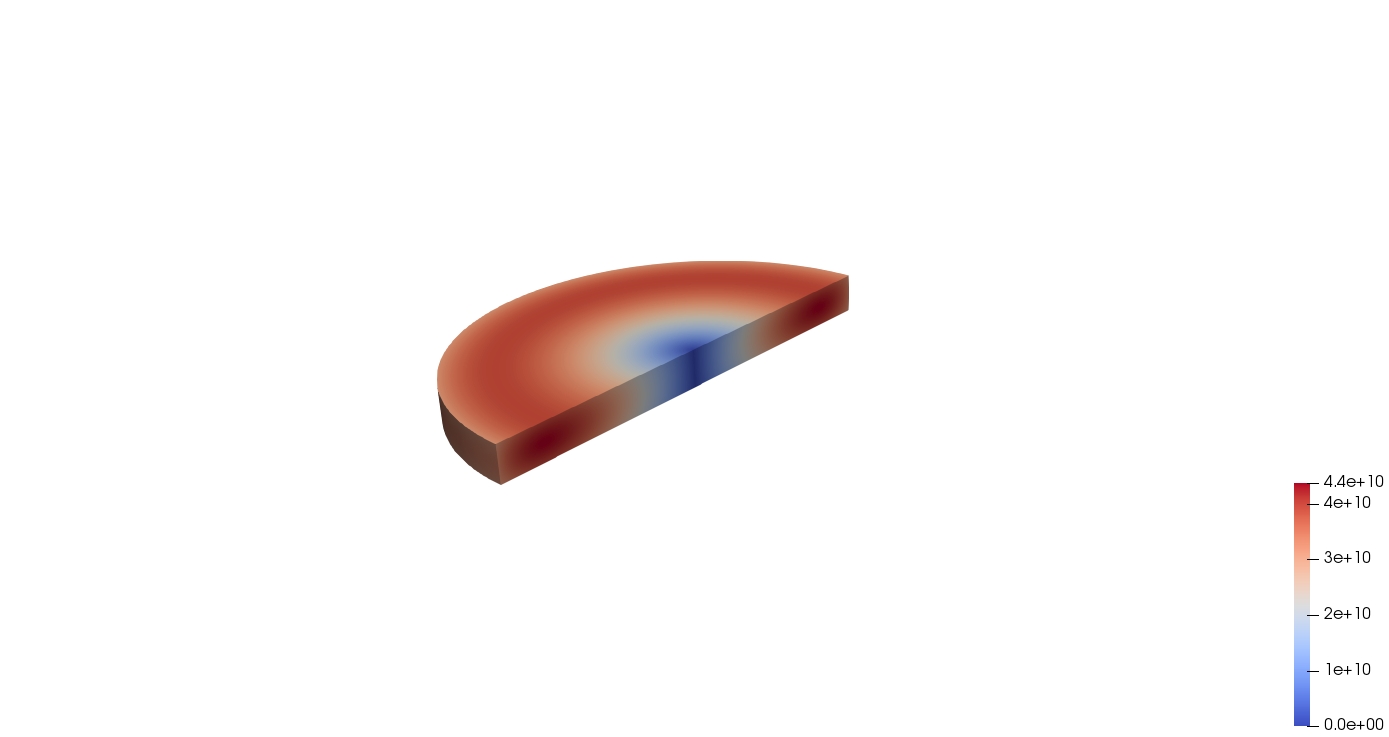} &
\includegraphics[width=0.5\textwidth, keepaspectratio]{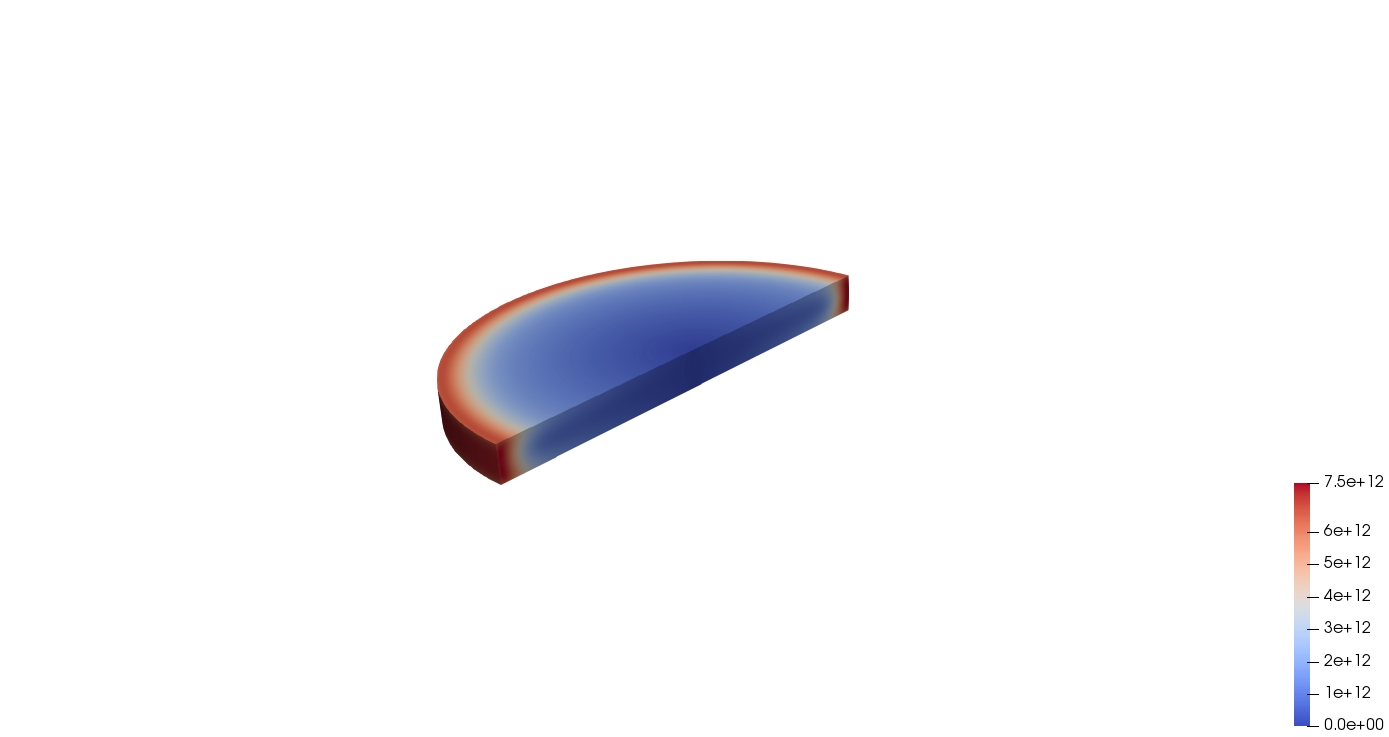} \\
\textrm{\footnotesize{(a) $| \text{Re}({\bm J}^e)|$ for $\omega=10^3$ rad/s.}} &
\textrm{\footnotesize{(b) $| \text{Re}({\bm J}^e )|$ for $\omega=10^5$ rad/s.}} 
\end{array}$\\
$\begin{array}{c}
\includegraphics[width=0.5\textwidth, keepaspectratio]{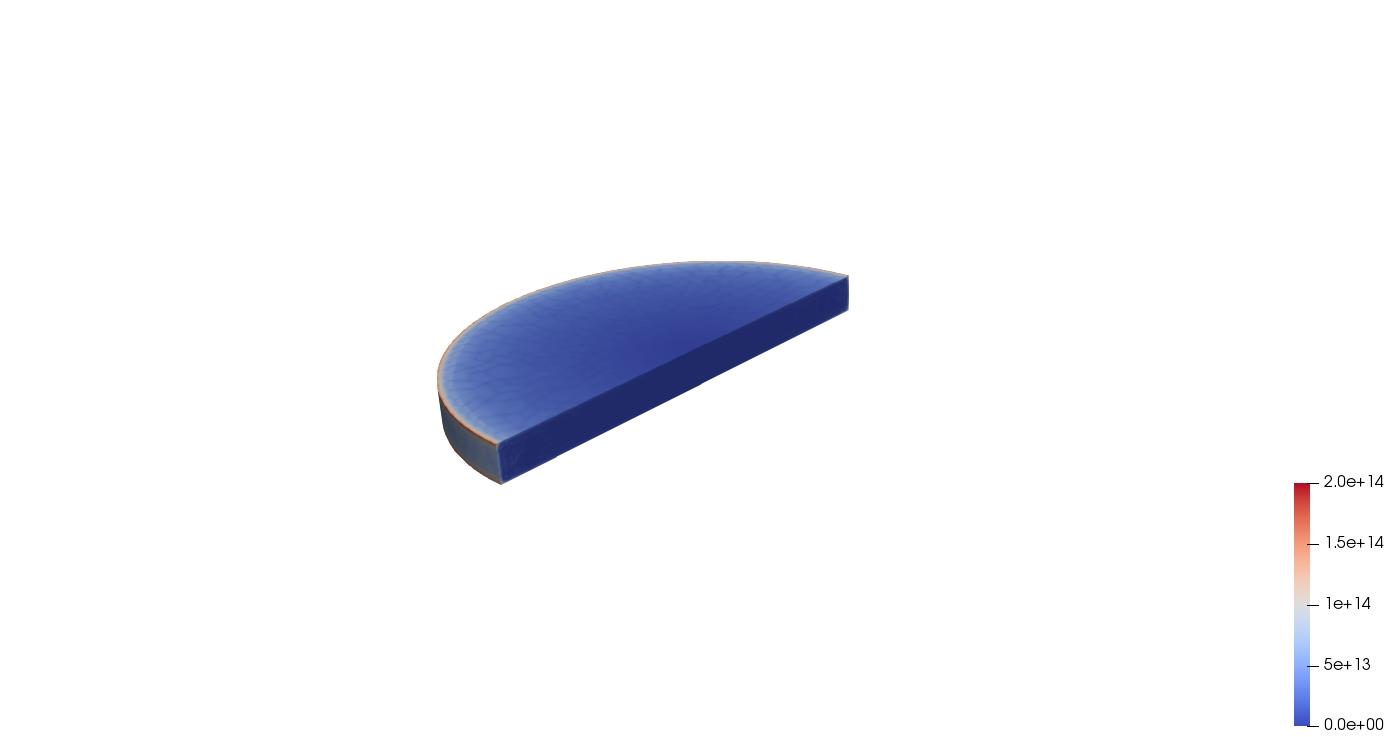} \\
\textrm{\footnotesize{(c) $| \text{Re}({\bm J}^e )|$ for $\omega=10^7$ rad/s.}}  
\end{array}$
\caption{British 1p coins: Contours of the eddy-currents $ {\bm J}^e =  \im \omega\sigma_*\bm{\theta}_3^{(1)}$ for different values of $\omega$, (a) $\omega=10^3$ rad/s, (b) $\omega=10^5$ rad/s, (c) $\omega=10^7$ rad/s in a cut through the coin, on the plane spanned by  $\bm{e}_1$ and $ \bm{e}_3$. }
\label{fig:con1p}
\end{center}
\end{figure}

\begin{figure}[!h]
\begin{center}
\includegraphics[width=0.5\textwidth, keepaspectratio]{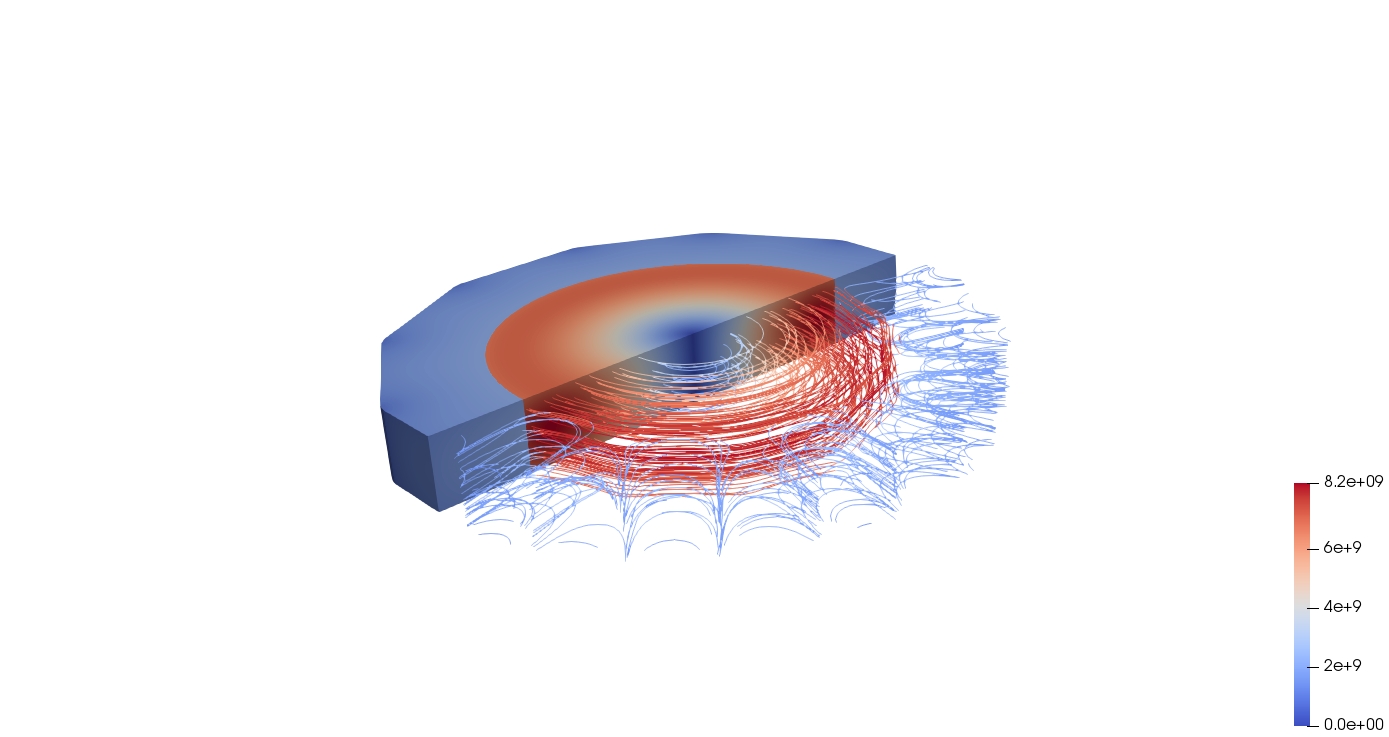}
\caption{British \pounds 1 coin: Contours  of  $| \text{Re}({\bm J}^e)| $ in a cut through the coin, on the plane spanned by $\bm{e}_1$ and $\bm{e}_3$, where  $ {\bm J}^e =  \im \omega\sigma_*\bm{\theta}_3^{(1)}$ are the eddy currents and showing the field lines corresponding to
$\text{Im}( {\bm J}^e)$ with $\omega = 10^3\text{ rad/s}$.}\label{fig:con1P}
\end{center}
\end{figure}

In order to compute the MPT spectral signature, the solution of a reduced order model is obtained at each frequency of interest. As the frequency increases, the skin depth reduces and the associated eddy currents become confined to a thin layer close to the surface of the conductor. In Figure~\ref{fig:con1p}, we show a cut through the 1p coin, on the plane spanned by $\bm{e}_1$ and 
$\bm{e}_3$,  in order to illustrate the eddy currents ${\bm J}^e =  \im \omega \sigma_*\bm{\theta}_3^{(1)}$ obtained at the frequencies $ \omega=10^3$ rad/s, $\omega=10^5$ rad/s and $\omega=10^7$ rad/s.
\noindent
In  Figure~\ref{fig:con1P}, we show a contour plot for the eddy-currents $\bm J^e = \im \omega \sigma_*\bm{\theta}_3^{(1)}$ in cut through the \pounds 1 coin, on a plane spanned by $\bm{e}_1$ and $\bm{e}_3$ for $\omega = 10^3 \text{rad/s}$. This figure also includes the field lines for $\text{Re}({\bm J}^e)$.
                  
For the coin models, each of the associated MPT frequency spectra have independent coefficients that are only associated with diagonal entries of the tensor.  Thus, we set $\sqrt{I_2 ( {\mathcal Z}[ \alpha B,\omega,  \sigma_*, \mu_r])}=0$ in each case.


\subsection{Threat items: TT-33 Semi-automatic pistol}
\begin{figure}[!h]
\begin{center}
\includegraphics[width=0.6\textwidth, keepaspectratio]{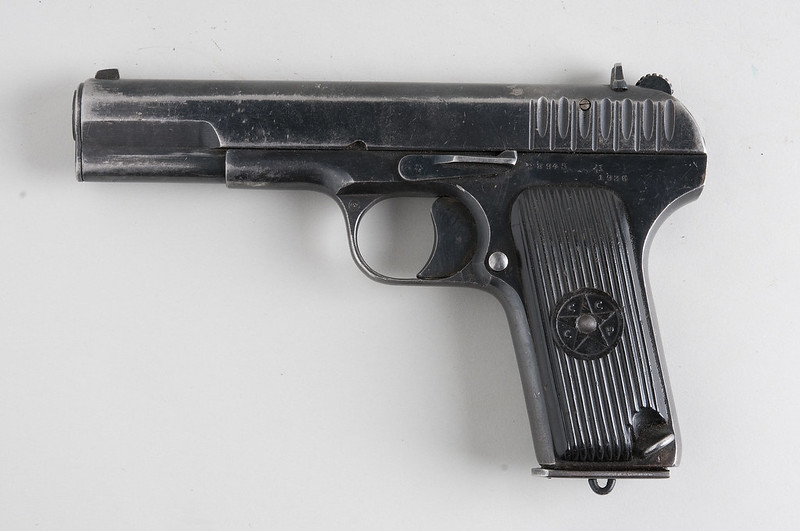}
\caption{Image of Tokarev TT-33 from \cite{tt33-image} reproduced with permission under Creative Commons License CC-BY}
\label{TT-33}
\end{center}
\end{figure}

In this section, we present the MPT spectral signature characterisations for components of an exemplar semi automatic pistol. 
We have chosen the Tokarev TT-33, shown in Figure~\ref{TT-33}, which was originally designed in the Soviet Union in the late 1920's, with production in the USSR between 1930-1954~\cite{TT-33History}. 
It has also been produced in other countries including China, Hungary, North Korea, Pakistan, Romania, Vietnam and Yugoslavia and exported to other nations around world. It is still used by the Bangladeshi and North Korean armed forces and the police in Pakistan often carry the pistol as a side arm. Under a different name, it is occasionally supplied to the police and armed forces in China~\cite{wikitt33}.
We chose to model this gun due to both it's simplicity and prevalence in conflict zones and less economically developed countries with about 1.7 million being produced in total~\cite{TT-33History}.

The starting point for our characterisation are the blueprints for the larger parts of a non-firing replica~\cite{blueprints}, which we expect to be sufficient for our purposes. We have chosen to focus our attention on the receiver (with the magazine and ammunition removed), the reason being that someone wishing to disguise an semi-automatic pistol might disassemble the main pieces  and  carry them separately  through a security control check. As the receiver is one of the larger components, it should be easiest to identify by a metal detector. The exact materials of the receiver are not known although it is likely to be made of a carbon steel alloy such as 1020 or 4140, which has a conductivity of around $\sigma_*=4.5 \times 10^6 \text{ S/m}$ to $\sigma_* =6.25 \times 10^6 \text{ S/m}$~\cite{tt33conductivity} and, for our simulations, we have chosen $\sigma_* =6.2 \times 10^6 \text{ S/m}$. Such steels are ferrous and exhibit a non-linear constitutive relationship between the magnetic flux density ${\bm B}$ and the magnetic field ${\bm H}$, but, if we restrict ourselves to low field strengths, where the relationship is linear,  $\mu_* = \mu_r \mu_0=|{\bm B}|/|{\bm H}|$ and the mathematical model developed in ~\cite{Ammari2014,LedgerLionheart2015} still applies. Values of $\mu_r $ obtained experimentally for different steels vary enormously (eg from $\mu_r=100$ to $\mu_r=600$ or larger) as often  $\mu_* = \mu_r \mu_0=|{\bm B}|/|{\bm H}|$ is applied when the curve is no longer straight. Numerical simulations using high values of  $\mu_r$ become increasingly challenging and so we have chosen $\mu_r=5$. 



\begin{figure}[!h]
\begin{center}
$\begin{array}{ccc}
 \includegraphics[width=0.25\textwidth, keepaspectratio]{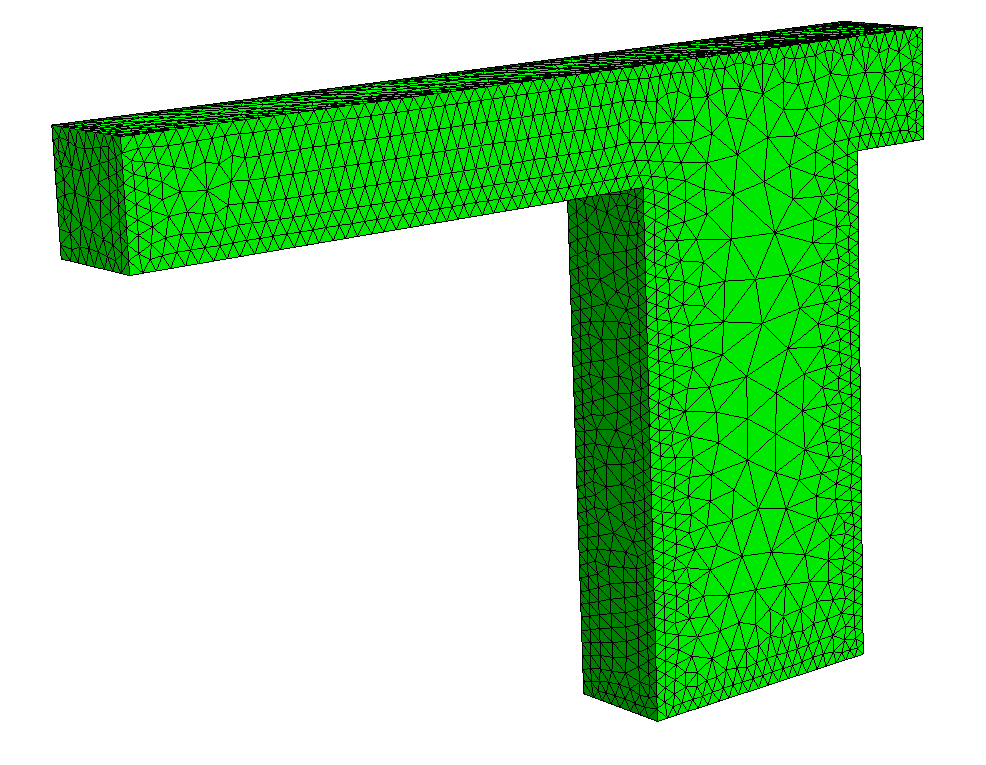} & \includegraphics[width=0.3\textwidth, keepaspectratio]{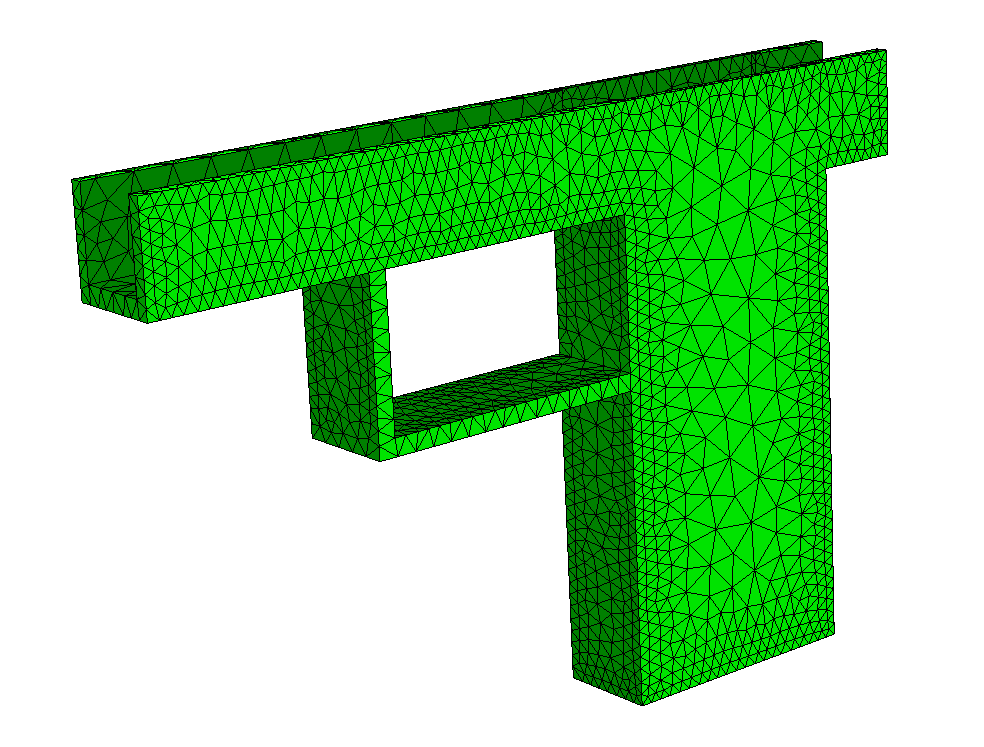} &
 \includegraphics[width=0.3\textwidth, keepaspectratio]{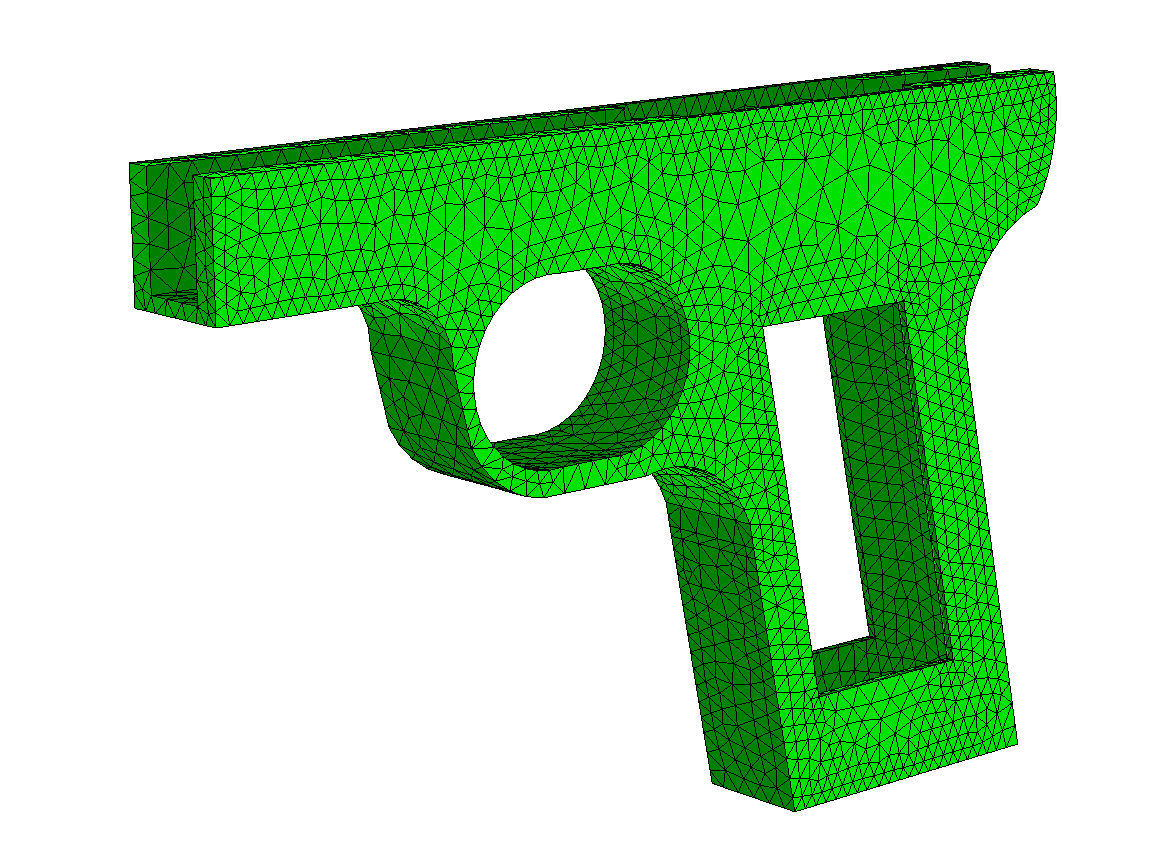} \\
 \text{(a)} & \text{(b)} & \text{(c)} 
 \end{array}$
 \end{center}
 \begin{center}
$ \begin{array}{cc}
 \includegraphics[width=0.3\textwidth, keepaspectratio]{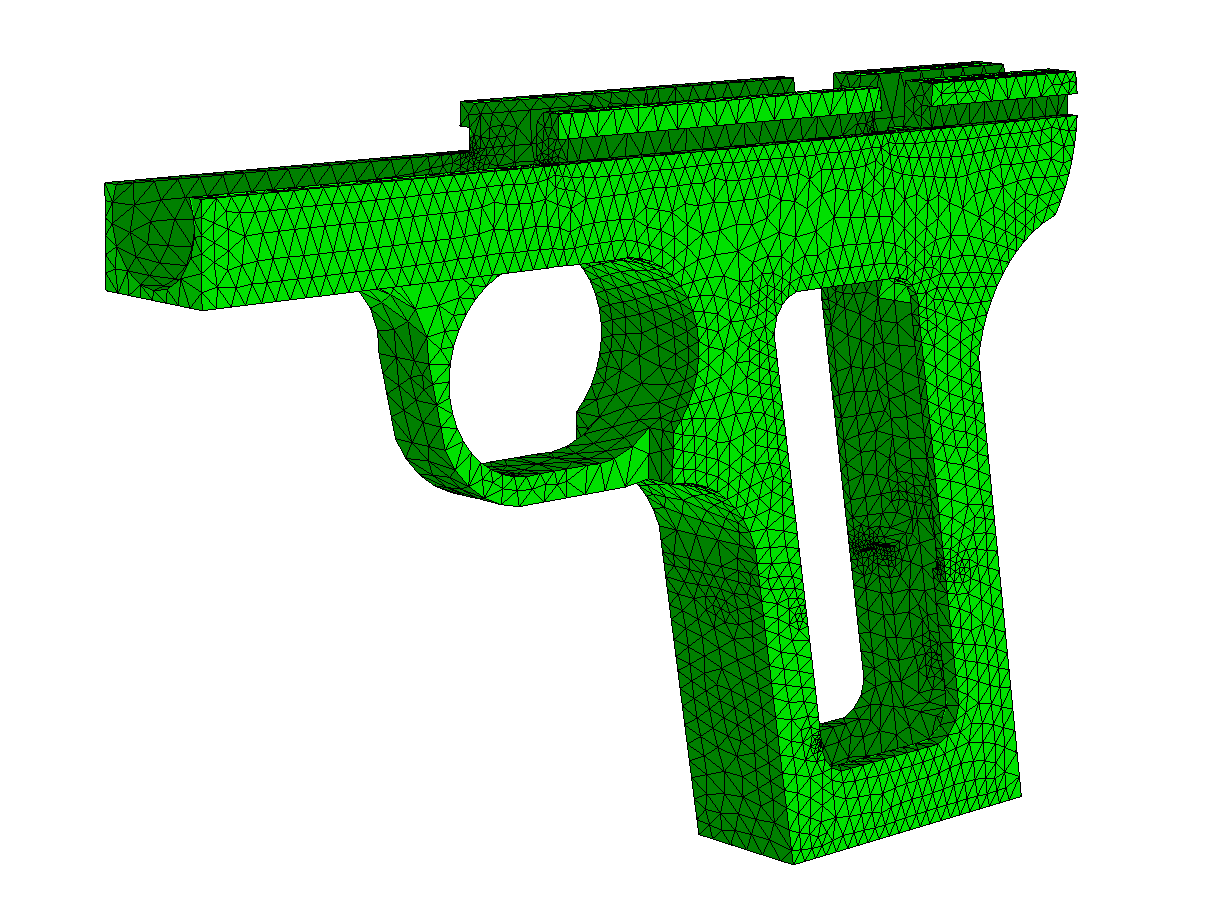} &
 \includegraphics[width=0.33\textwidth, keepaspectratio]{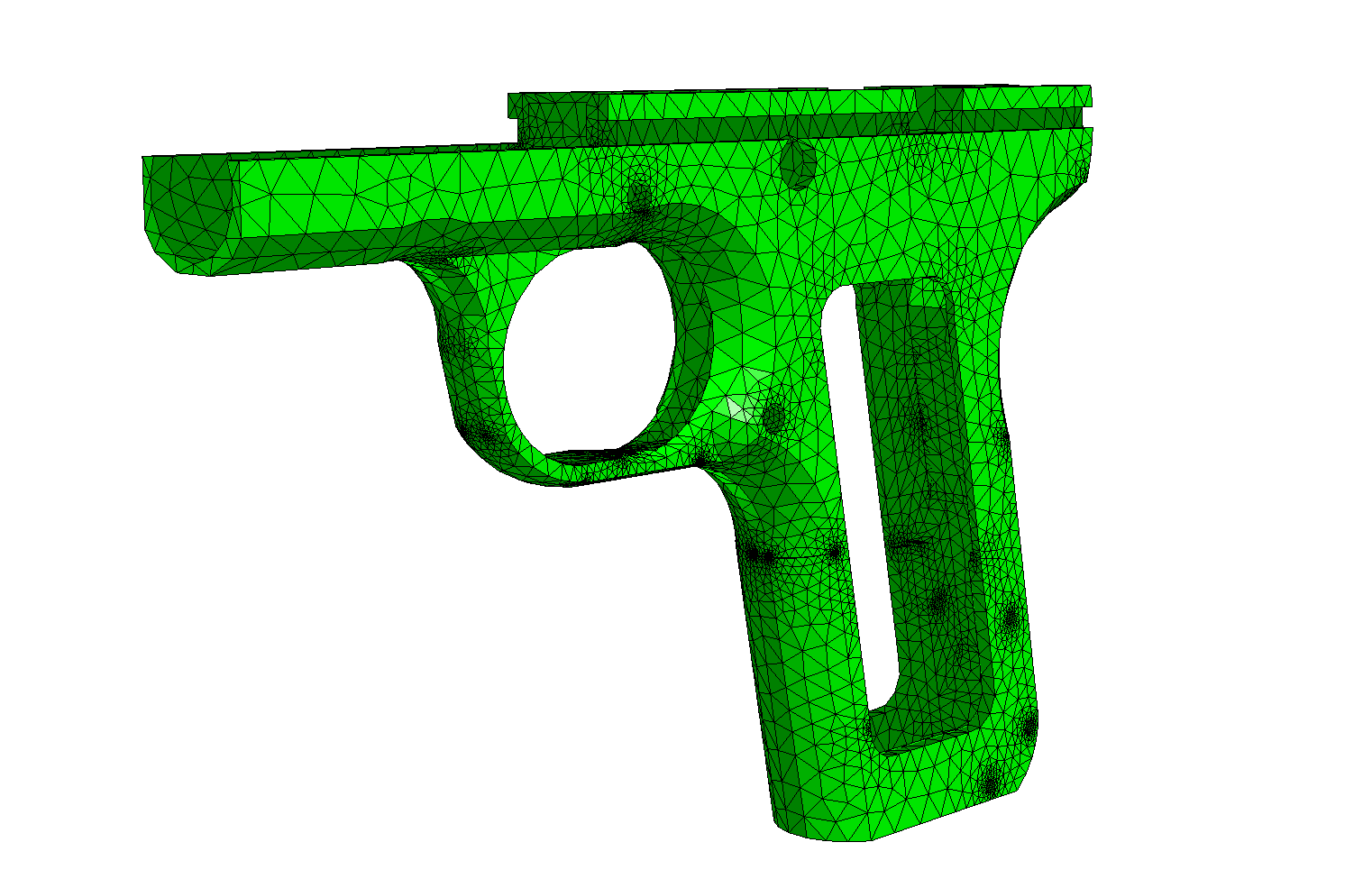} \\
 \text{(d)} & \text{(e)}
 \end{array}$
 \end{center}
\caption{Set of receiver models for TT-33 pistol: surface distribution of elements for (a) L-shape domain, (b) TT-33 with a trigger hole, (c) TT-33 with no internals, (d) TT-33 without chamfers and (e) TT-33 with chamfers} \label{fig:tt33meshes}
\end{figure}

\begin{figure}
\begin{center}
 \includegraphics[width=0.3\textwidth, keepaspectratio]{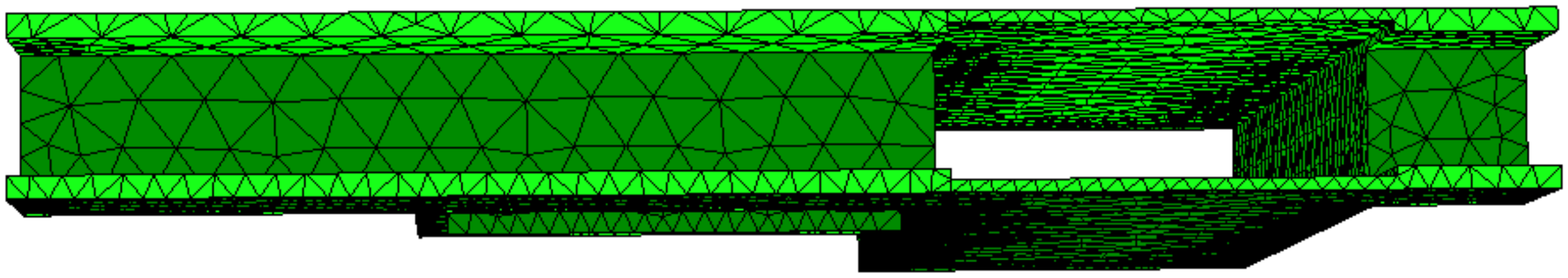}
\end{center}
\caption{Set of receiver models for TT-33 pistol: top view of TT-33 with a trigger hole} \label{fig:tt33meshestop}
\end{figure}

An extreme simplification of the receiver for TT-33 is to model it as simple L-shape made up of two rectangular regions glued together (the overall dimensions of the physical L-shape are $148\text{ mm} \times 17.5\text{ mm} \times 10.1 \text{ mm}$). A mesh discretising the { L-shape},  with overall dimensions $148 \times 17.5  \times 101 $, and the surrounding region out to a truncation boundary, in the form of a  box of dimensions $[-1000,1000]^3$, was generated with $h$-refinement towards the edges, containing $62\, 656$ unstructured tetrahedra with  $\alpha =0.001 \text{ m}$. We then considered a sequence of geometric improvements on the basic { L-shape model}, as shown in Figure~\ref{fig:tt33meshes}, which we refer to as { TT-33 with a trigger hole}, { TT-33 with no internals}, { TT-33 without chamfers} and  { TT-33 with chamfers}, having discretisations comprising of between $94 \, 092$ and $175\, 217$ tetrahedral elements, respectively. With the exception of the { L-shape}, all contain a model of the trigger guard (a loop of steel where the trigger would be placed) and assume that the magazine is removed,  as the top-view of { TT-33 with trigger hole} shown in Figure~\ref{fig:tt33meshestop} illustrates. { TT-33 with no internals} and { TT-33 without chamfers} offer further geometric improvements with { TT-33 with chamfers} being the closet to the actual blueprint and includes small holes in the receiver used to fix the other components of the pistol in place. Only the { L-shape} is simply connected with $\beta_1(B)=0$, the  { TT-33 with trigger hole} $\beta_1(B)=2$,  { TT-33 with no internals} and { TT-33 without chamfers} each have $\beta_1(B) =4$ while  { TT-33 without chamfers} has $\beta_1(B)=13$  each object has $\beta_0(B)=1$ and $\beta_2(B)=0$.

In each case,  $N = 13$ representative solution snapshots to full order problem at logarithmically spaced frequencies in the range $8\times 10^{-1} \le \omega \le 8 \times 10^{8} \text{ rad/s}$ were found to converge by using $p=4$ elements. Then, by applying the PODP approach described in Section~\ref{sect:rommethod} with a tolerance of $TOL=10^{-6}$, the MPT spectral signature for each of the receiver models was obtained. A detailed presentation of the results can be found in~\cite{thesisben}. 
With the exception of the model {with chamfers} each of the models of the receiver has an axis of symmetry in the ${\bm e}_2$ direction and so  there are $4$ independent coefficients each in $\tilde{\mathcal R}[\alpha B,\omega, \sigma_*, \mu_r   ]$ and  $ \mathcal{I}  [\alpha B,\omega, \sigma_*, \mu_r  ]$ corresponding to $( \tilde{\mathcal R}  )_{11}$, $(\tilde{\mathcal R}  )_{22}$, $(  \tilde{\mathcal R}  )_{33}$,  $(\tilde{\mathcal R} )_{13}=(\tilde {\mathcal R }  )_{31}$ at each frequency with similar for  $ \mathcal{I}  $. The model { with chamfers}, which lacks this symmetry, has 
$6$ independent coefficients each in $\tilde{\mathcal R}$ and  $ \mathcal{I} $ (at each frequency), which means that all coefficients of the tensor are independent of each other.
 
\begin{figure}[!h]
\centering
\hspace{-1.cm}
$\begin{array}{cc}
\includegraphics[scale=0.5]{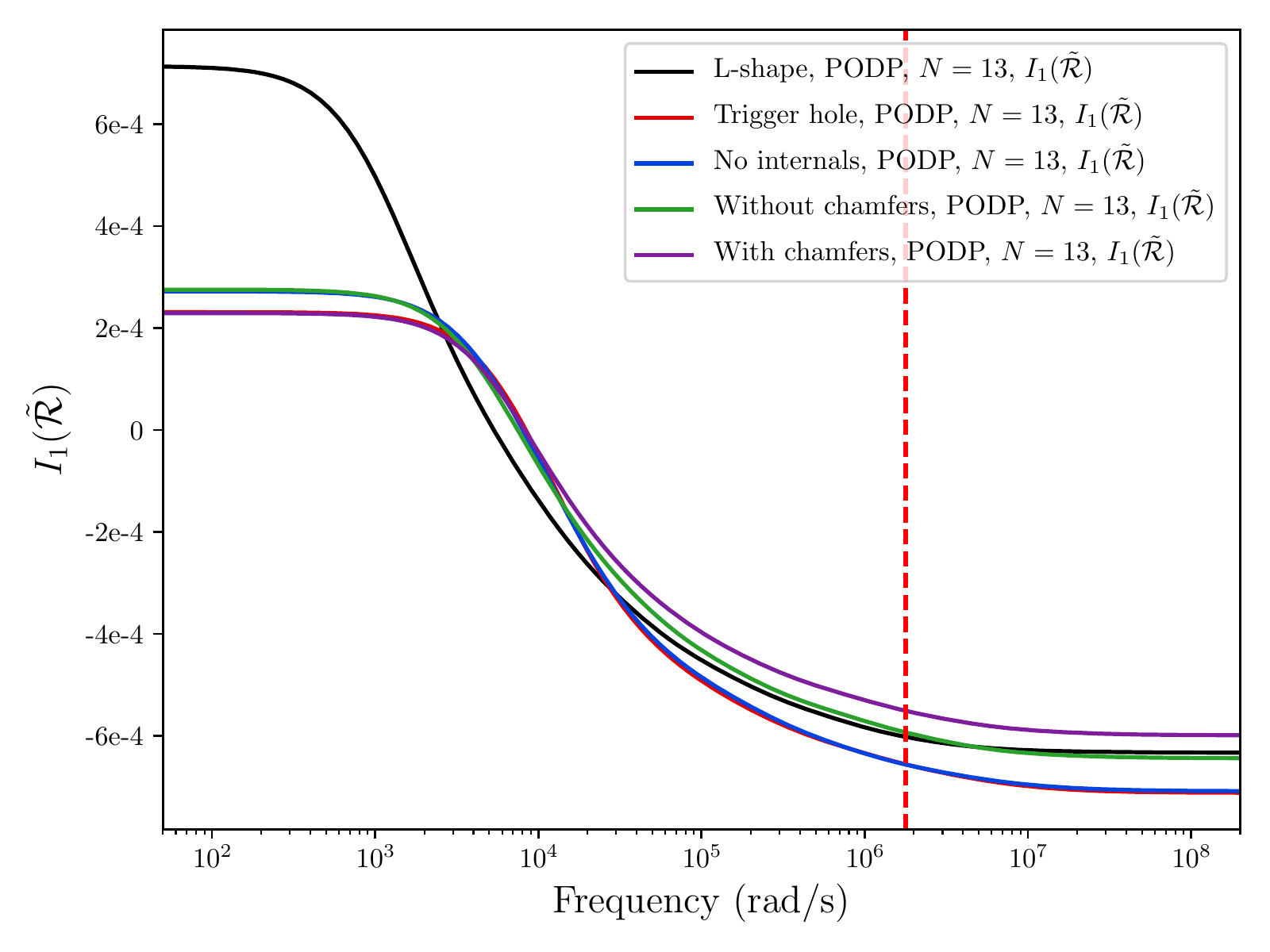} &
 \includegraphics[scale=0.5]{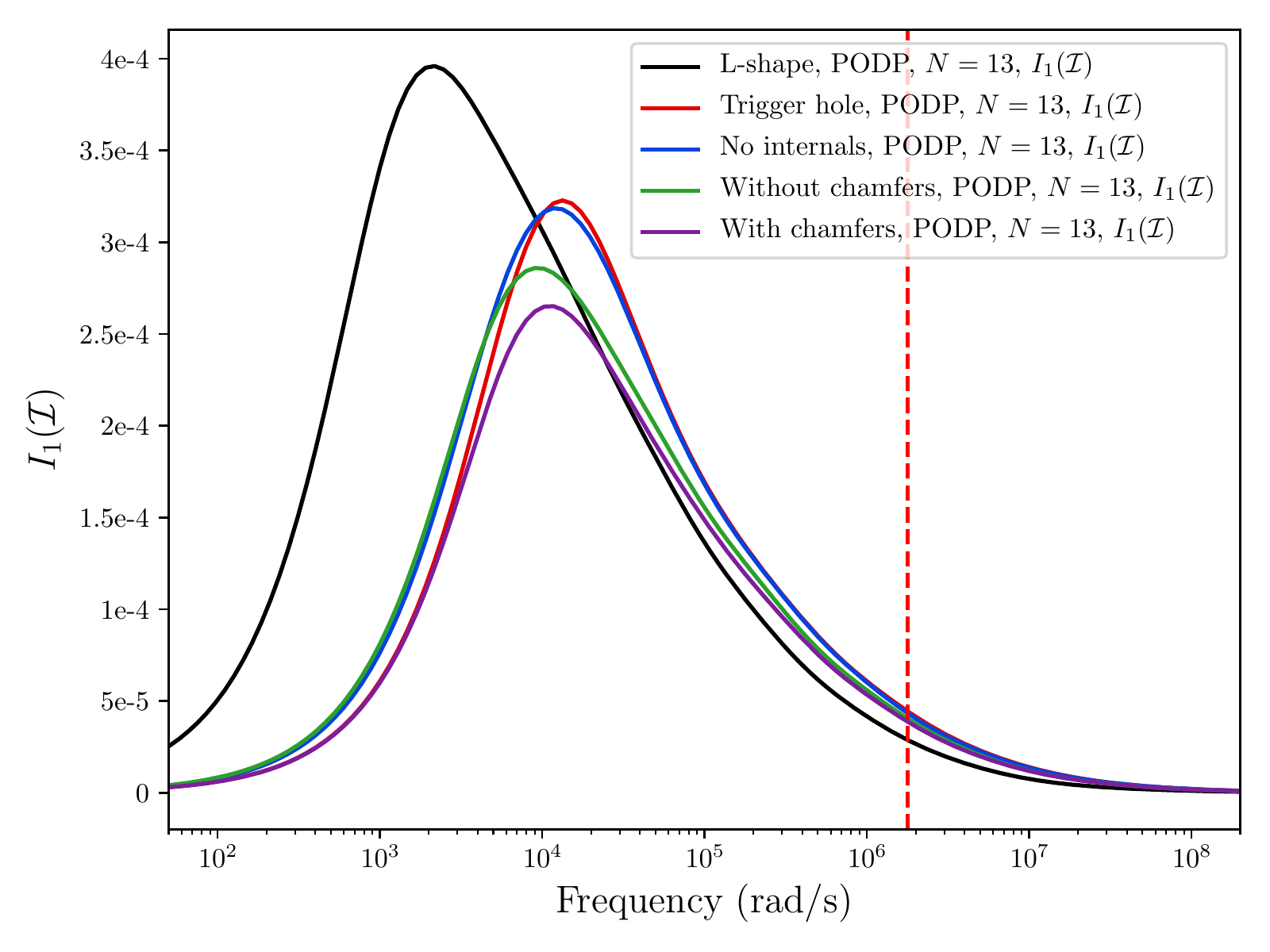}  \\
\text{(a) } I_{1} (\tilde{\mathcal{R}}  ) & 
\text{(b) } I_{1} ( \mathcal{I}  )  \\
\includegraphics[scale=0.5]{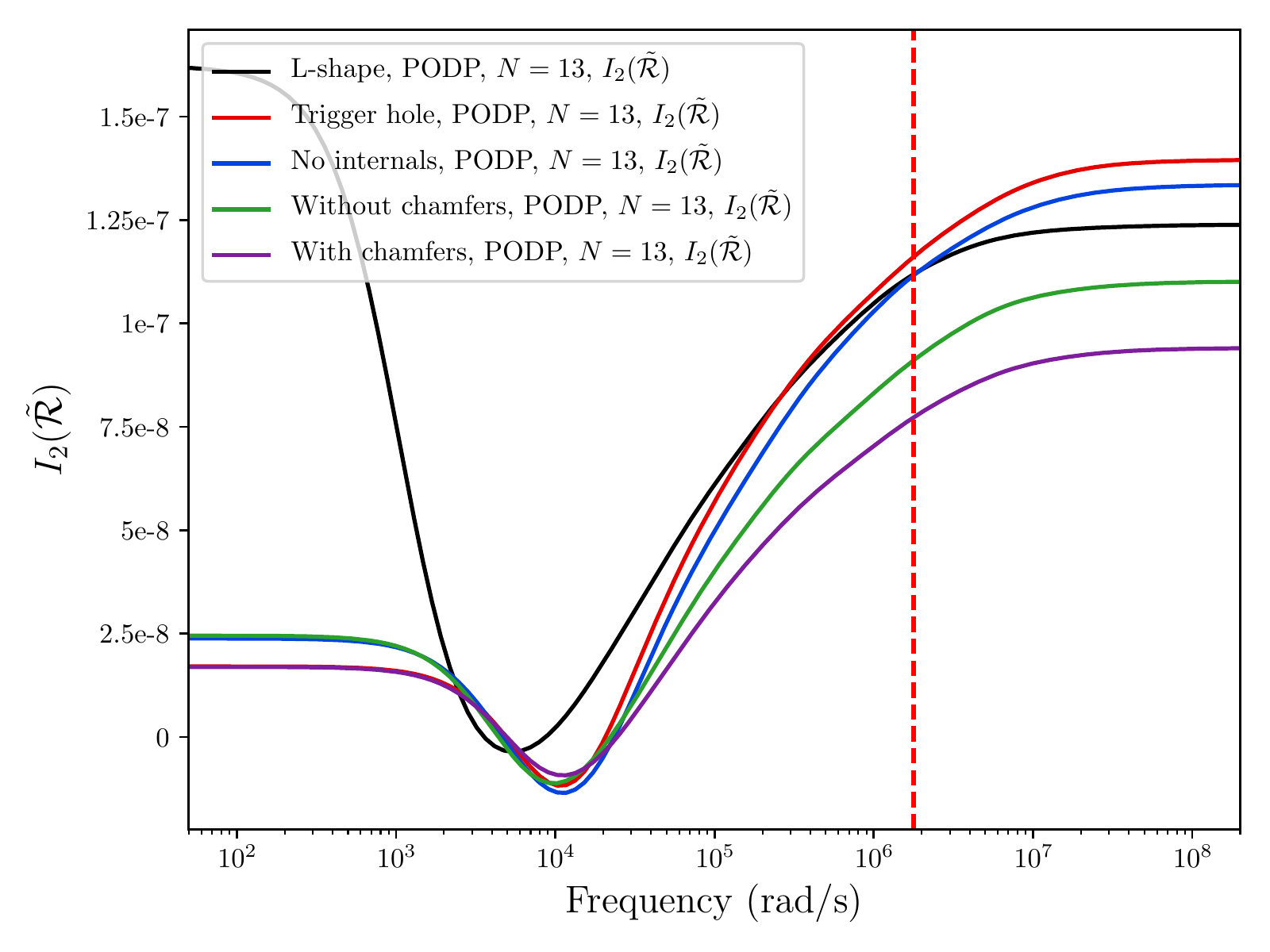} &
 \includegraphics[scale=0.5]{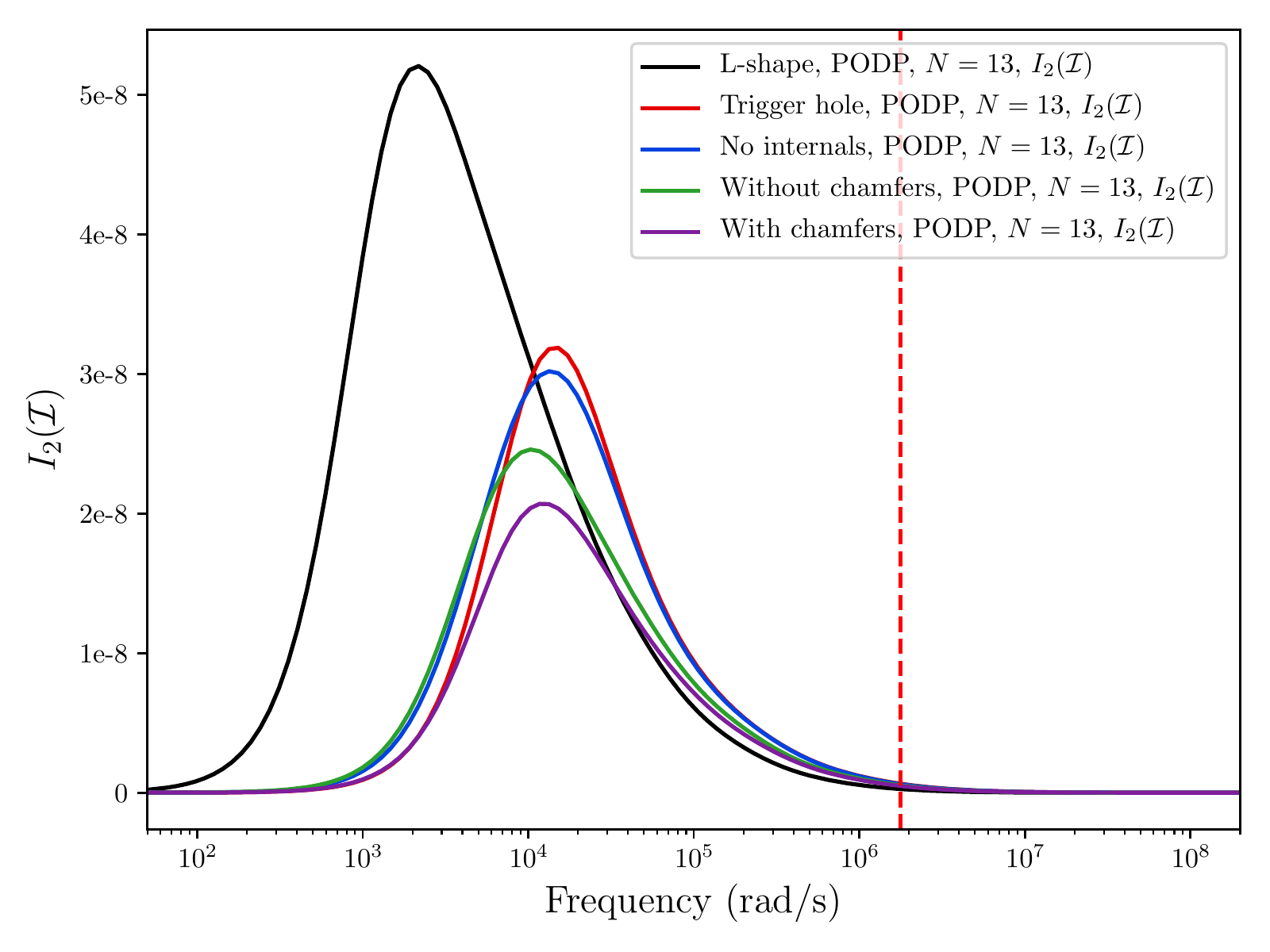}  \\
\text{(c) } I_{2}  (\tilde{\mathcal{R}} ) & 
\text{(d) } I_{2 } ( \mathcal{I}  )  \\
\includegraphics[scale=0.5]{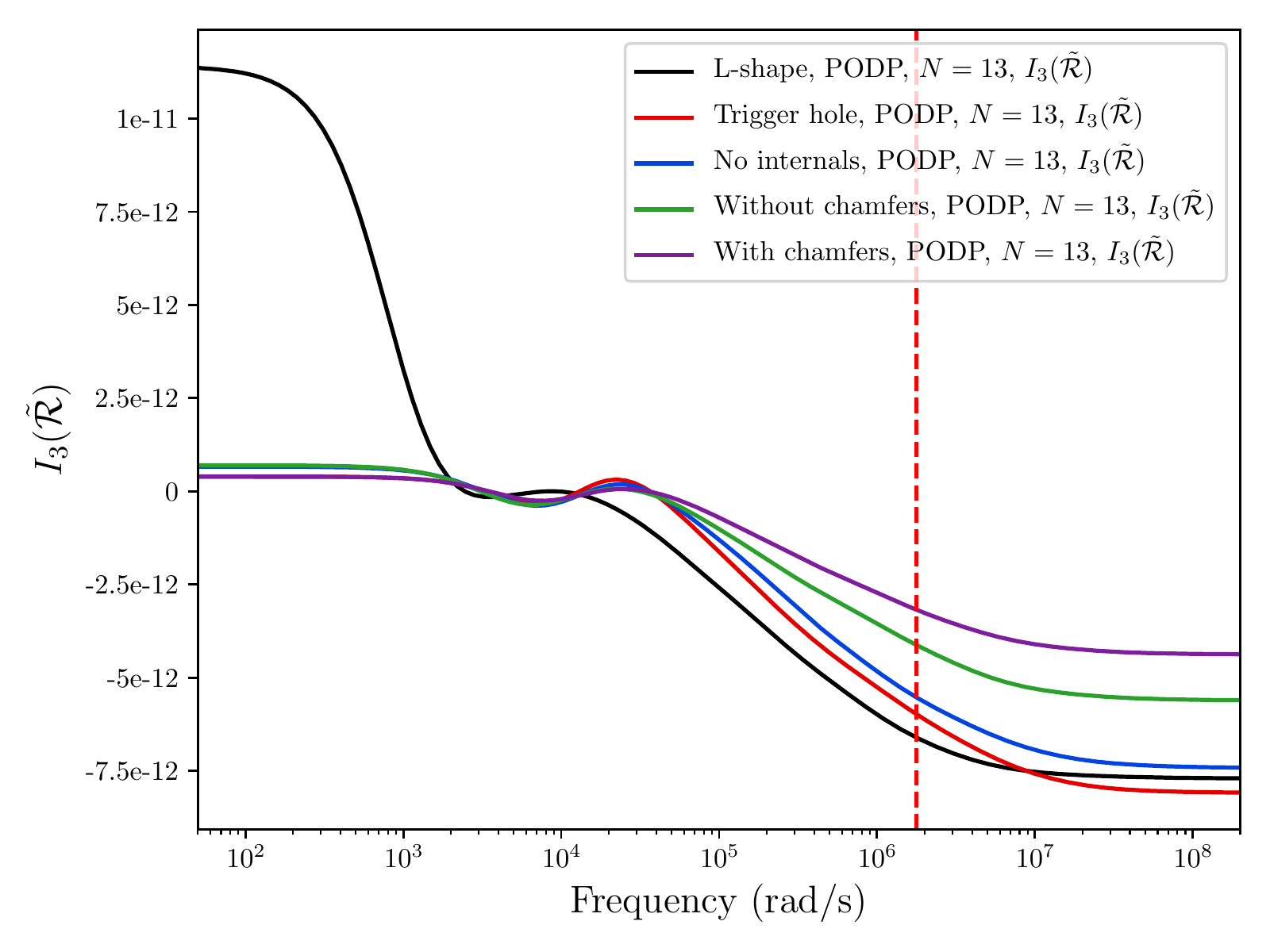} &
 \includegraphics[scale=0.5]{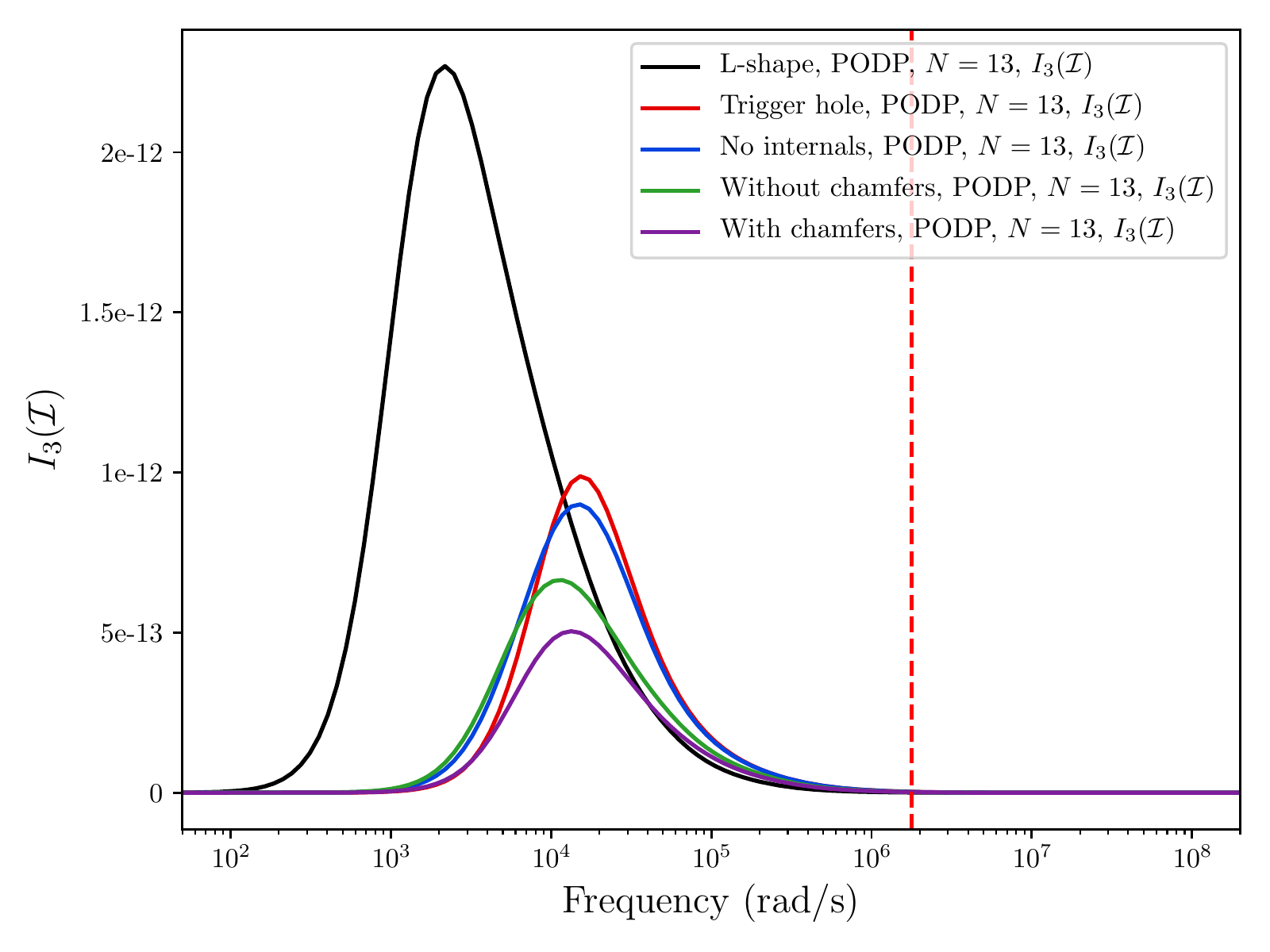}  \\
\text{(e) } I_{3}  ( \tilde{\mathcal{R}}  ) &
\text{(f) }  I_{3} ( \mathcal{I} )  
\end{array}$
  \caption{Set of receiver models for TT-33 pistol: Comparison of tensor invariants. (a) $I_{1}  ( \tilde{\mathcal{R}}  ) $, (b) $I_{1} ( \mathcal{I} ) $
  (c) $I_{2}  (\tilde{\mathcal{R}}  ) $, (d) $I_{2} ( \mathcal{I}  ) $,
  (e) $I_{3}  ( \tilde{\mathcal{R}} ) $ and (f)  $I_{3} ( \mathcal{I} ) $.}
        \label{fig:CompInvtt33}
\end{figure}

 We show a comparison of 
the  MPT spectra using the principal invariants $I_i$, $i=1,2,3$ for $\tilde{\mathcal R} [\alpha B,\omega,  \sigma_*,  \mu_r  ]$ and ${\mathcal I} [\alpha B,\omega, \sigma_*, \mu_r  ]$ that have been obtained using the PODP approach in Figure~\ref{fig:CompInvtt33}. 
In this figure, we have restricted consideration to frequencies such that $10^2 \le \omega \le 10^8 \text{ rad/s}$ in order to allow comparisons with the earlier key and coin results. In practice, the eddy current model brakes down at a frequency $\omega_{limit} < 2 \times 10^6 \text{ rad/s}$ for all the TT-33 models considered and so, in practice, higher frequencies are not relevant. 
 {While the $I_i({\mathcal I})$, $i=1,2,3$, invariants for the TT-33 models are similar to that for the keys and coins, the behaviour of $I_i(\tilde{\mathcal R} )$, $i=1,2,3$ 
are quite different due to $\mu_r \ne 1$. For $I_1( \tilde{\mathcal R}  )$, we see the curves are monotonically decreasing with $\log \omega$, but do not asymptote to $0$ for small $\omega$. For $I_2(\tilde{\mathcal R}  )$ and
$I_3(\tilde{\mathcal R} )$
 we see the curves are neither monotonically increasing or decreasing with $\log \omega$ and the curves do not asymptote to $0$ for small $\omega$ motivating that discrimination between the object is possible. The {L-shape} exhibits significant differences to the TT-33 models  with a different location of resonant peak in $I_i({\mathcal I})$, $i=1,2,3$ and a significantly different behaviour for $I_i(\tilde{\mathcal R})$, $i=1,2,3$. The results for the {TT--33 with a trigger hole} and {TT-33 with no internals} models are similar with further differences for the { TT-33 without chamfers} and {TT-33 with chamfers}. However, all these latter four cases exhibit a resonance peak of around $\omega = 10^4 \text{rad/s}$ for $I_i({\mathcal I})$, $i=1,2,3$. The magnitude of the resonance peak for $I_3({\mathcal I})$ decreases in sequence of the associated volume of the different TT-33 models.}

\begin{figure}[!h]
\centering
\hspace{-1.cm}
$\begin{array}{cc}
 \includegraphics[scale=0.5]{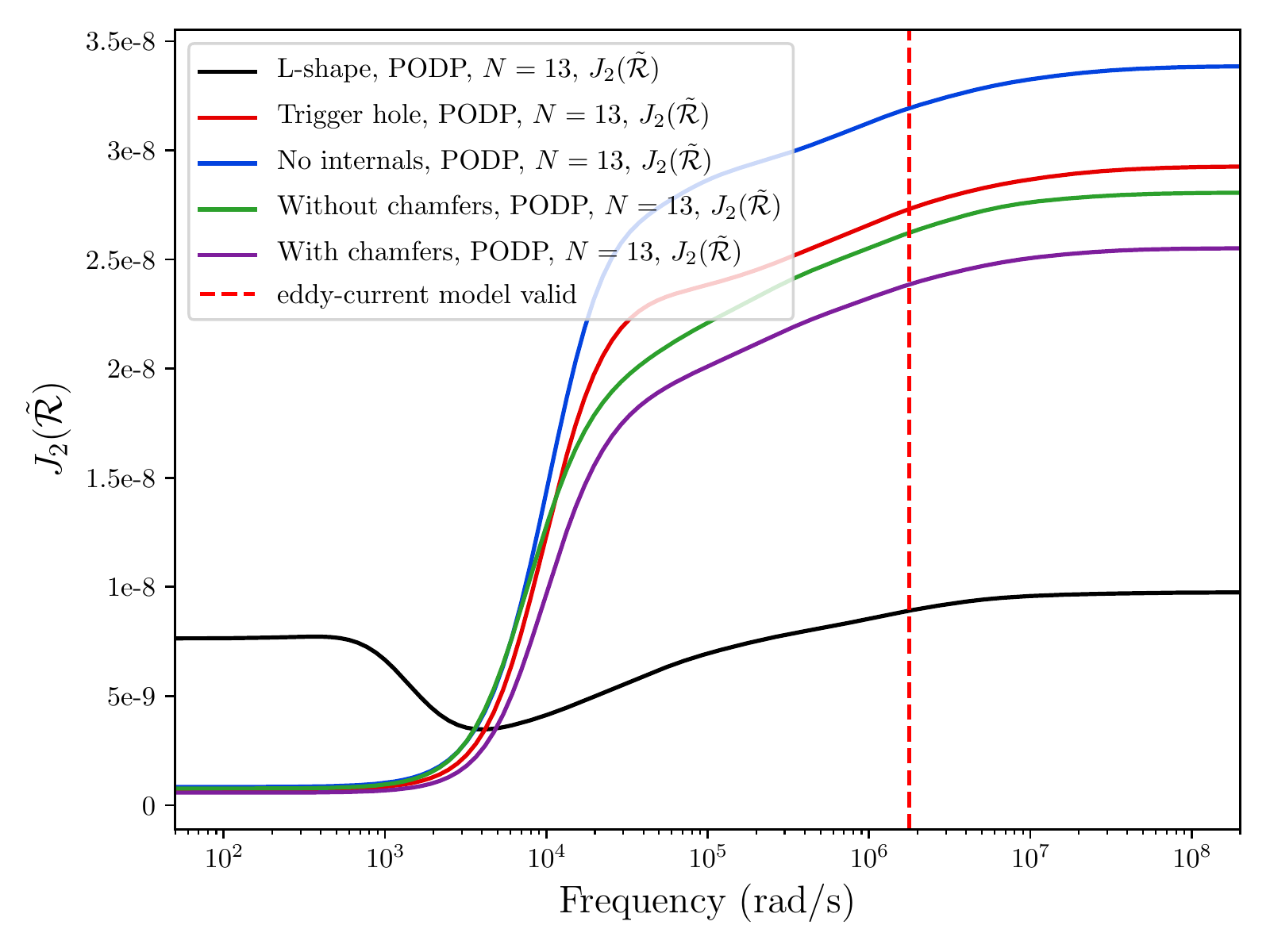} &
 \includegraphics[scale=0.5]{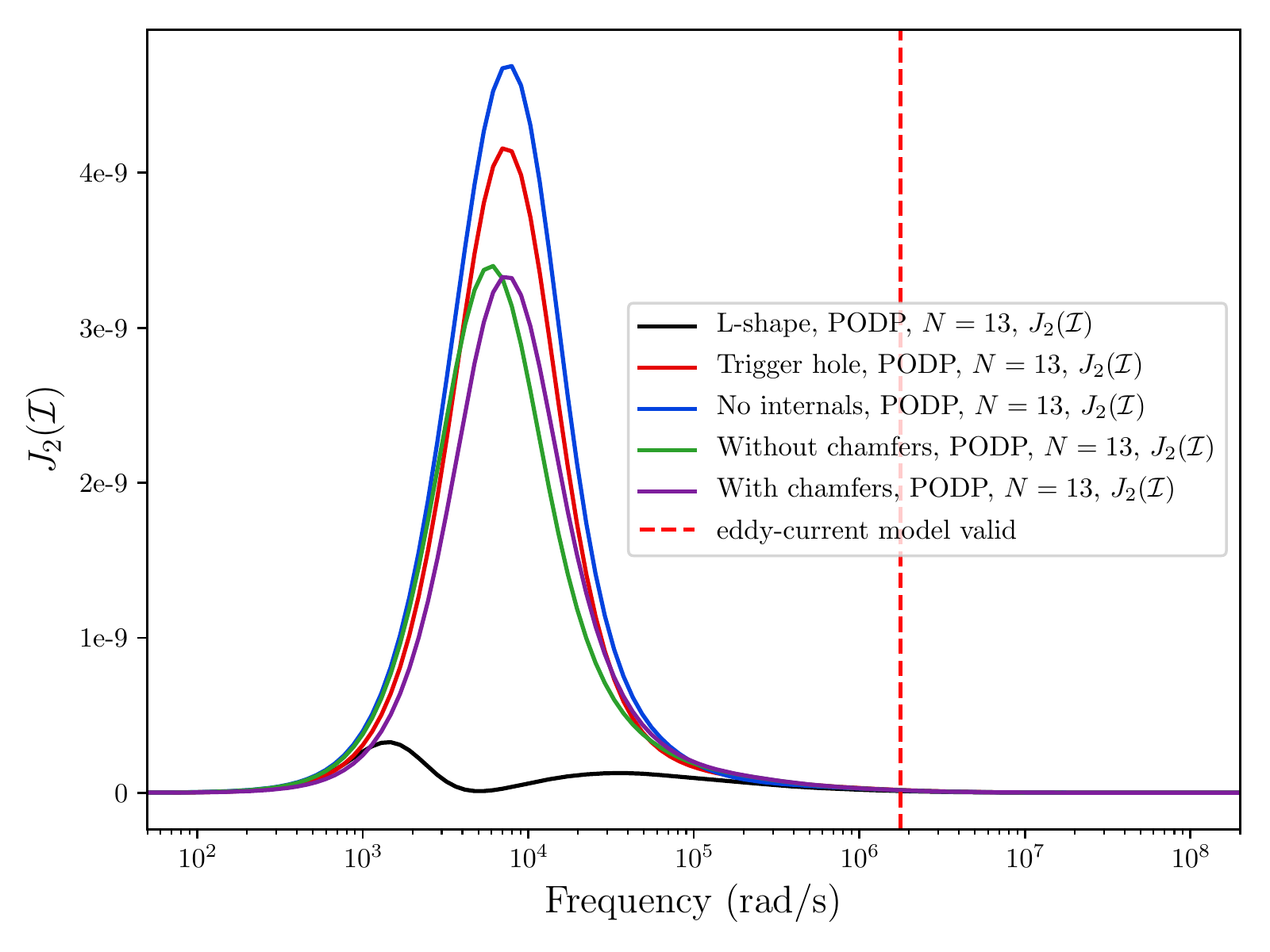}  \\
 text{(a) } J_{2} ( \tilde{\mathcal{R}} ) & 
\text{(b) } J_{2 } ( \mathcal{I}  )  \\
\includegraphics[scale=0.5]{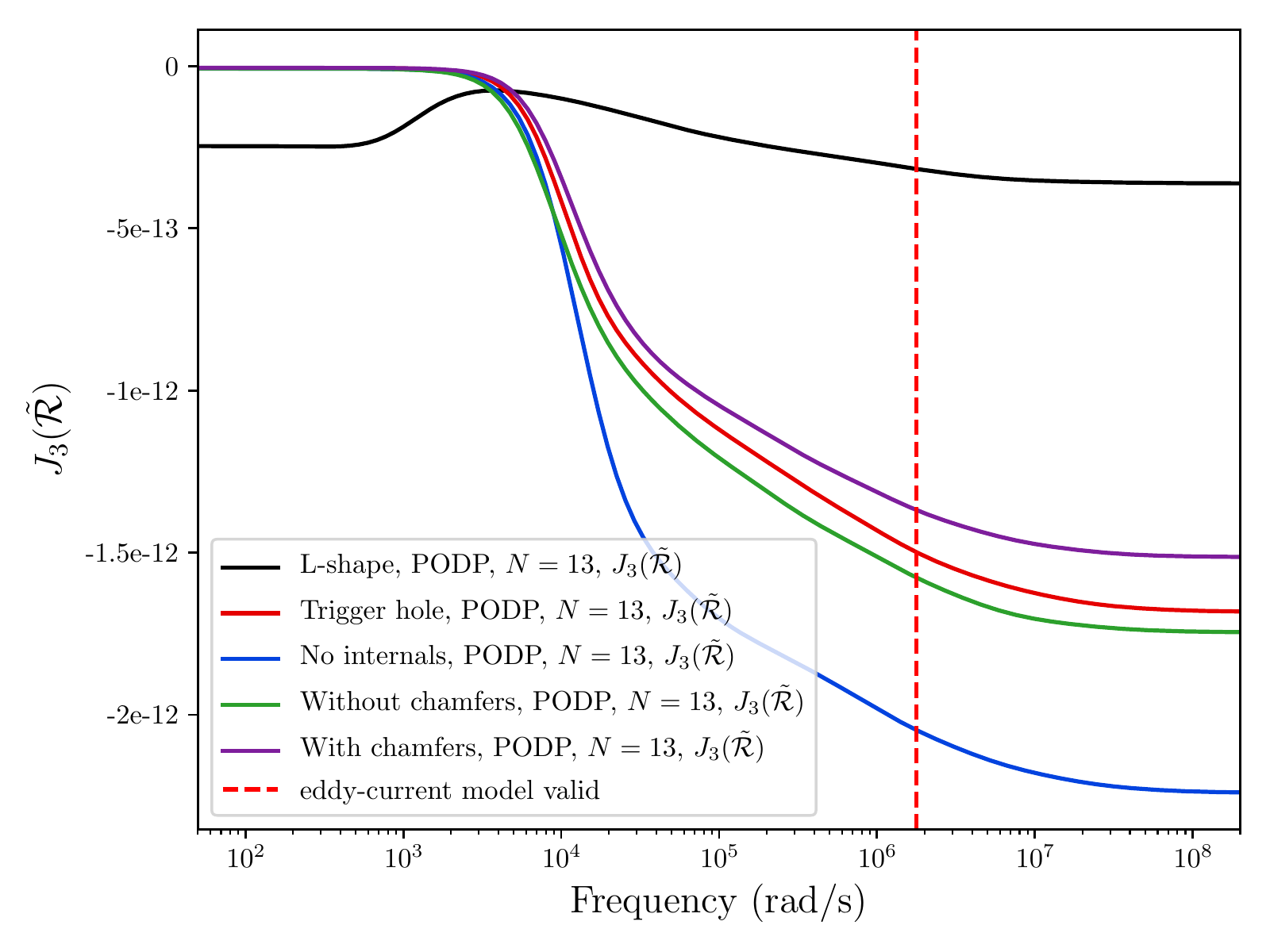} &
 \includegraphics[scale=0.5]{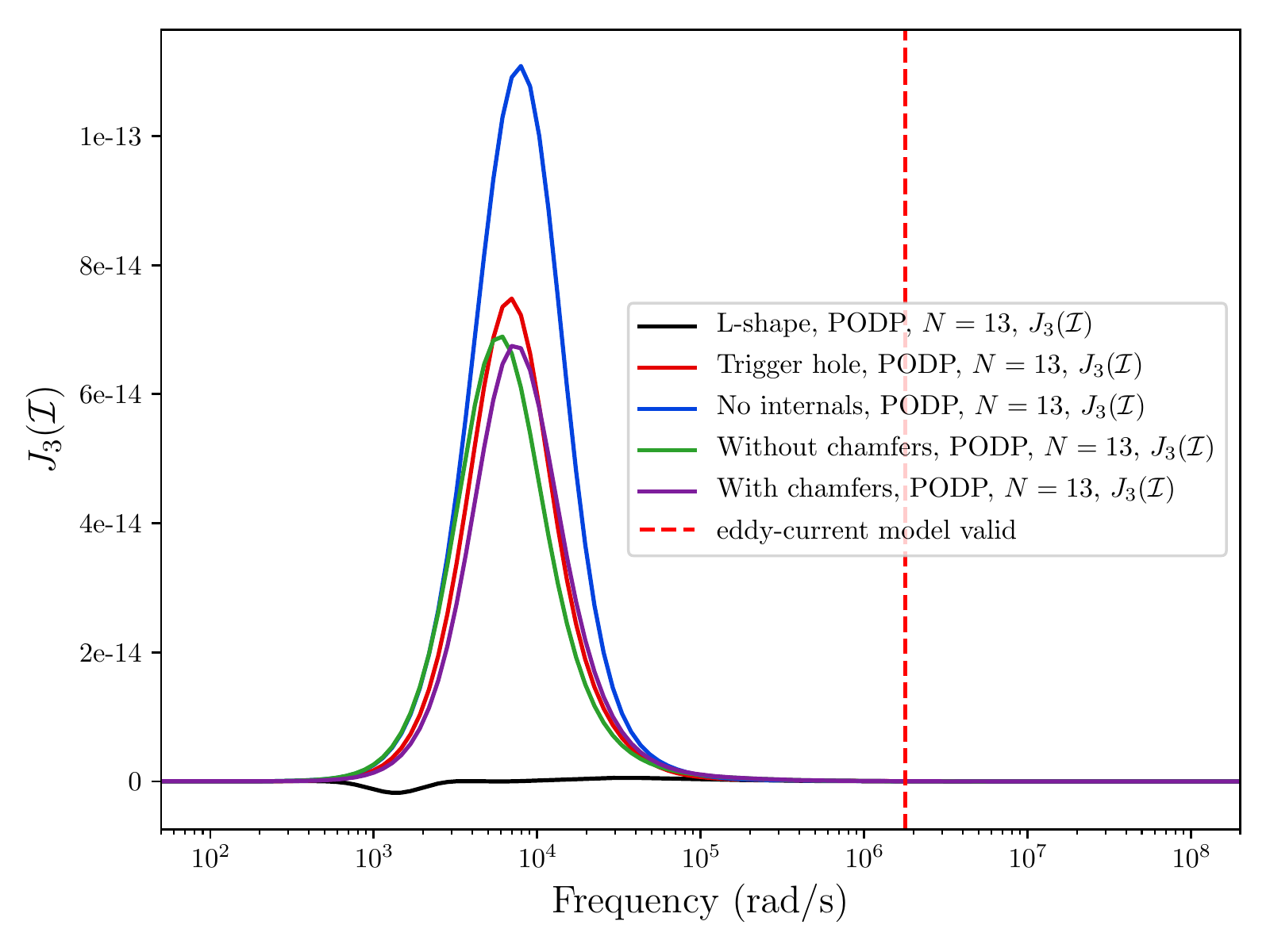}  \\
 \text{(c) } J_{3}  ( \tilde{\mathcal{R}} ) &
\text{(d) }  J_{3}( \mathcal{I} )  
\end{array}$
  \caption{Set of receiver models for TT-33 pistol:  Comparison of tensor invariants. 
  (a) $J_{2} ( \tilde{\mathcal{R}} ) $, (b) $J_{2} ( \mathcal{I} ) $,
  (c) $J_{3}  ( \tilde{\mathcal{R}} ) $ and (d)  $J_{3} ( \mathcal{I} ) $.}
        \label{fig:CompDevInvtt33}        
\end{figure}

The corresponding results obtained for the invariants $J_i$, $i=2,3$ for $\tilde{\mathcal R} [\alpha B,\omega , \sigma_*, \mu_r   ]$ and ${\mathcal I} [\alpha B,\omega, \sigma_*, \mu_r  ]$ that have been obtained using the PODP approach are shown in Figure~\ref{fig:CompDevInvtt33}. These results again illustrate the significant difference between the {L-shape} and the other models.
{With the exception of the {\em L-shape}, the results for $J_2(\tilde {\mathcal R}) $ show a monotonic increase with $\log \omega$, those for $J_3(\tilde {\mathcal R})$ show a monotonic decrease, $J_2({\mathcal I})$ and $J_3({\mathcal I})$ show a single local maximum. The  results for { TT-33 with and without chamfers} are similar with greater differences exhibited between the { TT-33 with no internals} and {TT-33 with trigger hole}.}
 The results shown in Figures~\ref{fig:CompInvtt33} and~\ref{fig:CompDevInvtt33} indicate the significant difference between the spectral signatures of the  { L-shape} and the more realistic models, which adds the credibility that the MPT spectral signature makes it possible to distinguish between a carpenters metallic set-square (which closely resembles an { L-shape}) and the receiver of a pistol, for example.

\begin{figure}[!h]
\centering
    \includegraphics[scale=0.5]{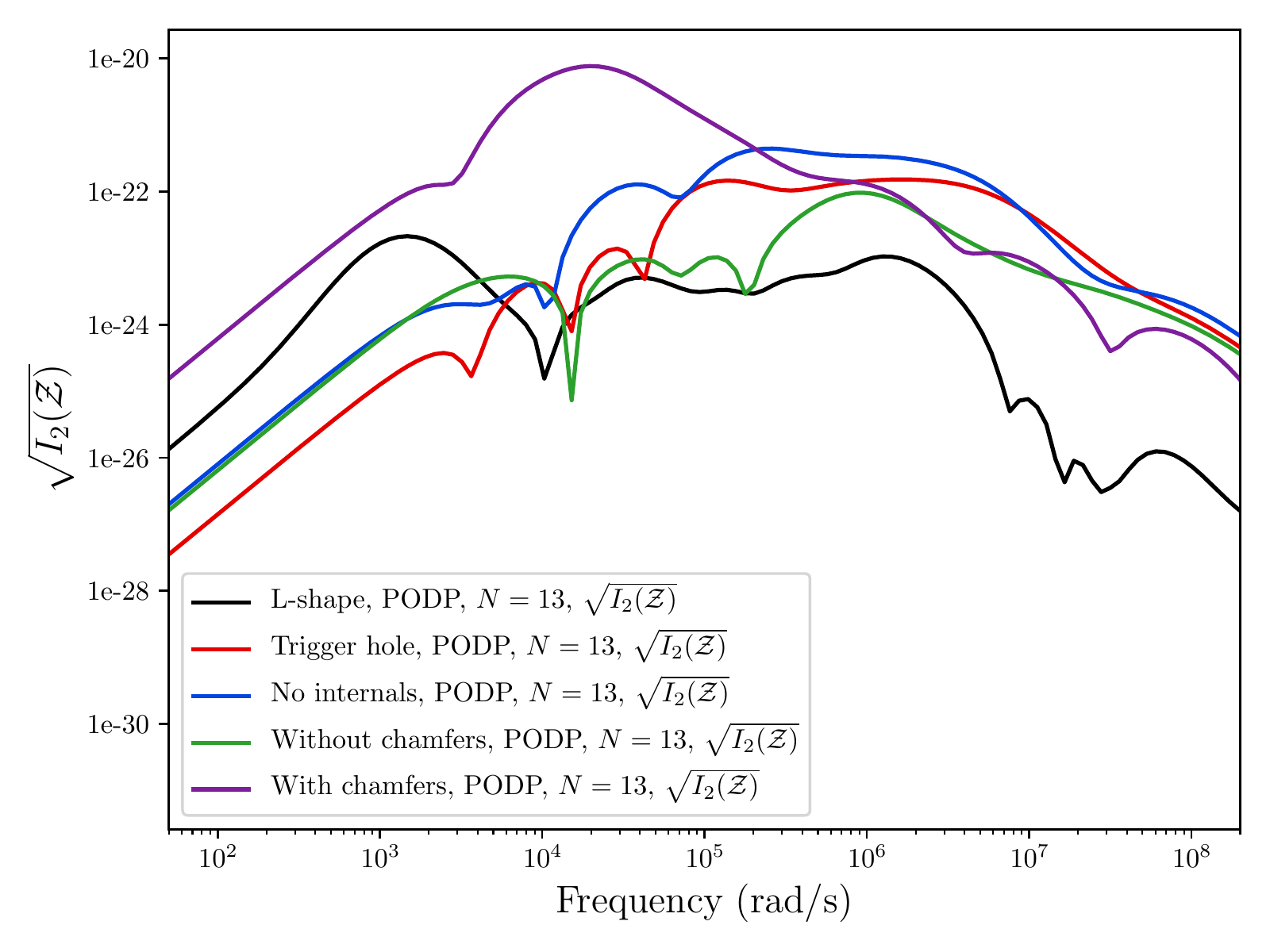} 
   \caption{Set of receiver models for TT-33 pistol: Comparison of the tensor invariant    $\sqrt{I_2 ( {\mathcal Z} )} $.}
        \label{fig:Mesh:zCurvett33}
\end{figure}

For the TT-33 models, each of the associated MPT frequency spectra have independent coefficients that are associated with both on and off diagonal entries of the tensor. The behaviour of $\sqrt{I_2 ( {\mathcal Z}[ \alpha B,\omega,  \sigma_*, \mu_r])} $ for the different models is shown in Figure~\ref{fig:Mesh:zCurvett33}.

\begin{figure}[!h]
\centering
\hspace{-1.cm}
$\begin{array}{cc}
\includegraphics[scale=0.15]{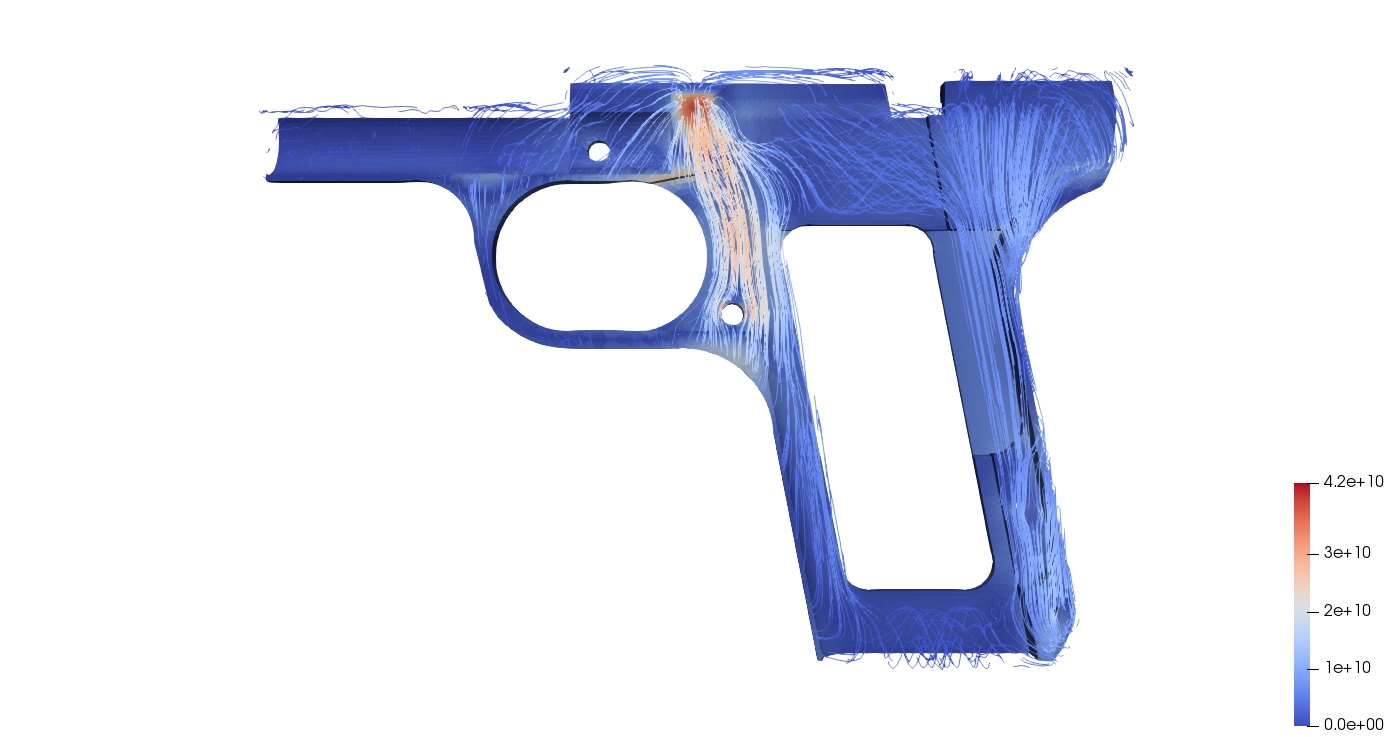} &
\includegraphics[scale=0.15]{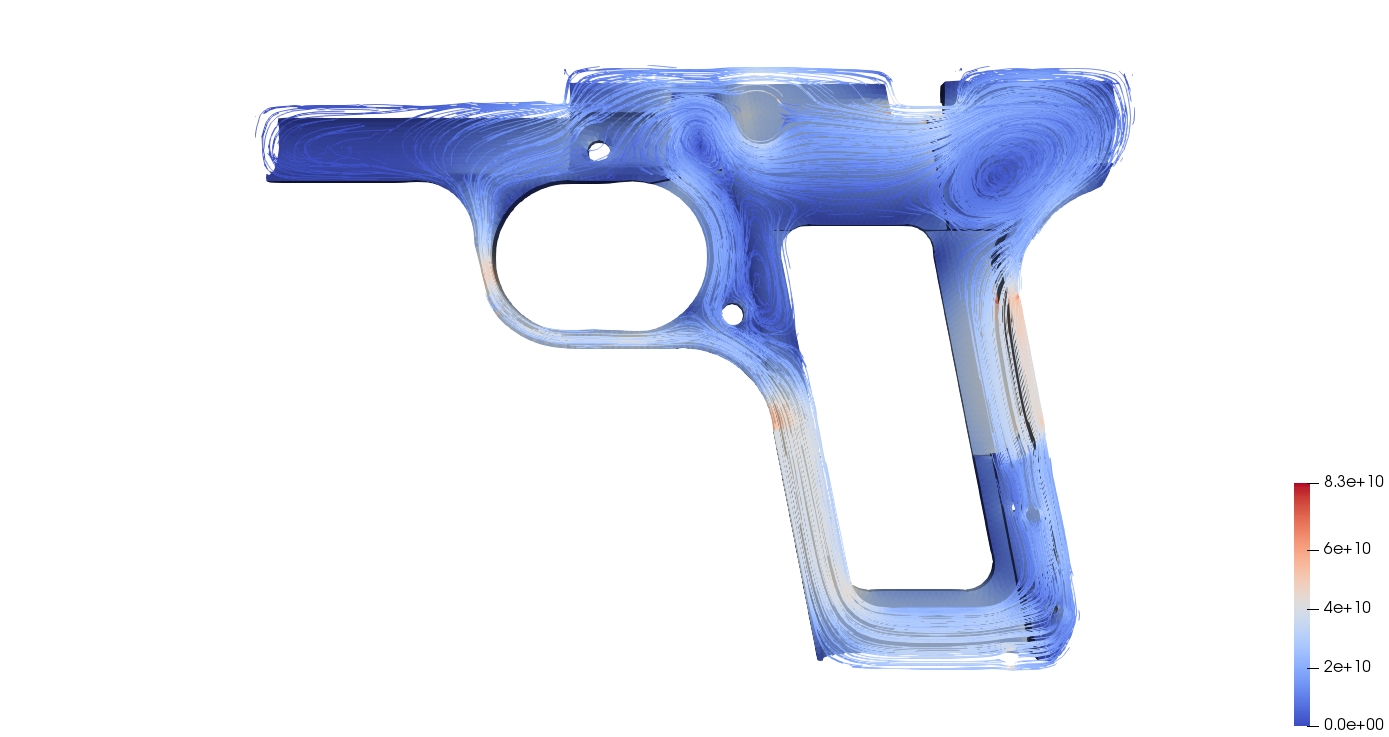} \\
\text{(a) } \bm{J}^e = \im \omega \sigma_* \bm{\theta}_1^{(1)} &  \text{(b) } \bm{J}^e = \im \omega \sigma_* \bm{\theta}_2^{(1)} 
\end{array}$\\
$\begin{array}{c}
\includegraphics[scale=0.15]{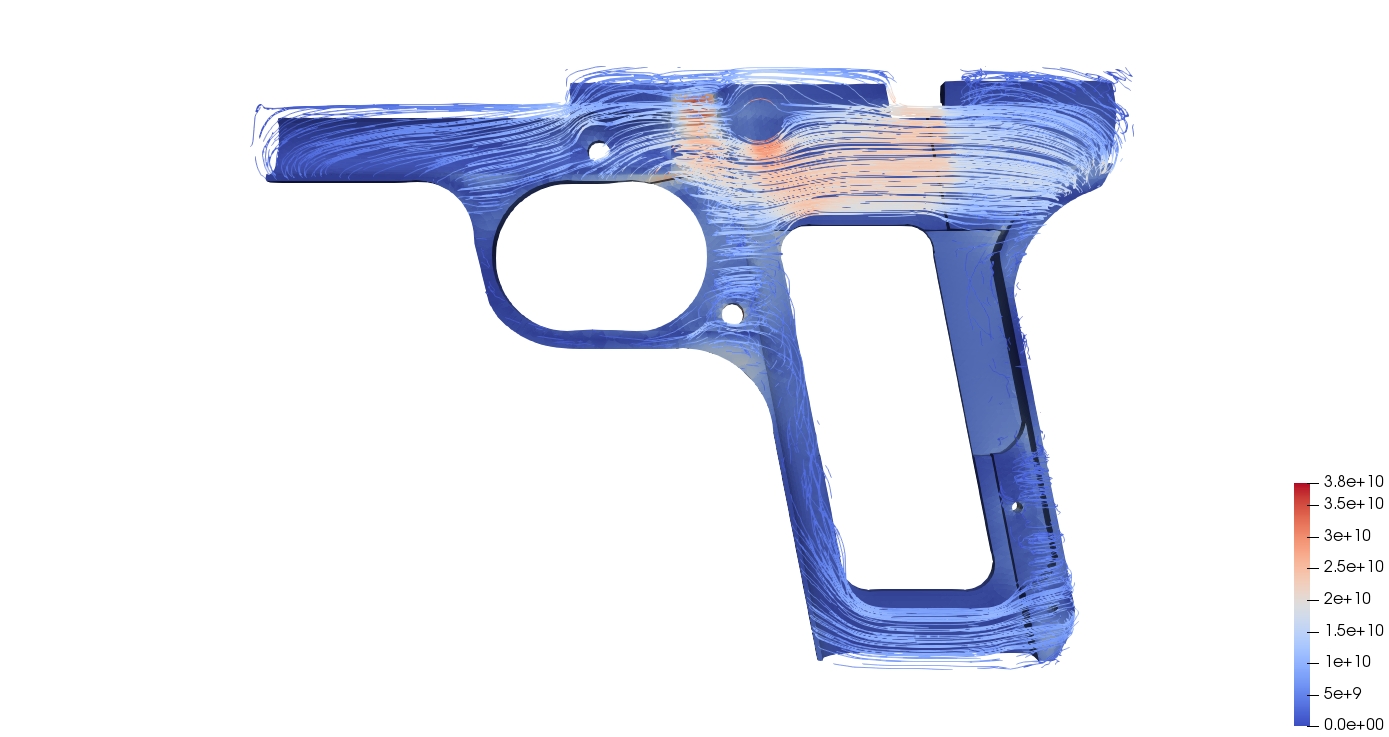} \\
\text{(c) } \bm{J}^e = \im \omega \sigma_* \bm{\theta}_3^{(1)}
\end{array}$
  \caption{Set of receiver models for TT-33 pistol: Contours of  $| \text{Re}({\bm J}^e)| $ and streamlines   for $ \text{Re}({\bm J}^e)$ on the plane spanned by $\bm{e}_1$ and $\bm{e}_3$ with $\xi_2=0$ and the {TT-33 with chamfers}  model,  showing (a) $ \bm{J}^e = \im \omega \sigma_* \bm{\theta}_1^{(1)}$, (b) $ \bm{J}^e = \im \omega \sigma_* \bm{\theta}_2^{(1)} $ and (c) $  \bm{J}^e = \im \omega \sigma_* \bm{\theta}_3^{(1)}$}
        \label{fig:t33contour}
\end{figure}

  Finally, in Figure~\ref{fig:t33contour} we show the contours of $| \text{Re}({\bm J}^e)| $ and field lines for  
 $ \text{Re}({\bm J}^e)$ on the plane spanned by $\bm{e}_1$ and $\bm{e}_3$ with $\xi_2=0$ and the {TT-33 with chamfers}  model for the situations where  $ {\bm J}^e =  \im \omega\sigma_*\bm{\theta}_1^{(1)}$, $ {\bm J}^e =  \im \omega\sigma_*\bm{\theta}_2^{(1)}$, $ {\bm J}^e =  \im \omega\sigma_*\bm{\theta}_3^{(1)}$ are the eddy currents corresponding to $\omega =10^3 \text{ rad/s}$.
 
\subsection{Threat items: Knives}
Knife crime in the U.K. is a persistent issue with 47\,000 offences involving a knife or sharp instrument in England and Wales from April 2018 - March 2019, 285 of which currently recorded as homicide \cite{Commons} with trend being a significant increase in the last 8-9 years. The early recognition of threat objects through metal detection may help to reduce the number of offences involving a sharp instrument. In this section, we present MPT spectral signature characterisations for exemplar knife models.

We will consider a set of 5 different knife models, which we name as { chef}, { cutlet}, { meat cleaver}, { Santoku} and {Wusthof}.
The {chef knife} is a model of a cheap chef knife, featuring a 20cm long, 5cm tall blade with a constant thickness of 1.5mm and a partial tang~\footnote{
The tang is the back portion of the blade, which  extends or connects to a handle, a full tang extends the full length of the handle while a partial tang only extends partially in to the handle~\cite{wikipediatang}}. This model is an example of a stamped knife would normally be constructed with a plastic handle and only the cutting edge of the knife would be sharpened to a point. Obviously, we have only modelled the metallic part of the knife. The {cutlet knife} is a model of a cheap cutlet, featuring a 11cm long, 2cm tall blade with a constant thickness of 1.25mm and a partial tang. This model is an example of a stamped knife and would normally be constructed with a plastic handle and only the cutting edge of the knife would be sharpened to a point.  The {cleaver} is a model of a meat cleaver featuring a 20cm long, 9cm tall blade with a thickness of 3mm over the majority of the blade with a double bevel 6.4 cm from the spin and a full tang with 3 rivets, which are each 2cm long, have radius 3mm and are spaced 45mm apart centre to centre. This model could be made using either the method of stamping or forging. The {Santoku} features a 17cm long, 4.5cm tall blade with a thickness of 1.275mm at its spine, which tapers over the height of the blade to a point at the cutting edge. It features a full tang with 3 rivets, which are each 2cm long, have radius 3.5mm and are spaced 42.5mm apart centre to centre. This model would normally be constructed from a single piece of steel with the two sides of the handle being made with either wood or a plastic material which are then both stuck and riveted to the steel. Finally, the {Wusthof} 
 has a 20cm long,  5cm tall blade with a thickness of 2mm at its spine, which tapers over the height of the blade to a point at the cutting edge. It features a full tang with 3 rivets, which are each 2cm long, have radius 3.5mm and are spaced 42.5mm apart centre to centre. In each case, the measurements quoted have been obtained by approximately measuring the dimensions of common household knives.
  The blade of the knives have been assumed to be made of 440 grade stainless steel, which has a relative permeability $\mu_r=62$ \cite{440bpermeability} and conductivity $\sigma_*=1.6\times10^6$ S/m\cite{tt33conductivity}, but modelled instead with a lower relative permeability $\mu_r=5$, and the rivets to be made of copper, which is non-magnetic having a relative permeability $\mu_r=1$ and a conductivity $\sigma_*=5.8\times10^7$ S/m \cite{tt33conductivity}.  Note that each of the knives are simply connected.
 
\begin{figure}
\begin{center}
$\begin{array}{ccc}
 \includegraphics[width=0.3\textwidth, keepaspectratio]{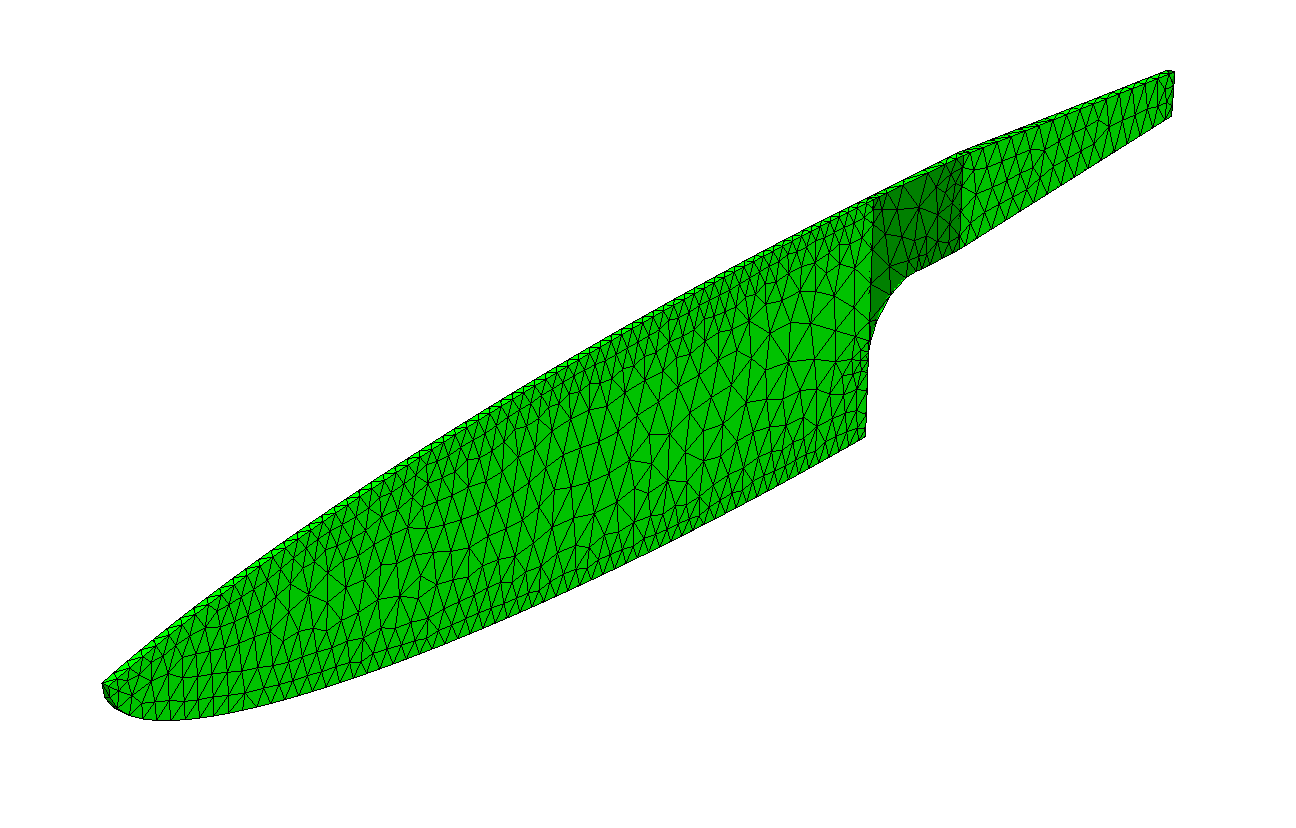} &
 \includegraphics[width=0.3\textwidth, keepaspectratio]{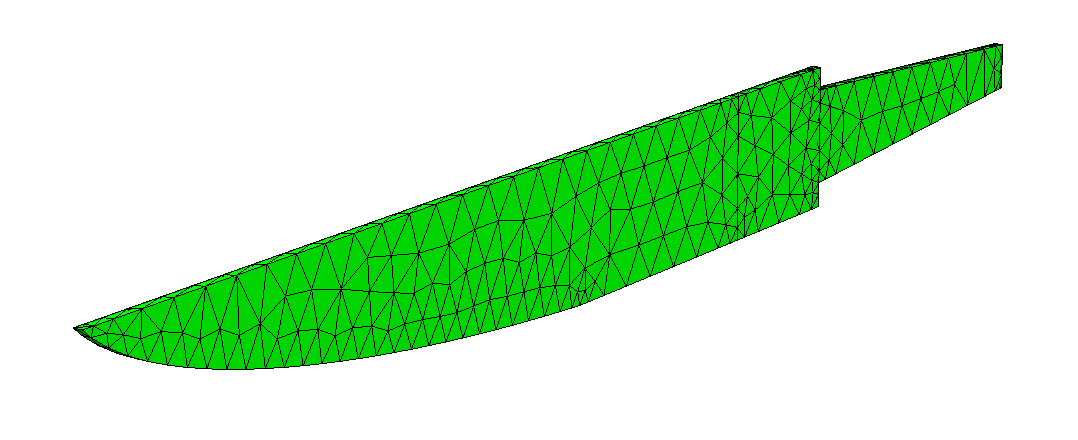} &
 \includegraphics[width=0.3\textwidth, keepaspectratio]{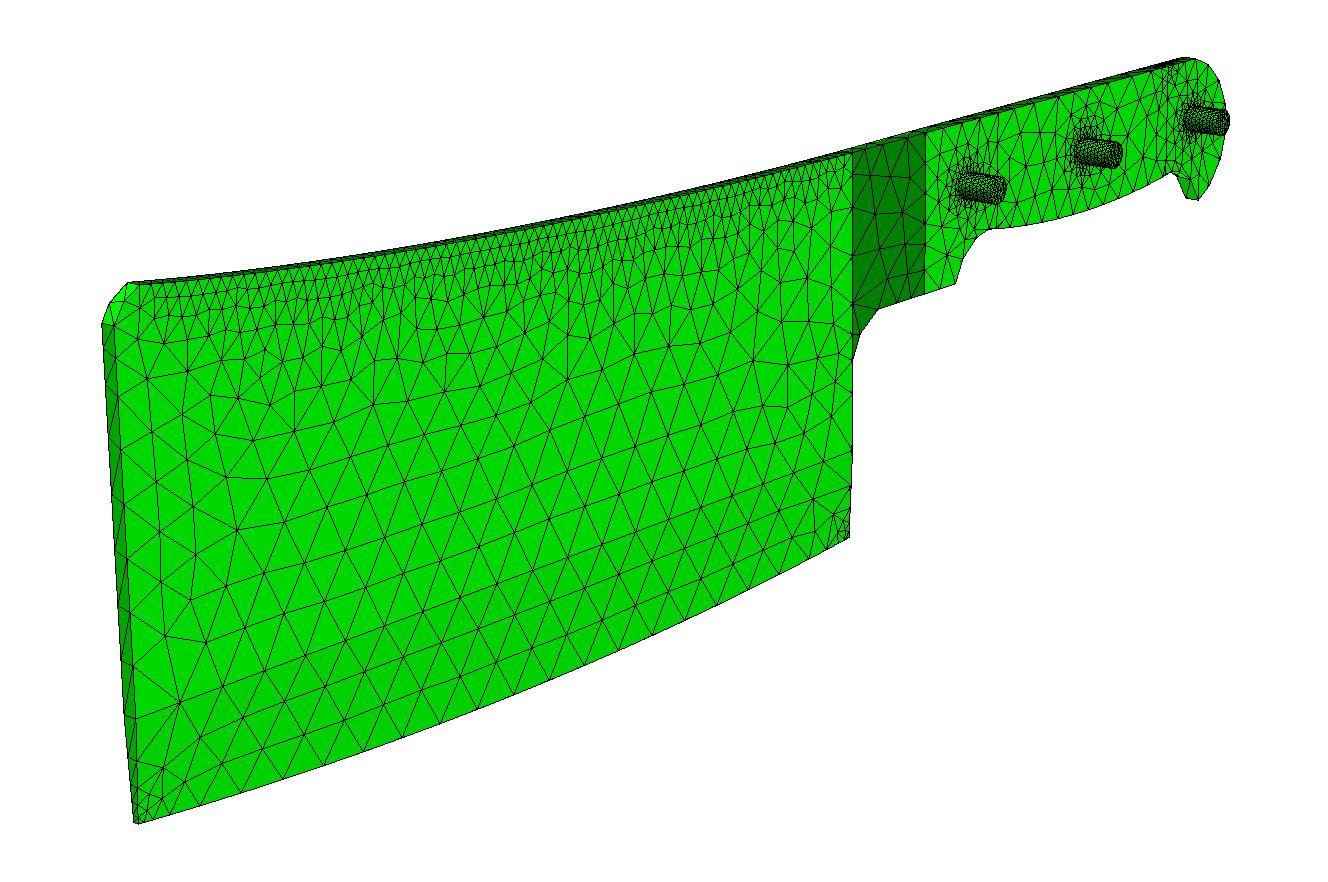} \\
 \text{(a)} & \text{(b)} & \text{(c)} 
 \end{array}$
 \end{center}
 \begin{center}
 $\begin{array}{cc}
 \includegraphics[width=0.3\textwidth, keepaspectratio]{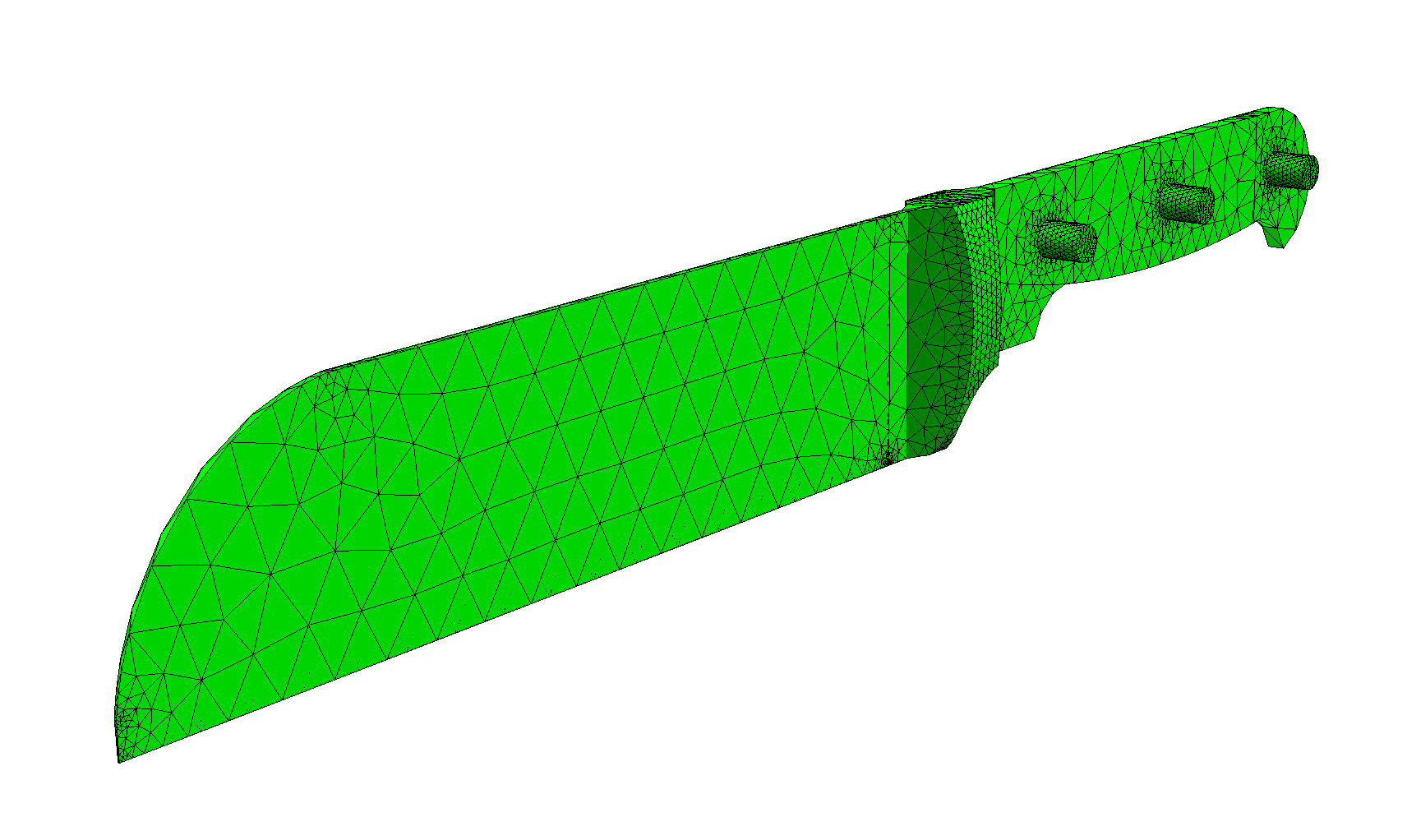} &
 \includegraphics[width=0.3\textwidth, keepaspectratio]{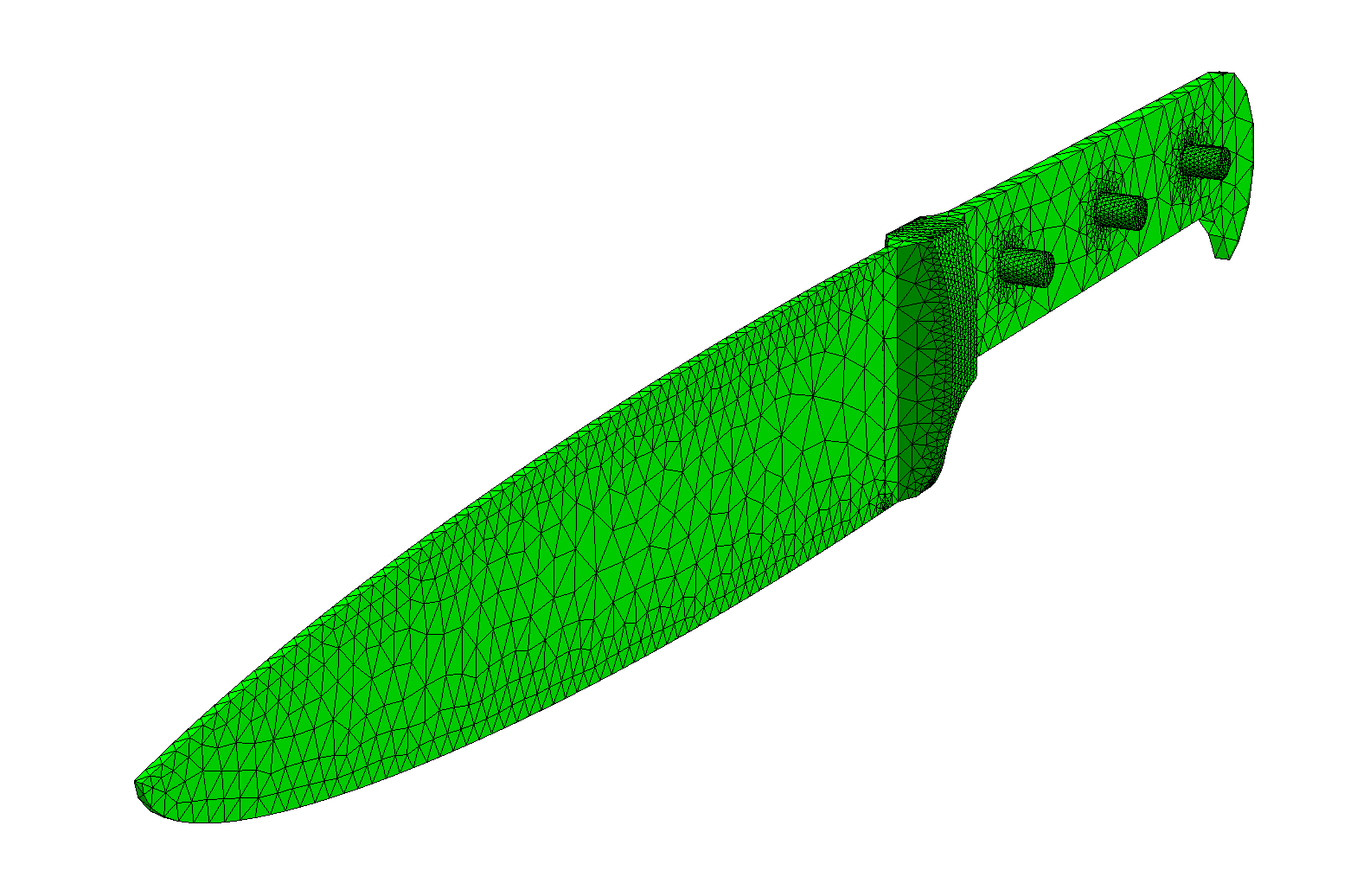} \\
 \text{(d)} & \text{(e)} 
 \end{array}$
 \end{center}
\caption{Set of knives: surface distribution of elements for (a) chef, (b) cutlet, (c) meat cleaver, (d) Santoku and (e) Wusthof} \label{fig:knifemeshes}
\end{figure}

A mesh of each of the geometries was generated assuming dimensionless units, the size parameter $\alpha =0.001$ m and by placing the knife configuration centrally in a box of dimensions $[-1000,1000]^3$ . The resulting meshes contain  $25\, 742$, $14\, 935$, $55\, 226$, $55\, 226 $ and $79\, 945$ unstructured tetrahedra for the {chef}, {cutlet}, {meat cleaver}, {Santoku} and {Wusthof} knives, respectively, and images of the distribution of elements on the surface of the object are reproduced in Figure~\ref{fig:knifemeshes}.
Each of the knives has been orientated so that the blade is parallel to the ${\bm e}_1$ direction and lies in the plane spanned by ${\bm e}_1$ and ${\bm e}_2$ with the knife configuration being symmetrical in the ${\bm e}_3$ direction.  Thus, there are $4$ independent coefficients each in $\tilde{\mathcal R}[\alpha B,\omega, \sigma_*, \mu_r  ]$ and  $ \mathcal{I}  [\alpha B,\omega, \sigma_*, \mu_r  ]$ corresponding to $(\tilde {\mathcal R}  )_{11}$, $( \tilde{\mathcal R} )_{22}$, $( \tilde{\mathcal R} )_{33}$,  $( \tilde{\mathcal R}  )_{13}=(\tilde{\mathcal R} )_{31}$ at each frequency with similar for $ \mathcal{I}  $.
  
\begin{figure}[!h]
\centering
\hspace{-1.cm}
$\begin{array}{cc}
\includegraphics[scale=0.5]{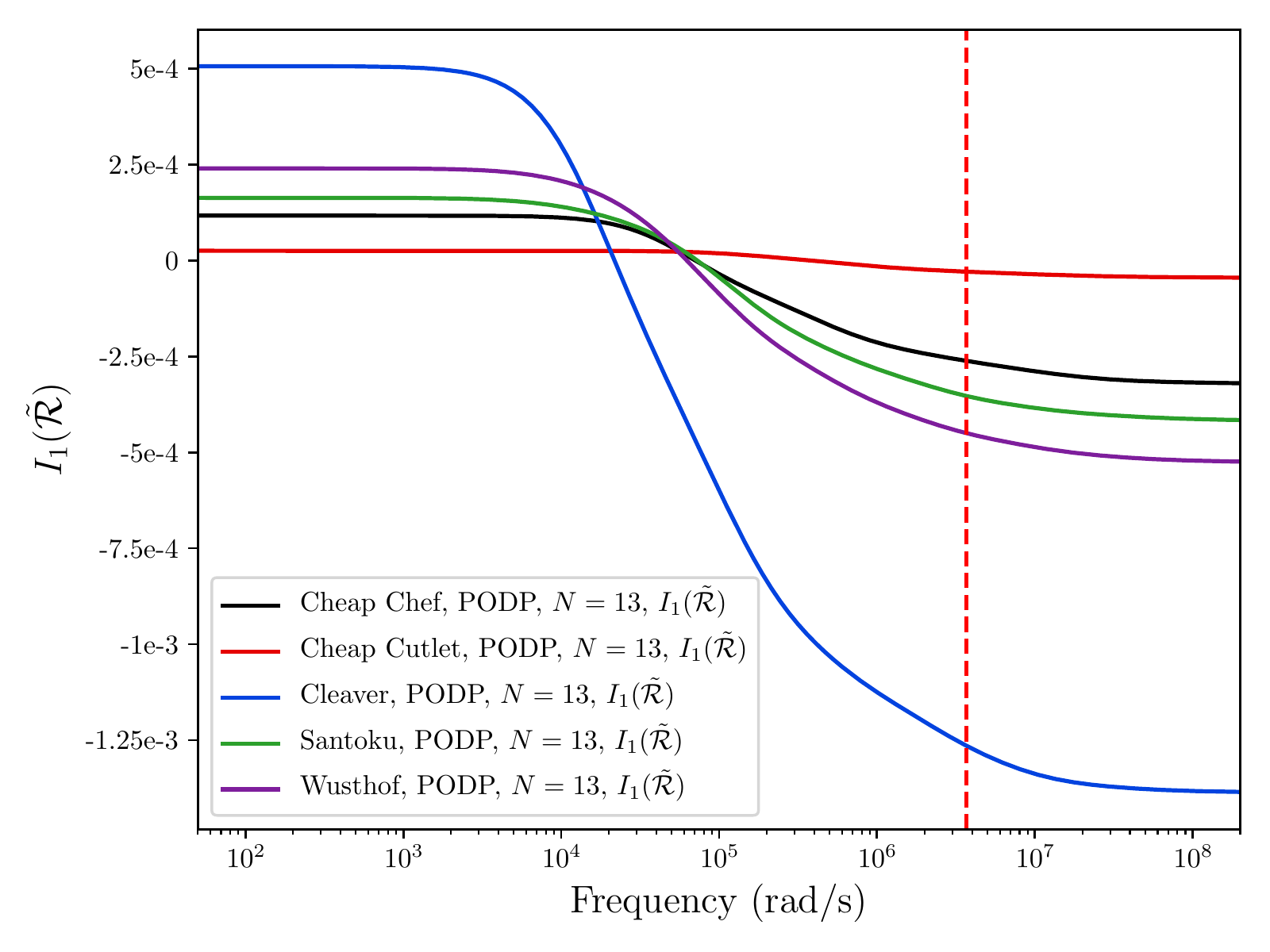} &
\includegraphics[scale=0.5]{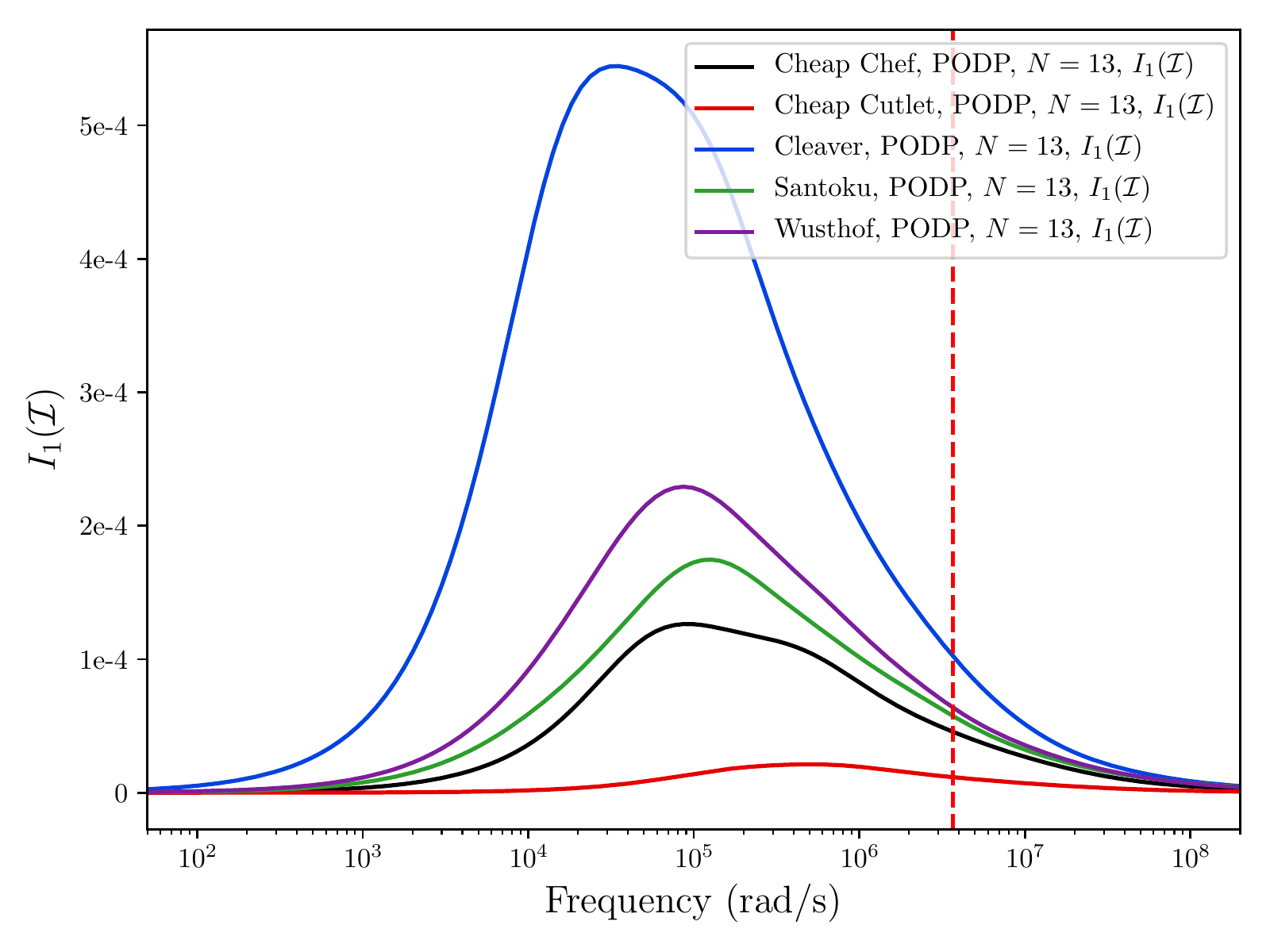}  \\
\text{(a) } I_{1}  ( \tilde{ \mathcal{R}} ) & 
\text{(b) } I_{1} ( \mathcal{I}  )  \\
\includegraphics[scale=0.5]{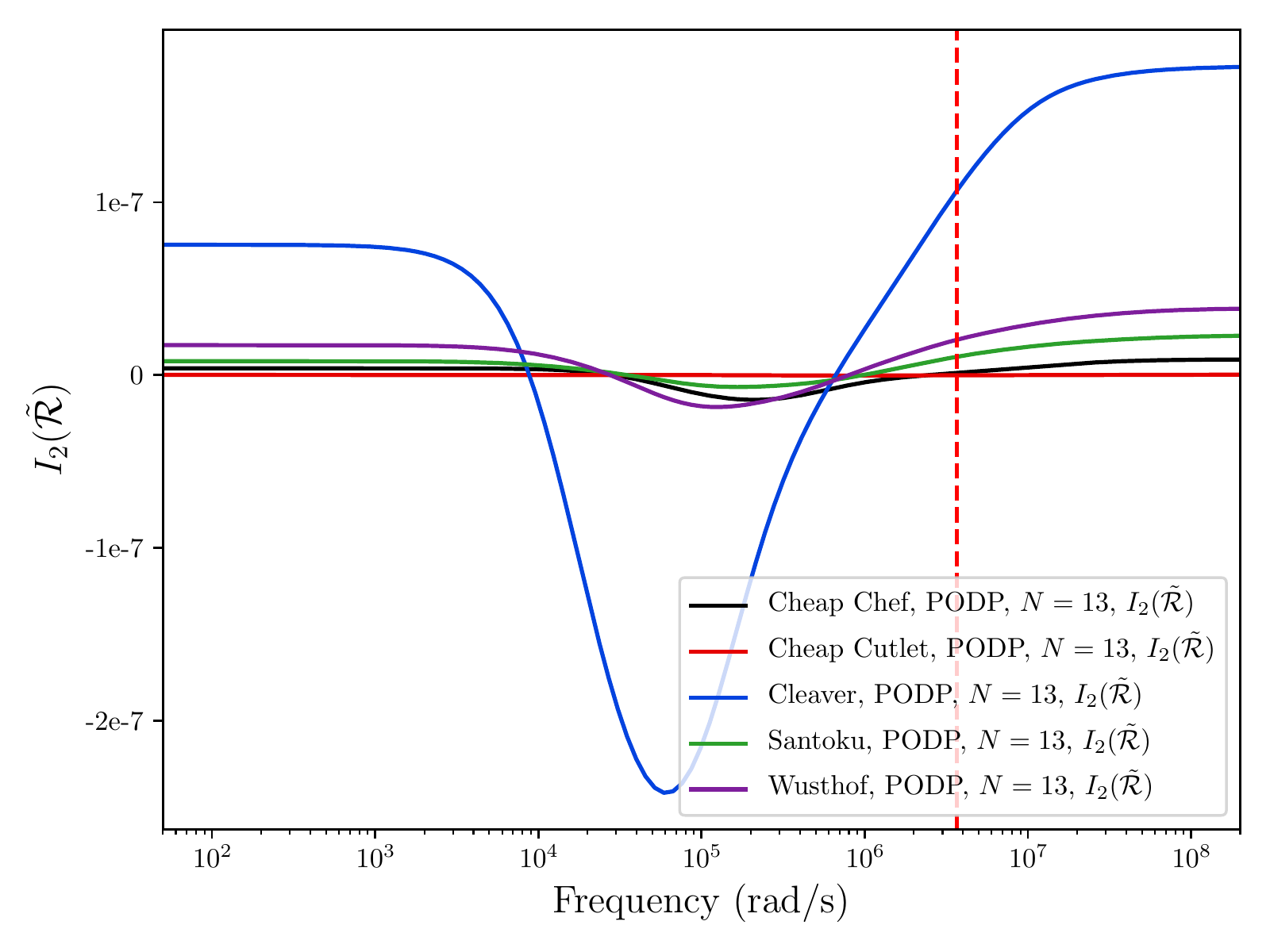} &
\includegraphics[scale=0.5]{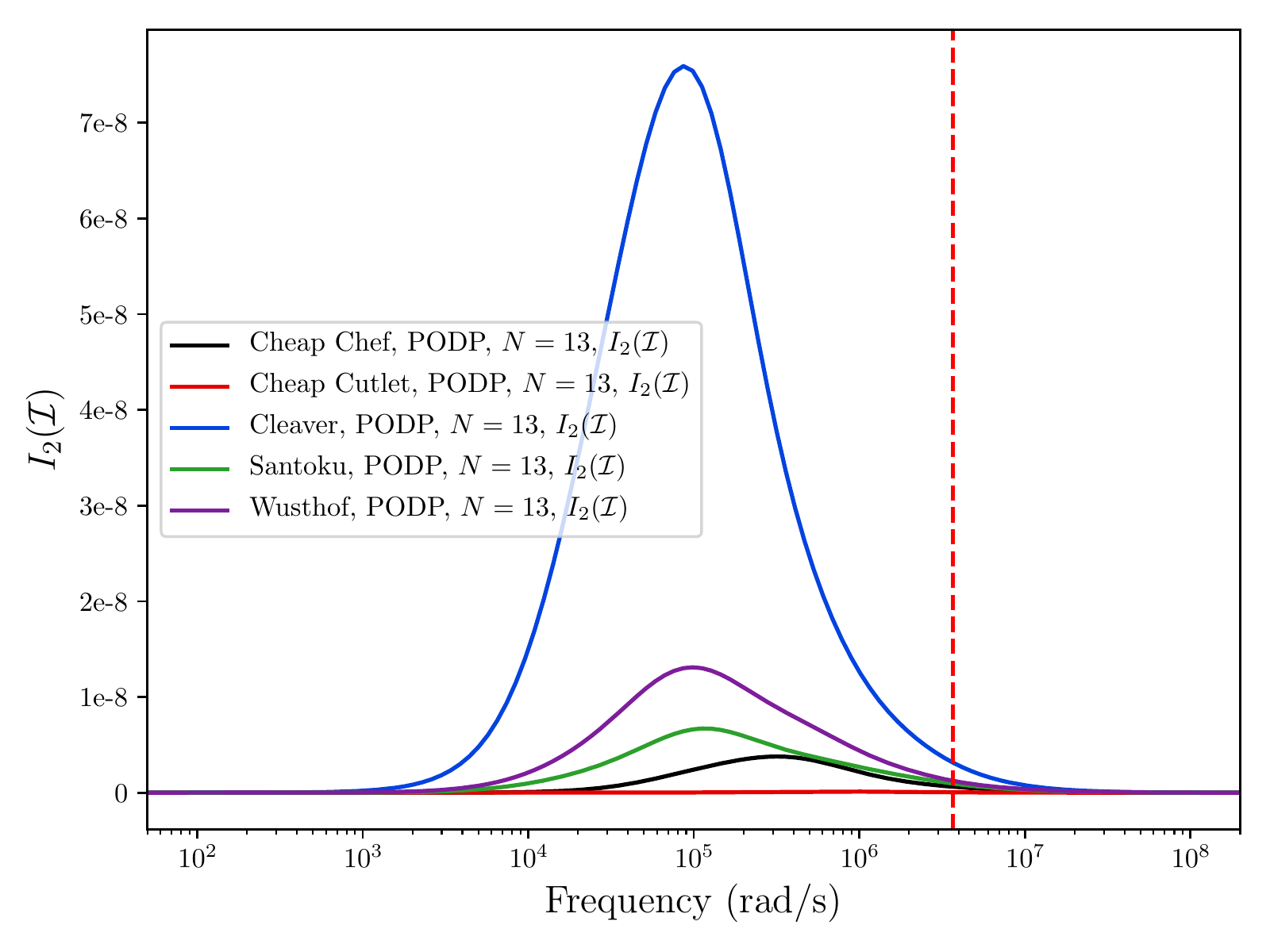}  \\
\text{(c) } I_{2}  ( \tilde{\mathcal{R}}  ) & 
\text{(d) } I_{2 } ( \mathcal{I}  )  \\
\includegraphics[scale=0.5]{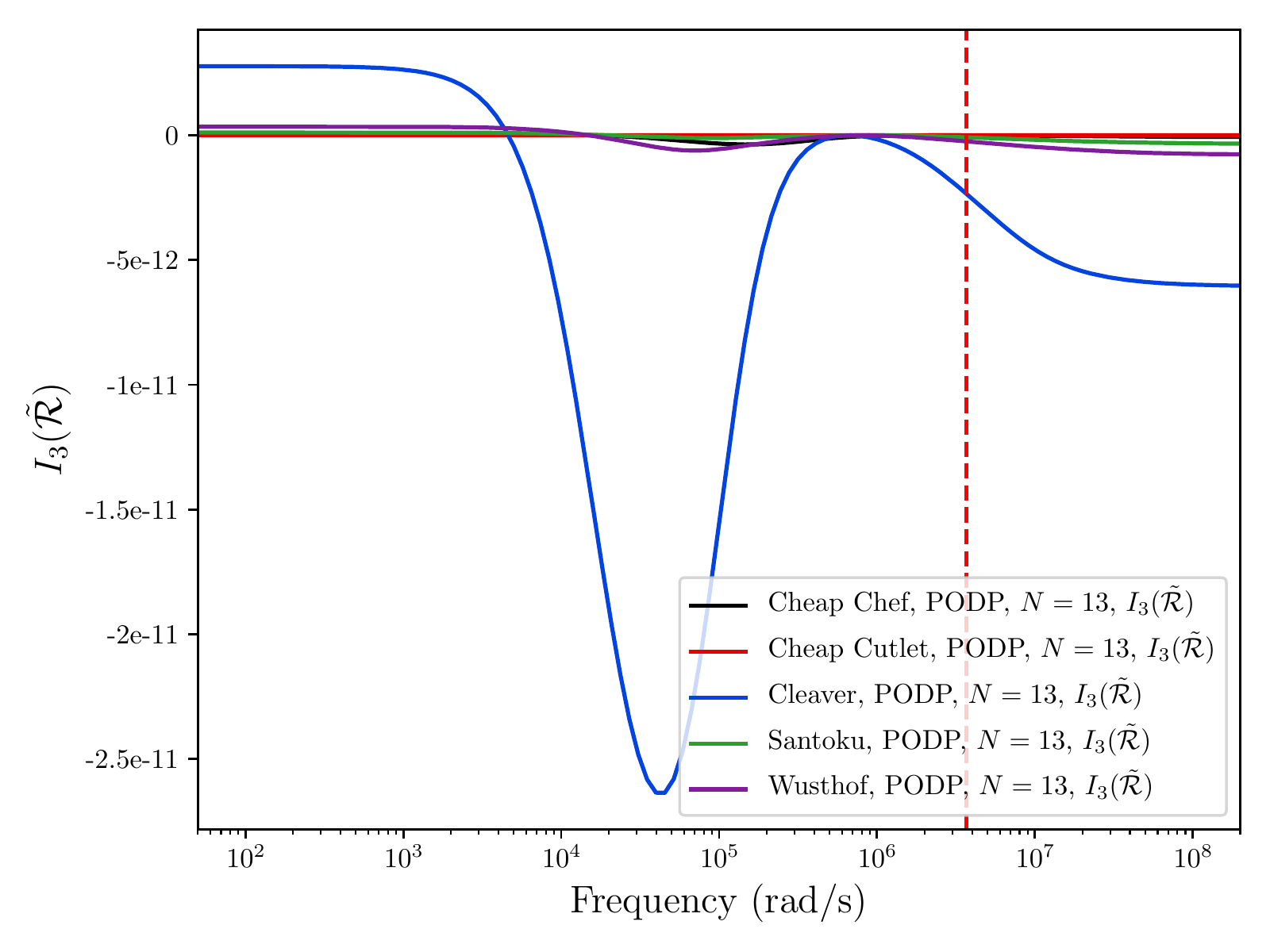} &
\includegraphics[scale=0.5]{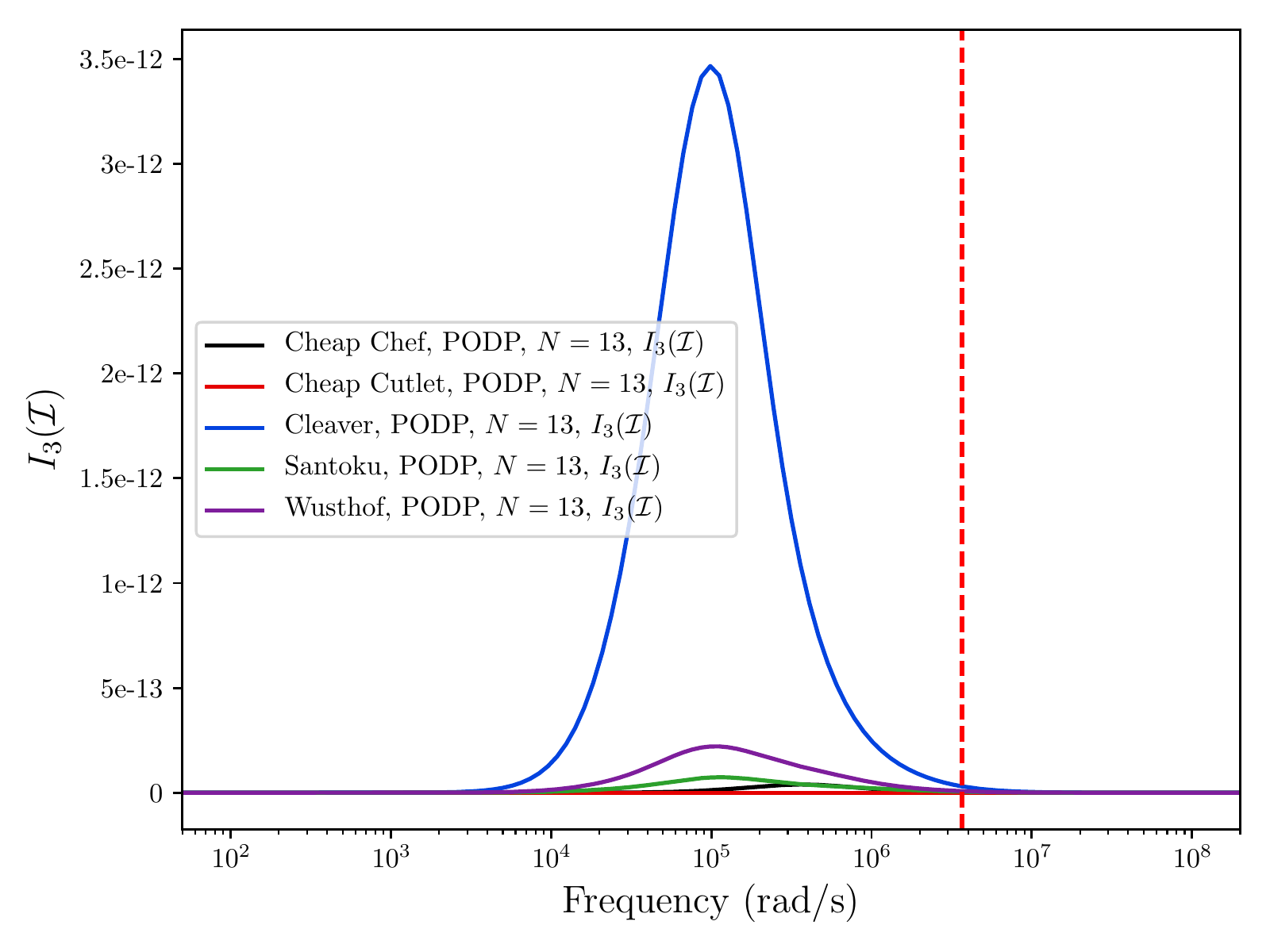}  \\
\text{(e) } I_{3}  ( \tilde{\mathcal{R}}  ) &
\text{(f) }  I_{3} ( \mathcal{I} )  
\end{array}$
  \caption{Set of knives: Comparison of tensor invariants. (a) $I_{1}  ( \tilde{\mathcal{R}} ) $, (b) $I_{1} ( \mathcal{I}  ) $
  (c) $I_{2}  ( \tilde{\mathcal{R}}  ) $, (d) $I_{2} ( \mathcal{I}  ) $,
  (e) $I_{3}  ( \tilde{\mathcal{R}}   ) $ and (f)  $I_{3} ( \mathcal{I} ) $.}
        \label{fig:CompInvknife}
\end{figure}

 The results obtained by using $N=13$  representative  full order solution snapshots at logarithmically spaced frequencies in the range $10^1 \le \omega \le 10^{10} $ rad/s were found to converge with $p=4 $ elements. Then, by applying the PODP algorithm described in Section~\ref{sect:rommethod} with a tolerance of $TOL=10^{-6}$, the MPT spectral signature for each of the receiver models was obtained. A detailed presentation of the results can be found in~\cite{thesisben}. 
 We show a comparison of 
the  MPT spectra using the principal invariants $I_i$, $i=1,2,3$ for $\tilde{\mathcal R} [\alpha B,\omega,\mu_r , \sigma_* ]$ and ${\mathcal I} [\alpha B,\omega,\mu_r , \sigma_* ]$ that have been obtained using the PODP approach in Figure~\ref{fig:CompInvknife}. In this figure, we have restricted consideration to frequencies such that $10^2 \le \omega \le 10^8 \text{rad/s}$ in order to allow comparisons with the earlier results. In practice, the eddy current model brakes down at a frequency $\omega_{limit} < 5 \times 10^6 \text{rad/s}$ for all the knives considered and so higher frequencies are not relevant. 

\begin{figure}[!h]
\centering
\hspace{-1.cm}
$\begin{array}{cc}
 \includegraphics[scale=0.5]{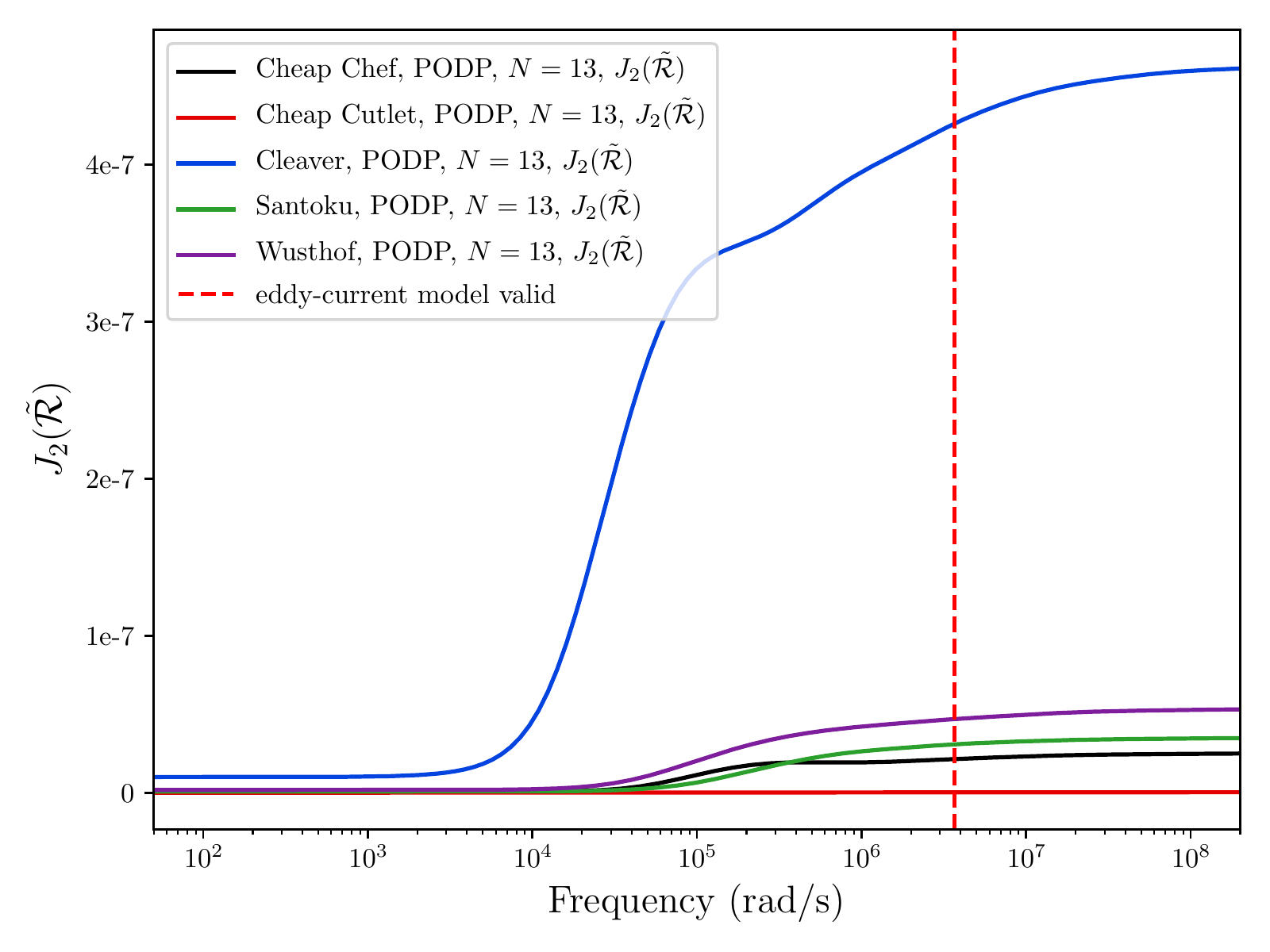} &
 \includegraphics[scale=0.5]{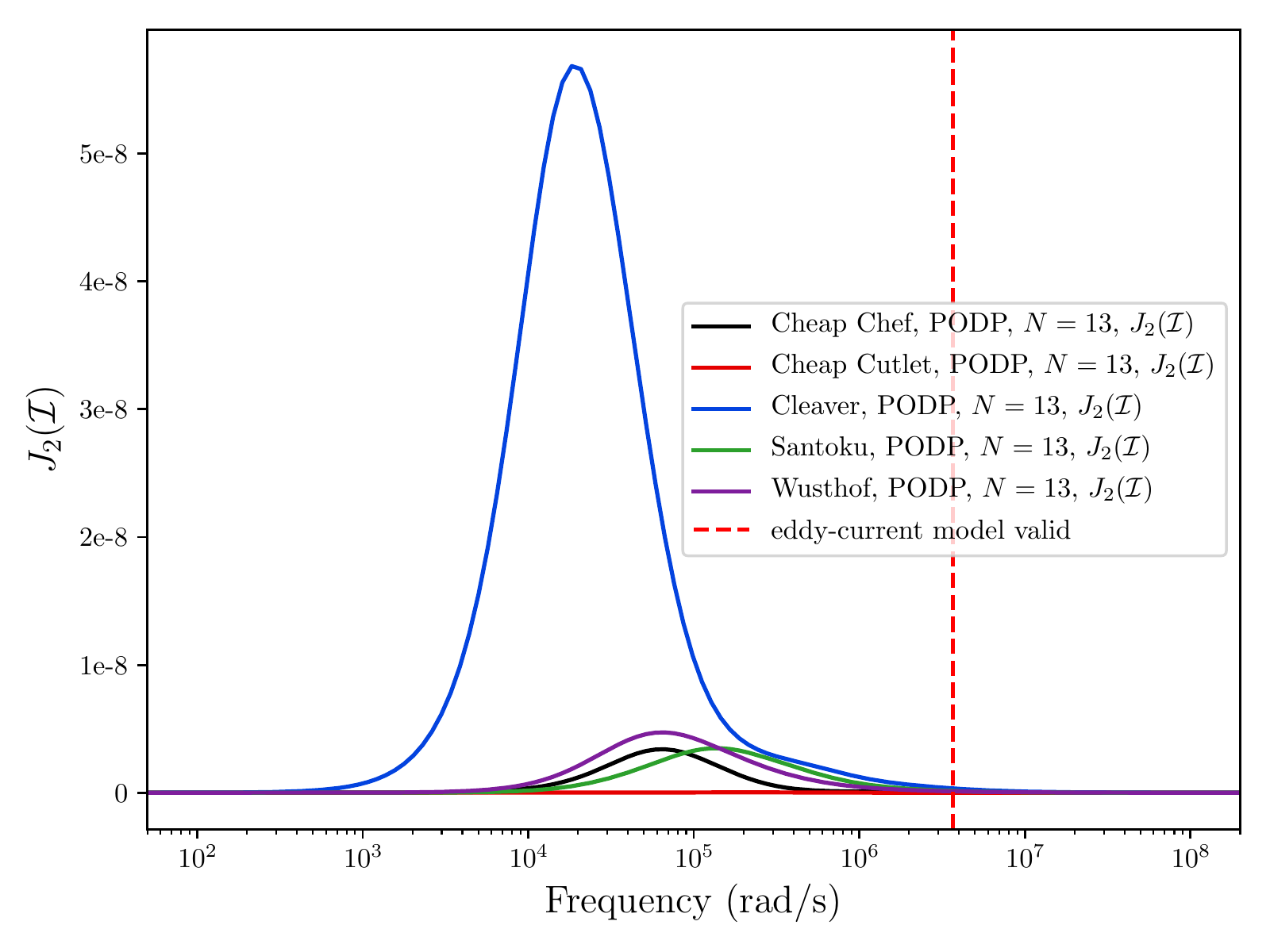}  \\
 \text{(a) } J_{2} ( \tilde{\mathcal{R}}  ) & 
\text{(b) } J_{2 }  ( \mathcal{I}  )  \\
 \includegraphics[scale=0.5]{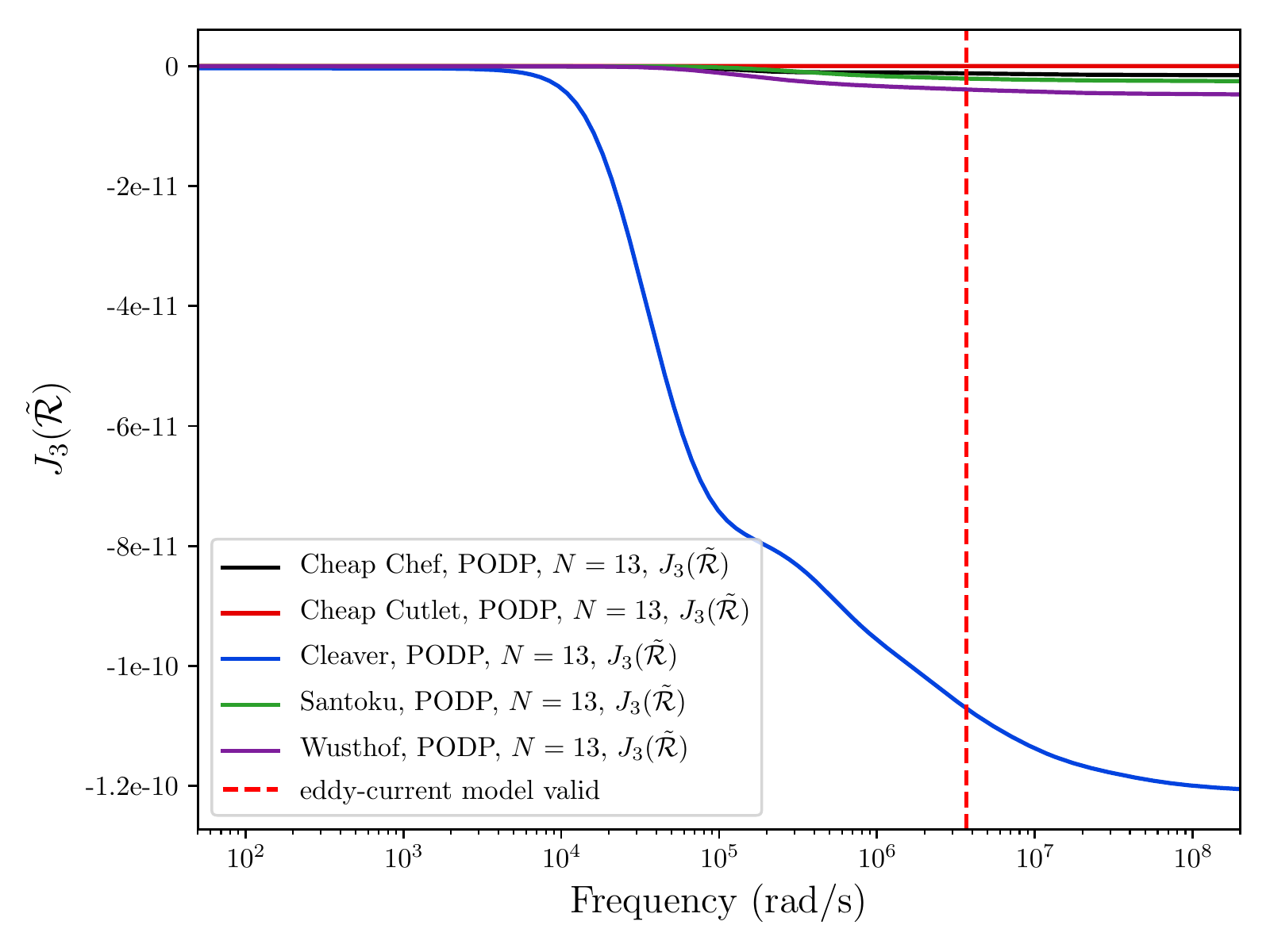} &
 \includegraphics[scale=0.5]{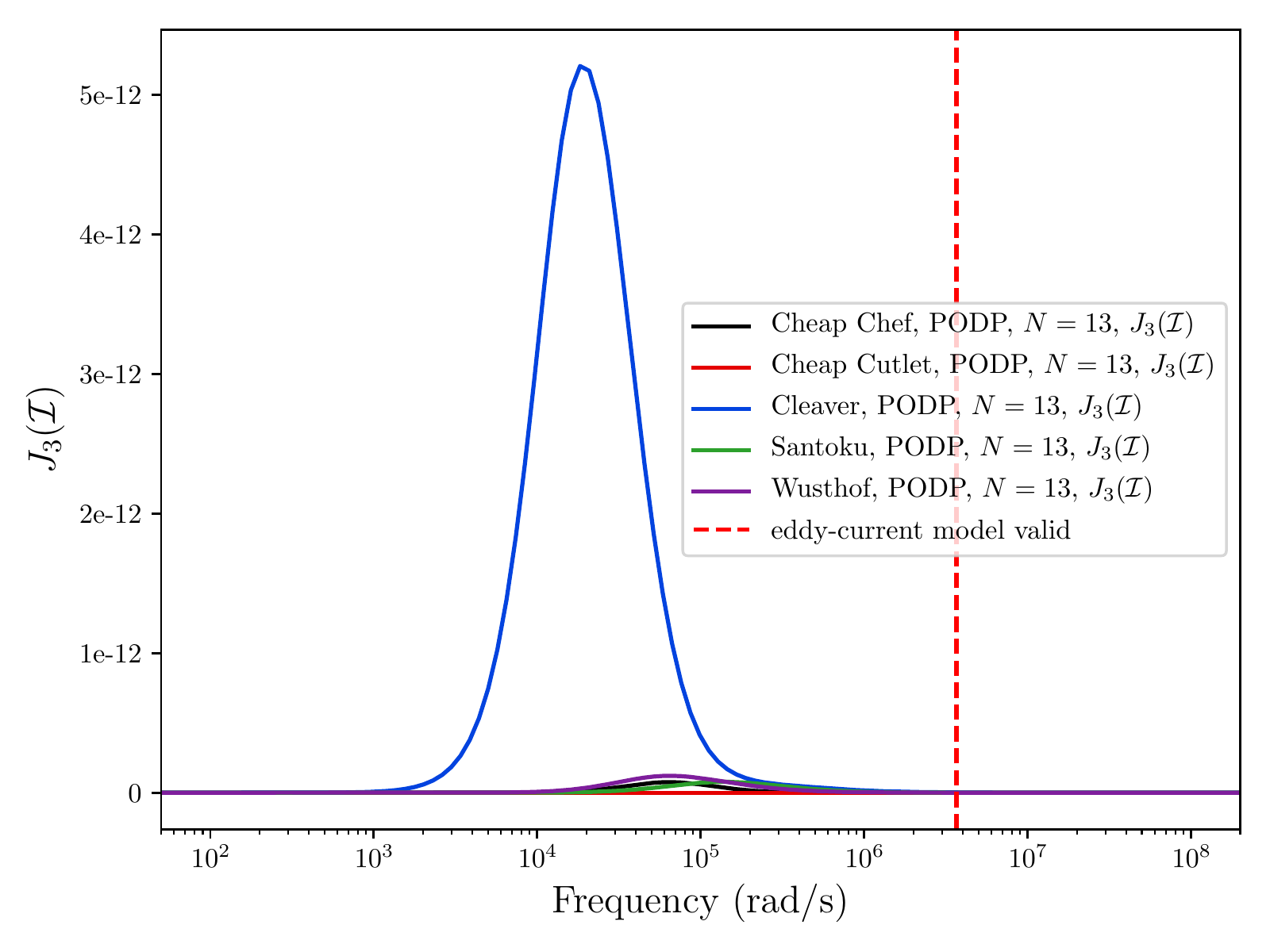}  \\
\text{(c) } J_{3}   (  \tilde{\mathcal{R}}  ) &
\text{(d) }  J_{3} ( \mathcal{I}  )  
\end{array}$
  \caption{Set of knives:  Comparison of tensor invariants. 
  (a) $J_{2}  ( \tilde{\mathcal{R}} ) $, (b) $J_{2} ( \mathcal{I} ) $,
  (c) $J_{3}  ( \tilde{\mathcal{R}}  ) $ and (d)  $J_{3} ( \mathcal{I} ) $.}
        \label{fig:CompDevInvknife}        
\end{figure}

{The results obtained for the different models shown in Figure~\ref{fig:CompInvknife} have some similarities to the TT-33 models in that $I_1(\tilde{\mathcal R})$ is monotonically increasing with $\log \omega$ and $I_2( \tilde{\mathcal R})$, $I_3(\tilde{\mathcal R})$ are not monotonically increasing or decreasing with $\log \omega$ and the curves $I_i({\mathcal I})$, $i=1,2,3$, each show a single local maximum with $\log \omega$. However, the characteristics of the curves is otherwise quite different, again motivating that discrimination between objects is possible. Comparing the different knife models, we see different behaviour of the invariants in each case.} On closer inspection of the eigenvalues of $\tilde{\mathcal R} $ and ${\mathcal I} $ (presented in~\cite{thesisben}) it is possible to observe multiple non-stationary points of inflection and multiple local maxima, respectively, particularly when considering the {Santoku}  and {Wusthof} knives, which are inhomogeneous. Although the {cleaver} also has inhomogeneous materials, the larger extent of material in the blade largely disguises these effects.
The corresponding results for the alternative invariants are shown in~\ref{fig:CompDevInvknife}, where again observe a significant difference between the {clever} and the other models.

\begin{figure}[!h]
\centering
    \includegraphics[scale=0.5]{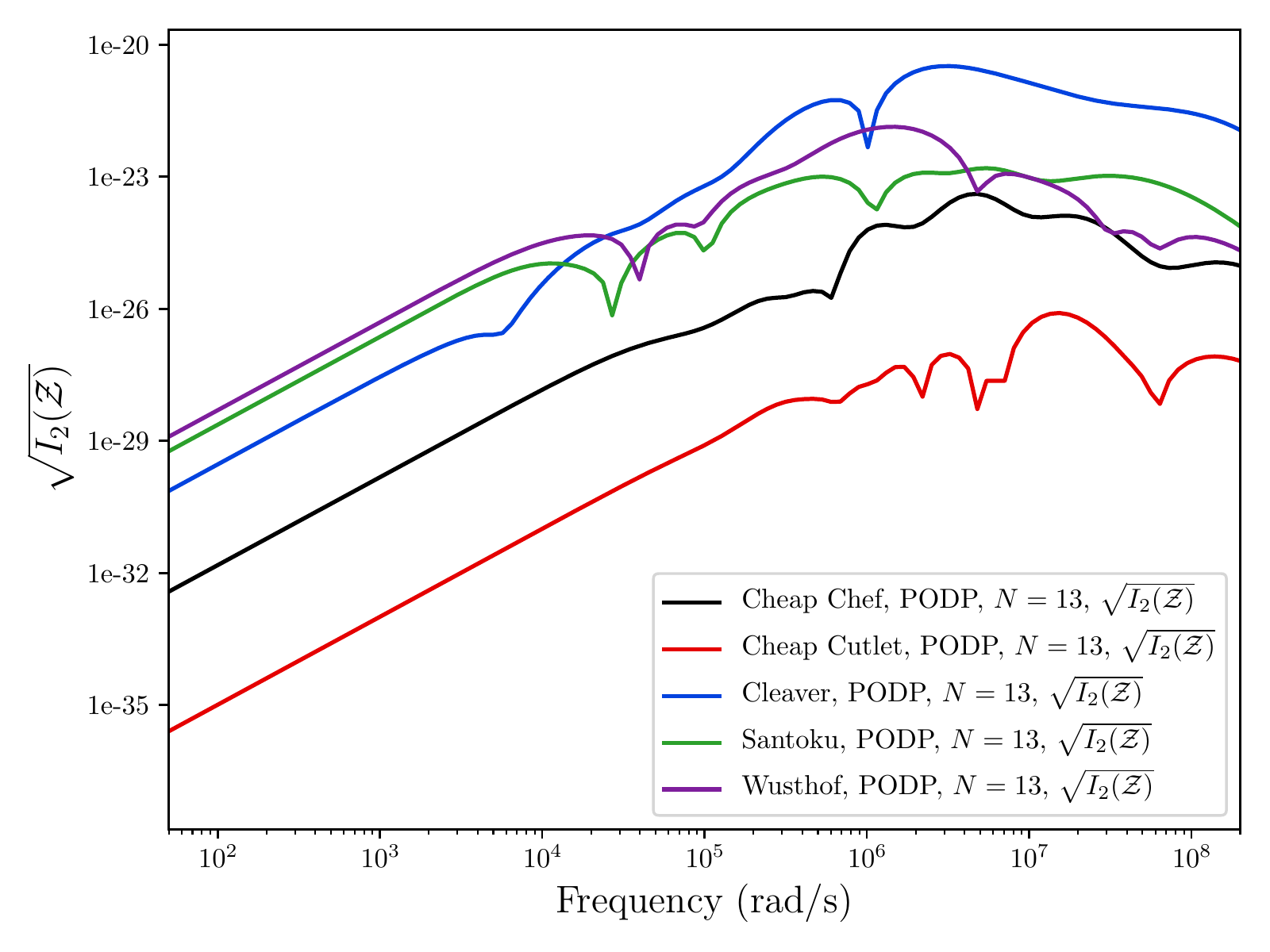}
  \caption{Set of knives: Comparison of the tensor invariant    $\sqrt{I_2 ( {\mathcal Z}[ \alpha B,\omega,  \sigma_*, \mu_r])} $.}
        \label{fig:Mesh:zCurveknife}
\end{figure}

For the knife models, each of the associated MPT frequency spectra have independent coefficients that are associated with both on and off diagonal entries of the tensor. The behaviour of $\sqrt{I_2 ( {\mathcal Z}[ \alpha B,\omega,  \sigma_*, \mu_r])} $ for the different models is shown in Figure~\ref{fig:Mesh:zCurveknife}.

\section{Conclusion}

In this paper we have demonstrated how a dictionary of object characterisation can be computed for training machine learning algorithms, with the end goal of being able to classify objects for metal detection.
We have focused on the MPT  characterisation of conducting objects and have shown, at a fixed frequency, that the characterisation provided is equivalent upto an an equivalent ellipsoid. But, by considering an MPT's spectral signature, in which the coefficients are obtained as a function of frequency, the characterisation becomes much richer, containing much more information, whereby the object is characterised by a series of different equivalent ellipsoids at each of the excitation frequencies considered. A series of alternative MPT spectral signature invariants have been provided, which are invariant to object rotation, and, although the {principal and alternative invariants contain the same information as a tensor's eigenvalues, they offer advantages by providing MPT spectral signatures that are not always sigmoid with $\log \omega$ when applied to $\tilde{\mathcal R}$ and have a single local maxima with $\log \omega$ for ${\mathcal I}$.}

 Finally, a series of practically motivated computational examples have been presented consisting of both non-threat objects (a range of brass keys and and set of British coins of different sizes and shapes and with different materials) and threat objects (a series of models of a the receiver part of a TT-33 semi-automatic pistol and a series of different knives). The MPT spectral signature of these different objects using the tensor invariants $I_i$, $i=1,2,3$ and $J_i$, $i=2,3$ applied to $\tilde{R}={\mathcal N}^0 +{\mathcal R}$ and ${\mathcal I}$ have shown quite different characteristics, illustrating the possibility to use them to discriminate between different objects.
  This set can easily extended to a much larger dictionary by combining these results with the previous scaling results obtained in~\cite{ben2020} that allow for changes in object size and object conductivities.

In future work we intend to compare the performance of different machine learning classifiers using the dictionary of MPT spectral signature characterisation described in this work.


\section*{Acknowledgements}

P.D. Ledger and A.A.S. Amad gratefully acknowledges the financial support received from EPSRC in the form of grant EP/R002134/1. B.A. Wilson gratefully acknowledges the financial support received from EPSRC in the form of a DTP studentship with project reference number 2129099. 
W.R.B. Lionheart gratefully acknowledges the  financial support received from EPSRC in the form of grant EP/R002177/1 and would like to thank the Royal Society for the financial support received from a Royal Society Wolfson Research Merit Award. EPSRC Data Statement: All data is provided in Section 5. This paper does not have any conflicts of interest.

\afterpage{\clearpage}
\bibliographystyle{acm}
\bibliography{paperbib}

\begin{thebibliography}{10}

\bibitem{conductivity}
Conductivity of metals sorted by resistivity.
\newblock
  \url{http://eddy-current.com/conductivity-of-metals-sorted-by-resistivity/}.
\newblock Date accessed 17th October 2020.

\bibitem{blueprints}
{TT33} blueprints.
\newblock
  \url{https://pdfslide.net/documents/tt-33-model-blueprints-drawings-of-main-parts.html}.
\newblock Date accessed 20th October 2020.

\bibitem{abdel}
{\sc Abdel-Rehim, O.~A., Davidson, J.~L., Marsh, L.~A., O'Toole, M.~D., and
  Peyton, A.~J.}
\newblock Magnetic polarizability tensor spectroscopy for low metal
  anti-personnel mine surrogates.
\newblock {\em IEEE Sensors 16\/} (2016), 3775--3783.

\bibitem{ambrus}
{\sc {Ambru\v{s}}, D., {Vasi\'c}, D., and Bilas, V.}
\newblock Robust estimation of metal target shape using time-domain
  electromagnetic induction data.
\newblock {\em IEEE Transactions Instrumentation and Measurement 65\/} (2016),
  795--807.

\bibitem{ammarishapeelectro}
{\sc Ammari, H., Boulier, T., Garnier, J., and Wang, H.}
\newblock Shape recognition and classification in electro-sensing.
\newblock {\em Proceedings of the National Academy of Sciences 111\/} (2014),
  11652--11657.

\bibitem{Ammari2014}
{\sc Ammari, H., Chen, J., Chen, Z., Garnier, J., and Volkov, D.}
\newblock Target detection and characterization from electromagnetic induction
  data.
\newblock {\em Journal de Math{\'e}matiques Pures et Appliqu{\'e}es 101(1)\/}
  (2014), 54--75.

\bibitem{Ammari2015}
{\sc Ammari, H., Chen, J., Chen, Z., Volkov, D., and Wang, H.}
\newblock Detection and classification from electromagnetic induction data.
\newblock {\em Journal of Computational Physics 301\/} (2015), 201--217.

\bibitem{ammarikangbook}
{\sc Ammari, H., and Kang, H.}
\newblock {\em Polarization and Moment Tensors with Applications to Inverse
  Problems and Effective Medium Theory}.
\newblock Springer-Verlag New York, 2007.

\bibitem{ammari2019}
{\sc Ammari, H., Putinar, M., Ruiz, M., Yu, S., and Zhang, H.}
\newblock Shape reconstruction of nanoparticles from their plasmonic
  resonances.
\newblock {\em Journal de Math{\'e}matiques Pures et Appliqu{\'e}es 122\/}
  (2019), 23--48.

\bibitem{tt33-image}
{\sc Antonosen, A.}
\newblock
  \url{https://www.flickr.com/photos/handvapensamlingen/6825679152/in/photolist-bpaoBj-eXwMWW-a664rz-2fhqR43-dQ7ipQ},
  2012.
\newblock Date Accessed 28th October 2020.

\bibitem{aospheroid}
{\sc Ao, C.~O., Braunisch, H., O'Neill, K., and Kong, J.~A.}
\newblock Quasi-magnetostatic solution for a conducting and permeable spheroid
  with arbitrary excitation.
\newblock {\em IEEE Transactions on Geoscience and Remote Sensing 40\/} (2002),
  887--897.

\bibitem{barrowes2004}
{\sc Barrowes, B.~E., O'Neill, K., Gregorcyzk, T., and Kong, J.~A.}
\newblock Broadband analytical magnetoquasistatic electromagnetic induction
  solution for a conducting and permeable spheroid.
\newblock {\em IEEE Transactions on Geoscience and Remote Sensing 42\/} (2004),
  2479--2489.

\bibitem{bishopbook}
{\sc Bishop, C.~M.}
\newblock {\em Pattern Recognition and Machine Learning}.
\newblock Springer, 2006.

\bibitem{bonet}
{\sc Bonet, J., and Wood, R.~D.}
\newblock {\em Nonlinear Continuum Mechanics for Finite Element Analysis}.
\newblock Cambridge University Press, 2010.

\bibitem{davidsoncoins}
{\sc Davidson, J.~L., Abdel-Rehim, O.~A., Hu, P., Marsh, L.~A., O'Toole, M.~D.,
  and Peyton, A.~J.}
\newblock On the magnetic polarizability tensor of us coinage.
\newblock {\em Measurement Science and Technology 29\/} (2018), 035501.

\bibitem{dekdouk}
{\sc Dekdouk, B., Ktistis, C., Marsh, L.~A., Armitage, D.~W., and Peyton,
  A.~J.}
\newblock Towards metal detection and identification for humanitarian demining
  using magnetic polarizability tensor spectroscopy.
\newblock {\em Measurement Science and Technology 26\/} (2015), 115501.

\bibitem{Commons}
{\sc G.~Allen, L.~A., Loft, P., and Bellis, A.}
\newblock Knife crime in england and wales.
\newblock Tech. Rep. SN4304, 2019.
\newblock Date Accessed 9th October 2020.

\bibitem{golub}
{\sc Golub, G.~H., and Loan, C.~F.~V.}
\newblock {\em Matrix Computations}.
\newblock JHU Press, 1996.

\bibitem{barrowes2008}
{\sc Gregorcyzk, T., Zhang, B., Kong, J., B.E.Barrowes, and O'Neill, K.}
\newblock Electromagnetic induction from highly permeable and conductive
  ellipsoids under arbitrary excitation: application to the detection of
  unexploded ordances.
\newblock {\em IEEE Transactions on Geoscience and Remote Sensing 46\/} (2008),
  1164--1176.

\bibitem{coinpermeability}
{\sc Gross, M.~R.}
\newblock Magnetic characteristics of non-magnetic metallic materials
  comparison of properties in strong and weak fields.
\newblock Tech. Rep. E.E.S. Report 4E(2)66904, U.S. Naval Engineering
  Experiment Station, Annapolis, Maryland, USA, 1951.

\bibitem{hesthaven2016}
{\sc Hesthaven, J.~S., Rozza, G., and Stamm, B.}
\newblock {\em Certified Reduced Basis Methods for Parametrized Partial
  Differential Equations}.
\newblock Springer, 2016.

\bibitem{coinconductivity}
{\sc Ho, C.~Y., Ackerman, M.~W., Wu, K.~Y., Havill, T.~N., Bogaard, R.~H.,
  Matula, R.~A., Oh, S.~G., and James, H.~M.}
\newblock Electrical resistivity of ten selected binary alloy systems.
\newblock {\em Journal of Physical and Chemical Reference Data 12\/} (1983),
  183--322.

\bibitem{wikitt33}
{\sc Image, T.~P.}
\newblock \url{https://en.wikipedia.org/wiki/TT_pistol}.
\newblock Date accessed 20th October 2020.

\bibitem{karimian2017}
{\sc Karimian, N., O'Toole, M.~D., and Peyton, A.~J.}
\newblock Electromagnetic tensor spectroscopy for sorting of shredded metallic
  scrap.
\newblock In {\em IEEE SENSORS 2017 - Conference Proceedings\/} (2017), IEEE.

\bibitem{taufiq2016}
{\sc Khairuddin, T.~A.~K., and Lionheart, W.~R.~B.}
\newblock Characterization of objects by electrosensing fish based on the first
  order polarization tensor.
\newblock {\em Bioinspiration and Biomimetics 11\/}, 055004.

\bibitem{taufiq2012}
{\sc Khairuddin, T.~A.~K., and Lionheart, W. R.~B.}
\newblock Do electro-sensing fish use the first order polarization tensor for
  object characterization?
\newblock In {\em 100 Years of Electrical Imaging}, p.~149.

\bibitem{taufiq}
{\sc Khairuddinm, T.~A.~K., and Lionheart, W.~R.~B.}
\newblock Fitting ellipsoids to objects by the first order polarization tensor.
\newblock {\em Malaya Journal of Matematik 4(1)\/} (2013), 44--53.

\bibitem{LedgerLionheart2015}
{\sc Ledger, P.~D., and Lionheart, W.~R.~B.}
\newblock Characterising the shape and material properties of hidden targets
  from magnetic induction data.
\newblock {\em IMA Journal of Applied Mathematics 80(6)\/} (2015), 1776--1798.

\bibitem{LedgerLionheart2016}
{\sc Ledger, P.~D., and Lionheart, W.~R.~B.}
\newblock Understanding the magnetic polarizability tensor.
\newblock {\em IEEE Transactions on Magnetics 52(5)\/} (2016), 6201216.

\bibitem{LedgerLionheart2018}
{\sc Ledger, P.~D., and Lionheart, W.~R.~B.}
\newblock An explicit formula for the magnetic polarizability tensor for object
  characterization.
\newblock {\em {IEEE} Transactions on Geoscience and Remote Sensing 56(6)\/}
  (2018), 3520--3533.

\bibitem{LedgerLionheart2018g}
{\sc Ledger, P.~D., and Lionheart, W.~R.~B.}
\newblock Generalised magnetic polarizability tensors.
\newblock {\em Mathematical Methods in the Applied Sciences 41\/} (2018),
  3175--3196.

\bibitem{LedgerLionheart2019}
{\sc Ledger, P.~D., and Lionheart, W.~R.~B.}
\newblock The spectral properties of the magnetic polarizability tensor for
  metallic object characterisation.
\newblock {\em Mathematical Methods in the Applied Sciences 43\/} (2020),
  78--113.

\bibitem{LedgerLionheartamad2019}
{\sc Ledger, P.~D., Lionheart, W.~R.~B., and Amad, A.~A.~S.}
\newblock Characterisation of multiple conducting permeable objects in metal
  detection by polarizability tensors.
\newblock {\em Mathematical Methods Applied Sciences 42(3)\/} (2019), 830--860.

\bibitem{ledgerzaglmayr2010}
{\sc Ledger, P.~D., and Zaglmayr, S.}
\newblock {$hp$}-finite element simulation of three-dimensional eddy current
  problems on multiply connected domains.
\newblock {\em Computer Methods in Applied Mechanics and Engineering 199\/}
  (2010), 3386--3401.

\bibitem{marsh2014b}
{\sc Makkonen, J., Marsh, L.~A., Vihonen, J., {J\"arvi}, A., Armitage, D.~W.,
  Visa, A., and Peyton, A.~J.}
\newblock {KNN} classification of metallic targets using the magnetic
  polarizability tensor.
\newblock {\em Measurement Science and Technology 25\/} (2014), 055105.

\bibitem{marsh2015}
{\sc Makkonen, J., Marsh, L.~A., Vihonen, J., {J\"arvi}, A., Armitage, D.~W.,
  Visa, A., and Peyton, A.~J.}
\newblock Improving reliability for classification of metallic objects using a
  {WTMD} portal.
\newblock {\em Measurement Science and Technology 26\/} (2015), 105103.

\bibitem{Makkonen2015}
{\sc Makkonen, J., Marsh, L.~A., Vihonen, J., {J\"arvi}, A., Armitage, D.~W.,
  Visa, A., and Peyton, A.~J.}
\newblock Improving the reliability for classification of metallic targets
  using a wtmd portal.
\newblock {\em Measurement Science and Technology 26\/} (2015), 105103.

\bibitem{marsh2013}
{\sc Marsh, L.~A., Ktisis, C., J{\"a}rvi, A., Armitage, D.~W., and Peyton,
  A.~J.}
\newblock Three-dimensional object location and inversion of the magnetic
  polarisability tensor at a single frequency using a walk-through metal
  detector.
\newblock {\em Measurement Science and Technology 24\/} (2013), 045102.

\bibitem{marsh2014}
{\sc Marsh, L.~A., Ktistis, C., {J\"arvi}, A., .Armitage, D.~W., and Peyton,
  A.~J.}
\newblock Determination of the magnetic polarizability tensor and three
  dimensional object location for multiple objects using a walk-through metal
  detector.
\newblock {\em Measurement Science and Technology 25\/} (2014), 055107.

\bibitem{royalmint}
{\sc Mint, R.}
\newblock
  \url{https://www.royalmint.com/discover/uk-coins/coin-design-and-specifications/}.
\newblock Date accessed 20th October 2020.

\bibitem{tt33conductivity}
{\sc Mitchell, B.~S.}
\newblock {\em An Introduction to Materials Engineering and Science: For
  Chemical and Materials Engineers}.
\newblock John Wiley {\&} Sons, 2004.

\bibitem{norton2001}
{\sc Norton, S.~J., and Won, I.~J.}
\newblock Identification of buried unexploded ordnance from broadband induction
  data.
\newblock {\em IEEE Transactions Geoscience Remote Sensing 39\/} (2001),
  2253--2261.

\bibitem{osborn}
{\sc Osborn, J.~A.}
\newblock Demagnetizing factors of the general ellipsoid.
\newblock {\em Physical Review 67\/} (1945), 351--357.

\bibitem{rehim2016}
{\sc Rehim, O.~A.~A., Davidson, J.~L., Marsh, L., O'Toole, M.~D., and Peyton,
  A.~J.}
\newblock Magnetic polarizability spectroscopy for low metal anti-personnel
  mine surrogates.
\newblock {\em IEEE Sensors Journal 16\/} (2016), 3775 -- 3783.

\bibitem{rehim2015}
{\sc Rehim, O.~A.~A., Davidson, J.~L., Marsh, L.~A., O'Toole, M.~D., Armitage,
  D., and Peyton, A.~J.}
\newblock Measurement system for determining the magnetic polarizability tensor
  of small metallic targets.
\newblock In {\em IEEE Sensor Application Symposium\/} (2015).

\bibitem{schmidteddycurrent}
{\sc {Schmidt}, K., {Sterz}, O., and {Hiptmair}, R.}
\newblock Estimating the eddy-current modeling error.
\newblock {\em IEEE Transactions on Magnetics 44}, 6 (2008), 686--689.

\bibitem{netgendet}
{\sc {Sch\"oberl}, J.}
\newblock Netgen - an advancing front 2d/3d-mesh generator based on abstract
  rules.
\newblock {\em Computing and Visualization in Science 1(1)\/} (1997), 41--52.

\bibitem{NGSolve}
{\sc {Sch\"oberl}, J.}
\newblock C++11 implementation of finite elements in ngsolve.
\newblock Tech. rep., ASC Report 30/2014, Institute for Analysis and Scientific
  Computing, Vienna University of Technology, 2014.

\bibitem{SchoberlZaglmayr2005}
{\sc {Sch\"{o}berl}, J., and Zaglmayr, S.}
\newblock High order {N\'{e}d\'{e}lec} elements with local complete sequence
  properties.
\newblock {\em COMPEL-The International Journal for Computation and Mathematics
  in Electrical and Electronic Engineering 24(2)\/} (2005), 374--384.

\bibitem{TT-33History}
{\sc Systems, W.}
\newblock Tokarev tt-33.
\newblock \url{https://weaponsystems.net/system/653-Tokarev+TT-33}.
\newblock Date accessed 20th October 2020.

\bibitem{440bpermeability}
{\sc Technology, C.}
\newblock Magnetic properties of stainless steels.
\newblock
  \url{https://www.carpentertechnology.com/en/alloy-techzone/technical-information/technical-articles/magnetic-properties-of-stainless-steels}.
\newblock Date Accessed 9th October 2020.

\bibitem{WoutervanVerre2019}
{\sc van Verre, W., {\"Ozde\"ger}, T., Gupta, A., Podd, F. J.~W., and Peyton,
  A.~J.}
\newblock Threat identification in humanitarian demining using machine learning
  and spectroscopic metal detection.
\newblock In {\em International Conference on Intelligent Data Engineering and
  Automated Learning (IDEAL)}. Springer, 2019, pp.~542--549.

\bibitem{wikipediatang}
{\sc Wikipedia}.
\newblock \url{https://en.wikipedia.org/wiki/Tang_(tools)}.
\newblock Date Accessed 9th October 2020.

\bibitem{thesisben}
{\sc Wilson, B.~A.}
\newblock {\em In preparation}.
\newblock PhD thesis, Swansea University, 2020.

\bibitem{ben2020}
{\sc Wilson, B.~A., and Ledger, P.~D.}
\newblock Efficient computation of the magnetic polarizability tensor spectral
  signature using pod.
\newblock {\em International Journal for Numerical Methods in Engineering\/}
  (2020).
\newblock Accepted, DOI:10.1002/nme.6606.

\bibitem{zaglmayrphd}
{\sc Zaglmayr, S.}
\newblock {\em High Order Finite Elements for Electromagnetic Field
  Computation}.
\newblock PhD thesis, Johannes Kepler University Linz, 2006.

\bibitem{zhao2014}
{\sc Zhao, Y., Yin, W., Ktistis, C., Butterworth, D., and Peyton, A.~J.}
\newblock On the low-frequency electromagnetic responses of in-line metal
  detectors to metal contaminants.
\newblock {\em IEEE Transactions on Instrumentation and Measurement 63\/}
  (2014), 3181--3189.

\bibitem{zhao2016}
{\sc Zhao, Y., Yin, W., Ktistis, C., Butterworth, D., and Peyton, A.~J.}
\newblock Determining the electromagnetic polarizability tensors of metal
  objects during in-line scanning.
\newblock {\em IEEE Transactions on Instrumentation and Measurement 65\/}
  (2016), 1172--1181.

\end{thebibliography}

\end{document}